\documentclass[reprint,superscriptaddress,amsmath,amssymb,aps,prx,longbibliography]{revtex4-1}

\usepackage{graphicx}
\usepackage{dcolumn}
\usepackage{bm}
\usepackage[colorlinks=true,urlcolor=blue,linkcolor=blue,citecolor=blue]{hyperref}
\usepackage[normalem]{ulem}
\usepackage{tikz}
\usetikzlibrary{quantikz}
\usepackage{amsmath}
\usepackage{amssymb}
\usepackage{bbold}
\DeclareMathOperator{\Tr}{Tr}
\usepackage{enumitem}
\usepackage{algpseudocode}
\usepackage{algorithm}
\usepackage{natbib}
\DeclareUnicodeCharacter{03B2}{\ensuremath{\beta}}
\DeclareUnicodeCharacter{2009}{\,}
 
\usepackage[all]{hypcap} 
\usepackage{braket}

\newcommand{\customref}[2]{\hyperref[#1]{\ref*{#1}#2}}

\usepackage{xcolor}

\definecolor{Ured}{HTML}{FF5C5C}
\definecolor{Ublue}{HTML}{ADD8E6}
\definecolor{Ugreen}{HTML}{198a11}

\newcommand{\xuparrow}[1]{%
  {\left\uparrow\vbox to #1{}\right.\kern-\nulldelimiterspace}
}

\renewcommand{\vec}[1]{\boldsymbol{#1}}

\newcommand{\up}{\uparrow}

\begin{document}

\title{Fundamental thresholds for computational and erasure errors via the coherent information}

\author{Luis Colmenarez}
\email{colmenarez@physik.rwth-aachen.de}
\affiliation{Institute for Quantum Information, RWTH Aachen University, 52056 Aachen, Germany}
\affiliation{Institute for Theoretical Nanoelectronics (PGI-2), Forschungszentrum Jülich, 52428 Jülich, Germany}

\author{Seyong Kim}
\affiliation{Department of Physics, Sejong University, 05006 Seoul, Republic of Korea}

\author{Markus M\"{u}ller}%
\affiliation{Institute for Quantum Information, RWTH Aachen University, 52056 Aachen, Germany}
\affiliation{Institute for Theoretical Nanoelectronics (PGI-2), Forschungszentrum Jülich, 52428 Jülich, Germany}

\date{\today}

\begin{abstract}

Quantum error correcting (QEC) codes protect quantum information against environmental noise. Computational errors caused by the environment change the quantum state within the qubit subspace, whereas quantum erasures correspond to the loss of qubits at known positions. Correcting either type of error involves different correction mechanisms, which makes studying the interplay between erasure and computational errors particularly challenging. 
In this work, we propose a framework based on the coherent information (CI) of the mixed-state density operator associated to noisy QEC codes, for treating both types of errors together. We show how to rigorously derive different families of statistical mechanics mappings for generic stabilizer QEC codes in the presence of both types of errors. We observe that the erasure errors enter as a classical average over fully depolarizing channels. Further, we show that computing the CI for erasure errors only can be done efficiently upon sampling over erasure configurations. 
We then test our approach on the 2D toric and color codes and compute optimal thresholds for erasure errors only, finding a $50\%$ threshold for both codes. This strengthens the notion that both codes share the same optimal thresholds. When considering both computational and erasure errors, the CI of small-size codes yields thresholds in very accurate agreement with established results that have been obtained in the thermodynamic limit. 
Next, we perform a similar analysis for a low-density parity-check (LDPC) code — the lift-connected surface code. We find a $50\%$ threshold under erasure errors alone and, for the first time, derive the exact statistical mechanics mappings in the presence of both computational and erasure errors.
We thereby further establish the CI as a practical tool for studying optimal thresholds for code classes beyond topological codes under realistic noise, and as a means for uncovering new relations between QEC codes and statistical physics models.

\end{abstract}
\maketitle

\section{Introduction}

Storing and manipulating quantum information in noisy quantum devices is one of the main challenges in the field of quantum technologies. In the framework of quantum computing, quantum error correction (QEC) \cite{lidar_quantum_2013,terhal_quantum_2015} is the main technique for protecting quantum information from noise in a scalable manner and is the key element on the route towards fault-tolerant quantum computing. QEC works by encoding one or several logical qubits into many noisy physical qubits and using the redundant degrees of freedom for the detection and subsequent removal of errors. This allows one to reduce the overall error rate of the logical qubit(s) compared to the bare physical qubits, provided the error rates of the faulty operations needed to operate the QEC code fall below code- and noise-model dependent critical threshold values \cite{aharonov_fault-tolerant_2008,knill_resilient_1998,kitaev_fault-tolerant_2003,shor_fault-tolerant_1996}. . There has been impressive recent progress in realizing error-corrected and fault-tolerantly operated logical qubits in a variety of physical platforms, including superconducting circuits \cite{google_quantum_ai_quantum_2024,google_quantum_ai_suppressing_2023,google_quantum_ai_exponential_2021, krinner_realizing_2022,zhao_realization_2022,andersen_repeated_2020,sivak_real-time_2023,gupta_encoding_2024,hetenyi_creating_2024,lacroix_scaling_2024}, trapped ions \cite{postler_demonstration_2022, ryan-anderson_implementing_2022,da_silva_demonstration_2024, berthusen_experiments_2024,ryan-anderson_realization_2021,ryan-anderson_high-fidelity_2024,pogorelov_experimental_2024,postler_demonstration_2024,huang_comparing_2024}, and neutral atoms \cite{bluvstein_logical_2024,bedalov_fault-tolerant_2024,reichardt_logical_2024,rodriguez_experimental_2024}. 
Fault-tolerant operation of logical qubits allows one to scale up the underlying QEC codes to larger distances, and thereby systematically suppress associated logical error rates. Recent experiments have demonstrated the suppression of logical error rates expected below threshold \cite{google_quantum_ai_quantum_2024,google_quantum_ai_suppressing_2023,da_silva_demonstration_2024,gupta_encoding_2024,hong_entangling_2024,sivak_real-time_2023}. 

The exact threshold value depends on the QEC code and decoding procedure used for recovering the logical information. Among all decoding strategies, Maximum Likelihood Decoding (MLD) stands out because it has the highest possible threshold, dubbed \emph{optimal threshold}. 
Then, a straightforward approach for obtaining optimal thresholds involves simulating QEC by sampling over all possible error processes and then computing the corrective operation using MLD. The latter requires, in general, to solve an exponentially hard problem \cite{iyer_hardness_2015,fuentes_degeneracy_2021} every time an error is corrected, thus severely limiting the size of the codes one can simulate. 
However, recently it has been pointed out that the coherent information (CI) \cite{schumacher_quantum_1996} of the mixed state associated to the noisy qubits forming the QEC code shows a discontinuity at the optimal threshold \cite{fan_diagnostics_2024,colmenarez_accurate_2024}. In fact, the CI captures the information left in the noisy mixed-state that can, in principle, be recovered~\cite{schumacher_quantum_1996}. Therefore, the CI provides a practical, generic tool for estimating optimal thresholds without resorting to numerical simulation of noisy QEC cycles and performing MLD.

Most QEC schemes are designed to tackle \emph{computational errors}, which are errors that change the state and/or phase of the physical qubits but preserve the qubits within the computational space. For instance, if the physical qubit consists of a ground and excited state of an atom, an amplitude damping channel that induces decay from the excited state to the ground state causes a computational error. The correction procedure for computational errors therefore amounts to identifying the operation generated by the noise that corrupts the logical information and reversing it by acting on the physical qubits.
However, correcting errors that take physical qubits out of the computational space, is gaining more attention. 
We consider the case of \textit{quantum erasures}, i.e.~when we \emph{know} on which qubit a transition to states outside the computational space occurred, e.g., by recording a photon emitted during the transition driving the quantum jump outside the computational space, or where qubits in a register have been lost entirely, e.g.~due to the loss of optically trapped atoms from tweezers or optical lattices~\cite{wu_erasure_2022, kang_quantum_2023, kubica_erasure_2023, vala_quantum_2005}. In the literature, detectable erasures are often called \emph{qubit losses} \cite{fujii_error_2012,  morley-short_loss-tolerant_2019, vala_quantum_2005, stace_thresholds_2009, vodola_twins_2018,amaro_analytical_2020} to highlight the difference from \emph{leakage} \cite{varbanov_leakage_2020, miao_overcoming_2023}, in which the transition outside the computational space goes undetected. This distinction is important because information about the place and time of erasure events can, in principle, be used in decoding. In our work, we focus on erasures at known positions; henceforth, throughout the manuscript, the terms qubit loss and erasure error will be used interchangeably.

Once one or more physical qubits are lost or erased, the corrective operation is equivalent to trying to retrieve the logical qubit from the remaining physical qubits. Several QEC schemes for the correction of erasure errors have been designed~\cite{grassl_codes_1997,amaro_analytical_2020,auger_fault-tolerance_2017,delfosse_almost-linear_2021,goto_soft-decision_2014,mclauchlan_accommodating_2024,morley-short_loss-tolerant_2019,nagayama_surface_2017,siegel_adaptive_2023,steinberg_far_2024,vodola_fundamental_2022}, including first experimental demonstrations of deterministic correction of qubit loss~\cite{stricker_experimental_2020,reichardt_logical_2024}. 
Let us note that optimal decoding of QEC codes in the presence of qubit loss is a fundamentally different problem compared to computational errors. For instance, in the case of the three-dimensional surface code, there is a linear-time maximum likelihood decoder (MLD) under qubit loss \cite{delfosse_linear-time_2020}, whereas efficient optimal decoding for computational errors has been demonstrated only for two-dimensional surface codes \cite{bravyi_efficient_2014,chubb_general_2021}. 
Another difference concerns the minimum number of physical qubits required to correct a single error on a qubit, which is four for qubit losses \cite{grassl_codes_1997} and five for an arbitrary computational error~\cite{laflamme_perfect_1996}.
Furthermore, optimal thresholds of QEC codes under erasure errors are usually higher than for computational errors \cite{stace_thresholds_2009}.
In fact, given the positive trade-off in correcting qubit losses instead of computational errors, it has been proposed that erasure conversion, i.e., converting computational errors into qubit losses, could significantly improve the overall performance of error correction protocols \cite{wu_erasure_2022, kang_quantum_2023, sahay_high-threshold_2023,quinn_high-fidelity_2024}.

Erasure thresholds are of great interest even beyond practical motivations. For instance, they impose restrictions on possible transversal logical gate sets in QEC stabilizer codes and guarantee the existence of a threshold for computational errors \cite{pastawski_fault-tolerant_2015}. Additionally, QEC codes under erasure errors are also used as toy models for the holographic bulk/boundary correspondence \cite{pastawski_holographic_2015}.
Regarding the approach to finding optimal thresholds, the standard procedure is to map the problem of reconstructing the logical operators from the remaining physical qubits to a percolation problem on a lattice defined by the code graph, i.e., stabilizers on nodes and physical qubits on edges \cite{dumer_thresholds_2015}. The percolation phase transition on the code graph usually provides tight threshold bounds~\cite{dennis_topological_2002, delfosse_upper_2013, kovalev_fault_2013, woolls_homology-changing_2022,dumer_thresholds_2015}
. In the case of the toric code, the equivalence between 2D bond percolation on a square lattice and optimal decoding is exact \cite{stace_thresholds_2009}.
However the latter might not always be true for more complex QEC codes, hence bounds beyond the percolation picture are often desired \cite{dumer_thresholds_2015}.  

On the other hand, it is less clear how to address the interplay between computational and erasure errors in determining optimal thresholds. In this case the goal is to remove the computational errors using only the information provided by the remaining physical qubits and the knowledge of the location of erased qubits. 
In Ref. \cite{ohzeki_error_2012}, the optimal thresholds of the toric code in the presence of both types of errors are obtained by mapping the relative probability distribution of errors and stabilizers to disordered spin models \cite{dennis_topological_2002}, however leaving out finite-size contributions. Then, in the thermodynamic limit, the phase transition between the ordered and disordered phase marks the optimal decoding threshold.
However, this methodology is difficult to extend beyond the realm of topological codes and lacks closed-form expressions that are useful for validating the mappings (we will return to this point later in the section). Therefore, a unified framework that encompasses qubit losses and computational errors on the remaining physical qubits is currently missing.

In this work, we use the CI of the noisy QEC code state as a general tool for studying both types of errors together. We derive analytical expressions for the CI in the form of classical statistical mechanics models for Calderbank-Steane-Shor (CSS) codes \cite{calderbank_good_1996, steane_multiple-particle_1997}. The corresponding mappings reflect the contribution and interplay of erasure and computational errors. Furthermore, they provide an insight into how erasure errors modify different families of statistical mechanics mappings in QEC.
Then, we focus on 2D topological codes, namely the toric and color code. 
On the one hand, we rigorously derive exact statistical mechanics mappings associated with the optimal decoding problem using the CI. We find that these mappings coincide with the mapping for the 2D toric code studied in Ref.~\cite{ohzeki_error_2012}, namely the diluted random-bond Ising model (RBIM) in which a qubit loss enters as a missing link at a given position.  However, this statistical mechanics model does not match the one studied in Ref.~\cite{stace_error_2010}, a diluted RBIM in which the erasures also modify the magnitude of the coupling near the location of the quantum erasure. Although there is no obvious connection between these two statistical mechanics models, both exhibit a phase transition at the same set of parameters, yielding the same thresholds as reported in Ref.~\cite{ohzeki_error_2012}.
The reason why these two models could describe the same optimal threshold (without a direct relation between them) is that the procedure outlined in Ref.~\cite{dennis_topological_2002}, namely mapping relative distributions of errors and stabilizers, does not provide a method for validating the mappings, which are only expected to describe optimal decoding in the thermodynamic limit.
In contrast, the CI provides exact analytical expressions valid for any code size. Therefore, the statistical mechanics models derived using the CI can be validated by ensuring that they faithfully reproduce the CI of the QEC code state. This illustrates the power of the CI in rigorously deriving statistical mechanics mappings for the optimal decoding problem of QEC codes.

On the other hand, we numerically compute the CI for different values of the error probability and find that the finite-size crossings of the CI in small distance codes yield very accurate approximations to the optimal error thresholds compared to rigorous solutions of the corresponding statistical mechanics models \cite{stace_thresholds_2009,stace_error_2010,ohzeki_error_2012}. 
This provides further evidence of the utility of the CI in estimating optimal thresholds of QEC codes from small code instances \cite{colmenarez_accurate_2024}. 
Let us note that in this work we do not intend to solve the statistical mechanics models and pinpoint their phase transition, as done in several other works \cite{dennis_topological_2002,chubb_statistical_2021}. Instead, we use the models as a mean to numerically compute the CI and extract optimal thresholds directly from it. 
To showcase the application of our method beyond topological codes, we perform a similar study on a low-density parity-check (LDPC) code—the lift-connected surface code~\cite{old_lift-connected_2024}. We find optimal thresholds for erasure errors only and derive statistical mechanics mappings for both computational and erasure errors. 
To the best of our knowledge, the models derived have not been studied previously in the literature. Our methods allow for a direct derivation of these mappings for LDPC codes and open the door to new families of statistical mechanics models beyond the realm of topological codes.

The structure of the paper is as follows. In Sec.~\ref{sec:summary} a summary of the main results of this work is presented. In Sec.~\ref{sec:background} the main concepts used are explained in detail. In Sec.~\ref{sec:results} we present the main derivations of our work in detail. In Sec.~\ref{sec:results_topo_codes} we show the numerical calculations of CI for surface and color codes. In Sec.~\ref{sec:lcs}, we apply the same methodology to the lift-connected surface code and present numerical results.
Finally, concluding remarks are presented in Sec.~\ref{sec:conclusions}.

\section{Summary of main results}\label{sec:summary}

In this section we summarize the main results of the paper. First, in Sec.~\ref{subsec:ci_erasure}, we describe how erasure errors reduce the CI of QEC stabilizer codes. Second, in Sec.~\ref{subsec:ci_comp} we recap how the CI of CSS codes under depolarizing noise is captured by the free energy cost of domain walls in certain families of statistical mechanics mappings. Third, in Sec.~\ref{subsec:ci_comp_and_erasure} we combine the two previous results and show how the CI captures both types of errors as two different but not independent contributions. Finally, in Sec.~\ref{subsec:toric_and_color}, we discuss mappings and optimal thresholds obtained for 2D topological codes using the methodology explained in the previous sections.

\begin{figure*}
    \centering
     \includegraphics[width=1.0 \linewidth]{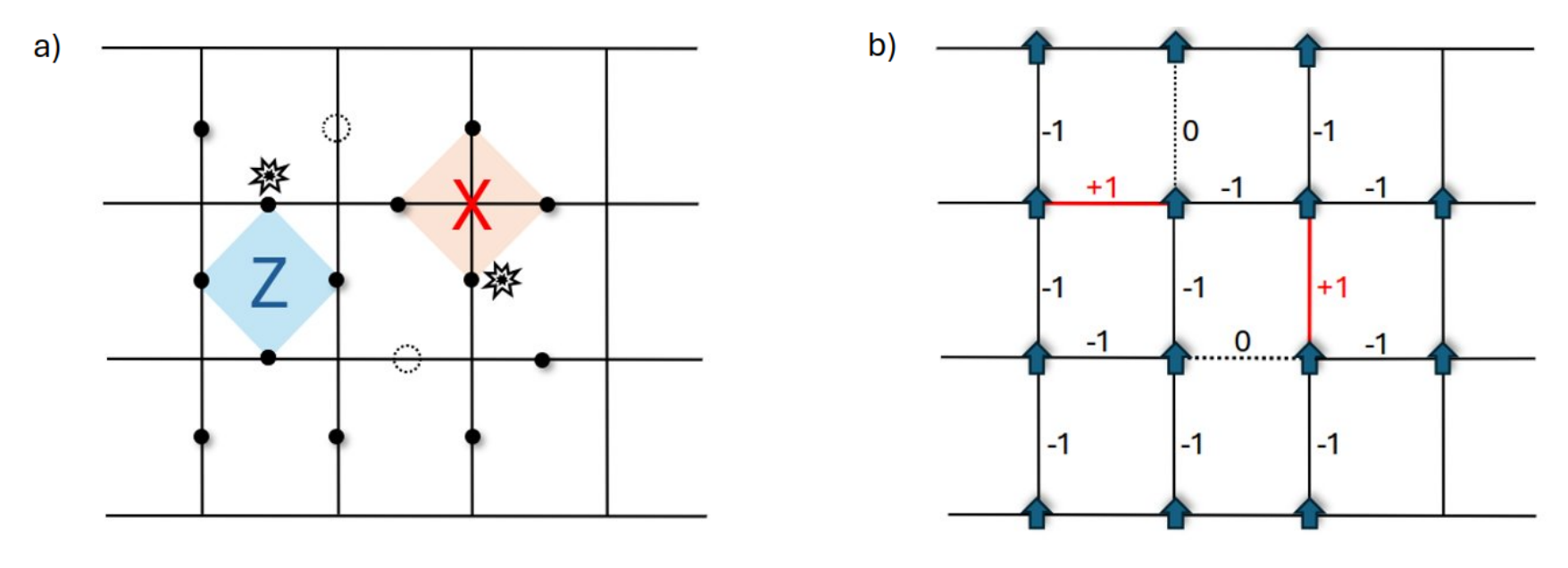}
    \caption{\textbf{a)} Toric code with $Z$ stabilizers defined on plaquettes and $X$ stabilizers centered on nodes of a square lattice. An error configuration is displayed: bit flip errors (stars next to qubits) and erasure errors as open circles replacing filled circles. \textbf{b)} Spin model corresponding to the error configuration shown in a). Spins are located at the center of the $X$ stabilizers, qubits without errors are denoted by a negative coupling. Bit flip and erasure errors are identified as positive and zero couplings, respectively.}
    \label{fig:spin_model_main}
\end{figure*}

\subsection{Coherent information and known erasure positions}\label{subsec:ci_erasure}

After $m$ physical qubits are erased, the logical information can be retrieved as long as the logical operators remain well-defined. A logical operator is said to be well-defined when at least one of its representatives, i.e., the logical operator up to stabilizer equivalences, does not have support on any of the $m$ erased qubits. Indeed, the coherent information (CI) of any QEC code $[[n,k,d]]$ effectively counts the number of logical operators that remain well-defined. The CI for an arbitrary configuration of erased qubits is given by
\begin{eqnarray}
\label{eq:ci_loss_conf}
I_l = (k-b_l-2c_l)\log 2,
\end{eqnarray}
where, $l=1,2,...,2^n$ is an index that denotes the specific locations of the erasures, which we call erasure configuration. $b_l$ is the number of logical qubits for which one logical generator, i.e either $X$ or $Z$ generator, becomes ill-defined, dubbed \emph{logical bits} in the following. $c_l$ is the number of \emph{lost logical qubits}, i.e., logical qubits for which both logical generators cannot be reconstructed on the remaining physical qubits. In summary, a logical qubit can either degrade into a classical bit or be completely lost, leading to the relation $k = k' + b_l + c_l$, where $k'$ is the number of remaining logical qubits, which holds for any $l$. Each erasure configuration $l$ occurs with a probability $P(l)$, and the total CI of the QEC code becomes
\begin{eqnarray}
\label{eq:average_ci_first}
I = \sum_{l=1}^{2^n} P(l) I_l = k \log 2-\langle b_l \rangle \log 2-2\langle c_l \rangle\log 2.
\end{eqnarray}

The brackets $\langle \dots \rangle$ denote averaging over erasure configurations $l$. Details of the derivation are shown in Sec.~\ref{subsec:loss_ci}. Computing $b_l$ and $c_l$ can be done via an algorithm based on Gaussian elimination of the parity check matrices of the code \cite{amaro_analytical_2020}, rendering the computation of $I_l$ for a given erasure configuation efficient. The challenging part of evaluating Eq.~\eqref{eq:average_ci_first} is sampling the probability distribution $P(l)$ over exponentially many erasure configurations. See Appendix~\ref{appendix:ci_erasure} for further details on how to compute Eq.~\eqref{eq:average_ci_first}.

\subsection{Coherent information mapping of CSS codes under depolarizing noise}\label{subsec:ci_comp}

\begin{table}
\begin{tabular}{||c|c|c|c||} 
 \hline
  Prob. & $\eta^x_\ell$ & $\eta^z_\ell$ & $C^{(1)}$ \\
 \hline
 $1-p$ & +1 & +1 & I \\ 
 \hline
 $p_x$ & -1 & +1 & $X$ \\
 \hline
 $p_y$ & -1 & -1 & $Y$ \\
 \hline
 $p_z$ & +1 & -1 & $Z$ \\ 
 \hline
\end{tabular}
\caption{Pauli error probabilities for the Pauli channel of Eq.~\eqref{eq:single_qubit_error_channel} on each site $\ell$ in term of the random variables $\eta^x_\ell$ and $\eta^z_\ell$. $C^{(1)}$ shows the connection to the error chains in Sec.~\ref{sec:results}. }
\label{table:chain_error}
\end{table}

The CI of $[[n,k,d]]$ CSS codes under the error channel $\mathcal{N} = \prod_i \mathcal{N}_i$ with
\begin{eqnarray}
\label{eq:single_qubit_error_channel}
\mathcal{N}_i(\rho) = (1-p)\rho + p_x X_i \rho X_i + p_y Y_i \rho Y_i + p_z Z_i \rho Z_i,
\end{eqnarray}
with $p = p_x+p_y+p_z$ and $i=1,2,...,n$, is given by: 
\begin{eqnarray}\label{eq:ci_first}
I  = k\log 2 - \overline{\log \left[\dfrac{\sum_{D} \mathcal{Z}_{D}}{\mathcal{Z}_0}\right]}
\end{eqnarray}
where $\mathcal{Z}_0$ is a partition function denoting the sum over all stabilizer operators, i.e. RBIM for the case of bit-flip noise and the toric code. The Hamiltonian associated to $\mathcal{Z}_0$ has the general form:  
\begin{eqnarray}\label{eq:hamiltonian_xyz_first}
&&H[\{\eta^x\},\{\eta^z\}] = \sum_{\ell} \eta^x_{\ell}\left(J_x-\dfrac{J_1}{2}\right)\dfrac{P^X_{\ell}}{2}+ \nonumber \\
&& \eta^z_{\ell}\left(J_z-\dfrac{J_1}{2}\right) \dfrac{P^Z_{\ell}}{2} +\eta^z_{\ell}\eta^x_{\ell}J_1 \dfrac{ P^Z_{\ell}  P^X_{\ell} }{4}.
\end{eqnarray}

Here, $\ell$ denotes the position of each physical qubit, see Fig.~\ref{fig:spin_model_main}. The binary variables $\eta^{x}_{\ell} = +1$ ($\eta^{z}_{\ell} = +1$) denote that the link $\ell$ has no $X$ ($Z$) error. Conversely, $\eta^{x}_{\ell} = -1$ ($\eta^{z}_{\ell} = -1$) indicates that the link $\ell$ is occupied by an $X$ ($Z$) error.  
The average $\overline{(...)} \equiv \sum_{\{ \eta^x \}, \{ \eta^z \}} P(\eta^x, \eta^z)$ runs over all possible error configurations, and $P(\eta^x, \eta^z)$ is given by the single-qubit error probabilities (see Table~\ref{table:chain_error}). 
The quantities $P^X_{\ell} = 1 - 2n^X_\ell$ and $P^Z_{\ell} = 1 - 2n^Z_\ell$ are related to the binary variables $n^{X,Z}_\ell = 0,1$, which reflect whether or not an error that does not change the stabilizer quantum numbers (i.e., generate a non-trivial syndrome) has support on link $\ell$. Imposing this condition is what gives rise to the celebrated spin models. $D = 0, \dots, 4^k - 1$ runs over the $4^k$ configurations of logical operators. 
Unlike the case of qubit loss in Section~\ref{subsec:ci_erasure}, computational errors may affect differently each logical operator. Therefore, one must consider all possible combinations independently.
For instance, for $k = 1$, those are $I$ (denoted as $D = 0$), $X_L$, $Z_L$, and $X_L Z_L$. The partition function $\mathcal{Z}_D$ has the same Hamiltonian as Eq.~\eqref{eq:hamiltonian_xyz_first}, with the exception that all links along the lines of logical $X$ ($Z$) operators are transformed as follows: (i) for $X_L$, $(\eta^x_\ell, \eta^z_\ell) \rightarrow (-\eta^x_\ell, \eta^z_\ell)$; (ii) for $Z_L$, $(\eta^x_\ell, \eta^z_\ell) \rightarrow (\eta^x_\ell, -\eta^z_\ell)$; and (iii) for $X_L Z_L$, $(\eta^x_\ell, \eta^z_\ell) \rightarrow (-\eta^x_\ell, -\eta^z_\ell)$. The coupling constants are defined as:

\begin{eqnarray}
e^{J_x} = \dfrac{1-p}{p_x},\quad, e^{J_z} = \dfrac{1-p}{p_z}, \quad e^{J_1} = \dfrac{(1-p)p_y}{p_x p_z}.
\end{eqnarray}

In this work, we are interested in two specific cases: (i) Uncorrelated bit and phase flip noise with probability $p_1$ and $p_2$, respectively: $p_x = p_1(1 - p_2)$, $p_z = p_2(1 - p_1)$, $p_y = p_1 p_2$, and $1 - p = (1 - p_1)(1 - p_2)$, yielding $J_1 = 0$. This allows us to write $\mathcal{Z} = \mathcal{Z}_1 \mathcal{Z}_2$ as two decoupled partition functions on $X$ and $Z$ errors. (ii) Depolarizing channel: $p_x = p_z = p_y = p/3$, which yields $e^{J_{x,z}} = \frac{3(1 - p)}{p}$ and $e^{J_1} = \frac{3(1 - p)}{p}$, a common case studied in the literature. It is important to note that Eq.~\eqref{eq:hamiltonian_xyz_first} can be generalized to in-homogeneous error rates by assigning site-dependent couplings.

As stated in previous works \cite{fan_diagnostics_2024,wang_intrinsic_2023,hauser_information_2024,su_tapestry_2024,lyons_understanding_2024,zhao_extracting_2023,lee_quantum_2023,bao_mixed-state_2023,sang_stability_2024,eckstein_robust_2024}, below threshold, the system is in a symmetry-broken macroscopic state. Therefore, inserting a domain wall $D$ (logical operator) incurs a high free energy cost, and thus $I \sim \log 2$. Above threshold, domain walls do not cost free energy because the macroscopic state is fully symmetric. Since Eq.~\eqref{eq:ci_first} is derived assuming only a CSS code (see Sec.~\ref{subsec:depo_ci}), the picture of domain wall cost applies to a wide variety of QEC codes.

\subsection{Coherent information mapping of CSS codes under depolarizing and erasure errors}\label{subsec:ci_comp_and_erasure}

\begin{table}
\begin{tabular}{||c |c| c||} 
 \hline
  Prob. & $\eta^X_{\ell}$ & $\eta^Z_{\ell}$ \\
 \hline
 $(1-p)(1-e)$ & +1 & +1  \\ 
 \hline
 $p_x(1-e)$ & -1 & +1 \\
 \hline
 $p_y(1-e)$ & -1 & -1 \\
 \hline
 $p_z (1-e)$ & +1 & -1 \\ 
 \hline
 $e$ & 0 & 0 \\
 \hline
\end{tabular}
\caption{Probabilities of each pair of variables $\eta^x$ and $\eta^Z$ on the same site $\ell$ on any of the spin models, where computational errors have probabilities $p_x,p_y,p_z$ and erasure errors have probability $e$.}
\label{table:chain_error_erasure}
\end{table}

When both single qubit erasure and computational errors are present, the CI for a fixed erasure configuration $l$ is written as 
\begin{eqnarray}\label{eq:ci_comp_and_erasure}
I_l  = (k-b_l-2c_l)\log 2 - \overline{\log \left[\dfrac{\sum_{D} \mathcal{Z}_{D_l}}{\mathcal{Z}_{0}}\right]}.
\end{eqnarray}
The average $\overline{(...)}$ is performed over the computational error configurations $\{\eta^x, \eta^z\}$. A lost qubit on site $\ell$ is represented as a missing ``link" with $\eta^x_\ell = \eta^z_\ell = 0$ (see Table~\ref{table:chain_error_erasure}), meaning that all erasure positions in the configuration $l$ have vanishing coupling. The actual CI is obtained after averaging over all loss configurations $l$ for a given probability distribution $P(l)$:

\begin{eqnarray}
\label{eq:average_ci_comp_flip}
I  & = & \sum_{l} P(l) I_l \nonumber \\
   & = & \left(k  -\langle b_l\rangle -2\langle c_l\rangle\right)\log 2 - \Biggr\langle \overline{\log \left[\dfrac{\sum_{D} \mathcal{Z}_{D_l}}{\mathcal{Z}_{0}}\right]}\Biggr\rangle.
\end{eqnarray}

The last equation shows the processes that reduce the CI: (i) qubit losses can degrade the logical qubit to a classical bit or completely destroy it, and (ii) the cost of computational errors and their interplay with qubit losses is modeled by the partition functions $\mathcal{Z}_{D_l}$, with missing bonds at the erased qubit positions. We can reason about these two processes as follows: erasure errors weaken the ability to protect quantum information by rendering some stabilizers ill-defined and effectively reducing the number of degrees of freedom used for error detection and correction. However, in the absence of computational errors, this does not affect the logical qubit unless at least one logical operator becomes ill-defined. 
In the presence of computational errors, there are fewer operators that commute with the well-defined stabilizers, but the partition functions $\mathcal{Z}_{D,l}$ still count the number of operators that commute with the well-defined stabilizers. 
Note that $D_l$ now denotes the set of operators that \emph{anti-commute} with the well-defined logical operators. For details, see Sec.~\ref{subsec:depo_ci_loss}.

\subsection{2D toric and color code under erasure and computational errors}\label{subsec:toric_and_color}

In this section we show the specific mappings for 2D topological codes under erasure and computational errors. Coming from the general mapping for CSS codes in Eq.~\eqref{eq:hamiltonian_xyz_first}, for the 2D toric code $n^X_{\ell} = (1-\sigma_i \sigma_j)/2$ and $n^Z_{\ell} = (1-\tau_n \tau_m)/2$, where $\sigma_i=\pm 1$ and $\tau_n = \pm 1 $ are classical spin variables living a square lattice and its dual, respectively (see Fig.~\ref{fig:depo_lattice_toric}). The $\sigma$ ($\tau$) spins represent the $X$($Z$) stabilizers, we then call $\sigma$ ($\tau$) the $X$($Z$) type spin. Hence we obtain the eight-vertex model \cite{bombin_strong_2012}: 
\begin{eqnarray}\label{eq:hamiltonian_xyz}
&&H_{TC} = \sum_{\langle i,j \rangle,\langle n,m \rangle } \eta^x_{ij}\left(J_x-\dfrac{J_1}{2}\right)\dfrac{\sigma_i\sigma_j}{2}+ \nonumber \\
&& \eta^z_{nm}\left(J_z-\dfrac{J_1}{2}\right) \dfrac{\tau_n\tau_m}{2}+\eta^z_{nm}\eta^x_{ij}J_1 \dfrac{\sigma_i\sigma_j\tau_n\tau_m}{4}.
\end{eqnarray}
For 2D color codes both spins $\sigma$ and $\tau$ live on the dual lattice with respect to the physical qubit lattice (see Fig.~\ref{fig:depo_lattice_color}). We then obtain a three-body coupled random-bond Ising model: 
\begin{eqnarray}\label{eq:hamiltonian_xyz}
&&H_{CC} = \sum_{\langle i,j,k \rangle,\langle n,m,o \rangle } \eta^x_{ijk}\left(J_x-\dfrac{J_1}{2}\right)\dfrac{\sigma_i\sigma_j\sigma_k}{2}+ \\
&& \eta^z_{nmo}\left(J_z-\dfrac{J_1}{2}\right) \dfrac{\tau_n\tau_m\tau_o}{2}+\eta^z_{nmo}\eta^x_{ijk}J_1 \dfrac{\sigma_i\sigma_j\sigma_k\tau_n\tau_m\tau_o}{4} \nonumber.
\end{eqnarray}

In general, the spins are located at the center of the stabilizers of the QEC code, and each spin interacts with the stabilizers with which physical qubits are shared. There are as many terms in the Hamiltonian as there are physical qubits.
The random variables $\{\eta^x\}$ and $\{\eta^z\}$ on each site $\ell$ have probabilities according to the computational and erasure error probabilities shown in Table~\ref{table:chain_error_erasure}. 

\subsubsection{Mapping the optimal threshold problem to statistical mechanics models}

The first case we study is uncorrelated bit and phase flip: after assigning $p_x= p_1(1-p_2)$, $p_z=p_2(1-p_1)$ and $p_y = p_1 p_2$ in which $p_1$ and $p_2$ are the bit and phase flip probability, respectively, we obtain $e^{J_x} = (1-p_1)/p_1$, $e^{J_z} = (1-p_2)/p_2$ and $J_1=0$. Therefore the $\sigma$ and $\tau$ spins get uncoupled and the CI can be written as 
\begin{eqnarray}
\label{eq:average_ci_comp_flip}
I & = & \left(k  -\langle b_l\rangle -2\langle c_l\rangle\right)\log 2 - \Biggr\langle \overline{\log \left[\dfrac{\sum_{D_X} \mathcal{Z}_{D_X,l}}{\mathcal{Z}_{0,l}}\right]}\Biggr\rangle \nonumber \\
&& - \Biggr\langle \overline{\log \left[\dfrac{\sum_{D_Z} \mathcal{Z}_{D_Z,l}}{\mathcal{Z}_{0,l}}\right]}\Biggr\rangle, 
\end{eqnarray}
where $D_{X(Z)}$ runs over the logical $X(Z)$ operators, the partition function $\mathcal{Z}_{D_{X(Z)},l}$ corresponds to the $Z(X)$ stabilizers. As a result, the contributions from $X$ and $Z$ errors are independent of each other. For the 2D toric code, this results in a \emph{diluted} random bond Ising model (RBIM), which has been previously studied in Ref.~\cite{ohzeki_error_2012}. From the point of view of the phase transition in the RBIM, the correctable phase is still denoted by the ferromagnetic phase, while the un-correctable phase is again the paramagnetic phase. However, now the paramagnetic phase is driven by two independent mechanisms: (i) anti-ferromagnetic links and finite temperature, which model computational errors, and (ii) missing links that model qubit losses. For 2D color codes, this choice of parameters yields a \emph{diluted} three-body random bond Ising model. To the best of our knowledge, this statistical mechanics model has not been studied before.

The second case is the isotropic depolarizing channel: choosing $p_x = p_y = p_z = p/3$ leads to a statistical mechanics model with a homogeneous coupling $J_x = J_z = J_1 = 3(1 - p)/p$ between spins of the same and different type. The diluted version of this model has been studied in \cite{ohzeki_error_2012} for the toric code. However, for the 2D color code, the model with both computational and erasure errors has not been studied so far.

\subsubsection{Optimal thresholds from numerical calculation of CI}

We numerically compute the CI and find optimal thresholds, see Sec.~\ref{sec:results_topo_codes}. For erasure errors only, we compute the thresholds by performing a finite-size scaling analysis. We find that both 2D toric and color codes have an optimal threshold of $50\%$. For 2D toric codes, this was previously known through the equivalence of this problem with 2D bond percolation \cite{stauffer_introduction_1992}. In principle, the equivalence between percolation on the code graph and erasure threshold does not hold for 2D color codes~\cite{vodola_twins_2018}; however, we find the same optimal threshold and scaling exponent as for the 2D toric code.

Under both computational and erasure errors, we find that the pseudo-thresholds, i.e. crossings between the CI of a single qubit and CI of finite distance codes, for small instances of the codes are in very good agreement with the known optimal thresholds of toric codes (which are summarised in Table~\ref{tab:thresholds}). This provides further evidence that the CI of small-distance codes accurately captures the asymptotic behavior of the QEC code \cite{colmenarez_accurate_2024}. We further explore the phase boundary of the correctability transition using the CI pseudo-threshold, as shown in Fig.~\ref{fig:phase_diagram}. We find that both 2D color and toric codes have almost identical boundaries for both types of computational errors.

\subsection{Lift-connected surface code}

Since the CI methodology developed in our work applies directly to low-density parity-check (LDPC) codes, we apply this machinery to the lift-connected surface code (LCS)~\cite{old_lift-connected_2024}. This code features a growing number of logical qubits, which is clearly distinct from topological codes.
As shown in Fig.~\ref{fig:lcs_code_figure}, the LCS code can be viewed as a periodic array of interconnected surface code sheets. Each surface code stabilizer is extended over neighboring surface code sheets. Despite the stabilizers having 3D-local connectivity, the structure of the logical operators does not resemble that of a topological code.
We numerically compute the coherent information (CI) for erasures only and find a $50\%$ threshold, along with a critical exponent that neither matches the 2D nor the 3D percolation transition. We then study the interplay between computational and erasure errors by deriving the corresponding statistical mechanics mapping and computing the CI numerically. The Hamiltonian of the model reads:
\begin{eqnarray}\label{eq:hamiltonian_lcs_first}
&&H_{LCS} = \sum_{\langle i,j \rangle,\langle n,m \rangle, q } \eta^x_{ij}\left(J_x-\dfrac{J_1}{2}\right)\dfrac{\sigma_{i,q}\sigma_{j,q}\sigma_{i,q+1}}{2} \nonumber \\
&& +\eta^z_{nm}\left(J_z-\dfrac{J_1}{2}\right) 
\dfrac{\tau_{n,q}\tau_{m,q}\tau_{m,q+1}}{2} \nonumber \\
&& +\eta^z_{nm}\eta^x_{ij}J_1 \dfrac{\sigma_{i,q}\sigma_{j,q}\sigma_{i,q+1}\tau_{n,q}\tau_{m,q}\tau_{m,q+1}}{4}.
\end{eqnarray}

The indices $\langle i, j \rangle$ and $\langle n, m \rangle$ denote qubit positions on the $q$-th surface code sheet. The two-spin interactions become three-spin interactions by coupling one spin from the same ``plaquette'' or ``star'' to a spin from the next surface code sheet. One instance of this model is depicted in Fig.~\ref{fig:lcs_d3}. 
This new family of statistical mechanics models is 3D-local and exhibits directionality in the interactions (see Appendix~\ref{sec:lcs_appendix} for further details). Furthermore, we perform a pseudo-threshold analysis for bit/phase-flip and depolarizing noise combined with erasure errors. We find thresholds comparable to those of the standard surface code, despite the growing number of logical qubits. 
We conclude that the LCS code offers similar levels of protection as the surface code, with the added benefit of supporting a growing number of logical qubits.

\section{Background}\label{sec:background}

\begin{figure*}
    \centering
    \includegraphics[width=1.0 \linewidth]{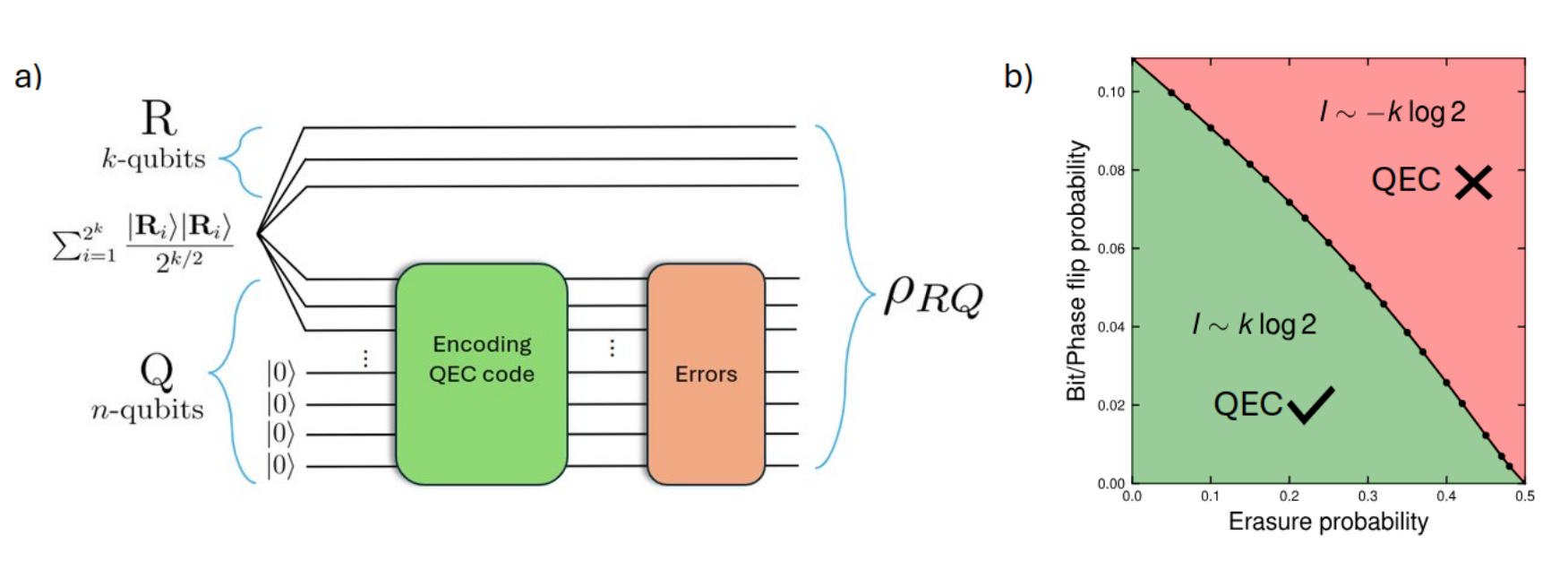}
    \caption{\textbf{a)} Coherent information setup. We start from a generalized Bell state $\sum_{i=1}^{2^k} | \vec{R}_i\rangle |\vec{R}_i \rangle /2^{k/2}$ between the $k$ reference qubits and identical $k$ seed qubits. Then we prepare the state $\sum_{i=1}^{2^k} | \vec{R}_i\rangle |\vec{L}_i \rangle /2^{k/2}$ where $|\vec{L}_i\rangle$ are the code words of a QEC code and send the state on $Q$ through an error channel. The CI is computed as $I=S(\rho_{Q})-S(\rho_{RQ})$.  \textbf{b)} Correctability phase diagram for the 2D color code under bit/phase flip and erasure errors. The black line is the same shown in Fig.~\ref{fig:phase_diagram} in Sec.~\ref{sec:results_topo_codes} for the color code. Below threshold (green region) the CI asymptotically approaches its maximum  value of $k\log 2$. Above threshold (red region) the CI approaches its minimum $-k\log 2$. }
    \label{fig:main_figure}
\end{figure*}

In this section, we present the background needed to derive the main results, as well as notation conventions. We recommend that the reader interested in the details of the derivations in Sec.~\ref{sec:results} first go through this section. In Sec.~\ref{subsec:stabilizer_codes}, we briefly introduce QEC stabilizer codes. Then, in Sec.~\ref{subsec:error_channels}, we describe the error channels investigated in this work. Finally, in Sec.~\ref{subsec:ci_definition}, we formally introduce the coherent information (CI) and explain how it will be used in this work.

\subsection{Stabilizer codes}\label{subsec:stabilizer_codes}

A QEC stabilizer code $[[n,k,d]]$ is defined by the $n-$qubit stabilizer group $\mathcal{S}_n$ \cite{gottesman_stabilizer_1997} whose $n-k$ generators $g_i$ mutually commute such that $[g_i,g_j]=0$ $\forall i,j$. 
A QEC code encodes $k$ logical qubits and the operators $g_i$ are $n-$qubit Pauli strings. We define the \emph{code space} as the subspace $\mathcal{C}$ of the Hilbert space for which $g_i = +1$ such that any $n-$qubit state $|\psi\rangle \in \mathcal{C}$ fulfills the condition $g_i |\psi\rangle = |\psi\rangle$ $\forall i$. We also have the set of logical operators $\{(O^x_i,O^{z}_i)\}$ with $i=1,..,k$, which satisfy the conditions $\{O^x_i,O^z_i\} = 0$ $\forall i$, $[O^x_i,O^z_j] = 0$ for $i\neq j$, $[O^z_i,g_j] = 0$ and $[O^x_i,g_j] = 0$ for any $i$ and $j$. The joint eigenstates of the set of operators $\{g_i\}\cup \{O^z_i\}$ form a complete basis of the $n-$qubit Hilbert space $\mathcal{H}$. The same applies for the set $\{g_i\}\cup \{O^x_i\}$. The code space $\mathcal{C}$ is spanned by $2^k$ basis vectors $|\boldsymbol{L}_i\rangle$ called \emph{code words}. 
By convention we choose all $|\boldsymbol{L}_i\rangle$ to be eigenstates of the logical operators $O^z_i$. The code distance $d$ is the minimum weight of the logical operators $\{(O^x_i,O^{z}_i)\}$. Sometimes we will denote $Z_L \equiv O^z_i $ and $X_L \equiv O^x_i $, specially when referring to QEC codes with $k=1$. 

In particular we are interested in codes for which the stabilizer generators $g_i$ are composed of either $X$ or $Z$ Pauli operators, i.e.~Calderbank-Shor-Steane (CSS) codes \cite{calderbank_good_1996,steane_multiple-particle_1997}. 
This class of codes allows the logical operators $\{O^x_i,O^z_i\}$ to be made of only $Z$ or $X$ Pauli operators. Within the class of CSS codes we are interested in topological codes \cite{kitaev_anyons_2006}. 
They are defined on $D-$dimensional lattices and encode a fixed number $k$ of logical qubits, i.e.~$k$ is independent of $n$ and only dependent on the topological properties of the manifold the lattice is embedded in. The stabilizer generators $\{g_i\}$ are defined on unit cells of the lattice and each qubit is shared by a constant number of stabilizer generators. In this work we focus on two-dimensional toric/surface and color code \cite{kitaev_quantum_1997,dennis_topological_2002,bombin_topological_2007} which are amongst today's leading contenders for the realization of scalable fault-tolerant quantum computing \cite{google_quantum_ai_quantum_2024,google_quantum_ai_suppressing_2023,google_quantum_ai_exponential_2021, krinner_realizing_2022,zhao_realization_2022,andersen_repeated_2020,postler_demonstration_2022, ryan-anderson_implementing_2022,da_silva_demonstration_2024, berthusen_experiments_2024,ryan-anderson_realization_2021,ryan-anderson_high-fidelity_2024,pogorelov_experimental_2024,postler_demonstration_2024,huang_comparing_2024,bluvstein_logical_2024,lacroix_scaling_2024,rodriguez_experimental_2024}. We refer to Appendix~\ref{app:qec_codes} for definitions and details of the topological codes studied in this work.
Besides topological codes, we study the lift-connected surface code~\cite{old_lift-connected_2024}, which belongs to the class of CSS low-density parity-check (LDPC) codes. This code is part of the hypergraph product code family and has the distinctive feature that it can be viewed as an array of interconnected surface codes. 
When the code parameters are appropriately chosen, the rate and distance of the code exceed those of disconnected surface codes.

\subsection{Error channels}\label{subsec:error_channels}

\subsubsection{Erasure channel}

The erasure channel on a single qubit state reads~\cite{bennett_capacities_1997} 
\begin{eqnarray}
\mathcal{E}_1(\rho) = (1-e)\rho + e \frac{\mathbb{1}}{2}. 
\end{eqnarray}

There is a probability $e$ that the state $\rho$ is replaced by a maximally mixed state. We are interested in the case when we \emph{know} the qubit is lost due to the action of the channel. 
Therefore, a more convenient way to describe this process is to define a CPTP map that attaches one ancilla bit to each Kraus operator of the channel such that 

\begin{eqnarray}\label{eq:single_qubit_erasure}
\mathcal{E}_1(\rho) = (1-e)\rho \otimes |\up\rangle \langle  \up| + e \frac{\mathbb{1}}{2} \otimes | \downarrow\rangle \langle \downarrow|
\end{eqnarray}

The ancilla bit in the state $|\up\rangle$ or $|\downarrow\rangle$ denotes the occurrence or not of the erasure event of the qubit attached to the given ancilla. 
As will be discussed in Sec.~\ref{subsec:loss_ci}, the erased degrees of freedom do not contribute to the CI. Now we generalize Eq.~\eqref{eq:single_qubit_erasure} to $n-$qubit states: 
\begin{eqnarray}\label{eq:n_qubit_erasure}
\mathcal{E}_n(\rho) = \sum_{l=1}^{2^n} P(l) \frac{1}{2^m}\mathbb{1}_{A_l} \otimes \Tr_{A_l}(\rho) \otimes |a_l\rangle . 
\end{eqnarray}
Here $A_l$ denotes the set of erased qubits and $l= 1,2,..,2^n$ runs over all possible configurations of erased qubits, $|a_l\rangle$ is the configuration of the ancilla bits guaranteed to satisfy $\langle a_{l'}|a_l\rangle = \delta_{ll'}$ and $m \equiv m(l)$ is the number of qubits in the erased region $A_l$. 
When erasures are independent on each qubit, e.g.~all qubits experience the same erasure operation given by Eq.~\eqref{eq:single_qubit_erasure} with the same probability $e$, then $P(l) = (1-e)^{1-m} e ^m$. In general a distance-$d$ QEC code can correct all erasure configurations with $m < d$. All erasure configurations for which at least one representative of each logical operator remains well-defined is fully correctable.

\subsubsection{Computational errors}

We also consider errors occurring independently on each qubit given by the Pauli
channel: 
\begin{eqnarray}\label{eq:depo_noise}
\mathcal{N}_i(\rho) = (1-p)\rho +p_x X_i \rho X_i + p_y Y_i \rho Y_i +  p_z Z_i \rho Z_i,
\end{eqnarray}
with $p=p_x+p_y+p_z$. The $n-$qubit error channel is then given by $\mathcal{N} = \prod_i \mathcal{N}_i$. Particularly we are interested in two choices of $(p_x,p_y,p_z)$: i) Symmetric uncorrelated bit and phase flip noise: $p_x=p_z= p_1(1-p_1)$ and $p_y = p_1^2$ with $p_1$ is the bit and phase flip probability; and ii) Symmetric depolarizing channel $p_x=p_z=p_y=p/3$. 

\subsection{Coherent information of QEC codes}\label{subsec:ci_definition}

The coherent information (CI) of a state $\rho_Q$ is defined as
\begin{eqnarray}
I = S(\rho_Q) - S(\rho_{RQ}),
\end{eqnarray}
where $S(\rho) = - \Tr \rho \log \rho$ is the von Neumann entropy of the state $\rho$ and $\rho_Q = \Tr_R(\rho_{RQ})$. This quantity was first introduced as a measure of the amount of quantum information transmitted by a quantum channel \cite{lloyd_capacity_1997} and as an indicator of the existence of a QEC protocol with maximum success probability \cite{schumacher_quantum_1996}. Recently, it has been shown that it can also signal optimal thresholds of QEC codes \cite{fan_diagnostics_2024,wang_intrinsic_2023,hauser_information_2024,su_tapestry_2024,lyons_understanding_2024,zhao_extracting_2023,colmenarez_accurate_2024}. The exact setting for studying optimal threhsolds of QEC codes is shown in Fig.~\ref{fig:main_figure}. 
We start with a generalized Bell pair between the reference system $R$ and the code space of an error correcting code $[[n,k,d]]$ on the system $Q$ of $n$ qubits. 
\begin{eqnarray}\label{eq:generalized_bell_pair}
|\psi_{RQ}\rangle = \frac{1}{2^{k/2}}\sum_{q=1}^{2^{k}} |\boldsymbol{R}_q,\boldsymbol{L}_q,\boldsymbol{S}\rangle,
\end{eqnarray}
%
where the state $|\boldsymbol{R}_q\rangle = |R_1 R_2 ... R_k\rangle$ is a basis state of the reference system. The reference system is composed of $k$ qubits, each of them denoted by the quantum number $R_i$. A ``code word" state $|\boldsymbol{L}_q,\boldsymbol{S} \rangle = |L_1 L_2 ... L_k, S_1,S_2,...,S_{n-k}\rangle$ is a basis $n-$qubit state for the code space of the QEC code living on the system $Q$. $L_i$ are the logical quantum numbers of the code space and $S_i = +1$ $\forall i$ are the stabilizer quantum numbers (syndromes). An alternative way to represent this state is 

\begin{eqnarray}\label{eq:projec_density_matrix}
\rho^0_{RQ} =  \prod_{j=1}^{2k}\left(\frac{1+O_{R_j}O_{L_j}}{2}\right)\prod_{i=1}^{n-k} \left(\frac{1+g_i}{2}\right).
\end{eqnarray}
In the last equation we wrote $\rho^0_{RQ} \equiv |\psi_{RQ}\rangle \langle \psi_{RQ}| $. The operators $O_{L_i}$ are the logical operators (the super-indices $x$ and $z$ are omitted in this section) of the code, hence $O_{R_i}$ are the respective images in the reference system. $g_i$ are the stabilizer generators of the QEC code. Since $[O_{R_j}O_{L_j}, O_{R_i}O_{L_i}]= 0$ and $[O_{R_j}O_{L_j}, g_i]=0$  then eigenstates of the operators  $O_{R_j}O_{L_j}$ and $g_i$ form a complete basis of the complete Hilbert space in the RQ system. Consequently, Eq.~\eqref{eq:projec_density_matrix} is a sum of projectors that uniquely determines the state $|\psi_{RQ}\rangle$. As an example, let us take $k=1$: in this case, there is only one pair of operators $Z_{R} Z_{L}$ and $X_{R} X_{L}$, therefore $|\psi_{RQ}\rangle = (|0,0,\boldsymbol{S} \rangle + |1,1,\boldsymbol{S}\rangle)/\sqrt{2} $ is a Bell pair between a single reference qubit and one logical qubit. 

After tracing out the reference system, the state $\rho^0_{Q} = \sum_{q=1}^{2^{k}} |\boldsymbol{L}_q,\boldsymbol{S}\rangle \langle \boldsymbol{L}_q,\boldsymbol{S}|/2^{k}$ is an incoherent superposition of codeword states, hence the state shown in Eq.~\eqref{eq:projec_density_matrix} has maximal CI of $I_0 = k\log 2$. 
After the state is exposed to noise, the CI might decrease $I\leq I_0$. The difference between the CI of a noiseless and a noisy state quantifies the amount of information that has leaked to the environment. In Sec. \ref{sec:results} we derive expressions for the exact CI obtained after the state $\rho^0_{RQ}$ is exposed to erasure and computational errors. In Sec.~\ref{sec:results_topo_codes} we use those expressions to estimate optimal thresholds of 2D topological codes. 

\section{Results}\label{sec:results}

\begin{figure}
    \centering
    \includegraphics[width=0.6\linewidth]{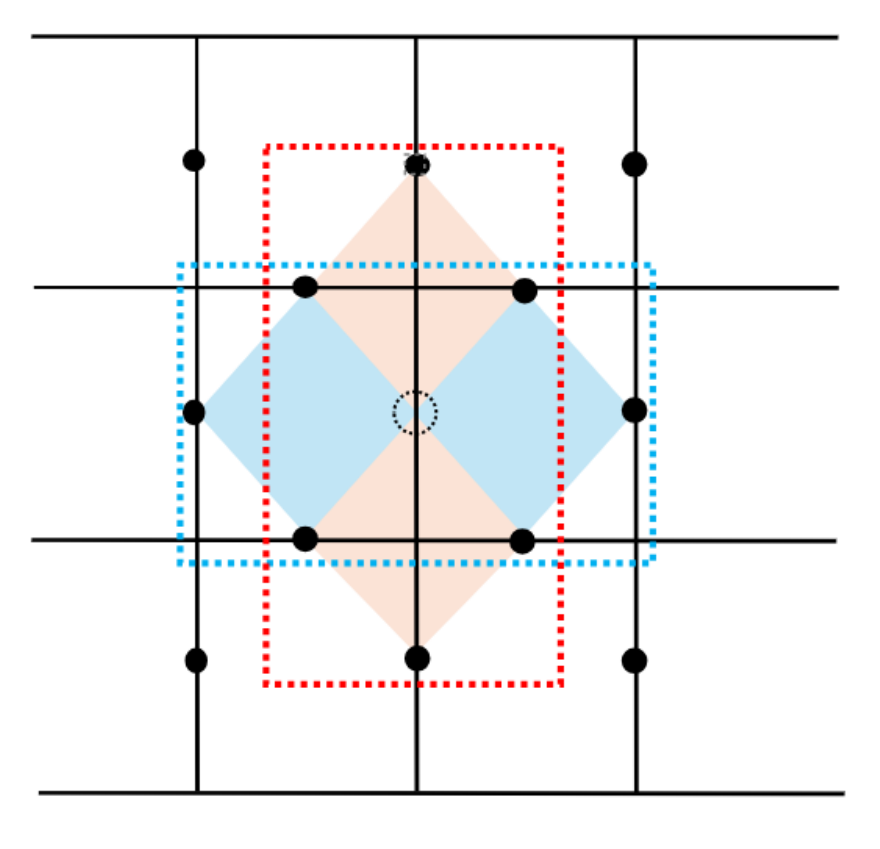}
    \caption{Formation of super-plaquettes and super-star operators after losing one qubit. Each qubit is shared between four stabilizers, two $X$ (red regions) and two $Z$ (blue regions). After erasure the product of stabilizers of the same species remains well-defined and forms what has been called super-plaquettes (enclosed by dashed blue line) and super-stars (enclosed by red dashed line) \cite{stace_thresholds_2009}. Let us note thateach of the former weight-3 stabilizers commutes with the super-plaquette and super-star operators, however its expectation value has been randomized by the qubit erasure.}
    \label{fig:super_plaquette}
\end{figure}

In this section, we describe in detail how to derive the mappings of the coherent information (CI) in the error configuration picture, both with and without qubit loss. In Sec.~\ref{subsec:loss_ci}, we present the exact expressions for the CI of stabilizer codes with erasure errors. Then, in Sec.~\ref{subsec:depo_ci}, we study computational errors by applying the methods used in Refs. \cite{wang_intrinsic_2023,hauser_information_2024,fan_diagnostics_2024,lyons_understanding_2024} to CSS codes and Pauli noise. In Sec.~\ref{subsec:depo_ci_loss}, we combine the methods and derivations explained in the two previous sections, and arrive at closed-form expressions for the CI of CSS codes under both computational and erasure errors. To simplify the notation across the section we choose to work with the base-two logarithm, such that $\log 2 = 1$.

\subsection{Coherent information under erasure errors}\label{subsec:loss_ci}

In this section, we compute the CI of QEC codes under the action of the erasure channel given by Eq.~\ref{eq:n_qubit_erasure}. Recall that the erasure operation traces out $m = m(l)$ physical qubits in region $A_l$, where $l$ denotes the configuration of erased qubits. For a fixed erasure configuration $l$, we then obtain the state (the ancilla bit states $|a_l\rangle$ are omitted from now on):

\begin{eqnarray}
\rho^l_{RQ} = \frac{1}{2^m}\mathbb{1}_{A_l} \otimes \Tr_{A_l}(|\psi_{RQ}\rangle \langle \psi_{RQ}|).
\end{eqnarray}
Each erased qubit leads to a small set of stabilizers that must be redefined, while some others become ill-defined. In total, only $n - k - 2m$ stabilizers remain well-defined after erasing $m$ qubits. The latter is a consequence of how projective measurements are treated in the stabilizer formalism \cite{nielsen_quantum_2010}. For the surface/toric code, an example can be seen in Fig.~\ref{fig:super_plaquette}. After erasure of one qubit, a total of four stabilizers is affected. Then each pair of $X$ and $Z$ stabilizers gets combined and becomes super-plaquette and super-star operators. Thus two out of four affected degrees of freedom remain well-defined. Furthermore, the fate of the logical operators also depends on the configuration $l$ of lost qubits. If at least one representative of the given logical operator still has support on the remaining physical qubits, then we say that such a logical operator remains well-defined. A logical operator becomes ill-defined once all representatives have support on at least one erased qubit. One example of the latter in the toric code is the case when erasure errors split the lattice into two disconnected parts, hence cutting the way of non-contractible loops, thereby impeding the existence of the respective logical (string) operators. To determine whether a logical operator $O_{L_i}$ remains well-defined, we use the procedure outlined in Ref. \cite{amaro_analytical_2020} (see Appendix~\ref{appendix:ci_erasure} for details). If $h$ logical operators are not recoverable, then one can write

\begin{eqnarray}
\rho^l_{RQ} =  \prod_{i=1}^{2k-h}\left(\frac{1+O_{R_i}O_{L_i}}{2}\right)\prod_{i=1}^{n-2m-k} \left(\frac{1+g'_i}{2}\right) \otimes \dfrac{\mathbb{1}}{2^{h+2m}}.\nonumber \\.
\end{eqnarray}
Here, $g'_i$ are the remaining well-defined stabilizers, $O_{L_i}$ denotes one remaining logical operator, and $O_{R_i}$ is its image in the reference system.

Now let us discuss how the generalized Bell pair is affected by qubit erasures. First, each pair of operators $O^x_{R_i}O^x_{L_i}$ and $O^z_{R_i}O^z_{L_i}$ generates a code space of size 2, denoted by a Bell pair $|00\rangle + |11\rangle$ (up to normalization), between one reference qubit and one logical qubit. In other words, the two quantum numbers of the Bell basis are fixed to $+1$. Importantly, the CI relies on preserving the Bell pairs, so in the following analysis, we consider a single Bell pair only. The intuition carries over for $k > 1$ and any QEC code. After some qubits are lost, we can distinguish three scenarios for each logical qubit:

(i) Both logical operators remain well-defined. Then, $Z_R Z_L = +1$ and $X_R X_L = +1$, preserving the Bell pair. Therefore, the whole logical qubit is preserved, and the CI does not decrease.

(ii) One logical operator becomes ill-defined. When $X_L$ is the only lost operator, the Bell quantum number $Z_R Z_L = +1$ is fixed. Thus, the state is now in an incoherent superposition of the two $+1$ eigenstates of $Z_R Z_L$, ending up with $|00\rangle \langle 00| + |11\rangle \langle 11|$ (up to normalization). Similarly, when $Z_L$ is lost, the Bell quantum number $X_R X_L = +1$ is fixed, so the state is now $|++\rangle \langle ++| + |--\rangle \langle --|$ (up to normalization). As a result, the logical qubit is degraded to a \emph{logical bit}, because the remaining logical operator is still able to transmit classical information.

(iii) Both $X_L$ and $Z_L$ logical operators are lost. In this case, we get an incoherent mixture of the type $|00\rangle \langle 00| + |11\rangle \langle 11| + |01\rangle \langle 01| + |10\rangle \langle 10|$. In this situation, neither classical nor quantum information can be transmitted, so we call it a \emph{lost qubit}.

In the light of the previous analysis, a state for which $k^{'}$ logical qubits are preserved, $b$ logical bits arise and $c$ logical qubits are lost can be written as 

\begin{widetext}
\begin{eqnarray}\label{eq:rho_rq_l}
\rho^{l}_{RQ} =  \frac{1}{2^{k'+b+m+2c} }\sum_{q=1,q'=1}^{2^{k'}} \sum_{\boldsymbol{S}'}^{2^m} \sum_{o=1}^{2^b} \sum_{p=1}^{2^{c} } \sum_{p'=1}^{2^{c} } |\boldsymbol{R}_q,\boldsymbol{L}_q,\boldsymbol{S}, \boldsymbol{S}',\boldsymbol{R}_o,\boldsymbol{L}_o, \boldsymbol{R}_p,\boldsymbol{L}_{p'}\rangle \langle \boldsymbol{R}_{q'},\boldsymbol{L}_{q'},\boldsymbol{S},\boldsymbol{S}', \boldsymbol{R}_o,\boldsymbol{L}_o, \boldsymbol{R}_p,\boldsymbol{L}_{p'}|, 
\end{eqnarray}
\end{widetext}
where $\boldsymbol{S}'$ are the ill-defined stabilizers and $\boldsymbol{S}$ are the respective well-defined stabilizers. The $m$ erased qubits are omitted because they do not contribute to the CI. The states $|\boldsymbol{R}_{\alpha}, \boldsymbol{L}_{\alpha}\rangle$ correspond to the basis states in the reference and code spaces. The choice of $\alpha$ indicates the type of logical qubit: an intact logical qubit is denoted by the subscript $q$, a logical bit has the subscript $o$, and a completely lost qubit has the subscript $p$. Note that the fate of the $k$ logical qubits is restricted to these three possibilities, hence the relation $k = k' + b + c$ holds for any erasure configuration.

From Eq.~\eqref{eq:rho_rq_l} we obtain the state after tracing out the reference system
\begin{widetext}
\begin{eqnarray}\label{eq:rho_q_l}
\rho^{l}_{Q} =  \frac{1}{2^{k'+b+m+c} }\sum_{q=1}^{2^{k'}} \sum_{\boldsymbol{S}'=\pm1}^{2^m} \sum_{o=1}^{2^b} \sum_{p=1}^{2^{c}} |\boldsymbol{L}_q,\boldsymbol{S}, \boldsymbol{S}',\boldsymbol{L}_o, \boldsymbol{L}_{p}\rangle \langle \boldsymbol{L}_{q},\boldsymbol{S},\boldsymbol{S}', \boldsymbol{L}_o, \boldsymbol{L}_{p}|.
\end{eqnarray}
\end{widetext}
The respective von Neumann entropies are:
\begin{eqnarray}
S(\rho^{l}_{RQ}) = m+b+ 2c,
\end{eqnarray}
\begin{eqnarray}
S(\rho^{l}_{Q}) = k'+m+b+ c  .
\end{eqnarray}
Hence
\begin{eqnarray}
\label{eq:average_ci}
I_l = k'-c = k-b-2c.
\end{eqnarray}

The CI contains the information about the lost and degraded logical qubits. For example, in the case of $k=1$, the logical qubit can either be degraded to a classical bit or completely lost. Therefore, only the cases $\{k=0, b=1, c=0\}$ or $\{k=0, b=0, c=1\}$ are possible. The stabilizers, whether ill-defined or not, do not contain any relevant information and therefore do not contribute to the CI.

The actual CI is then the average over erasure configurations:
\begin{eqnarray}
\label{eq:average_ci_2}
I = \sum_{l=1} P(l) I^l = k -\langle b_l \rangle -2\langle c_l \rangle.
\end{eqnarray}
Here, we explicitly write the dependence of $c$ and $b$ on the erasure configuration $l$. Additionally, $\langle \dots \rangle$ denotes the average over the probability of erasure configurations $P(l)$. 
Let us recall that the average in Eq.~\eqref{eq:average_ci_2} appears because the  states $\rho_{RQ}^l$ are orthogonal to each other due to the ancilla bit state $|a_l\rangle$ representing the knowledge of the erasure event.

The calculation of the CI in the presence of erasure errors reduces to sampling erasure configurations $l$ and counting the number of logical bits $b$ and lost logical qubits $c$. In Refs.~\cite{amaro_analytical_2020,vodola_twins_2018}, it was shown that the problem of determining whether a logical operator remains well-defined can be framed as Gaussian elimination on the parity check matrix of the QEC code. Therefore, computing the CI under erasure errors for a fixed erasure configuration $l$ scales polynomially with the number of physical qubits.

\subsection{Coherent information of CSS codes under computational errors}\label{subsec:depo_ci}

In this section, we compute the CI of QEC codes under depolarizing noise, as given by Eq.~\eqref{eq:depo_noise}. We consider CSS stabilizer codes $[[n, k, d]]$ without further assumptions. We start from the generalized Bell state Eq.~\eqref{eq:generalized_bell_pair} between the reference system $R$ and the code space on the system $Q$ of $n$ qubits. In general, the density matrix after applying the noise $\mathcal{N} = \prod \mathcal{N}_i$ to the state $\rho^0_{RQ} = |\psi_{RQ}\rangle \langle \psi_{RQ}|$ looks like:

\begin{eqnarray}
\rho_{RQ}  = \sum_{C}P(C) w_C\rho^0_{RQ}w_C,
\end{eqnarray}
where $w_C$ denotes an error chain $C$ of length $|C|$. Each error chain can be decomposed in terms of the $X$, $Y$, and $Z$ components as $C = C_x \circ C_y \circ C_z$, so the length of the error chain is given by $|C| = |C_x| + |C_y| + |C_z|$. The probability $P(C)$ is the product of having error chains of each kind: $P(C) = (1-p)^{n - |C|} \prod_{i = x, y, z} p_i^{|C_i|}$. In the following, we compute $S(\rho_{RQ})$ and $S(\rho_{Q})$ by first considering the trace of the $r$-th power of density matrices, $\Tr(\rho^r_{RQ})$.

\subsubsection{Calculation of $S(\rho_{RQ})$}\label{subsec:rho_rq_calculation_depo}

\begin{figure}
    \centering
    \includegraphics[width=\linewidth]{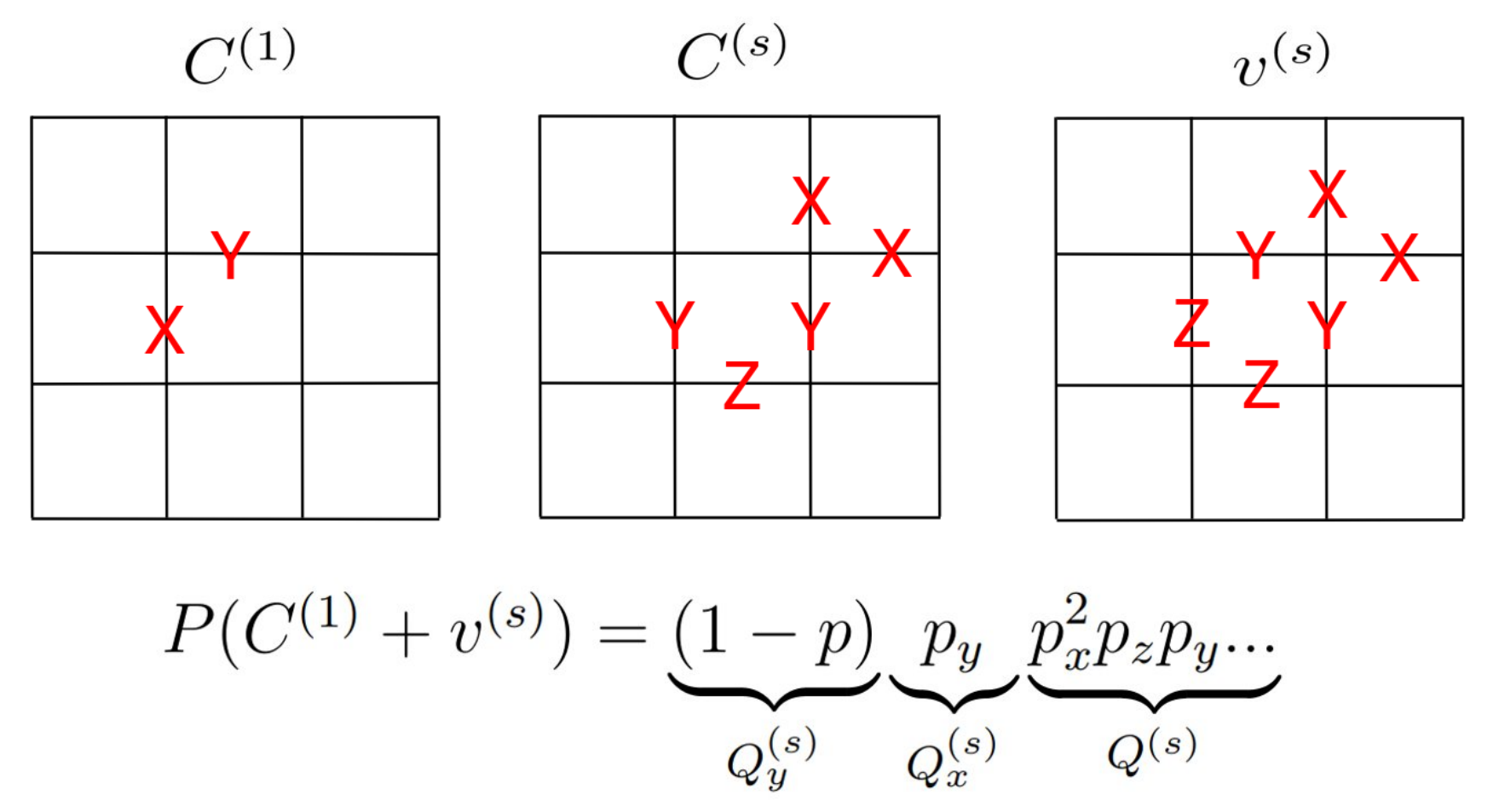}
    \caption{Error chains for the depolarizing noise model. Given a fixed error chain in the first replica $C^{(1)}$, the probability of each $C^{(s)}$ is parameterized according to the errors that appear in the first replica. The probability of having the $C^{(s)}$ depicted in the picture is parameterized  through the weighted probabilities $Q_x^{(s)}(l)$,$Q_y^{(s)}(l)$, shown in Eqs.~\eqref{eq:q_prob} and \eqref{eq:q_xyz}.}
    \label{fig:depo_mapping}
\end{figure}

Now we write $\rho_{RQ}^r$ as a product of $r$ replicas \cite{fan_diagnostics_2024} and take the trace:
\begin{eqnarray}
\rho_{RQ}^r  = \sum_{\{C^{(s)}\}}\prod_{s=1}^{r}P(C^{(s)}) w_{C^{(s)}}\rho^0_{RQ}w_{C^{(s)}},
\end{eqnarray}
\begin{eqnarray}
\Tr(\rho_{RQ}^r)  = \sum_{\{C^{(s)}\}}\Big[\prod_{s=1}^{r}P(C^{(s)}) \Big]\Tr\left(\prod_{s=1}^{r}w_{C^{(s)}}\rho^0_{RQ}w_{C^{(s)}}\right),\nonumber \\
\end{eqnarray}
\begin{widetext}
\begin{eqnarray}
\Tr(\rho_{RQ}^r) & = & \dfrac{1}{2^{kr}}\sum_{\{C^{(s)}\}} \sum_{\{q^{(s)},q^{'(s)}\}}\Big[\prod_{s=1}^{r}P(C^{(s)})\Big] \Tr\left(\prod_{s=1}^{r}w_{C^{(s)}}|\boldsymbol{R}_{q^{(s)}},\boldsymbol{L}_{q^{(s)}},\boldsymbol{S}\rangle\langle \boldsymbol{R}_{q^{'(s)}},\boldsymbol{L}_{q^{'(s)}},\boldsymbol{S}| w_{C^{(s)}}\right).
\end{eqnarray}
\end{widetext}
Here $\{q^{(s)}\}$ denotes replica quantum number configurations, e.g.~$q_1^{(1)}q_1^{(2)}...q_1^{(s)}$ is the configuration where all ``ket" replicas have the same logical quantum  number $q_1$. Since the errors do not act on the reference system, the reference quantum number can never be flipped and many aspects can be simplified. 
First, each replica contributes with a pair $q^{(s)},q'^{(s)}$ making up $2^{2rk}$ configurations. Second, there are $r-1$ inner products that force $q'^{(s)}=q^{(s+1)}$ for $s=1,..., r-1$ reducing the number of configurations to $2^{k(r-1)}$. Third, the first replica is left free thus contributing with a factor $2^k$. Once the first replica $s=1$ is fixed, let's say $w_{C^{(1)}}|\boldsymbol{R}_{q^{(1)}},\boldsymbol{L}_{q^{(1)}},\boldsymbol{S}\rangle = |\boldsymbol{R}_{q^{(1)}},\boldsymbol{L}_{q'^{(1)}},\boldsymbol{S'}\rangle $, the only non-vanishing contributions to the trace are those of the form $w_{C^{(s+1)}} |\boldsymbol{R}_{q^{(s)}},\boldsymbol{L}_{q^{(s)}},\boldsymbol{S}\rangle = |\boldsymbol{R}_{q^{(s)}},\boldsymbol{L}_{q^{'(s)}},\boldsymbol{S'}\rangle$ with $s=0,..,r-1$. Putting all together we arrive at
\begin{eqnarray}
\Tr(\rho_{RQ}^r)  = \sum_{\{C^{(s)}\}}\prod_{s=1}^{r-1}P(C^{(s)}) \langle\psi_0|w_{C^{(s)}}w_{C^{(s+1)}} |\psi_0\rangle, 
\end{eqnarray}
where each error chain fulfills the condition
\begin{eqnarray}
C^{(s+1)} = C^{(1)}+v^{(s)}, \quad s=1,2,...,r-1.
\end{eqnarray}
Here, $v^{(s)}$ stands for all error chains that do not flip any stabilizer quantum number $S_i$ and commute with all logical operators, i.e.~a member of the stabilizer group. In the toric code they have an interpretation in terms of homologically trivial loops \cite{dennis_topological_2002}. One can further simplify this expression as
\begin{eqnarray}
\Tr(\rho_{RQ}^r)  =  \sum_{C^{(1)}} P(C^{(1)}) \sum_{\{v^{(s)}\}} \prod_{s=1}^{r-1} P(C^{(1)}+v^{(s)}).
\end{eqnarray}
The error chains $C^{(s)}$ can contain $X$, $Z$ and $Y$ errors. Then we can express $P(C^{(1)}+v^{s})$ in terms of the binary variables $n^{x,z}_{v^{(s)}}(\ell)=0,1$ that denote whether the qubit $\ell$ has an $X$ or $Z$ error. Now the task is to parameterize the probability of each qubit $\ell$ in $C^{(s)}$ to be ``occupied" by any of the four Pauli operators (including the identity), see Fig.~\ref{fig:depo_mapping}. If $\ell \notin C^{(1)}$ then there is a probability $p_x$ that this site is occupied by an $X$ error, $p_z$ for $Z$, $p_y$ for $Y$ and $1-p$ of the link being unoccupied (i.e.~identity). Thus that qubit contributes with a probability 
\begin{eqnarray}\label{eq:q_prob}
Q^{(s)}(\ell) = p_x^{n_x(1-n_z)}p_z^{n_z(1-n_x)}p_y^{n_z n_x}(1-p)^{(1-n_x)(1-n_z)}. \nonumber \\
\end{eqnarray}
To avoid crowded notation we defined $n_{x,z} \equiv n^{x,z}_{v^{(s)}}(\ell) $. If $\ell \in C^{(1)}$ then one can identify three possibilities:
\begin{widetext}
\begin{eqnarray}\label{eq:q_xyz}
\begin{cases}
\ell \in C^{(1)}_x \rightarrow Q_x^{(s)}(\ell) =  p_x^{(1-n_x)(1-n_z)} p_z^{n_z n_x}p_y^{n_z(1-n_x)}(1-p)^{n_x(1-n_z)} \\
\ell \in C^{(1)}_y \rightarrow Q_y^{(s)}(\ell) = p_x^{n_z(1-n_x)} p_z^{n_x(1-n_z)}p_y^{(1-n_x)(1-n_z)}(1-p)^{n_x n_z} \\
\ell \in C^{(1)}_z \rightarrow Q_z^{(s)}(\ell) = p_x^{n_z n_x} p_z^{(1-n_x)(1-n_z)}p_y^{n_x(1-n_z)}(1-p)^{n_z(1-n_x)}
\end{cases}
\end{eqnarray}
\end{widetext}
Therefore we can write
\begin{widetext}
\begin{eqnarray}\label{eq:Qs}
P(C^{(1)}+v^{(s)}) & = \left[\prod_{\ell\in C_x^{(1)}} Q_x^{(s)}(\ell)\right] \left[\prod_{\ell\in C_y^{(1)}} Q_y^{(s)}(\ell)\right] \left[\prod_{\ell\in C_z^{(1)}} Q_z^{(s)}(\ell)\right] \left[\prod_{\ell \notin C^{(1)}}  Q^{(s)}(\ell) \right].
\end{eqnarray}
\end{widetext}
For each link $\ell$ we can define variables $P^{Z}_\ell$ and $P^{X}_\ell$ that are code-specific and related to $n^x$ and $n^z$ as
\begin{eqnarray}\label{eq:spin_def}
n^x_{v^{(s)}}(\ell) = \dfrac{1-P^X_\ell}{2}, \quad n^z_{v^{(s)}}(\ell) = \dfrac{1-P^Z_\ell}{2}. 
\end{eqnarray}
Here $P^X_{\ell}=\pm 1$ and $P^Z_{\ell}=\pm1$ and their description in terms of classical spin variables depends on the underlying space where the QEC code lives. In this section we leave this structure undetermined and retake it in Sec. \ref{sec:results_topo_codes} for the specific case of 2D toric and color code.  Now we rewrite Eqs. \eqref{eq:q_prob} and \eqref{eq:q_xyz} as 
\begin{widetext}
\begin{eqnarray}
Q^{(s)}(\ell) = \sqrt{p_x p_z}\left(\dfrac{(1-p)p_y}{p_x p_z}\right)^{1/4}\exp\left(J_x P^X_{\ell}/2\right) \exp\left(J_z P^Z_{\ell}/2\right) \exp\left[-J_1(P^X_{\ell}+P^Z_{\ell} - P^X_{\ell} P^Z_{\ell})/4\right], 
\end{eqnarray}
\begin{eqnarray}
Q_x^{(s)}(\ell) &= \sqrt{(1-p) p_y}&\left(\dfrac{(1-p)p_y}{p_x p_z}\right)^{-1/4}\exp\left(-J_x P^X_{\ell}/2\right) \exp\left[(J_z-J_1) P^Z_{\ell} /2\right] \exp\left[J_1(P^X_\ell+ P^Z_{\ell} - P^X_\ell P^Z_{\ell} )/4\right], \nonumber \\
\end{eqnarray}
\begin{eqnarray}
Q_y^{(s)}(\ell) = \sqrt{p_x p_z}\left(\dfrac{(1-p)p_y}{p_x p_z}\right)^{1/4} \exp\left[-(J_x-J_1)P^X_\ell/2\right] \exp\left[-(J_z-J_1)\tau_n \tau_m/2\right] \exp\left[-J_1(P^X_\ell+P^Z_\ell - P^X_\ell P^Z_\ell)/4\right], \nonumber \\\end{eqnarray}
\begin{eqnarray}
Q_z^{(s)}(\ell) = \sqrt{(1-p) p_y} \left(\dfrac{(1-p)p_y}{p_x p_z}\right)^{-1/4}\exp\left[(J_x-J_1)P^X_\ell/2\right] \exp\left(-J_z P^Z_\ell/2\right)\exp\left[J_1(P^X_\ell+P^Z_\ell- P^X_\ell P^Z_\ell)/4\right].
\end{eqnarray}
\end{widetext}
Now we can rewrite the whole $P(C^{(1)}+v^{(s)})$ as
\begin{eqnarray}
P(C^{(1)}+v^{(s)})  = \dfrac{1}{f}\left(p_x p_z p_y(1-p)\right)^{n/4}  e^{H[\{\eta^x\},\{\eta^z\}]} 
\end{eqnarray}
with (omitting the replica index $s$):
\begin{eqnarray}\label{eq:hamiltonian_xyz_2}
&&H[\{\eta^x\},\{\eta^z\}] = \sum_{\ell} \eta^x_{\ell}\left(J_x-\dfrac{J_1}{2}\right)\dfrac{P^X_{\ell}}{2}+ \nonumber \\
&& \eta^z_{\ell}\left(J_z-\dfrac{J_1}{2}\right) \dfrac{P^Z_{\ell}}{2} +\eta^z_{\ell}\eta^x_{\ell}J_1  \dfrac{P^Z_{\ell}  P^X_{\ell}}{4}.
\end{eqnarray}
The factor $f$ is code-dependent and reflects the symmetry of the resulting spin model. For instance, for the toric code, whose spin model is $\mathbb{Z}_2$-symmetric, $f=2$ because flipping all spins together leaves all $n^x_{v^{(s)}}(\ell)$ and $ n^z_{v^{(s)}}(\ell)$ invariant. As expected, the CI will not depend on the factor $f$. The coupling constants are defined as 
\begin{eqnarray}\label{eq:J_def}
e^{J_x} = \dfrac{1-p}{p_x},\quad, e^{J_z} = \dfrac{1-p}{p_z}, \quad e^{J_1} = \dfrac{(1-p)p_y}{p_x p_z}.
\end{eqnarray}
The whole expression for $\Tr(\rho_{RQ}^r)$ is then rewritten as
\begin{widetext}
\begin{eqnarray}
\Tr(\rho_{RQ}^r)  & = &  \dfrac{1}{f^{r-1}} \left[(1-p)p_xp_z p_y\right]^{(r-1)n/4}\sum_{\{\eta^x, \eta^z\}} P(\{\eta^x,\eta^z\}) \sum_{\{ P^{X(s)},P^{Z(s)}\} } \prod_{s=1}^{r-1} \exp\left(H(s)\right) \nonumber \\
                & = & \dfrac{1}{f^{r-1}}  \left[(1-p)p_xp_z p_y\right]^{(r-1)n/4} \sum_{\{\eta^x \eta^z\}} P(\{\eta^x \eta^z\}) \mathcal{Z}\left[\{\eta^x,\eta^z\}\right]^{r-1}.
\end{eqnarray}
\end{widetext}

Here, $\mathcal{Z}\left[\{\eta^x,\eta^z\}\right] = \sum_{\{ P^{X(s)},P^{Z(s)}\}} \exp\left[H(s)\right] $ is the partition function of a double spin model with the configuration $\{\eta^x, \eta^z\}$ of couplings. Each choice of $\{\eta^x, \eta^z\}$ represents one error chain and $P(\{\eta^x,\eta^z\})$ its probability. Now we take the limit $r\rightarrow 1$ using the identity 
\begin{eqnarray}\label{eq:entropy_identity}
S(\rho) =  - \lim_{r \to 1} \dfrac{\partial}{\partial r}  \Tr(\rho^r)
\end{eqnarray}
and obtain
\begin{eqnarray}\label{eq:entropy_depo_rho_rq}
S(\rho_{RQ})  =  -\dfrac{n}{4}\log\left[(1-p)p_x p_y p_z\right]+\log f-\overline{\log \mathcal{Z}},
\end{eqnarray}
where we have defined $\overline{(...)} = \sum_{\{\eta^x, \eta^z\}} P(\{\eta^x,\eta^z\}) (...)$ as the disorder average associated with the computational errors. 

\subsubsection{Calculation of $S(\rho_{Q})$}

Now we compute the $r-$th power of the state $\rho_Q=\Tr_R{(\rho_{RQ})}$ and take its trace,
\begin{eqnarray}
\rho_{Q}^r  = \sum_{\{C^{(s)}\}}\prod_{s=1}^{r}P(C^{(s)}) w_{C^{(s)}}\rho^0_{Q}w_{C^{(s)}},
\end{eqnarray}
\begin{eqnarray}
\Tr(\rho_{Q}^r)  = \sum_{\{C^{(s)}\}}\Big[\prod_{s=1}^{r}P(C^{(s)}) \Big]\Tr\left(\prod_{s=1}^{r}w_{C^{(s)}}\rho^0_{Q}w_{C^{(s)}}\right),
\end{eqnarray}
\begin{widetext}
\begin{eqnarray}
\Tr(\rho_{Q}^r) =  \dfrac{1}{2^{kr}}\sum_{\{C^{(s)}\}}\sum_{\{q^{(s)}\}}\Big[\prod_{s=1}^{r}P(C^{(s)})\Big]\Tr\left(\prod_{s=1}^{r}w_{C^{(s)}}|\boldsymbol{L}_{q^{(s)}},\boldsymbol{S}\rangle\langle \boldsymbol{L}_{q^{(s)}},\boldsymbol{S}| w_{C^{(s)}}\right).
\end{eqnarray}
\end{widetext}
In the last step, we replaced $\rho^0_Q = \sum_{q=1}^{2^k} |\boldsymbol{L}_{q}, \boldsymbol{S} \rangle \langle \boldsymbol{L}_{q}, \boldsymbol{S}| / 2^k$ as the mixture of all codewords in the QEC code. The summation $\sum_{\{q^{(s)}\}}$ runs over all codewords in the $r$ replicas. There are $2^{kr}$ configurations of $q^{(s)}$, and almost all of them have a non-vanishing contribution. Once the first replica $C^{(1)}$ is fixed, the configurations of $C^{(s)}$ with non-vanishing contributions satisfy the condition $C^{(s+1)} = C^{(1)} + v^{(s)} + D^{(s)}_l$, and one can write
\begin{widetext}
\begin{eqnarray}
\Tr(\rho_{Q}^r) =  \dfrac{1}{2^{k(r-1)}} \sum_{C^{(1)}} P(C^{(1)}) \prod_{s=1}^{r-1} \sum_{\{D^{(s)}\}} \sum_{\{v^{(s)}\}} P(C^{(1)}+v^{(s)}+D^{(s)}) .
\end{eqnarray}
\end{widetext}

Here, $v^{(s)}$ are the $X$ and $Z$ stabilizer operators, and $D^{(s)} = \prod_{i=1}^{k} X_{i}^{d^i_x} Z_{i}^{d^i_z}$ is a product of logical operators, where $d^i_{x,z} = 0, 1$ is a binary variable that denotes their absence or presence in the specific replica (the replica index in $d^i_{x,z}$ is omitted). In total, there are $4^k$ possible $D^{(s)}$, which represent all possible combinations of logical operators. For instance, for $k = 1$, there are four possibilities: $I$, $X_L$, $Z_L$, and $X_L Z_L$.
The summation $\sum_{\{v^{(s)}\}}$ becomes the summation over classical spin configurations, as discussed in Sec.~\ref{subsec:rho_rq_calculation_depo}. 
From the point of view of Eq.~\eqref{eq:Qs}, the logical operator $D^{(s)}$ can be absorbed into $C^{(1)}$, and therefore the summation $\sum_{\{D^{(s)}\}}$ becomes the summation over different partition functions for which the variables $\eta^x, \eta^z$ are flipped along the support of the respective logical operator. The rest of the derivation follows the same steps as for $\Tr(\rho^r_{RQ})$, which leads us to

\begin{widetext}
\begin{eqnarray}
\Tr(\rho_{Q}^r) &  = &  \dfrac{1}{f^{r-1}} \dfrac{1}{2^{k(r-1)}} \left[(1-p)p_xp_z p_y\right]^{(r-1)n/4}\sum_{\{\eta^x,\eta^z\}} P(\{\eta^x, \eta^z \}) \left( \sum_{\{D\}} \mathcal{Z}_D\right)^{r-1} \nonumber \\
                & = &  \dfrac{1}{f^{r-1}} \dfrac{1}{2^{k(r-1)}} \left[(1-p)p_xp_z p_y\right]^{(r-1)n/4} \overline{\left( \sum_{\{D\}} \mathcal{Z}_D\right)^{r-1}} .
\end{eqnarray}
\end{widetext}
In the partition function $\mathcal{Z}_D$ the variables $\eta^x$ and $\eta^z$ along the support of $D$ are transformed as follows: i) for $D$ containing a $X$ logical operator then $(\eta^x,\eta^z)\rightarrow(-\eta^x,\eta^z)$ ii) for $D$ containing a $Z$ logical operator then $(\eta^x,\eta^z)\rightarrow(\eta^x,-\eta^z)$ and iii) for $D$ containing both $X$ and $Z$ logical operators then $(\eta^x,\eta^z)\rightarrow(-\eta^x,-\eta^z)$. Now we take the limit $r\rightarrow 1$ using Eq.~\eqref{eq:entropy_identity} and get:
\begin{eqnarray}\label{eq:entropy_depo_rho_q}
S(\rho_{Q}) &= &  k +\log f  -\dfrac{n}{4}\log\left[(1-p)p_x p_y p_z\right] \nonumber \\
&&-\overline{\log  \left[\sum_{D} \mathcal{Z}_{D}\right]}
\end{eqnarray}
\subsubsection{Coherent information for generic Pauli noise}

Now we combine Eqs.~\eqref{eq:entropy_depo_rho_rq} and \eqref{eq:entropy_depo_rho_q} to write the CI for depolarizing noise:
\begin{eqnarray}\label{eq:ci_depo}
I  = k - \overline{\log \left[\dfrac{\sum_{D} \mathcal{Z}_{D}}{\mathcal{Z}_0}\right]}
\end{eqnarray}

where $\mathcal{Z}_0$ is the partition function without defects along the support of logical operators. Importantly, Eq.~\eqref{eq:ci_depo} does not make any assumptions about the underlying details of the QEC code. Neither dependencies on $k$ nor the underlying geometry of the code are specified. Therefore, we expect that the same approach applies to LDPC codes or topological codes in any spatial dimension. We highlight that the details of the code enter via the variables $P^X_\ell$ and $P^Z_\ell$ defined in Eq.~\eqref{eq:spin_def}. Thus, the work lies in properly parameterizing these variables in terms of classical spins. 
In this work, we specifically tackle the 2D toric and color codes in Sec.~\ref{sec:results_topo_codes} and the lift-connected surface code in Sec.~\ref{sec:lcs}. For simplicity, we assume identical Pauli channels on each qubit, but this can also be relaxed by allowing qubit-dependent couplings $J_x, J_z, J_1$ or considering correlated errors~\cite{chubb_statistical_2021}.

For topological codes, it has been observed that the phase transition in $\mathcal{Z}_0$ already pinpoints the optimal threshold \cite{dennis_topological_2002,huang_coherent_2024} and, hence, the singular behavior of the CI at the threshold \cite{fan_diagnostics_2024}. The reason is that fluctuations due to topological defects are exponentially suppressed in the thermodynamic limit, meaning that all terms $\mathcal{Z}_D / \mathcal{Z}_0$ vanish independently in the ordered phase. However, it remains to be seen whether this always holds for low-density parity-check (LDPC) codes \cite{breuckmann_quantum_2021}, whose growing number of logical qubits allows for many more classes of topological defects. These defects could potentially enhance the fluctuations in the free energy difference between phases \cite{placke_random-bond_2023,jiang_duality_2019}.

\subsection{Coherent information of CSS codes under computational and erasure errors}\label{subsec:depo_ci_loss}

We now turn to computing the CI for erasure and depolarizing errors together. We start from the state $\rho^l_{RQ}$ under a qubit erasure configuration $l$ where a total $m$ qubits are lost, given by Eq.~\ref{eq:rho_rq_l}. The computational errors enter in the usual way, 
\begin{eqnarray}
\rho^{l'}_{RQ}  = \sum_{C}P(C) w_C\rho^l_{RQ}w_C,
\end{eqnarray}
where $w_C$ denotes an error chain $C$ of length $|C|$ on the remaining $n-m$ data qubits. Each error chain can be decomposed as $C=C_x \circ C_y  \circ C_z$, therefore the length of the error chain is given by $|C| = |C_x|+|C_y|+|C_z|$. The probability $P(C)$ is the product of having error chains of each kind $P(C)=(1-p)^{n-m-|C|} \prod_{i=x,y,z}p_i^{|C_i|}$.

\subsubsection{Calculation of $S(\rho_{RQ})$}

Now we write $\rho_{RQ}^r$ (dropping for now the $l'$ index) as a product of $r$ replicas and take the trace,
\begin{eqnarray}
\rho_{RQ}^r  = \sum_{\{C^{(s)}\}}\prod_{s=1}^{r}P(C^{(s)})w_{C^{(s)}}\rho^l_{RQ}w_{C^{(s)}},
\end{eqnarray}
\begin{eqnarray}\label{eq:trace_rho_rq_l}
\Tr(\rho_{RQ}^r)  = \sum_{\{C^{(s)}\}}\Big[\prod_{s=1}^{r}P(C^{(s)})\Big]\Tr\left(\prod_{s=1}^{r}w_{C^{(s)}}\rho^l_{RQ}w_{C^{(s)}}\right). \nonumber \\
\end{eqnarray}

At this point we insert Eq.~\eqref{eq:rho_rq_l}, then the indices $q, o, p$ acquire a replica index $s$. The task is to identify the non-vanishing replica configurations of error chains $\{C^{(s)}\}$. Let us now walk through the restrictions imposed onto $\{C^{(s)}\}$ for each set of indices $q^{(s)}, o^{(s)}, p^{(s)}$ and the well-defined stabilizer generators $\boldsymbol{S}$ after fixing the first replica $C^{(1)}$:

\begin{itemize}
    \item Logical qubit quantum numbers $(\boldsymbol{R}_{q^{(s)}}, \boldsymbol{L}_{q^{(s)}})$ and well-defined stabilizers $\boldsymbol{S}$: both types of quantum numbers impose the constraint of not allowing $C^{(s)}$ for $s > 1$ to have a different logical operator and syndrome from $C^{(1)}$. Thus, $C^{(s)}$ can only contain error chains that produce the same syndrome as $C^{(1)}$ on the \emph{well-defined $Z$ and $X$ stabilizers}. As in the case without erasures, the total number of replica state configurations with a non-vanishing contribution is $2^{k'r}$.
    \item Ill-defined stabilizers $\boldsymbol{S'}$: their quantum numbers enter as identity operators. Therefore, for any combination $C^{(s)}$ of errors, each of them contributes with a factor of 2. Thus there are $2^m$ replica configurations with non-vanishing contributions. They do not impose any constraint on the error chains $C^{(s)}$.
    \item Logical bit quantum numbers $(\boldsymbol{R}_{o^{(s)}}, \boldsymbol{L}_{o^{(s)}})$: First, recall that the states $|\boldsymbol{R}_{o^{(s)}}, \boldsymbol{L}_{o^{(s)}}\rangle \langle \boldsymbol{R}_{o^{(s)}}, \boldsymbol{L}_{o^{(s)}}|$ can be either $(|00\rangle \langle 00|, |11\rangle \langle 11|)$, in the respective basis states of the logical qubit and associated reference qubit under consideration, when $O^x_{L_i}$ is lost, or $(|++\rangle \langle ++|, |--\rangle \langle --|)$ correspondingly when $O^z_{L_i}$ is lost. Thus, we decompose $b = b_x + b_z$, where $b_{x(z)}$ denotes the remaining bits in the $X(Z)$ basis. The reference quantum number forces all logical bit numbers $|o_R^{(s)}, o^{(s)}_L\rangle \langle o^{(s)}_R, o^{(s)}_L|$ to be the same. Therefore, the only non-vanishing contributions come from replica configurations with $o^{(s+1)} = o^{(s)}$ for $s = 1, \dots, r-1$. Hence, only $2^b$ configurations contribute to the trace. The constraints on $C^{(s)}$ are the same as for the logical qubits (all logical operators must be the same in all replicas).
    \item Lost qubits quantum numbers $(\boldsymbol{R}_{p^{(s)}},\boldsymbol{L}_{p^{'(s)}})$: the reference part $\boldsymbol{R}_{p^{(s)}}$ is forced the be equal in each replica. Since the logical operators for these logical qubits are gone, the quantum number $\boldsymbol{L}_{p^{'(s)}}$ cannot be flipped by an error chain. Therefore there are only $4^c$ configurations with non-vanishing contribution.  
    
\end{itemize}

In summary we obtain
\begin{eqnarray}
\Tr(\rho_{RQ}^r) & = & \frac{1}{2^{(b+m+2c)(r-1)} }\times\nonumber \\
&&\hspace{-1cm}\sum_{\{C^{(s)}\}}\prod_{s=1}^{r}P(C^{(s)}) \langle\psi_0|w_{C^{(s)}}w_{C^{(s+1)}} |\psi_0\rangle, 
\end{eqnarray}
where error chains $C^{(s)}$ are restricted to
\begin{eqnarray}
C^{(s+1)} = C^{(1)}+  v^{(s)}, \quad s=1,2,...,r-1.
\end{eqnarray}
Here, $\{v^{(s)}\}$ represent all operations that do not change \emph{any of the well-defined $Z$ and $X$ stabilizer} quantum numbers and commute with \emph{all} logical operators. In the case without erasures, $v^{(s)}$ is always a member of the stabilizer group. After erasing some qubits, the well-defined $X$ and $Z$ stabilizers and logical operators cannot generate all error chains $C$. However, we must consider every $v^{(s)}$, even if they are not themselves made up of the well-defined $X$ and $Z$ stabilizers. Therefore, we can write
\begin{eqnarray}
\Tr(\rho_{RQ}^r) & = &  \frac{1}{2^{(b+m+2c)(r-1)} }  \sum_{C^{(1)}} P(C^{(1)}) \times \nonumber \\
&& \sum_{\{v^{(s)}\}} \prod_{s=1}^{r-1} P(C^{(1)}+ v^{(s)}).
\end{eqnarray}
Now we must count all possible $v^{(s)}$. To do so, we define binary variables $n_{v^{(s)}}^{x}(\ell) = 0, 1$, $n_{v^{(s)}}^{z}(\ell) = 0, 1,$ to denote whether a link $\ell$ on the direct lattice is occupied in the error chain $v^{(s)}$ by $X$, $Y$, $Z$, or $I$ (not occupied). The procedure is the same as in Sec.~\ref{subsec:depo_ci}. Again, we express $P(C^{(1)} + v^{(s)})$ in terms of the binary variables $n^{x,z}_{v^{(s)}}(\ell) = 0, 1,$ and write down the probability for each single qubit error

\begin{widetext}
\begin{eqnarray}\label{eq:q_xyz_2}
\begin{cases}
\ell \notin C^{(1)} \rightarrow Q^{(s)}(\ell) = p_x^{n_x(1-n_z)}p_z^{n_z(1-n_x)}p_y^{n_z n_x}(1-p)^{(1-n_x)(1-n_z)} \\ 
\ell \in C^{(1)}_x \rightarrow Q_x^{(s)}(\ell) =  p_x^{(1-n_x)(1-n_z)} p_z^{n_z n_x}p_y^{n_z(1-n_x)}(1-p)^{n_x(1-n_z)} \\
\ell \in C^{(1)}_y \rightarrow Q_y^{(s)}(\ell) = p_x^{n_z(1-n_x)} p_z^{n_x(1-n_z)}p_y^{(1-n_x)(1-n_z)}(1-p)^{n_x n_z} \\
\ell \in C^{(1)}_z \rightarrow Q_z^{(s)}(\ell) = p_x^{n_z n_x} p_z^{(1-n_x)(1-n_z)}p_y^{n_x(1-n_z)}(1-p)^{n_z(1-n_x)}
\end{cases}
\end{eqnarray}
\end{widetext}

To avoid crowded notation we abbreviated $n_{x,z} \equiv n^{x,z}_{v^{(s)}}(\ell) $. The remainder of the calculation is similar to the case without erasures studied in Sec.~\ref{subsec:depo_ci}. The only difference is that now $\ell$ denotes a qubit on the complement of $A_l$ instead of any arbitrary qubit. We then obtain
\begin{widetext}
\begin{eqnarray}
\Tr(\rho_{RQ}^r)  & = &  \dfrac{\left[(1-p)p_xp_z p_y\right]^{(r-1)(n-m)/4}}{f^{r-1} 2^{(b+m+2c)(r-1)}}\sum_{\{\eta^x, \eta^z\}} P(\{\eta^x,\eta^z\}) \sum_{\{ P^{X(s)},P^{Z(s)}\} } \prod_{s=1}^{r-1} \exp\left(H_l(s)\right) \nonumber \\
                & = & \dfrac{\left[(1-p)p_xp_z p_y\right]^{(r-1)(n-m)/4}}{f^{r-1} 2^{(b+m+2c)(r-1)}} \sum_{\{\eta^x \eta^z\}} P(\{\eta^x \eta^z\}) \mathcal{Z}_l\left[\{\eta^x,\eta^z\}\right]^{r-1}.
\end{eqnarray}
\end{widetext}
The partition function $\mathcal{Z}_l\left[\{\eta^x, \eta^z\}\right] = \sum_{\{ P^{X(s)}, P^{Z(s)}\}} \exp\left[H_l(s)\right]$ is the partition function of a classical Hamiltonian given by Eq.~\eqref{eq:hamiltonian_xyz_2} with the configuration $\{\eta^x, \eta^z\}$ of couplings. Each choice of $\{\eta^x, \eta^z\}$ represents an error chain, and $P(\{\eta^x, \eta^z\})$ is its probability according to Table~\ref{table:chain_error_erasure}. The erasure errors enter as \emph{missing links} with probability $e$ in the statistical mechanics model derived in Sec.~\ref{subsec:depo_ci}. In general, erasure errors favor a fully symmetric equilibrium state by switching off the interaction on each link $\ell$ of the lattice. In Secs.~\ref{sec:results_topo_codes} and \ref{sec:lcs}, we further discuss this model for 2D toric and color codes and one class of LDPC code. Now, using the identity in Eq.~\eqref{eq:entropy_identity}, we obtain

\begin{eqnarray}\label{eq:entropy_depo_rho_rq_loss}
S(\rho_{RQ})  &=&  -\dfrac{n-m}{4}\log\left[(1-p)p_x p_y p_z\right]+\log f \nonumber \\
&&+ b+m+2c-\overline{\log \mathcal{Z}_l}.
\end{eqnarray}
We omitted the $l$ dependence on $b$, $c$ and $m$. 

\subsubsection{Calculation of $S(\rho_{Q})$}

Now we compute the trace of the $r-$th power of the state $\rho_Q=\Tr_R{(\rho_{RQ})}$ starting now from Eq.~\eqref{eq:rho_q_l}:
\begin{widetext}
\begin{eqnarray}
\Tr(\rho_{Q}^r) &=&  \frac{1}{2^{(k'+b+m+c)r}}\sum_{\{C^{(s)}\}}\sum_{\{q^{(s)}\}} \sum_{\{\boldsymbol{S}^{'(s)}\}} \sum_{\{o^{(s)}\}} \sum_{\{p^{(s)}\}}\Big[\prod_{s=1}^{r}P(C^{(s)})\Big]\Tr\Big(\prod_{s=1}^{r}w_{C^{(s)}}|\boldsymbol{L}_{q^{(s)}},\boldsymbol{S}^{(s)}, \boldsymbol{S}^{'(s)},\boldsymbol{L}_{o^{(s)}}, \boldsymbol{L}_{p^{(s)}}\rangle \nonumber \\
&&\langle \boldsymbol{L}_{q^{(s)}},\boldsymbol{S}^{(s)}, \boldsymbol{S}^{'(s)},\boldsymbol{L}_{o^{(s)}}, \boldsymbol{L}_{p^{(s)}}|w_{C^{(s)}}\Big).
\end{eqnarray}
\end{widetext}
The quantum numbers that enter as an identity contribute equally; the only degree of freedom to be considered is the well-defined stabilizers $\boldsymbol{S}$. After fixing the first replica $C^{(1)}$, the configuration of $C^{(s)}$ with non-vanishing contributions must satisfy the condition $C^{(s+1)} = C^{(1)} + v^{(s)} + D^{(s)}_l$, and one can write
\begin{widetext}
\begin{eqnarray}
\Tr(\rho_{Q}^r) =  \dfrac{1}{2^{(k'+m+b+c)(r-1)}} \sum_{C^{(1)}} P(C^{(1)}) \prod_{s=1}^{r-1} \sum_{\{D^{(s)}\}} \sum_{\{v^{(s)}\}} P(C^{(1)}+v^{(s)}+D^{(s)}_l) .
\end{eqnarray}
\end{widetext}
Similarly to the calculation of $\Tr{(\rho_{RQ})}$, here $v^{(s)}$ represents all operators that commute with the well-defined stabilizers. Also, $D^{(s)} = \prod_{i=1}^{k} X_{i}^{d^i_x} Z_{i}^{d^i_z}$ is a product of logical operators that \emph{anti-commute} with the well-defined logical operators. Clearly, for the $k'$ remaining logical qubits, both logical operators are included. For the $b$ logical bits, when $X$ ($Z$) is the well-defined logical operator, the logical $Z$ ($X$) is the one flipping the respective logical quantum number. The binary variables $d^i_{x,z} = 0, 1$ denote this set of logical operators. Hence, there are $2^{2k' + b}$ possible $D^{(s)}$, which represent all possible combinations of anti-commuting logical operators. The rest of the derivation proceeds as in the erasure-free case, and we then obtain

\begin{widetext}
\begin{eqnarray}
\Tr(\rho_{Q}^r) &  = &  \dfrac{1}{f^{r-1}} \dfrac{1}{2^{(k'+m+b+c)(r-1)}} \left[(1-p)p_xp_z p_y\right]^{(r-1)(n-m)/4}\sum_{\{\eta^x,\eta^z\}} P(\{\eta^x, \eta^z \}) \left( \sum_{\{D_l\}} \mathcal{Z}_{D_l}\right)^{r-1} \nonumber \\
                & = &  \dfrac{1}{f^{r-1}} \dfrac{1}{2^{(k'+m+b+c)(r-1)}} \left[(1-p)p_xp_z p_y\right]^{(r-1)(n-m)/4} \overline{\left( \sum_{\{D_l\}} \mathcal{Z}_{D_l}\right)^{r-1}} .
\end{eqnarray}
\end{widetext}
The partition function $\mathcal{Z}_{D_l}$ has flipped couplings $\eta^x$ and $\eta^z$ along the support of the anti-commuting set of logical operators $D_l$.  Now, as before, we take the limit $r\rightarrow 1$ using Eq.~\eqref{eq:entropy_identity} and obtain
\begin{eqnarray}\label{eq:entropy_depo_rho_q_loss}
&S(\rho_{Q}) =  &k'+m+b+c+\log f  - \nonumber \\
&\dfrac{n-m}{4}&\log\left[(1-p)p_x p_y p_z\right] -\overline{\log  \left[\sum_{D} \mathcal{Z}_{D}\right]}.
\end{eqnarray}

\begin{figure}
    \centering
    \includegraphics[width=0.8\linewidth]{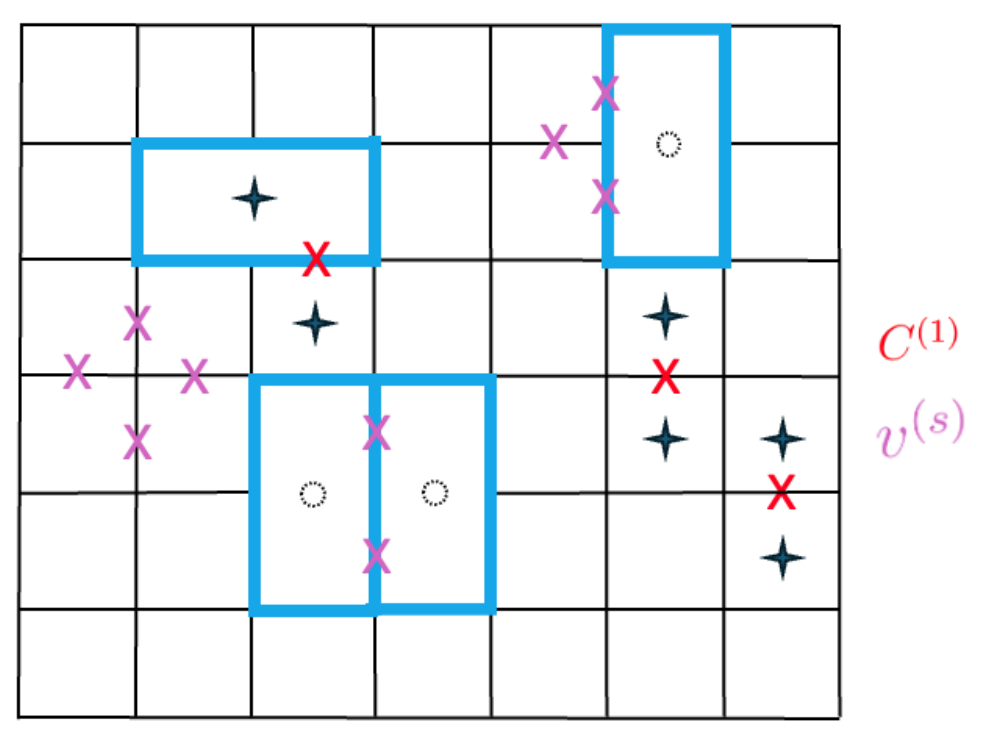}
    \caption{Toric code square lattice with qubits on the edges. Error chain $C^{(1)}$ in red and commuting error configuration $v^{(s)}$ in violet. The blue plaquettes are the super-plaquettes formed after erasures (denoted by open circles and missing edges of the square lattice). The syndrome generated by $C^{(1)}$ is shown as four-point stars in the center of the respective stabilizer plaquettes. The error chain $v^{(s)}$ does not create nor remove syndromes.}
    \label{fig:lattice_depo_loss}
\end{figure}

\subsubsection{Coherent information for Pauli noise and erasure errors}

Now we combine Eqs.~\eqref{eq:entropy_depo_rho_rq_loss} and \eqref{eq:entropy_depo_rho_q_loss} and get the CI for a fixed configuration $l$ of erased qubits, 
\begin{eqnarray}\label{eq:ci_depo_loss}
I_l  = k-b_l-2c_l - \overline{\log \left[\dfrac{\sum_{D_l} \mathcal{Z}_{D_l}}{\mathcal{Z}_0}\right]}, 
\end{eqnarray}
where we replaced $k'=k-b-c$. The averaged CI $I = \sum_l P(l)I_l$ is then written as
\begin{eqnarray}
\label{eq:average_ci_comp_loss}
I  = k  -\langle b_l\rangle -2\langle c_l\rangle - \Biggr\langle \overline{\log \left[\dfrac{\sum_{D} \mathcal{Z}_{D,l}}{\mathcal{Z}_{0,l}}\right]}\Biggr\rangle.
\end{eqnarray}
Let us note that $\overline{(...)}$ denotes the disorder average that arises from computational errors, and $\langle F \rangle = \sum_l P(l) F_l$ is the classical average over erased qubits. 

We can clearly distinguish two sources for the reduction of CI: (i) by means of the erasure errors the environment can effectively measure the logical operators and directly reduce the CI by integer multiples of $\log 2=1$. (ii) The computational errors act on the remaining physical qubits, with their effects quantified by the cost in free energy of inserting domain walls in the equilibrium state, as given by the partition function $\mathcal{Z}_0$ \cite{dennis_topological_2002,wang_confinement-higgs_2003}. The erasure errors drive the transition to a fully symmetric state by removing interactions locally, thereby lowering the cost of inserting domain walls.
This picture is general and does not make any assumptions about the underlying geometry and spatial structure of the QEC code. Moreover, since the details of the code enter via the variables $P^X_\ell$ and $P^Z_\ell$ defined in Eq.~\eqref{eq:spin_def}, the classical spin models derived are the same with and without erasure errors, except for the missing links in the latter. For simplicity, we assume identical Pauli channels on each qubit, but this could also be relaxed by allowing qubit-dependent couplings $J_x, J_z, J_1$ and considering spatially correlated noise \cite{chubb_statistical_2021}.

\section{2D topological codes}\label{sec:results_topo_codes}

\begin{figure}[h!]
    \centering
    \includegraphics[width=0.5\linewidth]{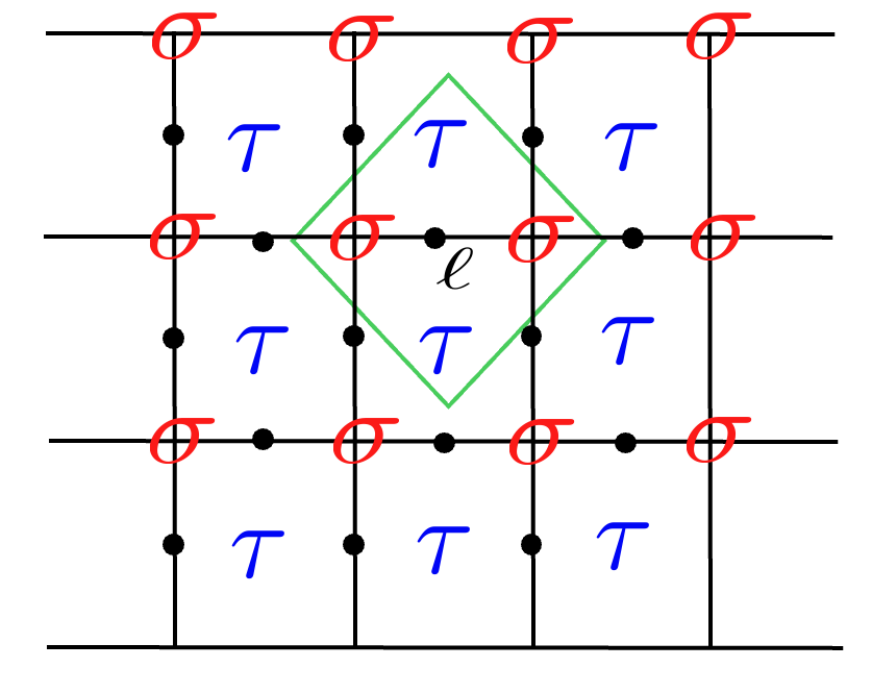}
    \caption{Spin model for the 2D toric code. Black circles denote physical qubits. $\sigma$ ($\tau$) are the spins counting the $X$($Z$) stabilizers. Each qubit $\ell$ is shared by two $X$ and $Z$ stabilizers. An $X$($Z$) error indicates a link between the two $\sigma$ ($\tau$) spins opposite to that site. A $Y$ error does so as a four spin interaction, two $\sigma$ and two $\tau$ around the site of the error (green diamond).}
    \label{fig:depo_lattice_toric}
\end{figure}

\begin{figure}[h!]
    \centering
    \includegraphics[width=0.55\linewidth]{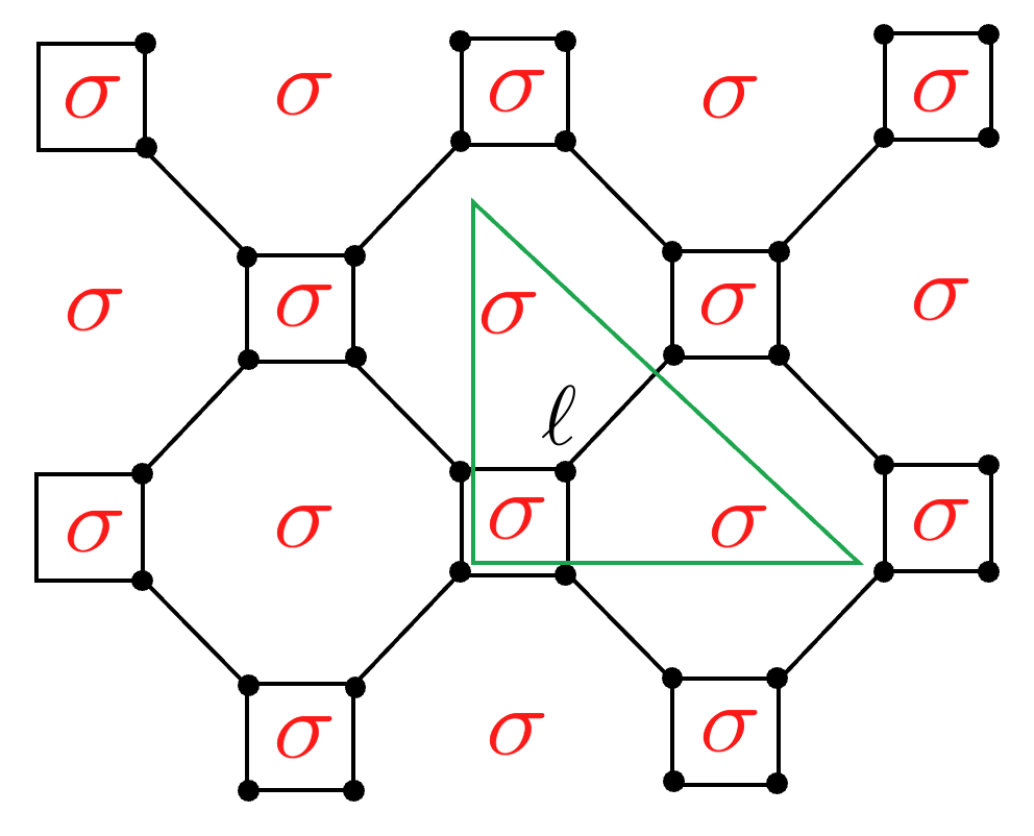}
    \caption{Spin model for the 4.8.8 color code. Black circles denote physical qubits. Only $\sigma$ spins counting $X$ stabilizers are shown. The $\tau$ spins live on the same positions as $\sigma$ spins. Each qubit $\ell$ is shared by three $X$ stabilizers. An $X$($Z$) error indicates a link between three $\sigma$ ($\tau$) spins. A $Y$ error does so as a six spin interaction, involving three $\sigma$ and three $\tau$.}
    \label{fig:depo_lattice_color}
\end{figure}

In this section, we present numerical calculations of the CI for the rotated surface code \cite{tomita_low-distance_2014} and the 4.8.8 octagonal color code \cite{bombin_topological_2006, bombin_topological_2007} (for details of the codes, see Appendix~\ref{app:qec_codes}). We calculate the CI exactly for code sizes up to $d=5$ in the presence of computational errors, and up to $d=17$ for erasure errors only.
For some choices of code and code distance, we calculate the CI via exact computation of partition functions derived in Sec.~\ref{subsec:depo_ci} and Sec.~\ref{subsec:depo_ci_loss}. For others, it is more convenient to directly use the mixed-state density matrix, as done in Ref.~\cite{colmenarez_accurate_2024}, which illustrates the practical complementarity of the two approaches. In two cases, namely (i) $d=5$ color code with depolarizing noise and erasure errors, and (ii) erasure errors at high distances, we use a stochastic approximation for the erasure errors, which involves sampling over qubit loss configurations (see Appendix~\ref{appendix:ci_erasure} for details).

\subsection{Erasure errors}

\begin{figure}
    \centering
    \includegraphics[width=1.0\columnwidth]{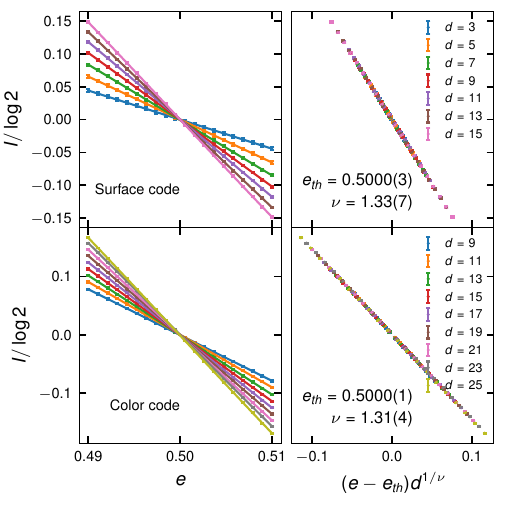}
    \caption{Coherent information of surface and 4.8.8 color code under erasure errors. For both code families we obtain a threshold of $e_{th}=0.5$ and a critical exponent close to  $\nu=4/3\approx 1.33$, which is the same as for the probability of having a percolating cluster in 2D bond percolation \cite{stauffer_introduction_1992}.}
    \label{fig:topo_erasure}
\end{figure}

We show in Fig.~\ref{fig:topo_erasure} the CI for the surface code and the 4.8.8 color code. The optimal threshold problem in the 2D toric code can be mapped to a bond percolation problem on a square lattice \cite{stace_thresholds_2009}, which yields a threshold of $e_{\text{th}} = 0.5$, saturating the fundamental limit set by the no-cloning theorem. We find the same threshold in our calculation with four-digit precision, as shown in Fig.~\ref{fig:topo_erasure}.
In terms of percolation theory, the CI, after renormalized and rescaled properly, can be viewed as the probability of the appearance of percolating clusters. At low enough erasure probability $e$, both logical string operators still have support on non-lost qubits in the lattice, and effectively can traverse the lattice. In contrast, at high $e$, they become confined within small, disconnected regions and therefore stop being well-defined. Therefore, we expect the CI and the percolating cluster probability to have the same scaling exponent. In fact, we find $\nu = 1.33(4)$ and $\nu = 1.4(1)$ for the surface code and the color code, respectively, which is, within statistical error bars, consistent with the $\nu = 4/3 \approx 1.33$ predicted by percolation theory \cite{stauffer_introduction_1992}.
A similar result for the toric code in the context of mixed-state topological order was obtained in Ref.~\cite{kuno_intrinsic_2024}.
Furthermore, since both codes have the same threshold and the same scaling exponent, we observe further evidence that the 2D toric and color codes have the same thresholds against quantum erasures in the code capacity setting \cite{dennis_topological_2002, bombin_strong_2012}.

\subsection{Computational and erasure errors}

\begin{figure}
    \centering
    \includegraphics[width=1.0\columnwidth]{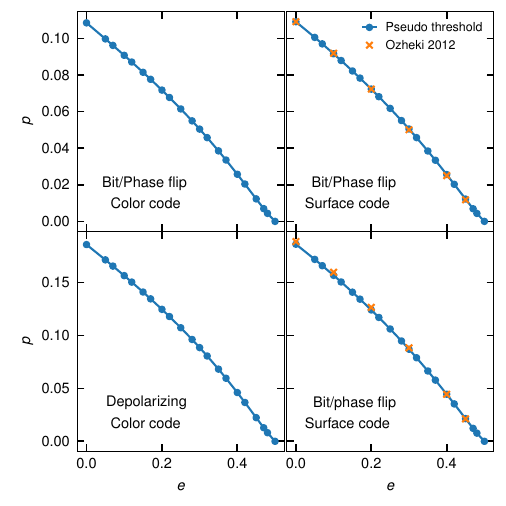}
    \caption{Phase boundary in the two-dimensional plane $(e,p)$ for quantum erasure occurring with probability $e$ and computational errors with probability $p$. The phase boundary has been extracted as the pseudo-thresholds of the $d=3$ surface code and 4.8.8 color code, respectively. For comparison, points marked by orange crosses are thresholds values obtained in Ref.~\cite{ohzeki_error_2012} for the toric code in the thermodynamic limit. The point $(0.5,0.0)$ is extracted from the data shown in Fig.~\ref{fig:topo_erasure}.}
    \label{fig:phase_diagram}
\end{figure}

\begin{figure}
    \centering
    \includegraphics[width=1.0\columnwidth]{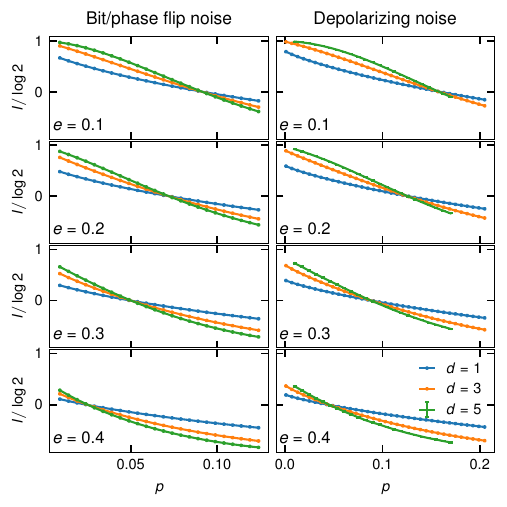}
    \caption{Coherent information of $d=3,5$ of the octagonal 4.8.8 color code and, for reference, a single physical qubit ($d=1$) for erasure probabilities $e=0.1,0.2,0.3,0.4$ as a function of phase/bit flip (left column) and depolarizing (right column) error probability $p$. For $d=5$ and depolarizing noise there are (very small) error bars due to sampling of erasure errors, see Appendix~\ref{appendix:ci_erasure}. }
    \label{fig:color_code_both}
\end{figure}

\begin{table}
\begin{center}
\begin{tabular}{ |c|c|c|c|c| } 
 \hline
 e & BF \cite{ohzeki_error_2012} & BF (this work) & Depol. \cite{ohzeki_error_2012} & Depol. (this work)  \\ 
 \hline
 0.0 & 0.10918 & 0.10913 & 0.18852 &  0.18605 \\ 
 \hline
 0.10 & 0.09189 & 0.09162 & 0.15960 & 0.15666  \\ 
 \hline
 0.20 & 0.07233 & 0.07230 & 0.12641 & 0.12397 \\ 
 \hline
 0.30 & 0.05009 & 0.05051 & 0.08815 & 0.08691  \\
 \hline
 0.40 & 0.02500 & 0.02561 & 0.04443 & 0.04444 \\ 
 \hline
 0.45 & 0.01179 & 0.01220 & 0.02117 & 0.02140  \\ 
 \hline
\end{tabular}
\end{center}
\caption{Comparison between pseudo-threshold values of coherent information and thresholds obtained by the spin glass duality mapping of Ref.~\cite{ohzeki_error_2012} for bit/phase flip (BF) and depolarizing (Depol.) noise in the toric code. The pseudo-threshold is computed as the crossing between the $[[9,1,3]]$ surface code and the single qubit coherent information. The statistical error bars can be estimated by the grid size prior to interpolation \cite{crossings}. For BF noise we use $\Delta p = 0.002$ and for depolarizing noise $\Delta p = 0.003$.   }
\label{tab:thresholds}
\end{table}

\begin{table}
\begin{center}
\begin{tabular}{ |c|c|c|c|c| } 
 \hline
 e & $(1,3)$ BF & $(3,5)$ BF & $(1,3)$ Depol. & $(3,5)$ Depol.  \\ 
 \hline
 0.0 & 0.10853 & 0.10842 & 0.18570 & 0.18629 \\ 
 \hline
 0.10 & 0.09077 & 0.09170 & 0.15639 & 0.1589(1) \\ 
 \hline
 0.20 & 0.07177 & 0.07246 & 0.12457 & 0.1265(4)\\ 
 \hline
 0.30 & 0.05495 &  0.05498 & 0.08847 & 0.0884(7) \\ 
 \hline
 0.40 & 0.03353 & 0.03317 & 0.04603 & 0.044(2) \\ 
 \hline
\end{tabular}
\end{center}
\caption{Pseudo thresholds of 4.8.8 color code as finite-size ($d=1,3,5$) crossings of the CI for the bit/phase flip (BF) and depolarizing (Depol.) with probability $p$ for fixed erasure probability (left most column). The columns $(1,3)$ show the crossing between the CI of a single qubit and the Steane code (distance-3 color code). The columns $(3,5)$ show the CI crossing between the Steane and the $d=5$ color code. Error bars for $d=5$ and $e>0$ result from erasure sampling. The error bars for the other crossings can be estimated by the difference between consecutive $p$ values prior to interpolation \cite{crossings}. For BF noise we use $\Delta p = 0.002$ and for depolarizing noise $\Delta p = 0.003$.    }
\label{tab:thresholds_color_bit_flip}
\end{table}

First, let us state what the statistical mechanics mappings are for the 2D toric and color code.  For the 2D toric code $P^{X}_\ell = \sigma_i \sigma_j$ and $P^{Z}_\ell = \tau_n \tau_m $ where $\langle i,j\rangle$ and $\langle n,m \rangle$ are points in a square lattice connected via the edge $\ell$ (see Fig.~\ref{fig:depo_lattice_toric}). The respective Hamiltonian then reads
\begin{eqnarray}\label{eq:hamiltonian_xyz_toric}
&&H_{TC} = \sum_{\langle i,j \rangle,\langle n,m \rangle } \eta^x_{ij}\left(J_x-\dfrac{J_1}{2}\right)\dfrac{\sigma_i\sigma_j}{2}+ \nonumber \\
&& \eta^z_{nm}\left(J_z-\dfrac{J_1}{2}\right) \dfrac{\tau_n\tau_m}{2}+\eta^z_{nm}\eta^x_{ij}J_1 \dfrac{\sigma_i\sigma_j\tau_n\tau_m}{4}.
\end{eqnarray}
For 2D color codes $P^{X}_\ell = \sigma_i \sigma_j \sigma_k$ and $P^{Z}_\ell = \tau_n \tau_m \tau_o $ where $\langle i,j,k \rangle$ and $\langle n,m,o \rangle$ denote the plaquettes that meet at the vertex $\ell$, see Fig.~\ref{fig:depo_lattice_color}. The Hamiltonian then reads
\begin{eqnarray}\label{eq:hamiltonian_xyz_color}
&&H_{CC} = \sum_{\langle i,j,k \rangle,\langle n,m,o \rangle } \eta^x_{ijk}\left(J_x-\dfrac{J_1}{2}\right)\dfrac{\sigma_i\sigma_j\sigma_k}{2}+  \\
&& \eta^z_{nmo}\left(J_z-\dfrac{J_1}{2}\right) \dfrac{\tau_n\tau_m\tau_o}{2}+\eta^z_{nmo}\eta^x_{ijk}J_1 \dfrac{\sigma_i\sigma_j\sigma_k\tau_n\tau_m\tau_o}{4}. \nonumber
\end{eqnarray}

Let us note that in Eqs.~\eqref{eq:hamiltonian_xyz_toric} and \eqref{eq:hamiltonian_xyz_color}, we assume periodic boundary conditions. Imposing open boundary conditions would only modify the physical qubits at the boundary, as they are shared among a reduced number of stabilizers. For example, in the rotated surface code, $P^{X}_\ell = \sigma_j$ when the qubit $\ell$ participates in only one $X$ stabilizer generator.

The probability distribution $P(\{\eta^x, \eta^z\})$ determines the presence or absence of erasure errors. Table~\ref{table:chain_error} shows the probabilities for computational errors only, while Table~\ref{table:chain_error_erasure} includes the effect of  erasures at known positions in the data qubits.
Among all possible choices of $p_x$, $p_y$, and $p_z$, we focus on the two most studied cases: (i) symmetric uncorrelated bit and phase flip noise, where $p_x = p_z = p_1(1 - p_1)$ and $p_y = p_1^2$, yielding $J_1 = 0$ and recovering two identical independent Hamiltonians for $X$ and $Z$ stabilizers. For the 2D toric code, these are two decoupled \emph{diluted} RBIMs living on shifted lattices, see Fig.~\ref{fig:depo_lattice_toric}. For 2D color codes, we obtain two decoupled three-body spin model living on the same lattice positions \cite{dennis_topological_2002, katzgraber_error_2009}. 
(ii) For the isotropic depolarizing channel, where $p_x = p_z = p_y = p/3$, it holds that $e^{J_{x,z}} = 3(1 - p) / p$ and $e^{J_1} = 3(1 - p) / p$, so all three interacting terms have the same strength.
For the toric code, this model is known as the eight-vertex model and has already been studied in the context of QEC and computational errors \cite{bombin_strong_2012}. In Ref.~\cite{ohzeki_error_2012}, the diluted RBIM and eight-vertex model are studied in the presence of both computational and erasure errors using a spin-glass duality technique. 

It is important to note that in Ref.~\cite{stace_thresholds_2009} the authors show a mapping for the 2D toric code under bit-flip noise and erasure errors in which the probability of ferromagnetic and anti-ferromagnetic links is renormalized by the presence of the erasures. This statistical mechanics model yields the same optimal threshold as the diluted RBIM, according to Ref.~\cite{ohzeki_error_2012}. However, to the best of our knowledge, the two models are not directly related.
At this point, we argue that the CI allows us to unambiguously derive the statistical mechanics mappings with the certainty that they must be correct as long as they faithfully reproduce the CI of the noisy code density matrix. In this sense, we propose the CI as a means to validate the statistical mechanics mappings derived using the approach in Refs.~\cite{dennis_topological_2002, ohzeki_error_2012, stace_error_2010}, and we conclude that the diluted RBIM ~faithfully captures the optimal decoding phase transition in the 2D toric code under bit-flip noise and erasure errors for arbitrary code sizes. 
It is worth to clarify that the latter does not exclude that some statistical mechanics models, which are not directly related to the one derived from the CI, can represent the optimal decoding phase transition \emph{only} in the thermodynamic limit. In other words, the advantage in deriving the statistical mechanics models from the CI is the built-in guarantee in representing the exact optimal decoding problem for arbitrary code sizes.

Next we compute the CI, by either diagonalizing the mixed-state density matrices or evaluating Eq.~\eqref{eq:average_ci_comp_loss}. 
In Fig.~\ref{fig:phase_diagram} we show thresholds in the plane $(e,p)$ for surface and 4.8.8 color code as crossings between the CI for $d=1$ (a single physical qubit) and the $d=3$ code. In the surface code case, the pseudo-thresholds we find are the same as the optimal thresholds from \cite{ohzeki_error_2012} within 2-4 digit precision, see Table~\ref{tab:thresholds}. This suggests that the CI is able to capture the asymptotic behavior of the code with high accuracy in small code instances. For the color code, due to the slightly smaller qubit numbers, we are able to compute the CI for $d=5$ as well and compare the crossings between $d=1$ and $d=3$ with crossings between $d=3$ and $d=5$ codes, see Fig.~\ref{fig:color_code_both} and Table~\ref{tab:thresholds_color_bit_flip}. 
We find in all cases examined and for both depolarizing and bit/phase flip noise that increasing the code distance from $d=3$ to $d=5$ only affects the threshold in the third digit. Therefore we are confident that the small-distance codes deliver accurate estimates for the thresholds when seen as crossings of the CI. This robustness against finite size effects is similar to the one observed in Ref.~\cite{ohzeki_error_2012} when increasing the size of the cluster treated with the duality equivalence.

\section{Lift-connected surface code}\label{sec:lcs}

\begin{figure}
    \centering
    \includegraphics[width=1.0\columnwidth]{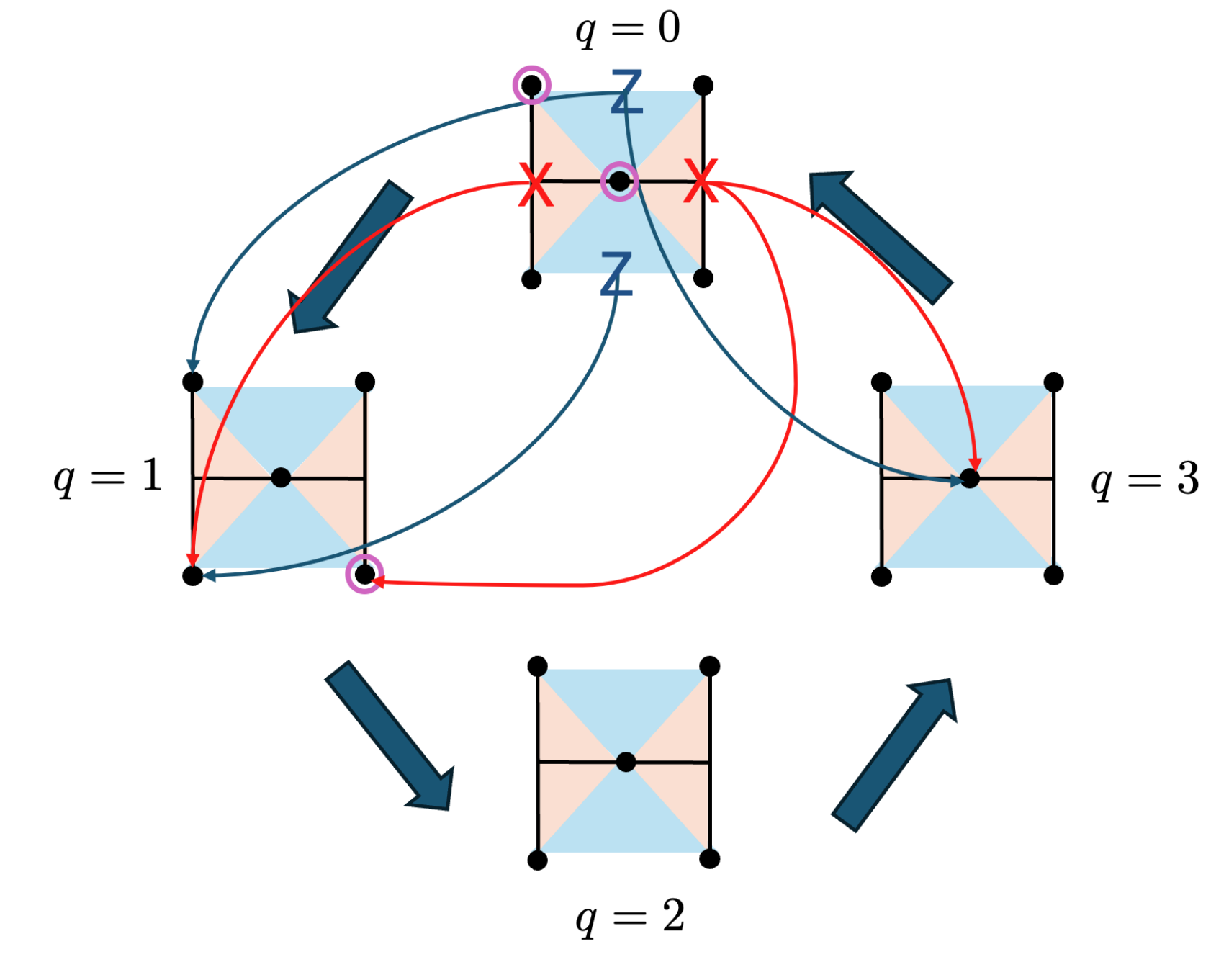}
    \caption{Lift-connected surface (LCS) code with parameters $L=4$ and $\ell=1$. The stabilizers associated to the $\ell+1$ surface code sheet $q=0$ are depicted. There are four stabilizers per surface code sheet. Each $X$ ($Z$) stabilizer is constructed by taking the qubits of the respective $\ell+1$ surface code plus the left (right) qubit on the $q-1$ sheet and the lower (upper) qubit on the $q+1$ sheet. The support of $X_L$ and $Z_L$ logical operators associated to the $q=0$ surface code sheet is shown highlighted in purple. The other stabilizers and logical operators can be visualized by rotating the indices $q \rightarrow (q+1)\mod L$.}
    \label{fig:lcs_code_figure}
\end{figure}

\begin{figure}
    \centering
    \includegraphics[width=1.0\columnwidth]{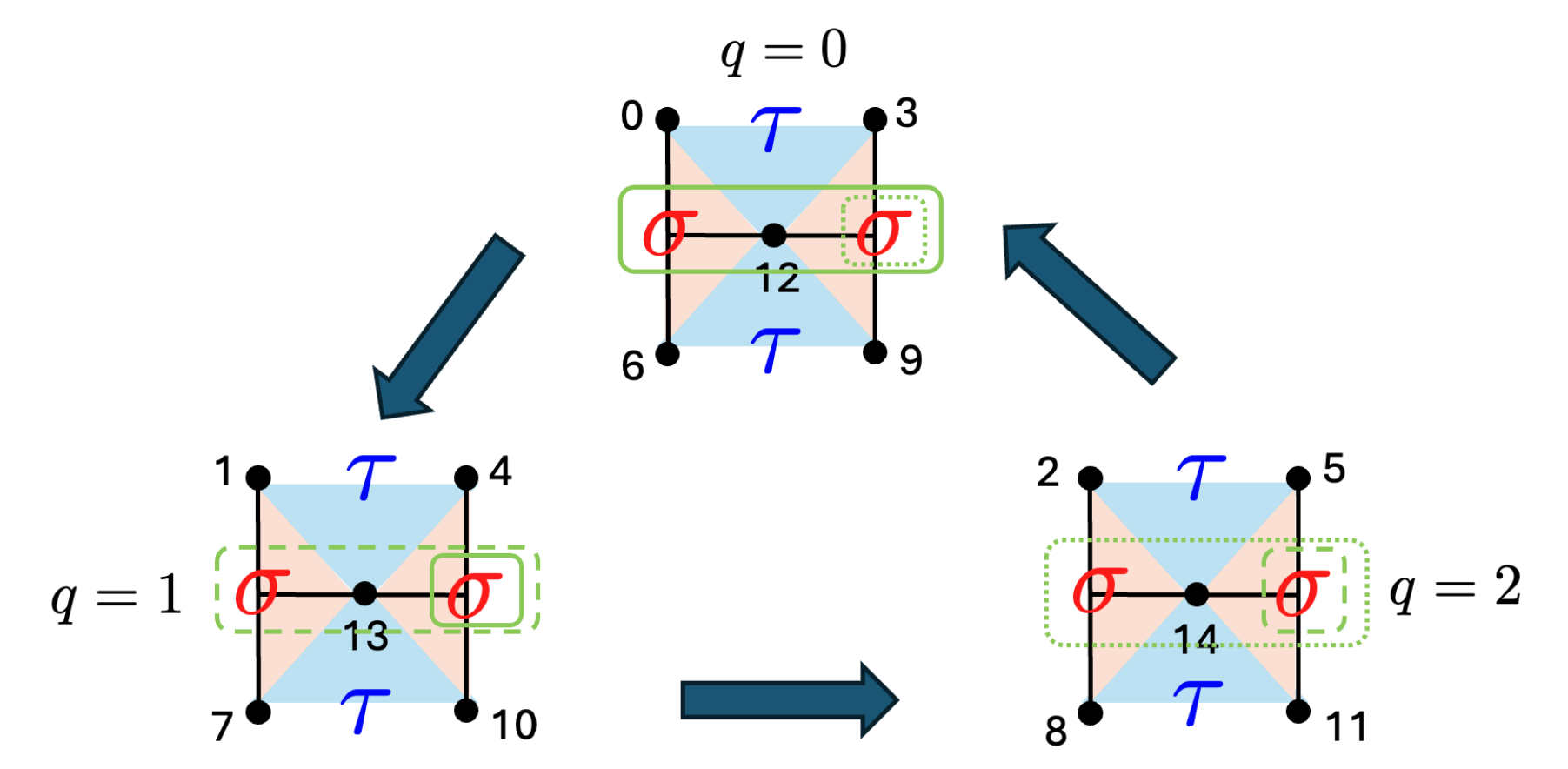}
    \caption{Depiction of $\sigma$ and $\tau$ spins and their location with respect to the qubits for the [[15,3,3]] lift-connected surface code. There are two $\sigma$ and $\tau$ spins on each surface code sheet. The weight-three spin interactions for the $\sigma$ spins are shown as green continuous, dashed and dotted lines. The third spin always comes from the $q+1$ surface code sheet on the right. See Appendix \ref{sec:lcs_appendix} for more details on the spin model.} 
    \label{fig:lcs_d3}
\end{figure}

In order to test our method beyond the realm of topological codes, we now study the lift-connected surface (LCS) code introduced in Ref.~\cite{old_lift-connected_2024}. To construct the code family, one takes two repetition codes of size $\ell+1$ and applies the lift-product~\cite{panteleev_degenerate_2021, panteleev_quantum_2022} with circulant matrices of size $L$. As explained in Ref.~\cite{old_lift-connected_2024}, this procedure yields an LDPC code $[[n,k,d]]$ with parameters

\begin{eqnarray}\label{eq:lcs}
   && n = [(\ell+1)^2+\ell^2] L \nonumber \\
   && k = L \nonumber \\
   && d = \min(L,2\ell+1).
\end{eqnarray}

This code is more easily interpreted as $L$ unrotated surface code sheets of linear size $\ell+1$ arranged in a periodic stack; see Fig.~\ref{fig:lcs_code_figure}. Each stabilizer of the standard surface code is now extended over neighboring surface code sheets, leading to weight-six $X$ and $Z$ stabilizers in the bulk and weight-four stabilizers at the boundary. The connectivity needed for measuring stabilizers is thus 3D-local. The logical operators are extended over multiple surface code sheets, and their minimum-weight representatives lie along the diagonals of each surface code sheet. Let us note that, since the logical operators are not defined on the boundaries, these are not topological codes in the same sense as, for example, 3D surface codes.

The two parameters $(L, \ell)$ are independent; therefore, it is convenient to choose a line in the $(L, \ell)$ plane to study thermodynamic properties. Here, we choose $L = 2\ell + 1$, which yields an ever-growing code distance $d$ and the minimum number of physical qubits for a fixed distance~\cite{old_lift-connected_2024}. Other choices that keep either $L$ or $\ell$ constant result in a bounded code distance $d$ (see Eq.~\eqref{eq:lcs}) which prevents the development of a finite threshold. The downside of the family $L = 2\ell + 1$ is its vanishing rate, $R = k/n = [(\ell+1)^2 + \ell^2]^{-1}$, as $\ell$ grows. We leave the study of finite-rate LDPC codes for future works.

In the remainder of this section, we present numerical calculations of the coherent information (CI) for the LCS code under two noise models: erasure errors only, and a combination of computational and erasure errors. For the former, we provide results up to $d = 17$, whereas for the latter, we are limited to $d = 3$ and $d = 1$. 
We also present, for the first time, the exact statistical mechanics mappings for LCS codes under bit/phase-flip and depolarizing noise, along with their direct extensions to include erasure errors. These mappings are used to compute the CI exactly in the case of combined bit/phase-flip and erasure errors. For depolarizing noise, we sample over erasure errors while treating the computational errors exactly.

\subsection{Erasure errors}

\begin{figure}
    \centering
    \includegraphics[width=1.0\columnwidth]{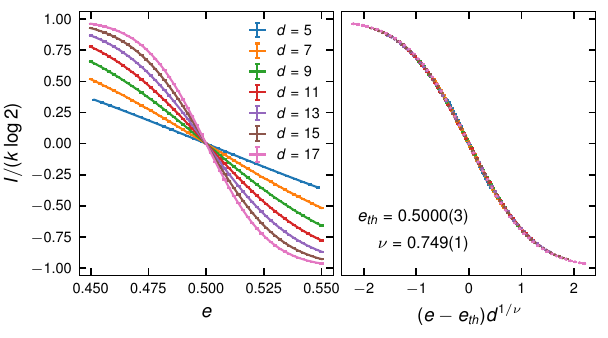}
    \caption{Coherent information of lift-connected surface code under erasure errors. We obtain a threshold of $e_{th}=0.5$ and a critical exponent $\nu\approx 0.75$, which  belongs  neither to the 3D nor the 2D percolation universality class \cite{stauffer_introduction_1992}.}
    \label{fig:lcs_erasure}
\end{figure}

In Fig.~\ref{fig:lcs_erasure}, we show the CI for the LCS code under erasure errors. We find that the optimal threshold is located at $e_{\text{th}} = 0.5$, which saturates the bound imposed by the no-cloning theorem~\cite{bennett_capacities_1997}. The exponent $\nu \approx 0.75$ does not match the exponents of 2D or 3D percolation~\cite{stauffer_introduction_1992}, reflecting the more complex nature of the stabilizers and logical operators compared to those in 2D and 3D topological codes.

\subsection{Computational and erasure errors}

\begin{figure}
    \centering
    \includegraphics[width=1.0\columnwidth]{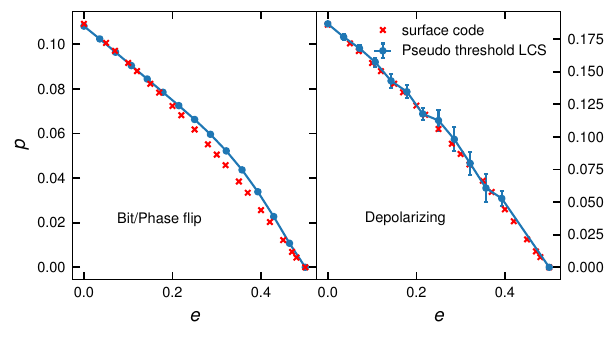}
    \caption{Pseudo-threshold, as crossing of the CI between $d=3$ and $d=1$ (single qubit) for various erasure probabilities $e$ and computational error probabilities $p$. In blue and red we show the pseudothreshold of the LCS surface code and the standard surface code, respectively. Both codes have very similar pseudo-thresholds under the same error processes.}
    \label{fig:lcs_phase_diagram}
\end{figure}

Let us now introduce the statistical mechanics mappings for the LCS codes. First, we observe that in the bulk, the stabilizers take the form $P^{X}_\ell = \sigma_{i,q} \sigma_{j,q} \sigma_{i,q+1}$ and $P^{Z}_\ell = \tau_{n,q} \tau_{m,q} \tau_{m,q+1}$, where the indices $(i,q)$ denote the 2D coordinates on the $q$-th surface code sheet. In other words, the two-spin terms within each surface code sheet are now lifted to three-spin interactions. The newly added spins are always at the same position $i$, but on neighboring surface code sheets. These three-spin terms are depicted in Fig.~\ref{fig:lcs_d3} for the example of a $[[15,3,3]]$ LCS code. The Hamiltonian is then written as

\begin{eqnarray}\label{eq:hamiltonian_lcs}
&&H_{LCS} = \sum_{\langle i,j \rangle,\langle n,m \rangle } \eta^x_{ij}\left(J_x-\dfrac{J_1}{2}\right)\dfrac{\sigma_{i,q}\sigma_{j,q}\sigma_{i,q+1}}{2} \nonumber \\
&& +\eta^z_{nm}\left(J_z-\dfrac{J_1}{2}\right) 
\dfrac{\tau_{n,q}\tau_{m,q}\tau_{m,q+1}}{2} \nonumber \\
&& +\eta^z_{nm}\eta^x_{ij}J_1 \dfrac{\sigma_{i,q}\sigma_{j,q}\sigma_{i,q+1}\tau_{n,q}\tau_{m,q}\tau_{m,q+1}}{4}.
\end{eqnarray}

Importantly, for the qubits located at the boundaries, some terms in the Hamiltonian will involve one- and two-spin interactions; see Appendix~\ref{sec:lcs_appendix} for the exact description of the $[[15,3,3]]$ Hamiltonian. To the best of our knowledge, this family of statistical mechanics models has not been studied in the literature. Notably, Eq.~\eqref{eq:hamiltonian_lcs} holds for any choice of $(L, \ell)$ and is therefore not limited to our case study of $L = 2\ell + 1$.

In Fig.~\ref{fig:lcs_phase_diagram}, we show a phase diagram for finite erasure and computational error probabilities (bit/phase-flip and depolarizing noise), defined by the crossing of the $d = 3$ and $d = 1$ CI
curves. Remarkably, the pseudo-thresholds of the LCS code are very close to those of the rotated surface code studied in Sec.~\ref{sec:results_topo_codes}. 
Together with the $50\%$ threshold found for erasures only, we note that the LCS code, in the code capacity setting, inherits most of the error correction capabilities of the 2D standard surface code, despite having a growing number of logical qubits.

\section{Conclusions}\label{sec:conclusions}

In this work, we have presented a framework for treating computational and erasure errors on the same footing in an exact manner. The method is based on the coherent information (CI) of the mixed-state density matrix. We have shown how to derive closed expressions for the CI of CSS codes under erasure and computational errors as families of statistical mechanics mappings. In general, erasure errors are introduced by removing links at the erasure positions. The CI shows two distinct contributions: one coming from the erased logical information, in the form of logical qubits degraded to logical bits—i.e., able to transmit only classical information—and lost logical qubits.
The second contribution arises from the interplay between computational and erasure errors, reflecting the optimal decoding problem restricted to the degrees of freedom unaffected by the erasures. 

Furthermore, we have successfully applied the proposed framework to 2D topological toric and color codes and developed three insights that are particularly worth to be pointed out: First, when considering erasure errors only, computing the CI for a single erasure configuration can be done efficiently (see Appendix~\ref{appendix:ci_erasure} for details). Therefore, most of the work then amounts to finding an effective way to sample erasure configurations. The same applies to any stabilizer QEC code with a constant number of logical qubits (i.e.~the number of encoded logical qubits does not grow with the number of physical qubits
), so we envision the CI as a practical, rigorous, and scalable tool for computing optimal thresholds of QEC stabilizer codes under erasure errors.

Second, we have observed that the finite crossings of the CI we obtain coincide within 2-4 digits of precision with the values previously reported in the literature as the optimal threshold. This is compatible with previous observations in Ref.~\cite{colmenarez_accurate_2024} for the same codes, where in previous work, however, only computational errors have been considered. Similarly, the 2D bond percolation \cite{stace_thresholds_2009} and the diluted RBIM \cite{ohzeki_error_2012} show the same thresholds as the crossings of the CI in the surface code.
We therefore argue that the crossings of the CI in small codes serve as accurate estimates of optimal thresholds, which is particularly useful when optimal thresholds are not known (as is, e.g., the case for 2D color codes with both erasure and computational errors). This claim is further supported by the recent exact equivalence found between the CI of the noisy toric code and the topological order parameter of Majorana fermions \cite{huang_coherent_2024}. Therefore, the robustness of the CI is related to the robustness of topological order parameters against disorder.

Third, we rigorously show how erasure errors modify the statistical mechanics mappings associated with computational errors. The key insight is that qubit erasures can be regarded as fully depolarizing channels acting at known positions. Hence, the mappings derived for computational errors are extended by adding an average over missing bonds, in addition to the bond disorder produced by the computational errors. This intuition also applies to mappings in the stabilizer configuration picture \cite{fan_diagnostics_2024, hauser_information_2024, lyons_understanding_2024}. There, a fully depolarizing channel translates to an infinite coupling, locking neighboring spins rigidly together (see Appendix \ref{appendix:stab_conf} for details).
Similarly, one can use the CI to establish connections between families of statistical mechanics models and to validate existing QEC mappings that lack closed-form expressions \cite{dennis_topological_2002, stace_error_2010, stace_thresholds_2009}.

We also showcase the generality of the CI approach by studying optimal thresholds for a low-density parity-check (LDPC) code: the lift-connected surface code~\cite{old_lift-connected_2024}. This code has a growing number of logical qubits, making it the first code with this feature to be systematically studied using CI. 
Under erasure errors only, we find a $50\%$ optimal threshold and a critical exponent that, so far, has not been related to any known percolation transition. Furthermore, we derive a family of statistical mechanics mappings describing the optimal decoding problem for this code and identify a 3D-local model exhibiting chiral spin interactions. 
An in-depth study of the statistical physics models arising from this code, potentially involving larger system sizes, is left for future work.
Finally, we compute the CI numerically and find pseudo-thresholds under combined computational and erasure errors that are comparable to those of the standard surface code. This indicates that the lift-connected surface code offers high levels of noise protection, despite having a growing number of logical qubits.

There are several routes to further explore erasure errors and the CI-based methodology to determine fundamental thresholds of QEC codes. In the absence of computational errors, one can now estimate optimal thresholds very reliably, making it feasible to extend this approach beyond 2D topological codes to higher dimensional codes or LDPC codes with finite-high rate. In both cases, erasure errors have not been extensively studied, and the equivalence of a QEC code graph and associated percolation picture is not always exact \cite{delfosse_upper_2013, kovalev_fault_2013, woolls_homology-changing_2022}.

In addition, one could study how the mappings are modified by coherent and/or correlated errors. A point of great interest is the exploration of new families of statistical mechanics models that can be obtained from LDPC codes with different features (e.g., finite rate), potentially uncovering new physical phenomena arising from exotic QEC codes.
Beyond the QEC perspective, mappings derived from quantum LDPC codes may also lead to new classes of ordered states and topological phenomena beyond the paradigm of Euclidean space~\cite{placke_topological_2024, roeck_ldpc_2024}.

\begin{acknowledgments}
We thank Davide Vodola for useful discussions. We thank Josias Old for fruitful discussions about the lift-connected surface code. L.C. and M.M. gratefully acknowledge funding by the U.S. ARO Grant No. W911NF-21-1-0007. M.M. furthermore acknowledges funding from the European Union’s Horizon Europe research and innovation programme under grant agreement No 101114305 (“MILLENION-SGA1” EU Project), and this research is also part of the Munich Quantum Valley (K-8), which is supported by the Bavarian state government with funds from the Hightech Agenda Bayern Plus. M.M. acknowledges funding from
the ERC Starting Grant QNets through Grant No. 804247, and from the European Union’s Horizon Europe research and innovation program under Grant Agreement No. 101046968 (BRISQ). 
L.C.~and M.M. also acknowledge support for the research that was sponsored by IARPA and the Army Research Office, under the Entangled Logical Qubits program, and was accomplished under Cooperative Agreement Number W911NF-23-2-0216. The views and conclusions contained in this document are those of the authors and should not be interpreted as representing the official policies, either expressed or implied, of IARPA, the Army Research Office, or the U.S. Government. The U.S. Government is authorized to reproduce and distribute reprints for Government purposes notwithstanding any copyright notation herein. M.M. acknowledges support from the Deutsche Forschungsgemeinschaft (DFG, German Research Foundation) under  Germany’s Excellence Strategy Cluster of Excellence Matter and Light for  Quantum Computing (ML4Q) EXC 2004/1 390534769.
S.K. is supported by the National Research Foundation of Korea under grant NRF-2021R1A2C1092701 funded by the Korean government (MEST) and by the Institute of Information \& Communication Technology Planning \& Evaluation grant funded by the Korean government (Ministry of Science and ICT) (IITP-2024-RS-2024-00437191). SK would like to thank IQI (RWTH Aachen University) and the Forschungszentrum J\"ulich for the support for the visit to the IQI in the initial phase of this work. The
authors gratefully acknowledge the computing time provided to them at the NHR Center NHR4CES at RWTH
Aachen University (Project No. p0020074). This is
funded by the Federal Ministry of Education and Research and the state governments participating on the
basis of the resolutions of the GWK for national high
performance computing at universities.

We used the python libraries LDPC \cite{Roffe_LDPC_Python_tools_2022}, NumPy \cite{harris2020array} and SciPy \cite{2020SciPy-NMeth}. 
\end{acknowledgments}

\bibliography{references,footnote}

\begin{thebibliography}{114}%
\makeatletter
\providecommand \@ifxundefined [1]{%
 \@ifx{#1\undefined}
}%
\providecommand \@ifnum [1]{%
 \ifnum #1\expandafter \@firstoftwo
 \else \expandafter \@secondoftwo
 \fi
}%
\providecommand \@ifx [1]{%
 \ifx #1\expandafter \@firstoftwo
 \else \expandafter \@secondoftwo
 \fi
}%
\providecommand \natexlab [1]{#1}%
\providecommand \enquote  [1]{``#1''}%
\providecommand \bibnamefont  [1]{#1}%
\providecommand \bibfnamefont [1]{#1}%
\providecommand \citenamefont [1]{#1}%
\providecommand \href@noop [0]{\@secondoftwo}%
\providecommand \href [0]{\begingroup \@sanitize@url \@href}%
\providecommand \@href[1]{\@@startlink{#1}\@@href}%
\providecommand \@@href[1]{\endgroup#1\@@endlink}%
\providecommand \@sanitize@url [0]{\catcode `\\12\catcode `\$12\catcode
  `\&12\catcode `\#12\catcode `\^12\catcode `\_12\catcode `\%12\relax}%
\providecommand \@@startlink[1]{}%
\providecommand \@@endlink[0]{}%
\providecommand \url  [0]{\begingroup\@sanitize@url \@url }%
\providecommand \@url [1]{\endgroup\@href {#1}{\urlprefix }}%
\providecommand \urlprefix  [0]{URL }%
\providecommand \Eprint [0]{\href }%
\providecommand \doibase [0]{http://dx.doi.org/}%
\providecommand \selectlanguage [0]{\@gobble}%
\providecommand \bibinfo  [0]{\@secondoftwo}%
\providecommand \bibfield  [0]{\@secondoftwo}%
\providecommand \translation [1]{[#1]}%
\providecommand \BibitemOpen [0]{}%
\providecommand \bibitemStop [0]{}%
\providecommand \bibitemNoStop [0]{.\EOS\space}%
\providecommand \EOS [0]{\spacefactor3000\relax}%
\providecommand \BibitemShut  [1]{\csname bibitem#1\endcsname}%
\let\auto@bib@innerbib\@empty
\bibitem [{\citenamefont {Lidar}\ and\ \citenamefont
  {Brun}(2013)}]{lidar_quantum_2013}%
  \BibitemOpen
  \bibinfo {editor} {\bibfnamefont {Daniel~A.}\ \bibnamefont {Lidar}}\ and\
  \bibinfo {editor} {\bibfnamefont {Todd~A.}\ \bibnamefont {Brun}},\ eds.,\
  \href
  {https://www.cambridge.org/core/books/quantum-error-correction/B51E8333050A0F9A67363254DC1EA15A}
  {\emph {\bibinfo {title} {Quantum {Error} {Correction}}}}\ (\bibinfo
  {publisher} {Cambridge University Press},\ \bibinfo {address} {Cambridge},\
  \bibinfo {year} {2013})\BibitemShut {NoStop}%
\bibitem [{\citenamefont {Terhal}(2015)}]{terhal_quantum_2015}%
  \BibitemOpen
  \bibfield  {author} {\bibinfo {author} {\bibfnamefont {Barbara~M.}\
  \bibnamefont {Terhal}},\ }\bibfield  {title} {\enquote {\bibinfo {title}
  {Quantum error correction for quantum memories},}\ }\href {\doibase
  10.1103/RevModPhys.87.307} {\bibfield  {journal} {\bibinfo  {journal}
  {Reviews of Modern Physics}\ }\textbf {\bibinfo {volume} {87}},\ \bibinfo
  {pages} {307--346} (\bibinfo {year} {2015})}\BibitemShut {NoStop}%
\bibitem [{\citenamefont {Aharonov}\ and\ \citenamefont
  {Ben-Or}(2008)}]{aharonov_fault-tolerant_2008}%
  \BibitemOpen
  \bibfield  {author} {\bibinfo {author} {\bibfnamefont {Dorit}\ \bibnamefont
  {Aharonov}}\ and\ \bibinfo {author} {\bibfnamefont {Michael}\ \bibnamefont
  {Ben-Or}},\ }\bibfield  {title} {\enquote {\bibinfo {title} {Fault-{Tolerant}
  {Quantum} {Computation} with {Constant} {Error} {Rate}},}\ }\href {\doibase
  10.1137/S0097539799359385} {\bibfield  {journal} {\bibinfo  {journal} {SIAM
  Journal on Computing}\ }\textbf {\bibinfo {volume} {38}},\ \bibinfo {pages}
  {1207--1282} (\bibinfo {year} {2008})}\BibitemShut {NoStop}%
\bibitem [{\citenamefont {Knill}\ \emph {et~al.}(1998)\citenamefont {Knill},
  \citenamefont {Laflamme},\ and\ \citenamefont
  {Zurek}}]{knill_resilient_1998}%
  \BibitemOpen
  \bibfield  {author} {\bibinfo {author} {\bibfnamefont {Emanuel}\ \bibnamefont
  {Knill}}, \bibinfo {author} {\bibfnamefont {Raymond}\ \bibnamefont
  {Laflamme}}, \ and\ \bibinfo {author} {\bibfnamefont {Wojciech~H.}\
  \bibnamefont {Zurek}},\ }\bibfield  {title} {\enquote {\bibinfo {title}
  {Resilient quantum computation: error models and thresholds},}\ }\href
  {\doibase 10.1098/rspa.1998.0166} {\bibfield  {journal} {\bibinfo  {journal}
  {Proceedings of the Royal Society of London. Series A: Mathematical, Physical
  and Engineering Sciences}\ }\textbf {\bibinfo {volume} {454}},\ \bibinfo
  {pages} {365--384} (\bibinfo {year} {1998})}\BibitemShut {NoStop}%
\bibitem [{\citenamefont {Kitaev}(2003)}]{kitaev_fault-tolerant_2003}%
  \BibitemOpen
  \bibfield  {author} {\bibinfo {author} {\bibfnamefont {A.~Yu.}\ \bibnamefont
  {Kitaev}},\ }\bibfield  {title} {\enquote {\bibinfo {title} {Fault-tolerant
  quantum computation by anyons},}\ }\href {\doibase
  10.1016/S0003-4916(02)00018-0} {\bibfield  {journal} {\bibinfo  {journal}
  {Annals of Physics}\ }\textbf {\bibinfo {volume} {303}},\ \bibinfo {pages}
  {2--30} (\bibinfo {year} {2003})}\BibitemShut {NoStop}%
\bibitem [{\citenamefont {Shor}(1996)}]{shor_fault-tolerant_1996}%
  \BibitemOpen
  \bibfield  {author} {\bibinfo {author} {\bibfnamefont {P.W.}\ \bibnamefont
  {Shor}},\ }\bibfield  {title} {\enquote {\bibinfo {title} {Fault-tolerant
  quantum computation},}\ }in\ \href {\doibase 10.1109/SFCS.1996.548464} {\emph
  {\bibinfo {booktitle} {Proceedings of 37th {Conference} on {Foundations} of
  {Computer} {Science}}}}\ (\bibinfo {year} {1996})\ pp.\ \bibinfo {pages}
  {56--65}\BibitemShut {NoStop}%
\bibitem [{\citenamefont {AI}(2024)}]{google_quantum_ai_quantum_2024}%
  \BibitemOpen
  \bibfield  {author} {\bibinfo {author} {\bibfnamefont {Google~Quantum}\
  \bibnamefont {AI}},\ }\href {\doibase 10.48550/arXiv.2408.13687} {\enquote
  {\bibinfo {title} {Quantum error correction below the surface code
  threshold},}\ } (\bibinfo {year} {2024}),\ \bibinfo {note}
  {arXiv:2408.13687}\BibitemShut {NoStop}%
\bibitem [{\citenamefont {{Google Quantum
  AI}}(2023)}]{google_quantum_ai_suppressing_2023}%
  \BibitemOpen
  \bibfield  {author} {\bibinfo {author} {\bibnamefont {{Google Quantum AI}}},\
  }\bibfield  {title} {\enquote {\bibinfo {title} {Suppressing quantum errors
  by scaling a surface code logical qubit},}\ }\href {\doibase
  10.1038/s41586-022-05434-1} {\bibfield  {journal} {\bibinfo  {journal}
  {Nature}\ }\textbf {\bibinfo {volume} {614}},\ \bibinfo {pages} {676--681}
  (\bibinfo {year} {2023})}\BibitemShut {NoStop}%
\bibitem [{\citenamefont {{Google Quantum
  AI}}(2021)}]{google_quantum_ai_exponential_2021}%
  \BibitemOpen
  \bibfield  {author} {\bibinfo {author} {\bibnamefont {{Google Quantum AI}}},\
  }\bibfield  {title} {\enquote {\bibinfo {title} {Exponential suppression of
  bit or phase errors with cyclic error correction},}\ }\href {\doibase
  10.1038/s41586-021-03588-y} {\bibfield  {journal} {\bibinfo  {journal}
  {Nature}\ }\textbf {\bibinfo {volume} {595}},\ \bibinfo {pages} {383--387}
  (\bibinfo {year} {2021})}\BibitemShut {NoStop}%
\bibitem [{\citenamefont {Krinner}\ \emph {et~al.}(2022)\citenamefont
  {Krinner}, \citenamefont {Lacroix}, \citenamefont {Remm}, \citenamefont
  {Di~Paolo}, \citenamefont {Genois}, \citenamefont {Leroux}, \citenamefont
  {Hellings}, \citenamefont {Lazar}, \citenamefont {Swiadek}, \citenamefont
  {Herrmann}, \citenamefont {Norris}, \citenamefont {Andersen}, \citenamefont
  {Müller}, \citenamefont {Blais}, \citenamefont {Eichler},\ and\
  \citenamefont {Wallraff}}]{krinner_realizing_2022}%
  \BibitemOpen
  \bibfield  {author} {\bibinfo {author} {\bibfnamefont {Sebastian}\
  \bibnamefont {Krinner}}, \bibinfo {author} {\bibfnamefont {Nathan}\
  \bibnamefont {Lacroix}}, \bibinfo {author} {\bibfnamefont {Ants}\
  \bibnamefont {Remm}}, \bibinfo {author} {\bibfnamefont {Agustin}\
  \bibnamefont {Di~Paolo}}, \bibinfo {author} {\bibfnamefont {Elie}\
  \bibnamefont {Genois}}, \bibinfo {author} {\bibfnamefont {Catherine}\
  \bibnamefont {Leroux}}, \bibinfo {author} {\bibfnamefont {Christoph}\
  \bibnamefont {Hellings}}, \bibinfo {author} {\bibfnamefont {Stefania}\
  \bibnamefont {Lazar}}, \bibinfo {author} {\bibfnamefont {Francois}\
  \bibnamefont {Swiadek}}, \bibinfo {author} {\bibfnamefont {Johannes}\
  \bibnamefont {Herrmann}}, \bibinfo {author} {\bibfnamefont {Graham~J.}\
  \bibnamefont {Norris}}, \bibinfo {author} {\bibfnamefont
  {Christian~Kraglund}\ \bibnamefont {Andersen}}, \bibinfo {author}
  {\bibfnamefont {Markus}\ \bibnamefont {Müller}}, \bibinfo {author}
  {\bibfnamefont {Alexandre}\ \bibnamefont {Blais}}, \bibinfo {author}
  {\bibfnamefont {Christopher}\ \bibnamefont {Eichler}}, \ and\ \bibinfo
  {author} {\bibfnamefont {Andreas}\ \bibnamefont {Wallraff}},\ }\bibfield
  {title} {\enquote {\bibinfo {title} {Realizing repeated quantum error
  correction in a distance-three surface code},}\ }\href {\doibase
  10.1038/s41586-022-04566-8} {\bibfield  {journal} {\bibinfo  {journal}
  {Nature}\ }\textbf {\bibinfo {volume} {605}},\ \bibinfo {pages} {669--674}
  (\bibinfo {year} {2022})}\BibitemShut {NoStop}%
\bibitem [{\citenamefont {Zhao}\ \emph {et~al.}(2022)\citenamefont {Zhao},
  \citenamefont {Ye}, \citenamefont {Huang}, \citenamefont {Zhang},
  \citenamefont {Wu}, \citenamefont {Guan}, \citenamefont {Zhu}, \citenamefont
  {Wei}, \citenamefont {He}, \citenamefont {Cao}, \citenamefont {Chen},
  \citenamefont {Chung}, \citenamefont {Deng}, \citenamefont {Fan},
  \citenamefont {Gong}, \citenamefont {Guo}, \citenamefont {Guo}, \citenamefont
  {Han}, \citenamefont {Li}, \citenamefont {Li}, \citenamefont {Li},
  \citenamefont {Liang}, \citenamefont {Lin}, \citenamefont {Qian},
  \citenamefont {Rong}, \citenamefont {Su}, \citenamefont {Sun}, \citenamefont
  {Wang}, \citenamefont {Wu}, \citenamefont {Xu}, \citenamefont {Ying},
  \citenamefont {Yu}, \citenamefont {Zha}, \citenamefont {Zhang}, \citenamefont
  {Huo}, \citenamefont {Lu}, \citenamefont {Peng}, \citenamefont {Zhu},\ and\
  \citenamefont {Pan}}]{zhao_realization_2022}%
  \BibitemOpen
  \bibfield  {author} {\bibinfo {author} {\bibfnamefont {Youwei}\ \bibnamefont
  {Zhao}}, \bibinfo {author} {\bibfnamefont {Yangsen}\ \bibnamefont {Ye}},
  \bibinfo {author} {\bibfnamefont {He-Liang}\ \bibnamefont {Huang}}, \bibinfo
  {author} {\bibfnamefont {Yiming}\ \bibnamefont {Zhang}}, \bibinfo {author}
  {\bibfnamefont {Dachao}\ \bibnamefont {Wu}}, \bibinfo {author} {\bibfnamefont
  {Huijie}\ \bibnamefont {Guan}}, \bibinfo {author} {\bibfnamefont {Qingling}\
  \bibnamefont {Zhu}}, \bibinfo {author} {\bibfnamefont {Zuolin}\ \bibnamefont
  {Wei}}, \bibinfo {author} {\bibfnamefont {Tan}\ \bibnamefont {He}}, \bibinfo
  {author} {\bibfnamefont {Sirui}\ \bibnamefont {Cao}}, \bibinfo {author}
  {\bibfnamefont {Fusheng}\ \bibnamefont {Chen}}, \bibinfo {author}
  {\bibfnamefont {Tung-Hsun}\ \bibnamefont {Chung}}, \bibinfo {author}
  {\bibfnamefont {Hui}\ \bibnamefont {Deng}}, \bibinfo {author} {\bibfnamefont
  {Daojin}\ \bibnamefont {Fan}}, \bibinfo {author} {\bibfnamefont {Ming}\
  \bibnamefont {Gong}}, \bibinfo {author} {\bibfnamefont {Cheng}\ \bibnamefont
  {Guo}}, \bibinfo {author} {\bibfnamefont {Shaojun}\ \bibnamefont {Guo}},
  \bibinfo {author} {\bibfnamefont {Lianchen}\ \bibnamefont {Han}}, \bibinfo
  {author} {\bibfnamefont {Na}~\bibnamefont {Li}}, \bibinfo {author}
  {\bibfnamefont {Shaowei}\ \bibnamefont {Li}}, \bibinfo {author}
  {\bibfnamefont {Yuan}\ \bibnamefont {Li}}, \bibinfo {author} {\bibfnamefont
  {Futian}\ \bibnamefont {Liang}}, \bibinfo {author} {\bibfnamefont {Jin}\
  \bibnamefont {Lin}}, \bibinfo {author} {\bibfnamefont {Haoran}\ \bibnamefont
  {Qian}}, \bibinfo {author} {\bibfnamefont {Hao}\ \bibnamefont {Rong}},
  \bibinfo {author} {\bibfnamefont {Hong}\ \bibnamefont {Su}}, \bibinfo
  {author} {\bibfnamefont {Lihua}\ \bibnamefont {Sun}}, \bibinfo {author}
  {\bibfnamefont {Shiyu}\ \bibnamefont {Wang}}, \bibinfo {author}
  {\bibfnamefont {Yulin}\ \bibnamefont {Wu}}, \bibinfo {author} {\bibfnamefont
  {Yu}~\bibnamefont {Xu}}, \bibinfo {author} {\bibfnamefont {Chong}\
  \bibnamefont {Ying}}, \bibinfo {author} {\bibfnamefont {Jiale}\ \bibnamefont
  {Yu}}, \bibinfo {author} {\bibfnamefont {Chen}\ \bibnamefont {Zha}}, \bibinfo
  {author} {\bibfnamefont {Kaili}\ \bibnamefont {Zhang}}, \bibinfo {author}
  {\bibfnamefont {Yong-Heng}\ \bibnamefont {Huo}}, \bibinfo {author}
  {\bibfnamefont {Chao-Yang}\ \bibnamefont {Lu}}, \bibinfo {author}
  {\bibfnamefont {Cheng-Zhi}\ \bibnamefont {Peng}}, \bibinfo {author}
  {\bibfnamefont {Xiaobo}\ \bibnamefont {Zhu}}, \ and\ \bibinfo {author}
  {\bibfnamefont {Jian-Wei}\ \bibnamefont {Pan}},\ }\bibfield  {title}
  {\enquote {\bibinfo {title} {Realization of an {Error}-{Correcting} {Surface}
  {Code} with {Superconducting} {Qubits}},}\ }\href {\doibase
  10.1103/PhysRevLett.129.030501} {\bibfield  {journal} {\bibinfo  {journal}
  {Physical Review Letters}\ }\textbf {\bibinfo {volume} {129}},\ \bibinfo
  {pages} {030501} (\bibinfo {year} {2022})}\BibitemShut {NoStop}%
\bibitem [{\citenamefont {Andersen}\ \emph {et~al.}(2020)\citenamefont
  {Andersen}, \citenamefont {Remm}, \citenamefont {Lazar}, \citenamefont
  {Krinner}, \citenamefont {Lacroix}, \citenamefont {Norris}, \citenamefont
  {Gabureac}, \citenamefont {Eichler},\ and\ \citenamefont
  {Wallraff}}]{andersen_repeated_2020}%
  \BibitemOpen
  \bibfield  {author} {\bibinfo {author} {\bibfnamefont {Christian~Kraglund}\
  \bibnamefont {Andersen}}, \bibinfo {author} {\bibfnamefont {Ants}\
  \bibnamefont {Remm}}, \bibinfo {author} {\bibfnamefont {Stefania}\
  \bibnamefont {Lazar}}, \bibinfo {author} {\bibfnamefont {Sebastian}\
  \bibnamefont {Krinner}}, \bibinfo {author} {\bibfnamefont {Nathan}\
  \bibnamefont {Lacroix}}, \bibinfo {author} {\bibfnamefont {Graham~J.}\
  \bibnamefont {Norris}}, \bibinfo {author} {\bibfnamefont {Mihai}\
  \bibnamefont {Gabureac}}, \bibinfo {author} {\bibfnamefont {Christopher}\
  \bibnamefont {Eichler}}, \ and\ \bibinfo {author} {\bibfnamefont {Andreas}\
  \bibnamefont {Wallraff}},\ }\bibfield  {title} {\enquote {\bibinfo {title}
  {Repeated quantum error detection in a surface code},}\ }\href {\doibase
  10.1038/s41567-020-0920-y} {\bibfield  {journal} {\bibinfo  {journal} {Nature
  Physics}\ }\textbf {\bibinfo {volume} {16}},\ \bibinfo {pages} {875--880}
  (\bibinfo {year} {2020})}\BibitemShut {NoStop}%
\bibitem [{\citenamefont {Sivak}\ \emph {et~al.}(2023)\citenamefont {Sivak},
  \citenamefont {Eickbusch}, \citenamefont {Royer}, \citenamefont {Singh},
  \citenamefont {Tsioutsios}, \citenamefont {Ganjam}, \citenamefont {Miano},
  \citenamefont {Brock}, \citenamefont {Ding}, \citenamefont {Frunzio},
  \citenamefont {Girvin}, \citenamefont {Schoelkopf},\ and\ \citenamefont
  {Devoret}}]{sivak_real-time_2023}%
  \BibitemOpen
  \bibfield  {author} {\bibinfo {author} {\bibfnamefont {V.~V.}\ \bibnamefont
  {Sivak}}, \bibinfo {author} {\bibfnamefont {A.}~\bibnamefont {Eickbusch}},
  \bibinfo {author} {\bibfnamefont {B.}~\bibnamefont {Royer}}, \bibinfo
  {author} {\bibfnamefont {S.}~\bibnamefont {Singh}}, \bibinfo {author}
  {\bibfnamefont {I.}~\bibnamefont {Tsioutsios}}, \bibinfo {author}
  {\bibfnamefont {S.}~\bibnamefont {Ganjam}}, \bibinfo {author} {\bibfnamefont
  {A.}~\bibnamefont {Miano}}, \bibinfo {author} {\bibfnamefont {B.~L.}\
  \bibnamefont {Brock}}, \bibinfo {author} {\bibfnamefont {A.~Z.}\ \bibnamefont
  {Ding}}, \bibinfo {author} {\bibfnamefont {L.}~\bibnamefont {Frunzio}},
  \bibinfo {author} {\bibfnamefont {S.~M.}\ \bibnamefont {Girvin}}, \bibinfo
  {author} {\bibfnamefont {R.~J.}\ \bibnamefont {Schoelkopf}}, \ and\ \bibinfo
  {author} {\bibfnamefont {M.~H.}\ \bibnamefont {Devoret}},\ }\bibfield
  {title} {\enquote {\bibinfo {title} {Real-time quantum error correction
  beyond break-even},}\ }\href {\doibase 10.1038/s41586-023-05782-6} {\bibfield
   {journal} {\bibinfo  {journal} {Nature}\ }\textbf {\bibinfo {volume}
  {616}},\ \bibinfo {pages} {50--55} (\bibinfo {year} {2023})}\BibitemShut
  {NoStop}%
\bibitem [{\citenamefont {Gupta}\ \emph {et~al.}(2024)\citenamefont {Gupta},
  \citenamefont {Sundaresan}, \citenamefont {Alexander}, \citenamefont {Wood},
  \citenamefont {Merkel}, \citenamefont {Healy}, \citenamefont {Hillenbrand},
  \citenamefont {Jochym-O’Connor}, \citenamefont {Wootton}, \citenamefont
  {Yoder}, \citenamefont {Cross}, \citenamefont {Takita},\ and\ \citenamefont
  {Brown}}]{gupta_encoding_2024}%
  \BibitemOpen
  \bibfield  {author} {\bibinfo {author} {\bibfnamefont {Riddhi~S.}\
  \bibnamefont {Gupta}}, \bibinfo {author} {\bibfnamefont {Neereja}\
  \bibnamefont {Sundaresan}}, \bibinfo {author} {\bibfnamefont {Thomas}\
  \bibnamefont {Alexander}}, \bibinfo {author} {\bibfnamefont {Christopher~J.}\
  \bibnamefont {Wood}}, \bibinfo {author} {\bibfnamefont {Seth~T.}\
  \bibnamefont {Merkel}}, \bibinfo {author} {\bibfnamefont {Michael~B.}\
  \bibnamefont {Healy}}, \bibinfo {author} {\bibfnamefont {Marius}\
  \bibnamefont {Hillenbrand}}, \bibinfo {author} {\bibfnamefont {Tomas}\
  \bibnamefont {Jochym-O’Connor}}, \bibinfo {author} {\bibfnamefont
  {James~R.}\ \bibnamefont {Wootton}}, \bibinfo {author} {\bibfnamefont
  {Theodore~J.}\ \bibnamefont {Yoder}}, \bibinfo {author} {\bibfnamefont
  {Andrew~W.}\ \bibnamefont {Cross}}, \bibinfo {author} {\bibfnamefont {Maika}\
  \bibnamefont {Takita}}, \ and\ \bibinfo {author} {\bibfnamefont
  {Benjamin~J.}\ \bibnamefont {Brown}},\ }\bibfield  {title} {\enquote
  {\bibinfo {title} {Encoding a magic state with beyond break-even fidelity},}\
  }\href {\doibase 10.1038/s41586-023-06846-3} {\bibfield  {journal} {\bibinfo
  {journal} {Nature}\ }\textbf {\bibinfo {volume} {625}},\ \bibinfo {pages}
  {259--263} (\bibinfo {year} {2024})}\BibitemShut {NoStop}%
\bibitem [{\citenamefont {Hetényi}\ and\ \citenamefont
  {Wootton}(2024)}]{hetenyi_creating_2024}%
  \BibitemOpen
  \bibfield  {author} {\bibinfo {author} {\bibfnamefont {Bence}\ \bibnamefont
  {Hetényi}}\ and\ \bibinfo {author} {\bibfnamefont {James~R.}\ \bibnamefont
  {Wootton}},\ }\href {http://arxiv.org/abs/2404.15989} {\enquote {\bibinfo
  {title} {Creating entangled logical qubits in the heavy-hex lattice with
  topological codes},}\ } (\bibinfo {year} {2024}),\ \bibinfo {note}
  {arXiv:2404.15989}\BibitemShut {NoStop}%
\bibitem [{\citenamefont {Lacroix}\ \emph {et~al.}(2024)\citenamefont
  {Lacroix}, \citenamefont {Bourassa}, \citenamefont {Heras}, \citenamefont
  {Zhang}, \citenamefont {Bausch}, \citenamefont {Senior}, \citenamefont
  {Edlich}, \citenamefont {Shutty}, \citenamefont {Sivak}, \citenamefont
  {Bengtsson}, \citenamefont {McEwen}, \citenamefont {Higgott}, \citenamefont
  {Kafri}, \citenamefont {Claes}, \citenamefont {Morvan}, \citenamefont {Chen},
  \citenamefont {Zalcman}, \citenamefont {Madhuk}, \citenamefont {Acharya},
  \citenamefont {Beni}, \citenamefont {Aigeldinger}, \citenamefont {Alcaraz},
  \citenamefont {Andersen}, \citenamefont {Ansmann}, \citenamefont {Arute},
  \citenamefont {Arya}, \citenamefont {Asfaw}, \citenamefont {Atalaya},
  \citenamefont {Babbush}, \citenamefont {Ballard}, \citenamefont {Bardin},
  \citenamefont {Bilmes}, \citenamefont {Blackwell}, \citenamefont {Bovaird},
  \citenamefont {Bowers}, \citenamefont {Brill}, \citenamefont {Broughton},
  \citenamefont {Browne}, \citenamefont {Buchea}, \citenamefont {Buckley},
  \citenamefont {Burger}, \citenamefont {Burkett}, \citenamefont {Bushnell},
  \citenamefont {Cabrera}, \citenamefont {Campero}, \citenamefont {Chang},
  \citenamefont {Chiaro}, \citenamefont {Chih}, \citenamefont {Cleland},
  \citenamefont {Cogan}, \citenamefont {Collins}, \citenamefont {Conner},
  \citenamefont {Courtney}, \citenamefont {Crook}, \citenamefont {Curtin},
  \citenamefont {Das}, \citenamefont {Demura}, \citenamefont {Lorenzo},
  \citenamefont {Paolo}, \citenamefont {Donohoe}, \citenamefont {Drozdov},
  \citenamefont {Dunsworth}, \citenamefont {Eickbusch}, \citenamefont {Elbag},
  \citenamefont {Elzouka}, \citenamefont {Erickson}, \citenamefont {Ferreira},
  \citenamefont {Burgos}, \citenamefont {Forati}, \citenamefont {Fowler},
  \citenamefont {Foxen}, \citenamefont {Ganjam}, \citenamefont {Garcia},
  \citenamefont {Gasca}, \citenamefont {Genois}, \citenamefont {Giang},
  \citenamefont {Gilboa}, \citenamefont {Gosula}, \citenamefont {Dau},
  \citenamefont {Graumann}, \citenamefont {Greene}, \citenamefont {Gross},
  \citenamefont {Ha}, \citenamefont {Habegger}, \citenamefont {Hansen},
  \citenamefont {Harrigan}, \citenamefont {Harrington}, \citenamefont {Heslin},
  \citenamefont {Heu}, \citenamefont {Hiltermann}, \citenamefont {Hilton},
  \citenamefont {Hong}, \citenamefont {Huang}, \citenamefont {Huff},
  \citenamefont {Huggins}, \citenamefont {Jeffrey}, \citenamefont {Jiang},
  \citenamefont {Jin}, \citenamefont {Joshi}, \citenamefont {Juhas},
  \citenamefont {Kabel}, \citenamefont {Kang}, \citenamefont {Karamlou},
  \citenamefont {Kechedzhi}, \citenamefont {Khaire}, \citenamefont {Khattar},
  \citenamefont {Khezri}, \citenamefont {Kim}, \citenamefont {Klimov},
  \citenamefont {Kobrin}, \citenamefont {Korotkov}, \citenamefont {Kostritsa},
  \citenamefont {Kreikebaum}, \citenamefont {Kurilovich}, \citenamefont
  {Landhuis}, \citenamefont {Lange-Dei}, \citenamefont {Langley}, \citenamefont
  {Laptev}, \citenamefont {Lau}, \citenamefont {Ledford}, \citenamefont {Lee},
  \citenamefont {Lester}, \citenamefont {Guevel}, \citenamefont {Li},
  \citenamefont {Li}, \citenamefont {Lill}, \citenamefont {Livingston},
  \citenamefont {Locharla}, \citenamefont {Lucero}, \citenamefont {Lundahl},
  \citenamefont {Lunt}, \citenamefont {Maloney}, \citenamefont {Mandrà},
  \citenamefont {Martin}, \citenamefont {Martin}, \citenamefont {Maxfield},
  \citenamefont {McClean}, \citenamefont {Meeks}, \citenamefont {Megrant},
  \citenamefont {Miao}, \citenamefont {Molavi}, \citenamefont {Molina},
  \citenamefont {Montazeri}, \citenamefont {Movassagh}, \citenamefont {Neill},
  \citenamefont {Newman}, \citenamefont {Nguyen}, \citenamefont {Nguyen},
  \citenamefont {Ni}, \citenamefont {Niu}, \citenamefont {Oas}, \citenamefont
  {Oliver}, \citenamefont {Orosco}, \citenamefont {Ottosson}, \citenamefont
  {Pizzuto}, \citenamefont {Potter}, \citenamefont {Pritchard}, \citenamefont
  {Quintana}, \citenamefont {Ramachandran}, \citenamefont {Reagor},
  \citenamefont {Resnick}, \citenamefont {Rhodes}, \citenamefont {Roberts},
  \citenamefont {Rosenberg}, \citenamefont {Rosenfeld}, \citenamefont {Rossi},
  \citenamefont {Roushan}, \citenamefont {Sankaragomathi}, \citenamefont
  {Schurkus}, \citenamefont {Shearn}, \citenamefont {Shorter}, \citenamefont
  {Shvarts}, \citenamefont {Small}, \citenamefont {Smith}, \citenamefont
  {Springer}, \citenamefont {Sterling}, \citenamefont {Suchard}, \citenamefont
  {Szasz}, \citenamefont {Sztein}, \citenamefont {Thor}, \citenamefont
  {Tomita}, \citenamefont {Torres}, \citenamefont {Torunbalci}, \citenamefont
  {Vaishnav}, \citenamefont {Vargas}, \citenamefont {Vdovichev}, \citenamefont
  {Vidal}, \citenamefont {Heidweiller}, \citenamefont {Waltman}, \citenamefont
  {Waltz}, \citenamefont {Wang}, \citenamefont {Ware}, \citenamefont {Weidel},
  \citenamefont {White}, \citenamefont {Wong}, \citenamefont {Woo},
  \citenamefont {Woodson}, \citenamefont {Xing}, \citenamefont {Yao},
  \citenamefont {Yeh}, \citenamefont {Ying}, \citenamefont {Yoo}, \citenamefont
  {Yosri}, \citenamefont {Young}, \citenamefont {Zhang}, \citenamefont {Zhu},
  \citenamefont {Zobrist}, \citenamefont {Neven}, \citenamefont {Kohli},
  \citenamefont {Davies}, \citenamefont {Boixo}, \citenamefont {Kelly},
  \citenamefont {Jones}, \citenamefont {Gidney},\ and\ \citenamefont
  {Satzinger}}]{lacroix_scaling_2024}%
  \BibitemOpen
  \bibfield  {author} {\bibinfo {author} {\bibfnamefont {Nathan}\ \bibnamefont
  {Lacroix}}, \bibinfo {author} {\bibfnamefont {Alexandre}\ \bibnamefont
  {Bourassa}}, \bibinfo {author} {\bibfnamefont {Francisco J.~H.}\ \bibnamefont
  {Heras}}, \bibinfo {author} {\bibfnamefont {Lei~M.}\ \bibnamefont {Zhang}},
  \bibinfo {author} {\bibfnamefont {Johannes}\ \bibnamefont {Bausch}}, \bibinfo
  {author} {\bibfnamefont {Andrew~W.}\ \bibnamefont {Senior}}, \bibinfo
  {author} {\bibfnamefont {Thomas}\ \bibnamefont {Edlich}}, \bibinfo {author}
  {\bibfnamefont {Noah}\ \bibnamefont {Shutty}}, \bibinfo {author}
  {\bibfnamefont {Volodymyr}\ \bibnamefont {Sivak}}, \bibinfo {author}
  {\bibfnamefont {Andreas}\ \bibnamefont {Bengtsson}}, \bibinfo {author}
  {\bibfnamefont {Matt}\ \bibnamefont {McEwen}}, \bibinfo {author}
  {\bibfnamefont {Oscar}\ \bibnamefont {Higgott}}, \bibinfo {author}
  {\bibfnamefont {Dvir}\ \bibnamefont {Kafri}}, \bibinfo {author}
  {\bibfnamefont {Jahan}\ \bibnamefont {Claes}}, \bibinfo {author}
  {\bibfnamefont {Alexis}\ \bibnamefont {Morvan}}, \bibinfo {author}
  {\bibfnamefont {Zijun}\ \bibnamefont {Chen}}, \bibinfo {author}
  {\bibfnamefont {Adam}\ \bibnamefont {Zalcman}}, \bibinfo {author}
  {\bibfnamefont {Sid}\ \bibnamefont {Madhuk}}, \bibinfo {author}
  {\bibfnamefont {Rajeev}\ \bibnamefont {Acharya}}, \bibinfo {author}
  {\bibfnamefont {Laleh~Aghababaie}\ \bibnamefont {Beni}}, \bibinfo {author}
  {\bibfnamefont {Georg}\ \bibnamefont {Aigeldinger}}, \bibinfo {author}
  {\bibfnamefont {Ross}\ \bibnamefont {Alcaraz}}, \bibinfo {author}
  {\bibfnamefont {Trond~I.}\ \bibnamefont {Andersen}}, \bibinfo {author}
  {\bibfnamefont {Markus}\ \bibnamefont {Ansmann}}, \bibinfo {author}
  {\bibfnamefont {Frank}\ \bibnamefont {Arute}}, \bibinfo {author}
  {\bibfnamefont {Kunal}\ \bibnamefont {Arya}}, \bibinfo {author}
  {\bibfnamefont {Abraham}\ \bibnamefont {Asfaw}}, \bibinfo {author}
  {\bibfnamefont {Juan}\ \bibnamefont {Atalaya}}, \bibinfo {author}
  {\bibfnamefont {Ryan}\ \bibnamefont {Babbush}}, \bibinfo {author}
  {\bibfnamefont {Brian}\ \bibnamefont {Ballard}}, \bibinfo {author}
  {\bibfnamefont {Joseph~C.}\ \bibnamefont {Bardin}}, \bibinfo {author}
  {\bibfnamefont {Alexander}\ \bibnamefont {Bilmes}}, \bibinfo {author}
  {\bibfnamefont {Sam}\ \bibnamefont {Blackwell}}, \bibinfo {author}
  {\bibfnamefont {Jenna}\ \bibnamefont {Bovaird}}, \bibinfo {author}
  {\bibfnamefont {Dylan}\ \bibnamefont {Bowers}}, \bibinfo {author}
  {\bibfnamefont {Leon}\ \bibnamefont {Brill}}, \bibinfo {author}
  {\bibfnamefont {Michael}\ \bibnamefont {Broughton}}, \bibinfo {author}
  {\bibfnamefont {David~A.}\ \bibnamefont {Browne}}, \bibinfo {author}
  {\bibfnamefont {Brett}\ \bibnamefont {Buchea}}, \bibinfo {author}
  {\bibfnamefont {Bob~B.}\ \bibnamefont {Buckley}}, \bibinfo {author}
  {\bibfnamefont {Tim}\ \bibnamefont {Burger}}, \bibinfo {author}
  {\bibfnamefont {Brian}\ \bibnamefont {Burkett}}, \bibinfo {author}
  {\bibfnamefont {Nicholas}\ \bibnamefont {Bushnell}}, \bibinfo {author}
  {\bibfnamefont {Anthony}\ \bibnamefont {Cabrera}}, \bibinfo {author}
  {\bibfnamefont {Juan}\ \bibnamefont {Campero}}, \bibinfo {author}
  {\bibfnamefont {Hung-Shen}\ \bibnamefont {Chang}}, \bibinfo {author}
  {\bibfnamefont {Ben}\ \bibnamefont {Chiaro}}, \bibinfo {author}
  {\bibfnamefont {Liang-Ying}\ \bibnamefont {Chih}}, \bibinfo {author}
  {\bibfnamefont {Agnetta~Y.}\ \bibnamefont {Cleland}}, \bibinfo {author}
  {\bibfnamefont {Josh}\ \bibnamefont {Cogan}}, \bibinfo {author}
  {\bibfnamefont {Roberto}\ \bibnamefont {Collins}}, \bibinfo {author}
  {\bibfnamefont {Paul}\ \bibnamefont {Conner}}, \bibinfo {author}
  {\bibfnamefont {William}\ \bibnamefont {Courtney}}, \bibinfo {author}
  {\bibfnamefont {Alexander~L.}\ \bibnamefont {Crook}}, \bibinfo {author}
  {\bibfnamefont {Ben}\ \bibnamefont {Curtin}}, \bibinfo {author}
  {\bibfnamefont {Sayan}\ \bibnamefont {Das}}, \bibinfo {author} {\bibfnamefont
  {Sean}\ \bibnamefont {Demura}}, \bibinfo {author} {\bibfnamefont {Laura~De}\
  \bibnamefont {Lorenzo}}, \bibinfo {author} {\bibfnamefont {Agustin~Di}\
  \bibnamefont {Paolo}}, \bibinfo {author} {\bibfnamefont {Paul}\ \bibnamefont
  {Donohoe}}, \bibinfo {author} {\bibfnamefont {Ilya}\ \bibnamefont {Drozdov}},
  \bibinfo {author} {\bibfnamefont {Andrew}\ \bibnamefont {Dunsworth}},
  \bibinfo {author} {\bibfnamefont {Alec}\ \bibnamefont {Eickbusch}}, \bibinfo
  {author} {\bibfnamefont {Aviv~Moshe}\ \bibnamefont {Elbag}}, \bibinfo
  {author} {\bibfnamefont {Mahmoud}\ \bibnamefont {Elzouka}}, \bibinfo {author}
  {\bibfnamefont {Catherine}\ \bibnamefont {Erickson}}, \bibinfo {author}
  {\bibfnamefont {Vinicius~S.}\ \bibnamefont {Ferreira}}, \bibinfo {author}
  {\bibfnamefont {Leslie~Flores}\ \bibnamefont {Burgos}}, \bibinfo {author}
  {\bibfnamefont {Ebrahim}\ \bibnamefont {Forati}}, \bibinfo {author}
  {\bibfnamefont {Austin~G.}\ \bibnamefont {Fowler}}, \bibinfo {author}
  {\bibfnamefont {Brooks}\ \bibnamefont {Foxen}}, \bibinfo {author}
  {\bibfnamefont {Suhas}\ \bibnamefont {Ganjam}}, \bibinfo {author}
  {\bibfnamefont {Gonzalo}\ \bibnamefont {Garcia}}, \bibinfo {author}
  {\bibfnamefont {Robert}\ \bibnamefont {Gasca}}, \bibinfo {author}
  {\bibfnamefont {Élie}\ \bibnamefont {Genois}}, \bibinfo {author}
  {\bibfnamefont {William}\ \bibnamefont {Giang}}, \bibinfo {author}
  {\bibfnamefont {Dar}\ \bibnamefont {Gilboa}}, \bibinfo {author}
  {\bibfnamefont {Raja}\ \bibnamefont {Gosula}}, \bibinfo {author}
  {\bibfnamefont {Alejandro~Grajales}\ \bibnamefont {Dau}}, \bibinfo {author}
  {\bibfnamefont {Dietrich}\ \bibnamefont {Graumann}}, \bibinfo {author}
  {\bibfnamefont {Alex}\ \bibnamefont {Greene}}, \bibinfo {author}
  {\bibfnamefont {Jonathan~A.}\ \bibnamefont {Gross}}, \bibinfo {author}
  {\bibfnamefont {Tan}\ \bibnamefont {Ha}}, \bibinfo {author} {\bibfnamefont
  {Steve}\ \bibnamefont {Habegger}}, \bibinfo {author} {\bibfnamefont {Monica}\
  \bibnamefont {Hansen}}, \bibinfo {author} {\bibfnamefont {Matthew~P.}\
  \bibnamefont {Harrigan}}, \bibinfo {author} {\bibfnamefont {Sean~D.}\
  \bibnamefont {Harrington}}, \bibinfo {author} {\bibfnamefont {Stephen}\
  \bibnamefont {Heslin}}, \bibinfo {author} {\bibfnamefont {Paula}\
  \bibnamefont {Heu}}, \bibinfo {author} {\bibfnamefont {Reno}\ \bibnamefont
  {Hiltermann}}, \bibinfo {author} {\bibfnamefont {Jeremy}\ \bibnamefont
  {Hilton}}, \bibinfo {author} {\bibfnamefont {Sabrina}\ \bibnamefont {Hong}},
  \bibinfo {author} {\bibfnamefont {Hsin-Yuan}\ \bibnamefont {Huang}}, \bibinfo
  {author} {\bibfnamefont {Ashley}\ \bibnamefont {Huff}}, \bibinfo {author}
  {\bibfnamefont {William~J.}\ \bibnamefont {Huggins}}, \bibinfo {author}
  {\bibfnamefont {Evan}\ \bibnamefont {Jeffrey}}, \bibinfo {author}
  {\bibfnamefont {Zhang}\ \bibnamefont {Jiang}}, \bibinfo {author}
  {\bibfnamefont {Xiaoxuan}\ \bibnamefont {Jin}}, \bibinfo {author}
  {\bibfnamefont {Chaitali}\ \bibnamefont {Joshi}}, \bibinfo {author}
  {\bibfnamefont {Pavol}\ \bibnamefont {Juhas}}, \bibinfo {author}
  {\bibfnamefont {Andreas}\ \bibnamefont {Kabel}}, \bibinfo {author}
  {\bibfnamefont {Hui}\ \bibnamefont {Kang}}, \bibinfo {author} {\bibfnamefont
  {Amir~H.}\ \bibnamefont {Karamlou}}, \bibinfo {author} {\bibfnamefont
  {Kostyantyn}\ \bibnamefont {Kechedzhi}}, \bibinfo {author} {\bibfnamefont
  {Trupti}\ \bibnamefont {Khaire}}, \bibinfo {author} {\bibfnamefont {Tanuj}\
  \bibnamefont {Khattar}}, \bibinfo {author} {\bibfnamefont {Mostafa}\
  \bibnamefont {Khezri}}, \bibinfo {author} {\bibfnamefont {Seon}\ \bibnamefont
  {Kim}}, \bibinfo {author} {\bibfnamefont {Paul~V.}\ \bibnamefont {Klimov}},
  \bibinfo {author} {\bibfnamefont {Bryce}\ \bibnamefont {Kobrin}}, \bibinfo
  {author} {\bibfnamefont {Alexander~N.}\ \bibnamefont {Korotkov}}, \bibinfo
  {author} {\bibfnamefont {Fedor}\ \bibnamefont {Kostritsa}}, \bibinfo {author}
  {\bibfnamefont {John~Mark}\ \bibnamefont {Kreikebaum}}, \bibinfo {author}
  {\bibfnamefont {Vladislav~D.}\ \bibnamefont {Kurilovich}}, \bibinfo {author}
  {\bibfnamefont {David}\ \bibnamefont {Landhuis}}, \bibinfo {author}
  {\bibfnamefont {Tiano}\ \bibnamefont {Lange-Dei}}, \bibinfo {author}
  {\bibfnamefont {Brandon~W.}\ \bibnamefont {Langley}}, \bibinfo {author}
  {\bibfnamefont {Pavel}\ \bibnamefont {Laptev}}, \bibinfo {author}
  {\bibfnamefont {Kim-Ming}\ \bibnamefont {Lau}}, \bibinfo {author}
  {\bibfnamefont {Justin}\ \bibnamefont {Ledford}}, \bibinfo {author}
  {\bibfnamefont {Kenny}\ \bibnamefont {Lee}}, \bibinfo {author} {\bibfnamefont
  {Brian~J.}\ \bibnamefont {Lester}}, \bibinfo {author} {\bibfnamefont
  {Loïck~Le}\ \bibnamefont {Guevel}}, \bibinfo {author} {\bibfnamefont
  {Wing~Yan}\ \bibnamefont {Li}}, \bibinfo {author} {\bibfnamefont {Yin}\
  \bibnamefont {Li}}, \bibinfo {author} {\bibfnamefont {Alexander~T.}\
  \bibnamefont {Lill}}, \bibinfo {author} {\bibfnamefont {William~P.}\
  \bibnamefont {Livingston}}, \bibinfo {author} {\bibfnamefont {Aditya}\
  \bibnamefont {Locharla}}, \bibinfo {author} {\bibfnamefont {Erik}\
  \bibnamefont {Lucero}}, \bibinfo {author} {\bibfnamefont {Daniel}\
  \bibnamefont {Lundahl}}, \bibinfo {author} {\bibfnamefont {Aaron}\
  \bibnamefont {Lunt}}, \bibinfo {author} {\bibfnamefont {Ashley}\ \bibnamefont
  {Maloney}}, \bibinfo {author} {\bibfnamefont {Salvatore}\ \bibnamefont
  {Mandrà}}, \bibinfo {author} {\bibfnamefont {Leigh~S.}\ \bibnamefont
  {Martin}}, \bibinfo {author} {\bibfnamefont {Orion}\ \bibnamefont {Martin}},
  \bibinfo {author} {\bibfnamefont {Cameron}\ \bibnamefont {Maxfield}},
  \bibinfo {author} {\bibfnamefont {Jarrod~R.}\ \bibnamefont {McClean}},
  \bibinfo {author} {\bibfnamefont {Seneca}\ \bibnamefont {Meeks}}, \bibinfo
  {author} {\bibfnamefont {Anthony}\ \bibnamefont {Megrant}}, \bibinfo {author}
  {\bibfnamefont {Kevin~C.}\ \bibnamefont {Miao}}, \bibinfo {author}
  {\bibfnamefont {Reza}\ \bibnamefont {Molavi}}, \bibinfo {author}
  {\bibfnamefont {Sebastian}\ \bibnamefont {Molina}}, \bibinfo {author}
  {\bibfnamefont {Shirin}\ \bibnamefont {Montazeri}}, \bibinfo {author}
  {\bibfnamefont {Ramis}\ \bibnamefont {Movassagh}}, \bibinfo {author}
  {\bibfnamefont {Charles}\ \bibnamefont {Neill}}, \bibinfo {author}
  {\bibfnamefont {Michael}\ \bibnamefont {Newman}}, \bibinfo {author}
  {\bibfnamefont {Anthony}\ \bibnamefont {Nguyen}}, \bibinfo {author}
  {\bibfnamefont {Murray}\ \bibnamefont {Nguyen}}, \bibinfo {author}
  {\bibfnamefont {Chia-Hung}\ \bibnamefont {Ni}}, \bibinfo {author}
  {\bibfnamefont {Murphy~Y.}\ \bibnamefont {Niu}}, \bibinfo {author}
  {\bibfnamefont {Logan}\ \bibnamefont {Oas}}, \bibinfo {author} {\bibfnamefont
  {William~D.}\ \bibnamefont {Oliver}}, \bibinfo {author} {\bibfnamefont
  {Raymond}\ \bibnamefont {Orosco}}, \bibinfo {author} {\bibfnamefont
  {Kristoffer}\ \bibnamefont {Ottosson}}, \bibinfo {author} {\bibfnamefont
  {Alex}\ \bibnamefont {Pizzuto}}, \bibinfo {author} {\bibfnamefont {Rebecca}\
  \bibnamefont {Potter}}, \bibinfo {author} {\bibfnamefont {Orion}\
  \bibnamefont {Pritchard}}, \bibinfo {author} {\bibfnamefont {Chris}\
  \bibnamefont {Quintana}}, \bibinfo {author} {\bibfnamefont {Ganesh}\
  \bibnamefont {Ramachandran}}, \bibinfo {author} {\bibfnamefont {Matthew~J.}\
  \bibnamefont {Reagor}}, \bibinfo {author} {\bibfnamefont {Rachel}\
  \bibnamefont {Resnick}}, \bibinfo {author} {\bibfnamefont {David~M.}\
  \bibnamefont {Rhodes}}, \bibinfo {author} {\bibfnamefont {Gabrielle}\
  \bibnamefont {Roberts}}, \bibinfo {author} {\bibfnamefont {Eliott}\
  \bibnamefont {Rosenberg}}, \bibinfo {author} {\bibfnamefont {Emma}\
  \bibnamefont {Rosenfeld}}, \bibinfo {author} {\bibfnamefont {Elizabeth}\
  \bibnamefont {Rossi}}, \bibinfo {author} {\bibfnamefont {Pedram}\
  \bibnamefont {Roushan}}, \bibinfo {author} {\bibfnamefont {Kannan}\
  \bibnamefont {Sankaragomathi}}, \bibinfo {author} {\bibfnamefont {Henry~F.}\
  \bibnamefont {Schurkus}}, \bibinfo {author} {\bibfnamefont {Michael~J.}\
  \bibnamefont {Shearn}}, \bibinfo {author} {\bibfnamefont {Aaron}\
  \bibnamefont {Shorter}}, \bibinfo {author} {\bibfnamefont {Vladimir}\
  \bibnamefont {Shvarts}}, \bibinfo {author} {\bibfnamefont {Spencer}\
  \bibnamefont {Small}}, \bibinfo {author} {\bibfnamefont {W.~Clarke}\
  \bibnamefont {Smith}}, \bibinfo {author} {\bibfnamefont {Sofia}\ \bibnamefont
  {Springer}}, \bibinfo {author} {\bibfnamefont {George}\ \bibnamefont
  {Sterling}}, \bibinfo {author} {\bibfnamefont {Jordan}\ \bibnamefont
  {Suchard}}, \bibinfo {author} {\bibfnamefont {Aaron}\ \bibnamefont {Szasz}},
  \bibinfo {author} {\bibfnamefont {Alex}\ \bibnamefont {Sztein}}, \bibinfo
  {author} {\bibfnamefont {Douglas}\ \bibnamefont {Thor}}, \bibinfo {author}
  {\bibfnamefont {Eifu}\ \bibnamefont {Tomita}}, \bibinfo {author}
  {\bibfnamefont {Alfredo}\ \bibnamefont {Torres}}, \bibinfo {author}
  {\bibfnamefont {M.~Mert}\ \bibnamefont {Torunbalci}}, \bibinfo {author}
  {\bibfnamefont {Abeer}\ \bibnamefont {Vaishnav}}, \bibinfo {author}
  {\bibfnamefont {Justin}\ \bibnamefont {Vargas}}, \bibinfo {author}
  {\bibfnamefont {Sergey}\ \bibnamefont {Vdovichev}}, \bibinfo {author}
  {\bibfnamefont {Guifre}\ \bibnamefont {Vidal}}, \bibinfo {author}
  {\bibfnamefont {Catherine~Vollgraff}\ \bibnamefont {Heidweiller}}, \bibinfo
  {author} {\bibfnamefont {Steven}\ \bibnamefont {Waltman}}, \bibinfo {author}
  {\bibfnamefont {Jonathan}\ \bibnamefont {Waltz}}, \bibinfo {author}
  {\bibfnamefont {Shannon~X.}\ \bibnamefont {Wang}}, \bibinfo {author}
  {\bibfnamefont {Brayden}\ \bibnamefont {Ware}}, \bibinfo {author}
  {\bibfnamefont {Travis}\ \bibnamefont {Weidel}}, \bibinfo {author}
  {\bibfnamefont {Theodore}\ \bibnamefont {White}}, \bibinfo {author}
  {\bibfnamefont {Kristi}\ \bibnamefont {Wong}}, \bibinfo {author}
  {\bibfnamefont {Bryan W.~K.}\ \bibnamefont {Woo}}, \bibinfo {author}
  {\bibfnamefont {Maddy}\ \bibnamefont {Woodson}}, \bibinfo {author}
  {\bibfnamefont {Cheng}\ \bibnamefont {Xing}}, \bibinfo {author}
  {\bibfnamefont {Z.~Jamie}\ \bibnamefont {Yao}}, \bibinfo {author}
  {\bibfnamefont {Ping}\ \bibnamefont {Yeh}}, \bibinfo {author} {\bibfnamefont
  {Bicheng}\ \bibnamefont {Ying}}, \bibinfo {author} {\bibfnamefont {Juhwan}\
  \bibnamefont {Yoo}}, \bibinfo {author} {\bibfnamefont {Noureldin}\
  \bibnamefont {Yosri}}, \bibinfo {author} {\bibfnamefont {Grayson}\
  \bibnamefont {Young}}, \bibinfo {author} {\bibfnamefont {Yaxing}\
  \bibnamefont {Zhang}}, \bibinfo {author} {\bibfnamefont {Ningfeng}\
  \bibnamefont {Zhu}}, \bibinfo {author} {\bibfnamefont {Nicholas}\
  \bibnamefont {Zobrist}}, \bibinfo {author} {\bibfnamefont {Hartmut}\
  \bibnamefont {Neven}}, \bibinfo {author} {\bibfnamefont {Pushmeet}\
  \bibnamefont {Kohli}}, \bibinfo {author} {\bibfnamefont {Alex}\ \bibnamefont
  {Davies}}, \bibinfo {author} {\bibfnamefont {Sergio}\ \bibnamefont {Boixo}},
  \bibinfo {author} {\bibfnamefont {Julian}\ \bibnamefont {Kelly}}, \bibinfo
  {author} {\bibfnamefont {Cody}\ \bibnamefont {Jones}}, \bibinfo {author}
  {\bibfnamefont {Craig}\ \bibnamefont {Gidney}}, \ and\ \bibinfo {author}
  {\bibfnamefont {Kevin~J.}\ \bibnamefont {Satzinger}},\ }\href {\doibase
  10.48550/arXiv.2412.14256} {\enquote {\bibinfo {title} {Scaling and logic in
  the color code on a superconducting quantum processor},}\ } (\bibinfo {year}
  {2024})\BibitemShut {NoStop}%
\bibitem [{\citenamefont {Postler}\ \emph {et~al.}(2022)\citenamefont
  {Postler}, \citenamefont {Heuβen}, \citenamefont {Pogorelov}, \citenamefont
  {Rispler}, \citenamefont {Feldker}, \citenamefont {Meth}, \citenamefont
  {Marciniak}, \citenamefont {Stricker}, \citenamefont {Ringbauer},
  \citenamefont {Blatt}, \citenamefont {Schindler}, \citenamefont {Müller},\
  and\ \citenamefont {Monz}}]{postler_demonstration_2022}%
  \BibitemOpen
  \bibfield  {author} {\bibinfo {author} {\bibfnamefont {Lukas}\ \bibnamefont
  {Postler}}, \bibinfo {author} {\bibfnamefont {Sascha}\ \bibnamefont
  {Heuβen}}, \bibinfo {author} {\bibfnamefont {Ivan}\ \bibnamefont
  {Pogorelov}}, \bibinfo {author} {\bibfnamefont {Manuel}\ \bibnamefont
  {Rispler}}, \bibinfo {author} {\bibfnamefont {Thomas}\ \bibnamefont
  {Feldker}}, \bibinfo {author} {\bibfnamefont {Michael}\ \bibnamefont {Meth}},
  \bibinfo {author} {\bibfnamefont {Christian~D.}\ \bibnamefont {Marciniak}},
  \bibinfo {author} {\bibfnamefont {Roman}\ \bibnamefont {Stricker}}, \bibinfo
  {author} {\bibfnamefont {Martin}\ \bibnamefont {Ringbauer}}, \bibinfo
  {author} {\bibfnamefont {Rainer}\ \bibnamefont {Blatt}}, \bibinfo {author}
  {\bibfnamefont {Philipp}\ \bibnamefont {Schindler}}, \bibinfo {author}
  {\bibfnamefont {Markus}\ \bibnamefont {Müller}}, \ and\ \bibinfo {author}
  {\bibfnamefont {Thomas}\ \bibnamefont {Monz}},\ }\bibfield  {title} {\enquote
  {\bibinfo {title} {Demonstration of fault-tolerant universal quantum gate
  operations},}\ }\href {https://www.nature.com/articles/s41586-022-04721-1}
  {\bibfield  {journal} {\bibinfo  {journal} {Nature}\ }\textbf {\bibinfo
  {volume} {605}},\ \bibinfo {pages} {675--680} (\bibinfo {year}
  {2022})}\BibitemShut {NoStop}%
\bibitem [{\citenamefont {Ryan-Anderson}\ \emph {et~al.}(2022)\citenamefont
  {Ryan-Anderson}, \citenamefont {Brown}, \citenamefont {Allman}, \citenamefont
  {Arkin}, \citenamefont {Asa-Attuah}, \citenamefont {Baldwin}, \citenamefont
  {Berg}, \citenamefont {Bohnet}, \citenamefont {Braxton}, \citenamefont
  {Burdick}, \citenamefont {Campora}, \citenamefont {Chernoguzov},
  \citenamefont {Esposito}, \citenamefont {Evans}, \citenamefont {Francois},
  \citenamefont {Gaebler}, \citenamefont {Gatterman}, \citenamefont {Gerber},
  \citenamefont {Gilmore}, \citenamefont {Gresh}, \citenamefont {Hall},
  \citenamefont {Hankin}, \citenamefont {Hostetter}, \citenamefont {Lucchetti},
  \citenamefont {Mayer}, \citenamefont {Myers}, \citenamefont {Neyenhuis},
  \citenamefont {Santiago}, \citenamefont {Sedlacek}, \citenamefont {Skripka},
  \citenamefont {Slattery}, \citenamefont {Stutz}, \citenamefont {Tait},
  \citenamefont {Tobey}, \citenamefont {Vittorini}, \citenamefont {Walker},\
  and\ \citenamefont {Hayes}}]{ryan-anderson_implementing_2022}%
  \BibitemOpen
  \bibfield  {author} {\bibinfo {author} {\bibfnamefont {C.}~\bibnamefont
  {Ryan-Anderson}}, \bibinfo {author} {\bibfnamefont {N.~C.}\ \bibnamefont
  {Brown}}, \bibinfo {author} {\bibfnamefont {M.~S.}\ \bibnamefont {Allman}},
  \bibinfo {author} {\bibfnamefont {B.}~\bibnamefont {Arkin}}, \bibinfo
  {author} {\bibfnamefont {G.}~\bibnamefont {Asa-Attuah}}, \bibinfo {author}
  {\bibfnamefont {C.}~\bibnamefont {Baldwin}}, \bibinfo {author} {\bibfnamefont
  {J.}~\bibnamefont {Berg}}, \bibinfo {author} {\bibfnamefont {J.~G.}\
  \bibnamefont {Bohnet}}, \bibinfo {author} {\bibfnamefont {S.}~\bibnamefont
  {Braxton}}, \bibinfo {author} {\bibfnamefont {N.}~\bibnamefont {Burdick}},
  \bibinfo {author} {\bibfnamefont {J.~P.}\ \bibnamefont {Campora}}, \bibinfo
  {author} {\bibfnamefont {A.}~\bibnamefont {Chernoguzov}}, \bibinfo {author}
  {\bibfnamefont {J.}~\bibnamefont {Esposito}}, \bibinfo {author}
  {\bibfnamefont {B.}~\bibnamefont {Evans}}, \bibinfo {author} {\bibfnamefont
  {D.}~\bibnamefont {Francois}}, \bibinfo {author} {\bibfnamefont {J.~P.}\
  \bibnamefont {Gaebler}}, \bibinfo {author} {\bibfnamefont {T.~M.}\
  \bibnamefont {Gatterman}}, \bibinfo {author} {\bibfnamefont {J.}~\bibnamefont
  {Gerber}}, \bibinfo {author} {\bibfnamefont {K.}~\bibnamefont {Gilmore}},
  \bibinfo {author} {\bibfnamefont {D.}~\bibnamefont {Gresh}}, \bibinfo
  {author} {\bibfnamefont {A.}~\bibnamefont {Hall}}, \bibinfo {author}
  {\bibfnamefont {A.}~\bibnamefont {Hankin}}, \bibinfo {author} {\bibfnamefont
  {J.}~\bibnamefont {Hostetter}}, \bibinfo {author} {\bibfnamefont
  {D.}~\bibnamefont {Lucchetti}}, \bibinfo {author} {\bibfnamefont
  {K.}~\bibnamefont {Mayer}}, \bibinfo {author} {\bibfnamefont
  {J.}~\bibnamefont {Myers}}, \bibinfo {author} {\bibfnamefont
  {B.}~\bibnamefont {Neyenhuis}}, \bibinfo {author} {\bibfnamefont
  {J.}~\bibnamefont {Santiago}}, \bibinfo {author} {\bibfnamefont
  {J.}~\bibnamefont {Sedlacek}}, \bibinfo {author} {\bibfnamefont
  {T.}~\bibnamefont {Skripka}}, \bibinfo {author} {\bibfnamefont
  {A.}~\bibnamefont {Slattery}}, \bibinfo {author} {\bibfnamefont {R.~P.}\
  \bibnamefont {Stutz}}, \bibinfo {author} {\bibfnamefont {J.}~\bibnamefont
  {Tait}}, \bibinfo {author} {\bibfnamefont {R.}~\bibnamefont {Tobey}},
  \bibinfo {author} {\bibfnamefont {G.}~\bibnamefont {Vittorini}}, \bibinfo
  {author} {\bibfnamefont {J.}~\bibnamefont {Walker}}, \ and\ \bibinfo {author}
  {\bibfnamefont {D.}~\bibnamefont {Hayes}},\ }\href
  {http://arxiv.org/abs/2208.01863} {\enquote {\bibinfo {title} {Implementing
  {Fault}-tolerant {Entangling} {Gates} on the {Five}-qubit {Code} and the
  {Color} {Code}},}\ } (\bibinfo {year} {2022}),\ \bibinfo {note}
  {arXiv:2208.01863}\BibitemShut {NoStop}%
\bibitem [{\citenamefont {da~Silva}\ \emph {et~al.}(2024)\citenamefont
  {da~Silva}, \citenamefont {Ryan-Anderson}, \citenamefont {Bello-Rivas},
  \citenamefont {Chernoguzov}, \citenamefont {Dreiling}, \citenamefont {Foltz},
  \citenamefont {Frachon}, \citenamefont {Gaebler}, \citenamefont {Gatterman},
  \citenamefont {Grans-Samuelsson}, \citenamefont {Hayes}, \citenamefont
  {Hewitt}, \citenamefont {Johansen}, \citenamefont {Lucchetti}, \citenamefont
  {Mills}, \citenamefont {Moses}, \citenamefont {Neyenhuis}, \citenamefont
  {Paz}, \citenamefont {Pino}, \citenamefont {Siegfried}, \citenamefont
  {Strabley}, \citenamefont {Sundaram}, \citenamefont {Tom}, \citenamefont
  {Wernli}, \citenamefont {Zanner}, \citenamefont {Stutz},\ and\ \citenamefont
  {Svore}}]{da_silva_demonstration_2024}%
  \BibitemOpen
  \bibfield  {author} {\bibinfo {author} {\bibfnamefont {M.~P.}\ \bibnamefont
  {da~Silva}}, \bibinfo {author} {\bibfnamefont {C.}~\bibnamefont
  {Ryan-Anderson}}, \bibinfo {author} {\bibfnamefont {J.~M.}\ \bibnamefont
  {Bello-Rivas}}, \bibinfo {author} {\bibfnamefont {A.}~\bibnamefont
  {Chernoguzov}}, \bibinfo {author} {\bibfnamefont {J.~M.}\ \bibnamefont
  {Dreiling}}, \bibinfo {author} {\bibfnamefont {C.}~\bibnamefont {Foltz}},
  \bibinfo {author} {\bibfnamefont {F.}~\bibnamefont {Frachon}}, \bibinfo
  {author} {\bibfnamefont {J.~P.}\ \bibnamefont {Gaebler}}, \bibinfo {author}
  {\bibfnamefont {T.~M.}\ \bibnamefont {Gatterman}}, \bibinfo {author}
  {\bibfnamefont {L.}~\bibnamefont {Grans-Samuelsson}}, \bibinfo {author}
  {\bibfnamefont {D.}~\bibnamefont {Hayes}}, \bibinfo {author} {\bibfnamefont
  {N.}~\bibnamefont {Hewitt}}, \bibinfo {author} {\bibfnamefont
  {J.}~\bibnamefont {Johansen}}, \bibinfo {author} {\bibfnamefont
  {D.}~\bibnamefont {Lucchetti}}, \bibinfo {author} {\bibfnamefont
  {M.}~\bibnamefont {Mills}}, \bibinfo {author} {\bibfnamefont {S.~A.}\
  \bibnamefont {Moses}}, \bibinfo {author} {\bibfnamefont {B.}~\bibnamefont
  {Neyenhuis}}, \bibinfo {author} {\bibfnamefont {A.}~\bibnamefont {Paz}},
  \bibinfo {author} {\bibfnamefont {J.}~\bibnamefont {Pino}}, \bibinfo {author}
  {\bibfnamefont {P.}~\bibnamefont {Siegfried}}, \bibinfo {author}
  {\bibfnamefont {J.}~\bibnamefont {Strabley}}, \bibinfo {author}
  {\bibfnamefont {A.}~\bibnamefont {Sundaram}}, \bibinfo {author}
  {\bibfnamefont {D.}~\bibnamefont {Tom}}, \bibinfo {author} {\bibfnamefont
  {S.~J.}\ \bibnamefont {Wernli}}, \bibinfo {author} {\bibfnamefont
  {M.}~\bibnamefont {Zanner}}, \bibinfo {author} {\bibfnamefont {R.~P.}\
  \bibnamefont {Stutz}}, \ and\ \bibinfo {author} {\bibfnamefont {K.~M.}\
  \bibnamefont {Svore}},\ }\href {\doibase 10.48550/arXiv.2404.02280} {\enquote
  {\bibinfo {title} {Demonstration of logical qubits and repeated error
  correction with better-than-physical error rates},}\ } (\bibinfo {year}
  {2024}),\ \bibinfo {note} {arXiv:2404.02280}\BibitemShut {NoStop}%
\bibitem [{\citenamefont {Berthusen}\ \emph {et~al.}(2024)\citenamefont
  {Berthusen}, \citenamefont {Dreiling}, \citenamefont {Foltz}, \citenamefont
  {Gaebler}, \citenamefont {Gatterman}, \citenamefont {Gresh}, \citenamefont
  {Hewitt}, \citenamefont {Mills}, \citenamefont {Moses}, \citenamefont
  {Neyenhuis}, \citenamefont {Siegfried},\ and\ \citenamefont
  {Hayes}}]{berthusen_experiments_2024}%
  \BibitemOpen
  \bibfield  {author} {\bibinfo {author} {\bibfnamefont {Noah}\ \bibnamefont
  {Berthusen}}, \bibinfo {author} {\bibfnamefont {Joan}\ \bibnamefont
  {Dreiling}}, \bibinfo {author} {\bibfnamefont {Cameron}\ \bibnamefont
  {Foltz}}, \bibinfo {author} {\bibfnamefont {John~P.}\ \bibnamefont
  {Gaebler}}, \bibinfo {author} {\bibfnamefont {Thomas~M.}\ \bibnamefont
  {Gatterman}}, \bibinfo {author} {\bibfnamefont {Dan}\ \bibnamefont {Gresh}},
  \bibinfo {author} {\bibfnamefont {Nathan}\ \bibnamefont {Hewitt}}, \bibinfo
  {author} {\bibfnamefont {Michael}\ \bibnamefont {Mills}}, \bibinfo {author}
  {\bibfnamefont {Steven~A.}\ \bibnamefont {Moses}}, \bibinfo {author}
  {\bibfnamefont {Brian}\ \bibnamefont {Neyenhuis}}, \bibinfo {author}
  {\bibfnamefont {Peter}\ \bibnamefont {Siegfried}}, \ and\ \bibinfo {author}
  {\bibfnamefont {David}\ \bibnamefont {Hayes}},\ }\href {\doibase
  10.48550/arXiv.2408.08865} {\enquote {\bibinfo {title} {Experiments with the
  {4D} {Surface} {Code} on a {QCCD} {Quantum} {Computer}},}\ } (\bibinfo {year}
  {2024}),\ \bibinfo {note} {arXiv:2408.08865}\BibitemShut {NoStop}%
\bibitem [{\citenamefont {Ryan-Anderson}\ \emph {et~al.}(2021)\citenamefont
  {Ryan-Anderson}, \citenamefont {Bohnet}, \citenamefont {Lee}, \citenamefont
  {Gresh}, \citenamefont {Hankin}, \citenamefont {Gaebler}, \citenamefont
  {Francois}, \citenamefont {Chernoguzov}, \citenamefont {Lucchetti},
  \citenamefont {Brown}, \citenamefont {Gatterman}, \citenamefont {Halit},
  \citenamefont {Gilmore}, \citenamefont {Gerber}, \citenamefont {Neyenhuis},
  \citenamefont {Hayes},\ and\ \citenamefont
  {Stutz}}]{ryan-anderson_realization_2021}%
  \BibitemOpen
  \bibfield  {author} {\bibinfo {author} {\bibfnamefont {C.}~\bibnamefont
  {Ryan-Anderson}}, \bibinfo {author} {\bibfnamefont {J. G.}\ \bibnamefont
  {Bohnet}}, \bibinfo {author} {\bibfnamefont {K.}~\bibnamefont {Lee}},
  \bibinfo {author} {\bibfnamefont {D.}~\bibnamefont {Gresh}}, \bibinfo
  {author} {\bibfnamefont {A.}~\bibnamefont {Hankin}}, \bibinfo {author}
  {\bibfnamefont {J. P.}\ \bibnamefont {Gaebler}}, \bibinfo {author}
  {\bibfnamefont {D.}~\bibnamefont {Francois}}, \bibinfo {author}
  {\bibfnamefont {A.}~\bibnamefont {Chernoguzov}}, \bibinfo {author}
  {\bibfnamefont {D.}~\bibnamefont {Lucchetti}}, \bibinfo {author}
  {\bibfnamefont {N. C.}\ \bibnamefont {Brown}}, \bibinfo {author}
  {\bibfnamefont {T. M.}\ \bibnamefont {Gatterman}}, \bibinfo {author}
  {\bibfnamefont {S. K.}\ \bibnamefont {Halit}}, \bibinfo {author}
  {\bibfnamefont {K.}~\bibnamefont {Gilmore}}, \bibinfo {author} {\bibfnamefont
  {J. A.}\ \bibnamefont {Gerber}}, \bibinfo {author} {\bibfnamefont
  {B.}~\bibnamefont {Neyenhuis}}, \bibinfo {author} {\bibfnamefont
  {D.}~\bibnamefont {Hayes}}, \ and\ \bibinfo {author} {\bibfnamefont
  {R. P.}\ \bibnamefont {Stutz}},\ }\bibfield  {title} {\enquote {\bibinfo
  {title} {Realization of {Real}-{Time} {Fault}-{Tolerant} {Quantum} {Error}
  {Correction}},}\ }\href {\doibase 10.1103/PhysRevX.11.041058} {\bibfield
  {journal} {\bibinfo  {journal} {Physical Review X}\ }\textbf {\bibinfo
  {volume} {11}},\ \bibinfo {pages} {041058} (\bibinfo {year}
  {2021})}\BibitemShut {NoStop}%
\bibitem [{\citenamefont {Ryan-Anderson}\ \emph {et~al.}(2024)\citenamefont
  {Ryan-Anderson}, \citenamefont {Brown}, \citenamefont {Baldwin},
  \citenamefont {Dreiling}, \citenamefont {Foltz}, \citenamefont {Gaebler},
  \citenamefont {Gatterman}, \citenamefont {Hewitt}, \citenamefont {Holliman},
  \citenamefont {Horst}, \citenamefont {Johansen}, \citenamefont {Lucchetti},
  \citenamefont {Mengle}, \citenamefont {Matheny}, \citenamefont {Matsuoka},
  \citenamefont {Mayer}, \citenamefont {Mills}, \citenamefont {Moses},
  \citenamefont {Neyenhuis}, \citenamefont {Pino}, \citenamefont {Siegfried},
  \citenamefont {Stutz}, \citenamefont {Walker},\ and\ \citenamefont
  {Hayes}}]{ryan-anderson_high-fidelity_2024}%
  \BibitemOpen
  \bibfield  {author} {\bibinfo {author} {\bibfnamefont {C.}~\bibnamefont
  {Ryan-Anderson}}, \bibinfo {author} {\bibfnamefont {N.~C.}\ \bibnamefont
  {Brown}}, \bibinfo {author} {\bibfnamefont {C.~H.}\ \bibnamefont {Baldwin}},
  \bibinfo {author} {\bibfnamefont {J.~M.}\ \bibnamefont {Dreiling}}, \bibinfo
  {author} {\bibfnamefont {C.}~\bibnamefont {Foltz}}, \bibinfo {author}
  {\bibfnamefont {J.~P.}\ \bibnamefont {Gaebler}}, \bibinfo {author}
  {\bibfnamefont {T.~M.}\ \bibnamefont {Gatterman}}, \bibinfo {author}
  {\bibfnamefont {N.}~\bibnamefont {Hewitt}}, \bibinfo {author} {\bibfnamefont
  {C.}~\bibnamefont {Holliman}}, \bibinfo {author} {\bibfnamefont {C.~V.}\
  \bibnamefont {Horst}}, \bibinfo {author} {\bibfnamefont {J.}~\bibnamefont
  {Johansen}}, \bibinfo {author} {\bibfnamefont {D.}~\bibnamefont {Lucchetti}},
  \bibinfo {author} {\bibfnamefont {T.}~\bibnamefont {Mengle}}, \bibinfo
  {author} {\bibfnamefont {M.}~\bibnamefont {Matheny}}, \bibinfo {author}
  {\bibfnamefont {Y.}~\bibnamefont {Matsuoka}}, \bibinfo {author}
  {\bibfnamefont {K.}~\bibnamefont {Mayer}}, \bibinfo {author} {\bibfnamefont
  {M.}~\bibnamefont {Mills}}, \bibinfo {author} {\bibfnamefont {S.~A.}\
  \bibnamefont {Moses}}, \bibinfo {author} {\bibfnamefont {B.}~\bibnamefont
  {Neyenhuis}}, \bibinfo {author} {\bibfnamefont {J.}~\bibnamefont {Pino}},
  \bibinfo {author} {\bibfnamefont {P.}~\bibnamefont {Siegfried}}, \bibinfo
  {author} {\bibfnamefont {R.~P.}\ \bibnamefont {Stutz}}, \bibinfo {author}
  {\bibfnamefont {J.}~\bibnamefont {Walker}}, \ and\ \bibinfo {author}
  {\bibfnamefont {D.}~\bibnamefont {Hayes}},\ }\href
  {http://arxiv.org/abs/2404.16728} {\enquote {\bibinfo {title} {High-fidelity
  and {Fault}-tolerant {Teleportation} of a {Logical} {Qubit} using
  {Transversal} {Gates} and {Lattice} {Surgery} on a {Trapped}-ion {Quantum}
  {Computer}},}\ } (\bibinfo {year} {2024}),\ \bibinfo {note}
  {arXiv:2404.16728}\BibitemShut {NoStop}%
\bibitem [{\citenamefont {Pogorelov}\ \emph {et~al.}(2024)\citenamefont
  {Pogorelov}, \citenamefont {Butt}, \citenamefont {Postler}, \citenamefont
  {Marciniak}, \citenamefont {Schindler}, \citenamefont {Müller},\ and\
  \citenamefont {Monz}}]{pogorelov_experimental_2024}%
  \BibitemOpen
  \bibfield  {author} {\bibinfo {author} {\bibfnamefont {Ivan}\ \bibnamefont
  {Pogorelov}}, \bibinfo {author} {\bibfnamefont {Friederike}\ \bibnamefont
  {Butt}}, \bibinfo {author} {\bibfnamefont {Lukas}\ \bibnamefont {Postler}},
  \bibinfo {author} {\bibfnamefont {Christian~D.}\ \bibnamefont {Marciniak}},
  \bibinfo {author} {\bibfnamefont {Philipp}\ \bibnamefont {Schindler}},
  \bibinfo {author} {\bibfnamefont {Markus}\ \bibnamefont {Müller}}, \ and\
  \bibinfo {author} {\bibfnamefont {Thomas}\ \bibnamefont {Monz}},\ }\href
  {http://arxiv.org/abs/2403.13732} {\enquote {\bibinfo {title} {Experimental
  fault-tolerant code switching},}\ } (\bibinfo {year} {2024}),\ \bibinfo
  {note} {arXiv:2403.13732}\BibitemShut {NoStop}%
\bibitem [{\citenamefont {Postler}\ \emph {et~al.}(2024)\citenamefont
  {Postler}, \citenamefont {Butt}, \citenamefont {Pogorelov}, \citenamefont
  {Marciniak}, \citenamefont {Heußen}, \citenamefont {Blatt}, \citenamefont
  {Schindler}, \citenamefont {Rispler}, \citenamefont {Müller},\ and\
  \citenamefont {Monz}}]{postler_demonstration_2024}%
  \BibitemOpen
  \bibfield  {author} {\bibinfo {author} {\bibfnamefont {Lukas}\ \bibnamefont
  {Postler}}, \bibinfo {author} {\bibfnamefont {Friederike}\ \bibnamefont
  {Butt}}, \bibinfo {author} {\bibfnamefont {Ivan}\ \bibnamefont {Pogorelov}},
  \bibinfo {author} {\bibfnamefont {Christian~D.}\ \bibnamefont {Marciniak}},
  \bibinfo {author} {\bibfnamefont {Sascha}\ \bibnamefont {Heußen}}, \bibinfo
  {author} {\bibfnamefont {Rainer}\ \bibnamefont {Blatt}}, \bibinfo {author}
  {\bibfnamefont {Philipp}\ \bibnamefont {Schindler}}, \bibinfo {author}
  {\bibfnamefont {Manuel}\ \bibnamefont {Rispler}}, \bibinfo {author}
  {\bibfnamefont {Markus}\ \bibnamefont {Müller}}, \ and\ \bibinfo {author}
  {\bibfnamefont {Thomas}\ \bibnamefont {Monz}},\ }\bibfield  {title} {\enquote
  {\bibinfo {title} {Demonstration of {Fault}-{Tolerant} {Steane} {Quantum}
  {Error} {Correction}},}\ }\href {\doibase 10.1103/PRXQuantum.5.030326}
  {\bibfield  {journal} {\bibinfo  {journal} {PRX Quantum}\ }\textbf {\bibinfo
  {volume} {5}},\ \bibinfo {pages} {030326} (\bibinfo {year}
  {2024})}\BibitemShut {NoStop}%
\bibitem [{\citenamefont {Huang}\ \emph
  {et~al.}(2024{\natexlab{a}})\citenamefont {Huang}, \citenamefont {Brown},\
  and\ \citenamefont {Cetina}}]{huang_comparing_2024}%
  \BibitemOpen
  \bibfield  {author} {\bibinfo {author} {\bibfnamefont {Shilin}\ \bibnamefont
  {Huang}}, \bibinfo {author} {\bibfnamefont {Kenneth~R.}\ \bibnamefont
  {Brown}}, \ and\ \bibinfo {author} {\bibfnamefont {Marko}\ \bibnamefont
  {Cetina}},\ }\bibfield  {title} {\enquote {\bibinfo {title} {Comparing {Shor}
  and {Steane} error correction using the {Bacon}-{Shor} code},}\ }\href
  {\doibase 10.1126/sciadv.adp2008} {\bibfield  {journal} {\bibinfo  {journal}
  {Science Advances}\ }\textbf {\bibinfo {volume} {10}},\ \bibinfo {pages}
  {eadp2008} (\bibinfo {year} {2024}{\natexlab{a}})}\BibitemShut {NoStop}%
\bibitem [{\citenamefont {Bluvstein}\ \emph {et~al.}(2024)\citenamefont
  {Bluvstein}, \citenamefont {Evered}, \citenamefont {Geim}, \citenamefont
  {Li}, \citenamefont {Zhou}, \citenamefont {Manovitz}, \citenamefont {Ebadi},
  \citenamefont {Cain}, \citenamefont {Kalinowski}, \citenamefont {Hangleiter},
  \citenamefont {Bonilla~Ataides}, \citenamefont {Maskara}, \citenamefont
  {Cong}, \citenamefont {Gao}, \citenamefont {Sales~Rodriguez}, \citenamefont
  {Karolyshyn}, \citenamefont {Semeghini}, \citenamefont {Gullans},
  \citenamefont {Greiner}, \citenamefont {Vuletić},\ and\ \citenamefont
  {Lukin}}]{bluvstein_logical_2024}%
  \BibitemOpen
  \bibfield  {author} {\bibinfo {author} {\bibfnamefont {Dolev}\ \bibnamefont
  {Bluvstein}}, \bibinfo {author} {\bibfnamefont {Simon~J.}\ \bibnamefont
  {Evered}}, \bibinfo {author} {\bibfnamefont {Alexandra~A.}\ \bibnamefont
  {Geim}}, \bibinfo {author} {\bibfnamefont {Sophie~H.}\ \bibnamefont {Li}},
  \bibinfo {author} {\bibfnamefont {Hengyun}\ \bibnamefont {Zhou}}, \bibinfo
  {author} {\bibfnamefont {Tom}\ \bibnamefont {Manovitz}}, \bibinfo {author}
  {\bibfnamefont {Sepehr}\ \bibnamefont {Ebadi}}, \bibinfo {author}
  {\bibfnamefont {Madelyn}\ \bibnamefont {Cain}}, \bibinfo {author}
  {\bibfnamefont {Marcin}\ \bibnamefont {Kalinowski}}, \bibinfo {author}
  {\bibfnamefont {Dominik}\ \bibnamefont {Hangleiter}}, \bibinfo {author}
  {\bibfnamefont {J.~Pablo}\ \bibnamefont {Bonilla~Ataides}}, \bibinfo {author}
  {\bibfnamefont {Nishad}\ \bibnamefont {Maskara}}, \bibinfo {author}
  {\bibfnamefont {Iris}\ \bibnamefont {Cong}}, \bibinfo {author} {\bibfnamefont
  {Xun}\ \bibnamefont {Gao}}, \bibinfo {author} {\bibfnamefont {Pedro}\
  \bibnamefont {Sales~Rodriguez}}, \bibinfo {author} {\bibfnamefont {Thomas}\
  \bibnamefont {Karolyshyn}}, \bibinfo {author} {\bibfnamefont {Giulia}\
  \bibnamefont {Semeghini}}, \bibinfo {author} {\bibfnamefont {Michael~J.}\
  \bibnamefont {Gullans}}, \bibinfo {author} {\bibfnamefont {Markus}\
  \bibnamefont {Greiner}}, \bibinfo {author} {\bibfnamefont {Vladan}\
  \bibnamefont {Vuletić}}, \ and\ \bibinfo {author} {\bibfnamefont
  {Mikhail~D.}\ \bibnamefont {Lukin}},\ }\bibfield  {title} {\enquote {\bibinfo
  {title} {Logical quantum processor based on reconfigurable atom arrays},}\
  }\href {\doibase 10.1038/s41586-023-06927-3} {\bibfield  {journal} {\bibinfo
  {journal} {Nature}\ }\textbf {\bibinfo {volume} {626}},\ \bibinfo {pages}
  {58--65} (\bibinfo {year} {2024})}\BibitemShut {NoStop}%
\bibitem [{\citenamefont {Bedalov}\ \emph {et~al.}(2024)\citenamefont
  {Bedalov}, \citenamefont {Blakely}, \citenamefont {Buttler}, \citenamefont
  {Carnahan}, \citenamefont {Chong}, \citenamefont {Chung}, \citenamefont
  {Cole}, \citenamefont {Goiporia}, \citenamefont {Gokhale}, \citenamefont
  {Heim}, \citenamefont {Hickman}, \citenamefont {Jones}, \citenamefont
  {Jones}, \citenamefont {Khalate}, \citenamefont {Kim}, \citenamefont {Kuper},
  \citenamefont {Lichtman}, \citenamefont {Lee}, \citenamefont {Mason},
  \citenamefont {Neff-Mallon}, \citenamefont {Noel}, \citenamefont {Omole},
  \citenamefont {Radnaev}, \citenamefont {Rines}, \citenamefont {Saffman},
  \citenamefont {Shabtai}, \citenamefont {Teo}, \citenamefont {Thotakura},
  \citenamefont {Tomesh},\ and\ \citenamefont
  {Tucker}}]{bedalov_fault-tolerant_2024}%
  \BibitemOpen
  \bibfield  {author} {\bibinfo {author} {\bibfnamefont {Matt~J.}\ \bibnamefont
  {Bedalov}}, \bibinfo {author} {\bibfnamefont {Matt}\ \bibnamefont {Blakely}},
  \bibinfo {author} {\bibfnamefont {Peter~D.}\ \bibnamefont {Buttler}},
  \bibinfo {author} {\bibfnamefont {Caitlin}\ \bibnamefont {Carnahan}},
  \bibinfo {author} {\bibfnamefont {Frederic~T.}\ \bibnamefont {Chong}},
  \bibinfo {author} {\bibfnamefont {Woo~Chang}\ \bibnamefont {Chung}}, \bibinfo
  {author} {\bibfnamefont {Dan~C.}\ \bibnamefont {Cole}}, \bibinfo {author}
  {\bibfnamefont {Palash}\ \bibnamefont {Goiporia}}, \bibinfo {author}
  {\bibfnamefont {Pranav}\ \bibnamefont {Gokhale}}, \bibinfo {author}
  {\bibfnamefont {Bettina}\ \bibnamefont {Heim}}, \bibinfo {author}
  {\bibfnamefont {Garrett~T.}\ \bibnamefont {Hickman}}, \bibinfo {author}
  {\bibfnamefont {Eric~B.}\ \bibnamefont {Jones}}, \bibinfo {author}
  {\bibfnamefont {Ryan~A.}\ \bibnamefont {Jones}}, \bibinfo {author}
  {\bibfnamefont {Pradnya}\ \bibnamefont {Khalate}}, \bibinfo {author}
  {\bibfnamefont {Jin-Sung}\ \bibnamefont {Kim}}, \bibinfo {author}
  {\bibfnamefont {Kevin~W.}\ \bibnamefont {Kuper}}, \bibinfo {author}
  {\bibfnamefont {Martin~T.}\ \bibnamefont {Lichtman}}, \bibinfo {author}
  {\bibfnamefont {Stephanie}\ \bibnamefont {Lee}}, \bibinfo {author}
  {\bibfnamefont {David}\ \bibnamefont {Mason}}, \bibinfo {author}
  {\bibfnamefont {Nathan~A.}\ \bibnamefont {Neff-Mallon}}, \bibinfo {author}
  {\bibfnamefont {Thomas~W.}\ \bibnamefont {Noel}}, \bibinfo {author}
  {\bibfnamefont {Victory}\ \bibnamefont {Omole}}, \bibinfo {author}
  {\bibfnamefont {Alexander~G.}\ \bibnamefont {Radnaev}}, \bibinfo {author}
  {\bibfnamefont {Rich}\ \bibnamefont {Rines}}, \bibinfo {author}
  {\bibfnamefont {Mark}\ \bibnamefont {Saffman}}, \bibinfo {author}
  {\bibfnamefont {Efrat}\ \bibnamefont {Shabtai}}, \bibinfo {author}
  {\bibfnamefont {Mariesa~H.}\ \bibnamefont {Teo}}, \bibinfo {author}
  {\bibfnamefont {Bharath}\ \bibnamefont {Thotakura}}, \bibinfo {author}
  {\bibfnamefont {Teague}\ \bibnamefont {Tomesh}}, \ and\ \bibinfo {author}
  {\bibfnamefont {Angela~K.}\ \bibnamefont {Tucker}},\ }\href {\doibase
  10.48550/arXiv.2412.07670} {\enquote {\bibinfo {title} {Fault-{Tolerant}
  {Operation} and {Materials} {Science} with {Neutral} {Atom} {Logical}
  {Qubits}},}\ } (\bibinfo {year} {2024}),\ \bibinfo {note}
  {arXiv:2412.07670}\BibitemShut {NoStop}%
\bibitem [{\citenamefont {Reichardt}\ \emph {et~al.}(2024)\citenamefont
  {Reichardt}, \citenamefont {Paetznick}, \citenamefont {Aasen}, \citenamefont
  {Basov}, \citenamefont {Bello-Rivas}, \citenamefont {Bonderson},
  \citenamefont {Chao}, \citenamefont {Dam}, \citenamefont {Hastings},
  \citenamefont {Paz}, \citenamefont {Silva}, \citenamefont {Sundaram},
  \citenamefont {Svore}, \citenamefont {Vaschillo}, \citenamefont {Wang},
  \citenamefont {Zanner}, \citenamefont {Cairncross}, \citenamefont {Chen},
  \citenamefont {Crow}, \citenamefont {Kim}, \citenamefont {Kindem},
  \citenamefont {King}, \citenamefont {McDonald}, \citenamefont {Norcia},
  \citenamefont {Ryou}, \citenamefont {Stone}, \citenamefont {Wadleigh},
  \citenamefont {Barnes}, \citenamefont {Battaglino}, \citenamefont
  {Bohdanowicz}, \citenamefont {Booth}, \citenamefont {Brown}, \citenamefont
  {Brown}, \citenamefont {Cassella}, \citenamefont {Coxe}, \citenamefont
  {Epstein}, \citenamefont {Feldkamp}, \citenamefont {Griger}, \citenamefont
  {Halperin}, \citenamefont {Heinz}, \citenamefont {Hummel}, \citenamefont
  {Jaffe}, \citenamefont {Jones}, \citenamefont {Kapit}, \citenamefont {Kotru},
  \citenamefont {Lauigan}, \citenamefont {Li}, \citenamefont {Marjanovic},
  \citenamefont {Megidish}, \citenamefont {Meredith}, \citenamefont {Morshead},
  \citenamefont {Muniz}, \citenamefont {Narayanaswami}, \citenamefont
  {Nishiguchi}, \citenamefont {Paule}, \citenamefont {Pawlak}, \citenamefont
  {Pudenz}, \citenamefont {Pérez}, \citenamefont {Simon}, \citenamefont
  {Smull}, \citenamefont {Stack}, \citenamefont {Urbanek}, \citenamefont
  {Veerdonk}, \citenamefont {Vendeiro}, \citenamefont {Weverka}, \citenamefont
  {Wilkason}, \citenamefont {Wu}, \citenamefont {Xie}, \citenamefont
  {Zalys-Geller}, \citenamefont {Zhang},\ and\ \citenamefont
  {Bloom}}]{reichardt_logical_2024}%
  \BibitemOpen
  \bibfield  {author} {\bibinfo {author} {\bibfnamefont {Ben~W.}\ \bibnamefont
  {Reichardt}}, \bibinfo {author} {\bibfnamefont {Adam}\ \bibnamefont
  {Paetznick}}, \bibinfo {author} {\bibfnamefont {David}\ \bibnamefont
  {Aasen}}, \bibinfo {author} {\bibfnamefont {Ivan}\ \bibnamefont {Basov}},
  \bibinfo {author} {\bibfnamefont {Juan~M.}\ \bibnamefont {Bello-Rivas}},
  \bibinfo {author} {\bibfnamefont {Parsa}\ \bibnamefont {Bonderson}}, \bibinfo
  {author} {\bibfnamefont {Rui}\ \bibnamefont {Chao}}, \bibinfo {author}
  {\bibfnamefont {Wim~van}\ \bibnamefont {Dam}}, \bibinfo {author}
  {\bibfnamefont {Matthew~B.}\ \bibnamefont {Hastings}}, \bibinfo {author}
  {\bibfnamefont {Andres}\ \bibnamefont {Paz}}, \bibinfo {author}
  {\bibfnamefont {Marcus P.~da}\ \bibnamefont {Silva}}, \bibinfo {author}
  {\bibfnamefont {Aarthi}\ \bibnamefont {Sundaram}}, \bibinfo {author}
  {\bibfnamefont {Krysta~M.}\ \bibnamefont {Svore}}, \bibinfo {author}
  {\bibfnamefont {Alexander}\ \bibnamefont {Vaschillo}}, \bibinfo {author}
  {\bibfnamefont {Zhenghan}\ \bibnamefont {Wang}}, \bibinfo {author}
  {\bibfnamefont {Matt}\ \bibnamefont {Zanner}}, \bibinfo {author}
  {\bibfnamefont {William~B.}\ \bibnamefont {Cairncross}}, \bibinfo {author}
  {\bibfnamefont {Cheng-An}\ \bibnamefont {Chen}}, \bibinfo {author}
  {\bibfnamefont {Daniel}\ \bibnamefont {Crow}}, \bibinfo {author}
  {\bibfnamefont {Hyosub}\ \bibnamefont {Kim}}, \bibinfo {author}
  {\bibfnamefont {Jonathan~M.}\ \bibnamefont {Kindem}}, \bibinfo {author}
  {\bibfnamefont {Jonathan}\ \bibnamefont {King}}, \bibinfo {author}
  {\bibfnamefont {Michael}\ \bibnamefont {McDonald}}, \bibinfo {author}
  {\bibfnamefont {Matthew~A.}\ \bibnamefont {Norcia}}, \bibinfo {author}
  {\bibfnamefont {Albert}\ \bibnamefont {Ryou}}, \bibinfo {author}
  {\bibfnamefont {Mark}\ \bibnamefont {Stone}}, \bibinfo {author}
  {\bibfnamefont {Laura}\ \bibnamefont {Wadleigh}}, \bibinfo {author}
  {\bibfnamefont {Katrina}\ \bibnamefont {Barnes}}, \bibinfo {author}
  {\bibfnamefont {Peter}\ \bibnamefont {Battaglino}}, \bibinfo {author}
  {\bibfnamefont {Thomas~C.}\ \bibnamefont {Bohdanowicz}}, \bibinfo {author}
  {\bibfnamefont {Graham}\ \bibnamefont {Booth}}, \bibinfo {author}
  {\bibfnamefont {Andrew}\ \bibnamefont {Brown}}, \bibinfo {author}
  {\bibfnamefont {Mark~O.}\ \bibnamefont {Brown}}, \bibinfo {author}
  {\bibfnamefont {Kayleigh}\ \bibnamefont {Cassella}}, \bibinfo {author}
  {\bibfnamefont {Robin}\ \bibnamefont {Coxe}}, \bibinfo {author}
  {\bibfnamefont {Jeffrey~M.}\ \bibnamefont {Epstein}}, \bibinfo {author}
  {\bibfnamefont {Max}\ \bibnamefont {Feldkamp}}, \bibinfo {author}
  {\bibfnamefont {Christopher}\ \bibnamefont {Griger}}, \bibinfo {author}
  {\bibfnamefont {Eli}\ \bibnamefont {Halperin}}, \bibinfo {author}
  {\bibfnamefont {Andre}\ \bibnamefont {Heinz}}, \bibinfo {author}
  {\bibfnamefont {Frederic}\ \bibnamefont {Hummel}}, \bibinfo {author}
  {\bibfnamefont {Matthew}\ \bibnamefont {Jaffe}}, \bibinfo {author}
  {\bibfnamefont {Antonia M.~W.}\ \bibnamefont {Jones}}, \bibinfo {author}
  {\bibfnamefont {Eliot}\ \bibnamefont {Kapit}}, \bibinfo {author}
  {\bibfnamefont {Krish}\ \bibnamefont {Kotru}}, \bibinfo {author}
  {\bibfnamefont {Joseph}\ \bibnamefont {Lauigan}}, \bibinfo {author}
  {\bibfnamefont {Ming}\ \bibnamefont {Li}}, \bibinfo {author} {\bibfnamefont
  {Jan}\ \bibnamefont {Marjanovic}}, \bibinfo {author} {\bibfnamefont {Eli}\
  \bibnamefont {Megidish}}, \bibinfo {author} {\bibfnamefont {Matthew}\
  \bibnamefont {Meredith}}, \bibinfo {author} {\bibfnamefont {Ryan}\
  \bibnamefont {Morshead}}, \bibinfo {author} {\bibfnamefont {Juan~A.}\
  \bibnamefont {Muniz}}, \bibinfo {author} {\bibfnamefont {Sandeep}\
  \bibnamefont {Narayanaswami}}, \bibinfo {author} {\bibfnamefont {Ciro}\
  \bibnamefont {Nishiguchi}}, \bibinfo {author} {\bibfnamefont {Timothy}\
  \bibnamefont {Paule}}, \bibinfo {author} {\bibfnamefont {Kelly~A.}\
  \bibnamefont {Pawlak}}, \bibinfo {author} {\bibfnamefont {Kristen~L.}\
  \bibnamefont {Pudenz}}, \bibinfo {author} {\bibfnamefont {David~Rodríguez}\
  \bibnamefont {Pérez}}, \bibinfo {author} {\bibfnamefont {Jon}\ \bibnamefont
  {Simon}}, \bibinfo {author} {\bibfnamefont {Aaron}\ \bibnamefont {Smull}},
  \bibinfo {author} {\bibfnamefont {Daniel}\ \bibnamefont {Stack}}, \bibinfo
  {author} {\bibfnamefont {Miroslav}\ \bibnamefont {Urbanek}}, \bibinfo
  {author} {\bibfnamefont {René J. M. van~de}\ \bibnamefont {Veerdonk}},
  \bibinfo {author} {\bibfnamefont {Zachary}\ \bibnamefont {Vendeiro}},
  \bibinfo {author} {\bibfnamefont {Robert~T.}\ \bibnamefont {Weverka}},
  \bibinfo {author} {\bibfnamefont {Thomas}\ \bibnamefont {Wilkason}}, \bibinfo
  {author} {\bibfnamefont {Tsung-Yao}\ \bibnamefont {Wu}}, \bibinfo {author}
  {\bibfnamefont {Xin}\ \bibnamefont {Xie}}, \bibinfo {author} {\bibfnamefont
  {Evan}\ \bibnamefont {Zalys-Geller}}, \bibinfo {author} {\bibfnamefont
  {Xiaogang}\ \bibnamefont {Zhang}}, \ and\ \bibinfo {author} {\bibfnamefont
  {Benjamin~J.}\ \bibnamefont {Bloom}},\ }\href {\doibase
  10.48550/arXiv.2411.11822} {\enquote {\bibinfo {title} {Logical computation
  demonstrated with a neutral atom quantum processor},}\ } (\bibinfo {year}
  {2024}),\ \bibinfo {note} {arXiv:2411.11822}\BibitemShut {NoStop}%
\bibitem [{\citenamefont {Rodriguez}\ \emph {et~al.}(2024)\citenamefont
  {Rodriguez}, \citenamefont {Robinson}, \citenamefont {Jepsen}, \citenamefont
  {He}, \citenamefont {Duckering}, \citenamefont {Zhao}, \citenamefont {Wu},
  \citenamefont {Campo}, \citenamefont {Bagnall}, \citenamefont {Kwon},
  \citenamefont {Karolyshyn}, \citenamefont {Weinberg}, \citenamefont {Cain},
  \citenamefont {Evered}, \citenamefont {Geim}, \citenamefont {Kalinowski},
  \citenamefont {Li}, \citenamefont {Manovitz}, \citenamefont {Amato-Grill},
  \citenamefont {Basham}, \citenamefont {Bernstein}, \citenamefont {Braverman},
  \citenamefont {Bylinskii}, \citenamefont {Choukri}, \citenamefont {DeAngelo},
  \citenamefont {Fang}, \citenamefont {Fieweger}, \citenamefont {Frederick},
  \citenamefont {Haines}, \citenamefont {Hamdan}, \citenamefont {Hammett},
  \citenamefont {Hsu}, \citenamefont {Hu}, \citenamefont {Huber}, \citenamefont
  {Jia}, \citenamefont {Kedar}, \citenamefont {Kornjača}, \citenamefont {Liu},
  \citenamefont {Long}, \citenamefont {Lopatin}, \citenamefont {Lopes},
  \citenamefont {Luo}, \citenamefont {Macrì}, \citenamefont {Marković},
  \citenamefont {Martínez-Martínez}, \citenamefont {Meng}, \citenamefont
  {Ostermann}, \citenamefont {Ostroumov}, \citenamefont {Paquette},
  \citenamefont {Qiang}, \citenamefont {Shofman}, \citenamefont {Singh},
  \citenamefont {Singh}, \citenamefont {Sinha}, \citenamefont {Thoreen},
  \citenamefont {Wan}, \citenamefont {Wang}, \citenamefont {Waxman-Lenz},
  \citenamefont {Wong}, \citenamefont {Wurtz}, \citenamefont {Zhdanov},
  \citenamefont {Zheng}, \citenamefont {Greiner}, \citenamefont {Keesling},
  \citenamefont {Gemelke}, \citenamefont {Vuletić}, \citenamefont {Kitagawa},
  \citenamefont {Wang}, \citenamefont {Bluvstein}, \citenamefont {Lukin},
  \citenamefont {Lukin}, \citenamefont {Zhou},\ and\ \citenamefont
  {Cantú}}]{rodriguez_experimental_2024}%
  \BibitemOpen
  \bibfield  {author} {\bibinfo {author} {\bibfnamefont {Pedro~Sales}\
  \bibnamefont {Rodriguez}}, \bibinfo {author} {\bibfnamefont {John~M.}\
  \bibnamefont {Robinson}}, \bibinfo {author} {\bibfnamefont {Paul~Niklas}\
  \bibnamefont {Jepsen}}, \bibinfo {author} {\bibfnamefont {Zhiyang}\
  \bibnamefont {He}}, \bibinfo {author} {\bibfnamefont {Casey}\ \bibnamefont
  {Duckering}}, \bibinfo {author} {\bibfnamefont {Chen}\ \bibnamefont {Zhao}},
  \bibinfo {author} {\bibfnamefont {Kai-Hsin}\ \bibnamefont {Wu}}, \bibinfo
  {author} {\bibfnamefont {Joseph}\ \bibnamefont {Campo}}, \bibinfo {author}
  {\bibfnamefont {Kevin}\ \bibnamefont {Bagnall}}, \bibinfo {author}
  {\bibfnamefont {Minho}\ \bibnamefont {Kwon}}, \bibinfo {author}
  {\bibfnamefont {Thomas}\ \bibnamefont {Karolyshyn}}, \bibinfo {author}
  {\bibfnamefont {Phillip}\ \bibnamefont {Weinberg}}, \bibinfo {author}
  {\bibfnamefont {Madelyn}\ \bibnamefont {Cain}}, \bibinfo {author}
  {\bibfnamefont {Simon~J.}\ \bibnamefont {Evered}}, \bibinfo {author}
  {\bibfnamefont {Alexandra~A.}\ \bibnamefont {Geim}}, \bibinfo {author}
  {\bibfnamefont {Marcin}\ \bibnamefont {Kalinowski}}, \bibinfo {author}
  {\bibfnamefont {Sophie~H.}\ \bibnamefont {Li}}, \bibinfo {author}
  {\bibfnamefont {Tom}\ \bibnamefont {Manovitz}}, \bibinfo {author}
  {\bibfnamefont {Jesse}\ \bibnamefont {Amato-Grill}}, \bibinfo {author}
  {\bibfnamefont {James~I.}\ \bibnamefont {Basham}}, \bibinfo {author}
  {\bibfnamefont {Liane}\ \bibnamefont {Bernstein}}, \bibinfo {author}
  {\bibfnamefont {Boris}\ \bibnamefont {Braverman}}, \bibinfo {author}
  {\bibfnamefont {Alexei}\ \bibnamefont {Bylinskii}}, \bibinfo {author}
  {\bibfnamefont {Adam}\ \bibnamefont {Choukri}}, \bibinfo {author}
  {\bibfnamefont {Robert}\ \bibnamefont {DeAngelo}}, \bibinfo {author}
  {\bibfnamefont {Fang}\ \bibnamefont {Fang}}, \bibinfo {author} {\bibfnamefont
  {Connor}\ \bibnamefont {Fieweger}}, \bibinfo {author} {\bibfnamefont {Paige}\
  \bibnamefont {Frederick}}, \bibinfo {author} {\bibfnamefont {David}\
  \bibnamefont {Haines}}, \bibinfo {author} {\bibfnamefont {Majd}\ \bibnamefont
  {Hamdan}}, \bibinfo {author} {\bibfnamefont {Julian}\ \bibnamefont
  {Hammett}}, \bibinfo {author} {\bibfnamefont {Ning}\ \bibnamefont {Hsu}},
  \bibinfo {author} {\bibfnamefont {Ming-Guang}\ \bibnamefont {Hu}}, \bibinfo
  {author} {\bibfnamefont {Florian}\ \bibnamefont {Huber}}, \bibinfo {author}
  {\bibfnamefont {Ningyuan}\ \bibnamefont {Jia}}, \bibinfo {author}
  {\bibfnamefont {Dhruv}\ \bibnamefont {Kedar}}, \bibinfo {author}
  {\bibfnamefont {Milan}\ \bibnamefont {Kornjača}}, \bibinfo {author}
  {\bibfnamefont {Fangli}\ \bibnamefont {Liu}}, \bibinfo {author}
  {\bibfnamefont {John}\ \bibnamefont {Long}}, \bibinfo {author} {\bibfnamefont
  {Jonathan}\ \bibnamefont {Lopatin}}, \bibinfo {author} {\bibfnamefont {Pedro
  L.~S.}\ \bibnamefont {Lopes}}, \bibinfo {author} {\bibfnamefont {Xiu-Zhe}\
  \bibnamefont {Luo}}, \bibinfo {author} {\bibfnamefont {Tommaso}\ \bibnamefont
  {Macrì}}, \bibinfo {author} {\bibfnamefont {Ognjen}\ \bibnamefont
  {Marković}}, \bibinfo {author} {\bibfnamefont {Luis~A.}\ \bibnamefont
  {Martínez-Martínez}}, \bibinfo {author} {\bibfnamefont {Xianmei}\
  \bibnamefont {Meng}}, \bibinfo {author} {\bibfnamefont {Stefan}\ \bibnamefont
  {Ostermann}}, \bibinfo {author} {\bibfnamefont {Evgeny}\ \bibnamefont
  {Ostroumov}}, \bibinfo {author} {\bibfnamefont {David}\ \bibnamefont
  {Paquette}}, \bibinfo {author} {\bibfnamefont {Zexuan}\ \bibnamefont
  {Qiang}}, \bibinfo {author} {\bibfnamefont {Vadim}\ \bibnamefont {Shofman}},
  \bibinfo {author} {\bibfnamefont {Anshuman}\ \bibnamefont {Singh}}, \bibinfo
  {author} {\bibfnamefont {Manuj}\ \bibnamefont {Singh}}, \bibinfo {author}
  {\bibfnamefont {Nandan}\ \bibnamefont {Sinha}}, \bibinfo {author}
  {\bibfnamefont {Henry}\ \bibnamefont {Thoreen}}, \bibinfo {author}
  {\bibfnamefont {Noel}\ \bibnamefont {Wan}}, \bibinfo {author} {\bibfnamefont
  {Yiping}\ \bibnamefont {Wang}}, \bibinfo {author} {\bibfnamefont {Daniel}\
  \bibnamefont {Waxman-Lenz}}, \bibinfo {author} {\bibfnamefont {Tak}\
  \bibnamefont {Wong}}, \bibinfo {author} {\bibfnamefont {Jonathan}\
  \bibnamefont {Wurtz}}, \bibinfo {author} {\bibfnamefont {Andrii}\
  \bibnamefont {Zhdanov}}, \bibinfo {author} {\bibfnamefont {Laurent}\
  \bibnamefont {Zheng}}, \bibinfo {author} {\bibfnamefont {Markus}\
  \bibnamefont {Greiner}}, \bibinfo {author} {\bibfnamefont {Alexander}\
  \bibnamefont {Keesling}}, \bibinfo {author} {\bibfnamefont {Nathan}\
  \bibnamefont {Gemelke}}, \bibinfo {author} {\bibfnamefont {Vladan}\
  \bibnamefont {Vuletić}}, \bibinfo {author} {\bibfnamefont {Takuya}\
  \bibnamefont {Kitagawa}}, \bibinfo {author} {\bibfnamefont {Sheng-Tao}\
  \bibnamefont {Wang}}, \bibinfo {author} {\bibfnamefont {Dolev}\ \bibnamefont
  {Bluvstein}}, \bibinfo {author} {\bibfnamefont {Mikhail~D.}\ \bibnamefont
  {Lukin}}, \bibinfo {author} {\bibfnamefont {Alexander}\ \bibnamefont
  {Lukin}}, \bibinfo {author} {\bibfnamefont {Hengyun}\ \bibnamefont {Zhou}}, \
  and\ \bibinfo {author} {\bibfnamefont {Sergio~H.}\ \bibnamefont {Cantú}},\
  }\href {\doibase 10.48550/arXiv.2412.15165} {\enquote {\bibinfo {title}
  {Experimental {Demonstration} of {Logical} {Magic} {State} {Distillation}},}\
  } (\bibinfo {year} {2024}),\ \bibinfo {note} {arXiv:2412.15165}\BibitemShut
  {NoStop}%
\bibitem [{\citenamefont {Hong}\ \emph {et~al.}(2024)\citenamefont {Hong},
  \citenamefont {Durso-Sabina}, \citenamefont {Hayes},\ and\ \citenamefont
  {Lucas}}]{hong_entangling_2024}%
  \BibitemOpen
  \bibfield  {author} {\bibinfo {author} {\bibfnamefont {Yifan}\ \bibnamefont
  {Hong}}, \bibinfo {author} {\bibfnamefont {Elijah}\ \bibnamefont
  {Durso-Sabina}}, \bibinfo {author} {\bibfnamefont {David}\ \bibnamefont
  {Hayes}}, \ and\ \bibinfo {author} {\bibfnamefont {Andrew}\ \bibnamefont
  {Lucas}},\ }\href {\doibase 10.48550/arXiv.2406.02666} {\enquote {\bibinfo
  {title} {Entangling four logical qubits beyond break-even in a nonlocal
  code},}\ } (\bibinfo {year} {2024}),\ \bibinfo {note}
  {arXiv:2406.02666}\BibitemShut {NoStop}%
\bibitem [{\citenamefont {Iyer}\ and\ \citenamefont
  {Poulin}(2015)}]{iyer_hardness_2015}%
  \BibitemOpen
  \bibfield  {author} {\bibinfo {author} {\bibfnamefont {Pavithran}\
  \bibnamefont {Iyer}}\ and\ \bibinfo {author} {\bibfnamefont {David}\
  \bibnamefont {Poulin}},\ }\bibfield  {title} {\enquote {\bibinfo {title}
  {Hardness of {Decoding} {Quantum} {Stabilizer} {Codes}},}\ }\href {\doibase
  10.1109/TIT.2015.2422294} {\bibfield  {journal} {\bibinfo  {journal} {IEEE
  Transactions on Information Theory}\ }\textbf {\bibinfo {volume} {61}},\
  \bibinfo {pages} {5209--5223} (\bibinfo {year} {2015})}\BibitemShut {NoStop}%
\bibitem [{\citenamefont {Fuentes}\ \emph {et~al.}(2021)\citenamefont
  {Fuentes}, \citenamefont {Etxezarreta~Martinez}, \citenamefont {Crespo},\
  and\ \citenamefont {Garcia-Frías}}]{fuentes_degeneracy_2021}%
  \BibitemOpen
  \bibfield  {author} {\bibinfo {author} {\bibfnamefont {Patricio}\
  \bibnamefont {Fuentes}}, \bibinfo {author} {\bibfnamefont {Josu}\
  \bibnamefont {Etxezarreta~Martinez}}, \bibinfo {author} {\bibfnamefont
  {Pedro~M.}\ \bibnamefont {Crespo}}, \ and\ \bibinfo {author} {\bibfnamefont
  {Javier}\ \bibnamefont {Garcia-Frías}},\ }\bibfield  {title} {\enquote
  {\bibinfo {title} {Degeneracy and {Its} {Impact} on the {Decoding} of
  {Sparse} {Quantum} {Codes}},}\ }\href {\doibase 10.1109/ACCESS.2021.3089829}
  {\bibfield  {journal} {\bibinfo  {journal} {IEEE Access}\ }\textbf {\bibinfo
  {volume} {9}},\ \bibinfo {pages} {89093--89119} (\bibinfo {year}
  {2021})}\BibitemShut {NoStop}%
\bibitem [{\citenamefont {Schumacher}\ and\ \citenamefont
  {Nielsen}(1996)}]{schumacher_quantum_1996}%
  \BibitemOpen
  \bibfield  {author} {\bibinfo {author} {\bibfnamefont {Benjamin}\
  \bibnamefont {Schumacher}}\ and\ \bibinfo {author} {\bibfnamefont {M.~A.}\
  \bibnamefont {Nielsen}},\ }\bibfield  {title} {\enquote {\bibinfo {title}
  {Quantum data processing and error correction},}\ }\href {\doibase
  10.1103/PhysRevA.54.2629} {\bibfield  {journal} {\bibinfo  {journal}
  {Physical Review A}\ }\textbf {\bibinfo {volume} {54}},\ \bibinfo {pages}
  {2629--2635} (\bibinfo {year} {1996})}\BibitemShut {NoStop}%
\bibitem [{\citenamefont {Fan}\ \emph {et~al.}(2024)\citenamefont {Fan},
  \citenamefont {Bao}, \citenamefont {Altman},\ and\ \citenamefont
  {Vishwanath}}]{fan_diagnostics_2024}%
  \BibitemOpen
  \bibfield  {author} {\bibinfo {author} {\bibfnamefont {Ruihua}\ \bibnamefont
  {Fan}}, \bibinfo {author} {\bibfnamefont {Yimu}\ \bibnamefont {Bao}},
  \bibinfo {author} {\bibfnamefont {Ehud}\ \bibnamefont {Altman}}, \ and\
  \bibinfo {author} {\bibfnamefont {Ashvin}\ \bibnamefont {Vishwanath}},\
  }\bibfield  {title} {\enquote {\bibinfo {title} {Diagnostics of
  {Mixed}-{State} {Topological} {Order} and {Breakdown} of {Quantum}
  {Memory}},}\ }\href {\doibase 10.1103/PRXQuantum.5.020343} {\bibfield
  {journal} {\bibinfo  {journal} {PRX Quantum}\ }\textbf {\bibinfo {volume}
  {5}},\ \bibinfo {pages} {020343} (\bibinfo {year} {2024})}\BibitemShut
  {NoStop}%
\bibitem [{\citenamefont {Colmenarez}\ \emph {et~al.}(2024)\citenamefont
  {Colmenarez}, \citenamefont {Huang}, \citenamefont {Diehl},\ and\
  \citenamefont {Müller}}]{colmenarez_accurate_2024}%
  \BibitemOpen
  \bibfield  {author} {\bibinfo {author} {\bibfnamefont {Luis}\ \bibnamefont
  {Colmenarez}}, \bibinfo {author} {\bibfnamefont {Ze-Min}\ \bibnamefont
  {Huang}}, \bibinfo {author} {\bibfnamefont {Sebastian}\ \bibnamefont
  {Diehl}}, \ and\ \bibinfo {author} {\bibfnamefont {Markus}\ \bibnamefont
  {Müller}},\ }\bibfield  {title} {\enquote {\bibinfo {title} {Accurate
  optimal quantum error correction thresholds from coherent information},}\
  }\href {\doibase 10.1103/PhysRevResearch.6.L042014} {\bibfield  {journal}
  {\bibinfo  {journal} {Physical Review Research}\ }\textbf {\bibinfo {volume}
  {6}},\ \bibinfo {pages} {L042014} (\bibinfo {year} {2024})}\BibitemShut
  {NoStop}%
\bibitem [{\citenamefont {Wu}\ \emph {et~al.}(2022)\citenamefont {Wu},
  \citenamefont {Kolkowitz}, \citenamefont {Puri},\ and\ \citenamefont
  {Thompson}}]{wu_erasure_2022}%
  \BibitemOpen
  \bibfield  {author} {\bibinfo {author} {\bibfnamefont {Yue}\ \bibnamefont
  {Wu}}, \bibinfo {author} {\bibfnamefont {Shimon}\ \bibnamefont {Kolkowitz}},
  \bibinfo {author} {\bibfnamefont {Shruti}\ \bibnamefont {Puri}}, \ and\
  \bibinfo {author} {\bibfnamefont {Jeff~D.}\ \bibnamefont {Thompson}},\
  }\bibfield  {title} {\enquote {\bibinfo {title} {Erasure conversion for
  fault-tolerant quantum computing in alkaline earth {Rydberg} atom arrays},}\
  }\href {\doibase 10.1038/s41467-022-32094-6} {\bibfield  {journal} {\bibinfo
  {journal} {Nature Communications}\ }\textbf {\bibinfo {volume} {13}},\
  \bibinfo {pages} {4657} (\bibinfo {year} {2022})}\BibitemShut {NoStop}%
\bibitem [{\citenamefont {Kang}\ \emph {et~al.}(2023)\citenamefont {Kang},
  \citenamefont {Campbell},\ and\ \citenamefont {Brown}}]{kang_quantum_2023}%
  \BibitemOpen
  \bibfield  {author} {\bibinfo {author} {\bibfnamefont {Mingyu}\ \bibnamefont
  {Kang}}, \bibinfo {author} {\bibfnamefont {Wesley~C.}\ \bibnamefont
  {Campbell}}, \ and\ \bibinfo {author} {\bibfnamefont {Kenneth~R.}\
  \bibnamefont {Brown}},\ }\bibfield  {title} {\enquote {\bibinfo {title}
  {Quantum {Error} {Correction} with {Metastable} {States} of {Trapped} {Ions}
  {Using} {Erasure} {Conversion}},}\ }\href {\doibase
  10.1103/PRXQuantum.4.020358} {\bibfield  {journal} {\bibinfo  {journal} {PRX
  Quantum}\ }\textbf {\bibinfo {volume} {4}},\ \bibinfo {pages} {020358}
  (\bibinfo {year} {2023})}\BibitemShut {NoStop}%
\bibitem [{\citenamefont {Kubica}\ \emph {et~al.}(2023)\citenamefont {Kubica},
  \citenamefont {Haim}, \citenamefont {Vaknin}, \citenamefont {Levine},
  \citenamefont {Brandão},\ and\ \citenamefont
  {Retzker}}]{kubica_erasure_2023}%
  \BibitemOpen
  \bibfield  {author} {\bibinfo {author} {\bibfnamefont {Aleksander}\
  \bibnamefont {Kubica}}, \bibinfo {author} {\bibfnamefont {Arbel}\
  \bibnamefont {Haim}}, \bibinfo {author} {\bibfnamefont {Yotam}\ \bibnamefont
  {Vaknin}}, \bibinfo {author} {\bibfnamefont {Harry}\ \bibnamefont {Levine}},
  \bibinfo {author} {\bibfnamefont {Fernando}\ \bibnamefont {Brandão}}, \ and\
  \bibinfo {author} {\bibfnamefont {Alex}\ \bibnamefont {Retzker}},\ }\bibfield
   {title} {\enquote {\bibinfo {title} {Erasure {Qubits}: {Overcoming} the
  {$T_1$} {Limit} in {Superconducting} {Circuits}},}\ }\href {\doibase
  10.1103/PhysRevX.13.041022} {\bibfield  {journal} {\bibinfo  {journal}
  {Physical Review X}\ }\textbf {\bibinfo {volume} {13}},\ \bibinfo {pages}
  {041022} (\bibinfo {year} {2023})}\BibitemShut {NoStop}%
\bibitem [{\citenamefont {Vala}\ \emph {et~al.}(2005)\citenamefont {Vala},
  \citenamefont {Whaley},\ and\ \citenamefont {Weiss}}]{vala_quantum_2005}%
  \BibitemOpen
  \bibfield  {author} {\bibinfo {author} {\bibfnamefont {J.}~\bibnamefont
  {Vala}}, \bibinfo {author} {\bibfnamefont {K.~B.}\ \bibnamefont {Whaley}}, \
  and\ \bibinfo {author} {\bibfnamefont {D.~S.}\ \bibnamefont {Weiss}},\
  }\bibfield  {title} {\enquote {\bibinfo {title} {Quantum error correction of
  a qubit loss in an addressable atomic system},}\ }\href {\doibase
  10.1103/PhysRevA.72.052318} {\bibfield  {journal} {\bibinfo  {journal}
  {Physical Review A}\ }\textbf {\bibinfo {volume} {72}},\ \bibinfo {pages}
  {052318} (\bibinfo {year} {2005})}\BibitemShut {NoStop}%
\bibitem [{\citenamefont {Fujii}\ and\ \citenamefont
  {Tokunaga}(2012)}]{fujii_error_2012}%
  \BibitemOpen
  \bibfield  {author} {\bibinfo {author} {\bibfnamefont {Keisuke}\ \bibnamefont
  {Fujii}}\ and\ \bibinfo {author} {\bibfnamefont {Yuuki}\ \bibnamefont
  {Tokunaga}},\ }\bibfield  {title} {\enquote {\bibinfo {title} {Error and loss
  tolerances of surface codes with general lattice structures},}\ }\href
  {\doibase 10.1103/PhysRevA.86.020303} {\bibfield  {journal} {\bibinfo
  {journal} {Physical Review A}\ }\textbf {\bibinfo {volume} {86}},\ \bibinfo
  {pages} {020303} (\bibinfo {year} {2012})}\BibitemShut {NoStop}%
\bibitem [{\citenamefont {Morley-Short}\ \emph {et~al.}(2019)\citenamefont
  {Morley-Short}, \citenamefont {Gimeno-Segovia}, \citenamefont {Rudolph},\
  and\ \citenamefont {Cable}}]{morley-short_loss-tolerant_2019}%
  \BibitemOpen
  \bibfield  {author} {\bibinfo {author} {\bibfnamefont {Sam}\ \bibnamefont
  {Morley-Short}}, \bibinfo {author} {\bibfnamefont {Mercedes}\ \bibnamefont
  {Gimeno-Segovia}}, \bibinfo {author} {\bibfnamefont {Terry}\ \bibnamefont
  {Rudolph}}, \ and\ \bibinfo {author} {\bibfnamefont {Hugo}\ \bibnamefont
  {Cable}},\ }\bibfield  {title} {\enquote {\bibinfo {title} {Loss-tolerant
  teleportation on large stabilizer states},}\ }\href {\doibase
  10.1088/2058-9565/aaf6c4} {\bibfield  {journal} {\bibinfo  {journal} {Quantum
  Science and Technology}\ }\textbf {\bibinfo {volume} {4}},\ \bibinfo {pages}
  {025014} (\bibinfo {year} {2019})}\BibitemShut {NoStop}%
\bibitem [{\citenamefont {Stace}\ \emph {et~al.}(2009)\citenamefont {Stace},
  \citenamefont {Barrett},\ and\ \citenamefont
  {Doherty}}]{stace_thresholds_2009}%
  \BibitemOpen
  \bibfield  {author} {\bibinfo {author} {\bibfnamefont {Thomas~M.}\
  \bibnamefont {Stace}}, \bibinfo {author} {\bibfnamefont {Sean~D.}\
  \bibnamefont {Barrett}}, \ and\ \bibinfo {author} {\bibfnamefont {Andrew~C.}\
  \bibnamefont {Doherty}},\ }\bibfield  {title} {\enquote {\bibinfo {title}
  {Thresholds for {Topological} {Codes} in the {Presence} of {Loss}},}\ }\href
  {\doibase 10.1103/PhysRevLett.102.200501} {\bibfield  {journal} {\bibinfo
  {journal} {Physical Review Letters}\ }\textbf {\bibinfo {volume} {102}},\
  \bibinfo {pages} {200501} (\bibinfo {year} {2009})}\BibitemShut {NoStop}%
\bibitem [{\citenamefont {Vodola}\ \emph {et~al.}(2018)\citenamefont {Vodola},
  \citenamefont {Amaro}, \citenamefont {Martin-Delgado},\ and\ \citenamefont
  {Müller}}]{vodola_twins_2018}%
  \BibitemOpen
  \bibfield  {author} {\bibinfo {author} {\bibfnamefont {Davide}\ \bibnamefont
  {Vodola}}, \bibinfo {author} {\bibfnamefont {David}\ \bibnamefont {Amaro}},
  \bibinfo {author} {\bibfnamefont {Miguel~Angel}\ \bibnamefont
  {Martin-Delgado}}, \ and\ \bibinfo {author} {\bibfnamefont {Markus}\
  \bibnamefont {Müller}},\ }\bibfield  {title} {\enquote {\bibinfo {title}
  {Twins {Percolation} for {Qubit} {Losses} in {Topological} {Color}
  {Codes}},}\ }\href {\doibase 10.1103/PhysRevLett.121.060501} {\bibfield
  {journal} {\bibinfo  {journal} {Physical Review Letters}\ }\textbf {\bibinfo
  {volume} {121}},\ \bibinfo {pages} {060501} (\bibinfo {year}
  {2018})}\BibitemShut {NoStop}%
\bibitem [{\citenamefont {Amaro}\ \emph {et~al.}(2020)\citenamefont {Amaro},
  \citenamefont {Bennett}, \citenamefont {Vodola},\ and\ \citenamefont
  {Müller}}]{amaro_analytical_2020}%
  \BibitemOpen
  \bibfield  {author} {\bibinfo {author} {\bibfnamefont {David}\ \bibnamefont
  {Amaro}}, \bibinfo {author} {\bibfnamefont {Jemma}\ \bibnamefont {Bennett}},
  \bibinfo {author} {\bibfnamefont {Davide}\ \bibnamefont {Vodola}}, \ and\
  \bibinfo {author} {\bibfnamefont {Markus}\ \bibnamefont {Müller}},\
  }\bibfield  {title} {\enquote {\bibinfo {title} {Analytical percolation
  theory for topological color codes under qubit loss},}\ }\href {\doibase
  10.1103/PhysRevA.101.032317} {\bibfield  {journal} {\bibinfo  {journal}
  {Physical Review A}\ }\textbf {\bibinfo {volume} {101}},\ \bibinfo {pages}
  {032317} (\bibinfo {year} {2020})}\BibitemShut {NoStop}%
\bibitem [{\citenamefont {Varbanov}\ \emph {et~al.}(2020)\citenamefont
  {Varbanov}, \citenamefont {Battistel}, \citenamefont {Tarasinski},
  \citenamefont {Ostroukh}, \citenamefont {O’Brien}, \citenamefont
  {DiCarlo},\ and\ \citenamefont {Terhal}}]{varbanov_leakage_2020}%
  \BibitemOpen
  \bibfield  {author} {\bibinfo {author} {\bibfnamefont {Boris~Mihailov}\
  \bibnamefont {Varbanov}}, \bibinfo {author} {\bibfnamefont {Francesco}\
  \bibnamefont {Battistel}}, \bibinfo {author} {\bibfnamefont {Brian~Michael}\
  \bibnamefont {Tarasinski}}, \bibinfo {author} {\bibfnamefont
  {Viacheslav~Petrovych}\ \bibnamefont {Ostroukh}}, \bibinfo {author}
  {\bibfnamefont {Thomas~Eugene}\ \bibnamefont {O’Brien}}, \bibinfo {author}
  {\bibfnamefont {Leonardo}\ \bibnamefont {DiCarlo}}, \ and\ \bibinfo {author}
  {\bibfnamefont {Barbara~Maria}\ \bibnamefont {Terhal}},\ }\bibfield  {title}
  {\enquote {\bibinfo {title} {Leakage detection for a transmon-based surface
  code},}\ }\href {\doibase 10.1038/s41534-020-00330-w} {\bibfield  {journal}
  {\bibinfo  {journal} {npj Quantum Information}\ }\textbf {\bibinfo {volume}
  {6}},\ \bibinfo {pages} {1--13} (\bibinfo {year} {2020})}\BibitemShut
  {NoStop}%
\bibitem [{\citenamefont {Miao}\ \emph {et~al.}(2023)\citenamefont {Miao},
  \citenamefont {McEwen}, \citenamefont {Atalaya}, \citenamefont {Kafri},
  \citenamefont {Pryadko}, \citenamefont {Bengtsson}, \citenamefont {Opremcak},
  \citenamefont {Satzinger}, \citenamefont {Chen}, \citenamefont {Klimov},
  \citenamefont {Quintana}, \citenamefont {Acharya}, \citenamefont {Anderson},
  \citenamefont {Ansmann}, \citenamefont {Arute}, \citenamefont {Arya},
  \citenamefont {Asfaw}, \citenamefont {Bardin}, \citenamefont {Bourassa},
  \citenamefont {Bovaird}, \citenamefont {Brill}, \citenamefont {Buckley},
  \citenamefont {Buell}, \citenamefont {Burger}, \citenamefont {Burkett},
  \citenamefont {Bushnell}, \citenamefont {Campero}, \citenamefont {Chiaro},
  \citenamefont {Collins}, \citenamefont {Conner}, \citenamefont {Crook},
  \citenamefont {Curtin}, \citenamefont {Debroy}, \citenamefont {Demura},
  \citenamefont {Dunsworth}, \citenamefont {Erickson}, \citenamefont {Fatemi},
  \citenamefont {Ferreira}, \citenamefont {Burgos}, \citenamefont {Forati},
  \citenamefont {Fowler}, \citenamefont {Foxen}, \citenamefont {Garcia},
  \citenamefont {Giang}, \citenamefont {Gidney}, \citenamefont {Giustina},
  \citenamefont {Gosula}, \citenamefont {Dau}, \citenamefont {Gross},
  \citenamefont {Hamilton}, \citenamefont {Harrington}, \citenamefont {Heu},
  \citenamefont {Hilton}, \citenamefont {Hoffmann}, \citenamefont {Hong},
  \citenamefont {Huang}, \citenamefont {Huff}, \citenamefont {Iveland},
  \citenamefont {Jeffrey}, \citenamefont {Jiang}, \citenamefont {Jones},
  \citenamefont {Kelly}, \citenamefont {Kim}, \citenamefont {Kostritsa},
  \citenamefont {Kreikebaum}, \citenamefont {Landhuis}, \citenamefont {Laptev},
  \citenamefont {Laws}, \citenamefont {Lee}, \citenamefont {Lester},
  \citenamefont {Lill}, \citenamefont {Liu}, \citenamefont {Locharla},
  \citenamefont {Lucero}, \citenamefont {Martin}, \citenamefont {Megrant},
  \citenamefont {Mi}, \citenamefont {Montazeri}, \citenamefont {Morvan},
  \citenamefont {Naaman}, \citenamefont {Neeley}, \citenamefont {Neill},
  \citenamefont {Nersisyan}, \citenamefont {Newman}, \citenamefont {Ng},
  \citenamefont {Nguyen}, \citenamefont {Nguyen}, \citenamefont {Potter},
  \citenamefont {Rocque}, \citenamefont {Roushan}, \citenamefont
  {Sankaragomathi}, \citenamefont {Schurkus}, \citenamefont {Schuster},
  \citenamefont {Shearn}, \citenamefont {Shorter}, \citenamefont {Shutty},
  \citenamefont {Shvarts}, \citenamefont {Skruzny}, \citenamefont {Smith},
  \citenamefont {Sterling}, \citenamefont {Szalay}, \citenamefont {Thor},
  \citenamefont {Torres}, \citenamefont {White}, \citenamefont {Woo},
  \citenamefont {Yao}, \citenamefont {Yeh}, \citenamefont {Yoo}, \citenamefont
  {Young}, \citenamefont {Zalcman}, \citenamefont {Zhu}, \citenamefont
  {Zobrist}, \citenamefont {Neven}, \citenamefont {Smelyanskiy}, \citenamefont
  {Petukhov}, \citenamefont {Korotkov}, \citenamefont {Sank},\ and\
  \citenamefont {Chen}}]{miao_overcoming_2023}%
  \BibitemOpen
  \bibfield  {author} {\bibinfo {author} {\bibfnamefont {Kevin~C.}\
  \bibnamefont {Miao}}, \bibinfo {author} {\bibfnamefont {Matt}\ \bibnamefont
  {McEwen}}, \bibinfo {author} {\bibfnamefont {Juan}\ \bibnamefont {Atalaya}},
  \bibinfo {author} {\bibfnamefont {Dvir}\ \bibnamefont {Kafri}}, \bibinfo
  {author} {\bibfnamefont {Leonid~P.}\ \bibnamefont {Pryadko}}, \bibinfo
  {author} {\bibfnamefont {Andreas}\ \bibnamefont {Bengtsson}}, \bibinfo
  {author} {\bibfnamefont {Alex}\ \bibnamefont {Opremcak}}, \bibinfo {author}
  {\bibfnamefont {Kevin~J.}\ \bibnamefont {Satzinger}}, \bibinfo {author}
  {\bibfnamefont {Zijun}\ \bibnamefont {Chen}}, \bibinfo {author}
  {\bibfnamefont {Paul~V.}\ \bibnamefont {Klimov}}, \bibinfo {author}
  {\bibfnamefont {Chris}\ \bibnamefont {Quintana}}, \bibinfo {author}
  {\bibfnamefont {Rajeev}\ \bibnamefont {Acharya}}, \bibinfo {author}
  {\bibfnamefont {Kyle}\ \bibnamefont {Anderson}}, \bibinfo {author}
  {\bibfnamefont {Markus}\ \bibnamefont {Ansmann}}, \bibinfo {author}
  {\bibfnamefont {Frank}\ \bibnamefont {Arute}}, \bibinfo {author}
  {\bibfnamefont {Kunal}\ \bibnamefont {Arya}}, \bibinfo {author}
  {\bibfnamefont {Abraham}\ \bibnamefont {Asfaw}}, \bibinfo {author}
  {\bibfnamefont {Joseph~C.}\ \bibnamefont {Bardin}}, \bibinfo {author}
  {\bibfnamefont {Alexandre}\ \bibnamefont {Bourassa}}, \bibinfo {author}
  {\bibfnamefont {Jenna}\ \bibnamefont {Bovaird}}, \bibinfo {author}
  {\bibfnamefont {Leon}\ \bibnamefont {Brill}}, \bibinfo {author}
  {\bibfnamefont {Bob~B.}\ \bibnamefont {Buckley}}, \bibinfo {author}
  {\bibfnamefont {David~A.}\ \bibnamefont {Buell}}, \bibinfo {author}
  {\bibfnamefont {Tim}\ \bibnamefont {Burger}}, \bibinfo {author}
  {\bibfnamefont {Brian}\ \bibnamefont {Burkett}}, \bibinfo {author}
  {\bibfnamefont {Nicholas}\ \bibnamefont {Bushnell}}, \bibinfo {author}
  {\bibfnamefont {Juan}\ \bibnamefont {Campero}}, \bibinfo {author}
  {\bibfnamefont {Ben}\ \bibnamefont {Chiaro}}, \bibinfo {author}
  {\bibfnamefont {Roberto}\ \bibnamefont {Collins}}, \bibinfo {author}
  {\bibfnamefont {Paul}\ \bibnamefont {Conner}}, \bibinfo {author}
  {\bibfnamefont {Alexander~L.}\ \bibnamefont {Crook}}, \bibinfo {author}
  {\bibfnamefont {Ben}\ \bibnamefont {Curtin}}, \bibinfo {author}
  {\bibfnamefont {Dripto~M.}\ \bibnamefont {Debroy}}, \bibinfo {author}
  {\bibfnamefont {Sean}\ \bibnamefont {Demura}}, \bibinfo {author}
  {\bibfnamefont {Andrew}\ \bibnamefont {Dunsworth}}, \bibinfo {author}
  {\bibfnamefont {Catherine}\ \bibnamefont {Erickson}}, \bibinfo {author}
  {\bibfnamefont {Reza}\ \bibnamefont {Fatemi}}, \bibinfo {author}
  {\bibfnamefont {Vinicius~S.}\ \bibnamefont {Ferreira}}, \bibinfo {author}
  {\bibfnamefont {Leslie~Flores}\ \bibnamefont {Burgos}}, \bibinfo {author}
  {\bibfnamefont {Ebrahim}\ \bibnamefont {Forati}}, \bibinfo {author}
  {\bibfnamefont {Austin~G.}\ \bibnamefont {Fowler}}, \bibinfo {author}
  {\bibfnamefont {Brooks}\ \bibnamefont {Foxen}}, \bibinfo {author}
  {\bibfnamefont {Gonzalo}\ \bibnamefont {Garcia}}, \bibinfo {author}
  {\bibfnamefont {William}\ \bibnamefont {Giang}}, \bibinfo {author}
  {\bibfnamefont {Craig}\ \bibnamefont {Gidney}}, \bibinfo {author}
  {\bibfnamefont {Marissa}\ \bibnamefont {Giustina}}, \bibinfo {author}
  {\bibfnamefont {Raja}\ \bibnamefont {Gosula}}, \bibinfo {author}
  {\bibfnamefont {Alejandro~Grajales}\ \bibnamefont {Dau}}, \bibinfo {author}
  {\bibfnamefont {Jonathan~A.}\ \bibnamefont {Gross}}, \bibinfo {author}
  {\bibfnamefont {Michael~C.}\ \bibnamefont {Hamilton}}, \bibinfo {author}
  {\bibfnamefont {Sean~D.}\ \bibnamefont {Harrington}}, \bibinfo {author}
  {\bibfnamefont {Paula}\ \bibnamefont {Heu}}, \bibinfo {author} {\bibfnamefont
  {Jeremy}\ \bibnamefont {Hilton}}, \bibinfo {author} {\bibfnamefont
  {Markus~R.}\ \bibnamefont {Hoffmann}}, \bibinfo {author} {\bibfnamefont
  {Sabrina}\ \bibnamefont {Hong}}, \bibinfo {author} {\bibfnamefont {Trent}\
  \bibnamefont {Huang}}, \bibinfo {author} {\bibfnamefont {Ashley}\
  \bibnamefont {Huff}}, \bibinfo {author} {\bibfnamefont {Justin}\ \bibnamefont
  {Iveland}}, \bibinfo {author} {\bibfnamefont {Evan}\ \bibnamefont {Jeffrey}},
  \bibinfo {author} {\bibfnamefont {Zhang}\ \bibnamefont {Jiang}}, \bibinfo
  {author} {\bibfnamefont {Cody}\ \bibnamefont {Jones}}, \bibinfo {author}
  {\bibfnamefont {Julian}\ \bibnamefont {Kelly}}, \bibinfo {author}
  {\bibfnamefont {Seon}\ \bibnamefont {Kim}}, \bibinfo {author} {\bibfnamefont
  {Fedor}\ \bibnamefont {Kostritsa}}, \bibinfo {author} {\bibfnamefont
  {John~Mark}\ \bibnamefont {Kreikebaum}}, \bibinfo {author} {\bibfnamefont
  {David}\ \bibnamefont {Landhuis}}, \bibinfo {author} {\bibfnamefont {Pavel}\
  \bibnamefont {Laptev}}, \bibinfo {author} {\bibfnamefont {Lily}\ \bibnamefont
  {Laws}}, \bibinfo {author} {\bibfnamefont {Kenny}\ \bibnamefont {Lee}},
  \bibinfo {author} {\bibfnamefont {Brian~J.}\ \bibnamefont {Lester}}, \bibinfo
  {author} {\bibfnamefont {Alexander~T.}\ \bibnamefont {Lill}}, \bibinfo
  {author} {\bibfnamefont {Wayne}\ \bibnamefont {Liu}}, \bibinfo {author}
  {\bibfnamefont {Aditya}\ \bibnamefont {Locharla}}, \bibinfo {author}
  {\bibfnamefont {Erik}\ \bibnamefont {Lucero}}, \bibinfo {author}
  {\bibfnamefont {Steven}\ \bibnamefont {Martin}}, \bibinfo {author}
  {\bibfnamefont {Anthony}\ \bibnamefont {Megrant}}, \bibinfo {author}
  {\bibfnamefont {Xiao}\ \bibnamefont {Mi}}, \bibinfo {author} {\bibfnamefont
  {Shirin}\ \bibnamefont {Montazeri}}, \bibinfo {author} {\bibfnamefont
  {Alexis}\ \bibnamefont {Morvan}}, \bibinfo {author} {\bibfnamefont {Ofer}\
  \bibnamefont {Naaman}}, \bibinfo {author} {\bibfnamefont {Matthew}\
  \bibnamefont {Neeley}}, \bibinfo {author} {\bibfnamefont {Charles}\
  \bibnamefont {Neill}}, \bibinfo {author} {\bibfnamefont {Ani}\ \bibnamefont
  {Nersisyan}}, \bibinfo {author} {\bibfnamefont {Michael}\ \bibnamefont
  {Newman}}, \bibinfo {author} {\bibfnamefont {Jiun~How}\ \bibnamefont {Ng}},
  \bibinfo {author} {\bibfnamefont {Anthony}\ \bibnamefont {Nguyen}}, \bibinfo
  {author} {\bibfnamefont {Murray}\ \bibnamefont {Nguyen}}, \bibinfo {author}
  {\bibfnamefont {Rebecca}\ \bibnamefont {Potter}}, \bibinfo {author}
  {\bibfnamefont {Charles}\ \bibnamefont {Rocque}}, \bibinfo {author}
  {\bibfnamefont {Pedram}\ \bibnamefont {Roushan}}, \bibinfo {author}
  {\bibfnamefont {Kannan}\ \bibnamefont {Sankaragomathi}}, \bibinfo {author}
  {\bibfnamefont {Henry~F.}\ \bibnamefont {Schurkus}}, \bibinfo {author}
  {\bibfnamefont {Christopher}\ \bibnamefont {Schuster}}, \bibinfo {author}
  {\bibfnamefont {Michael~J.}\ \bibnamefont {Shearn}}, \bibinfo {author}
  {\bibfnamefont {Aaron}\ \bibnamefont {Shorter}}, \bibinfo {author}
  {\bibfnamefont {Noah}\ \bibnamefont {Shutty}}, \bibinfo {author}
  {\bibfnamefont {Vladimir}\ \bibnamefont {Shvarts}}, \bibinfo {author}
  {\bibfnamefont {Jindra}\ \bibnamefont {Skruzny}}, \bibinfo {author}
  {\bibfnamefont {W.~Clarke}\ \bibnamefont {Smith}}, \bibinfo {author}
  {\bibfnamefont {George}\ \bibnamefont {Sterling}}, \bibinfo {author}
  {\bibfnamefont {Marco}\ \bibnamefont {Szalay}}, \bibinfo {author}
  {\bibfnamefont {Douglas}\ \bibnamefont {Thor}}, \bibinfo {author}
  {\bibfnamefont {Alfredo}\ \bibnamefont {Torres}}, \bibinfo {author}
  {\bibfnamefont {Theodore}\ \bibnamefont {White}}, \bibinfo {author}
  {\bibfnamefont {Bryan W.~K.}\ \bibnamefont {Woo}}, \bibinfo {author}
  {\bibfnamefont {Z.~Jamie}\ \bibnamefont {Yao}}, \bibinfo {author}
  {\bibfnamefont {Ping}\ \bibnamefont {Yeh}}, \bibinfo {author} {\bibfnamefont
  {Juhwan}\ \bibnamefont {Yoo}}, \bibinfo {author} {\bibfnamefont {Grayson}\
  \bibnamefont {Young}}, \bibinfo {author} {\bibfnamefont {Adam}\ \bibnamefont
  {Zalcman}}, \bibinfo {author} {\bibfnamefont {Ningfeng}\ \bibnamefont {Zhu}},
  \bibinfo {author} {\bibfnamefont {Nicholas}\ \bibnamefont {Zobrist}},
  \bibinfo {author} {\bibfnamefont {Hartmut}\ \bibnamefont {Neven}}, \bibinfo
  {author} {\bibfnamefont {Vadim}\ \bibnamefont {Smelyanskiy}}, \bibinfo
  {author} {\bibfnamefont {Andre}\ \bibnamefont {Petukhov}}, \bibinfo {author}
  {\bibfnamefont {Alexander~N.}\ \bibnamefont {Korotkov}}, \bibinfo {author}
  {\bibfnamefont {Daniel}\ \bibnamefont {Sank}}, \ and\ \bibinfo {author}
  {\bibfnamefont {Yu}~\bibnamefont {Chen}},\ }\bibfield  {title} {\enquote
  {\bibinfo {title} {Overcoming leakage in quantum error correction},}\ }\href
  {\doibase 10.1038/s41567-023-02226-w} {\bibfield  {journal} {\bibinfo
  {journal} {Nature Physics}\ }\textbf {\bibinfo {volume} {19}},\ \bibinfo
  {pages} {1780--1786} (\bibinfo {year} {2023})}\BibitemShut {NoStop}%
\bibitem [{\citenamefont {Grassl}\ \emph {et~al.}(1997)\citenamefont {Grassl},
  \citenamefont {Beth},\ and\ \citenamefont {Pellizzari}}]{grassl_codes_1997}%
  \BibitemOpen
  \bibfield  {author} {\bibinfo {author} {\bibfnamefont {M.}~\bibnamefont
  {Grassl}}, \bibinfo {author} {\bibfnamefont {Th.}\ \bibnamefont {Beth}}, \
  and\ \bibinfo {author} {\bibfnamefont {T.}~\bibnamefont {Pellizzari}},\
  }\bibfield  {title} {\enquote {\bibinfo {title} {Codes for the quantum
  erasure channel},}\ }\href {\doibase 10.1103/PhysRevA.56.33} {\bibfield
  {journal} {\bibinfo  {journal} {Physical Review A}\ }\textbf {\bibinfo
  {volume} {56}},\ \bibinfo {pages} {33--38} (\bibinfo {year}
  {1997})}\BibitemShut {NoStop}%
\bibitem [{\citenamefont {Auger}\ \emph {et~al.}(2017)\citenamefont {Auger},
  \citenamefont {Anwar}, \citenamefont {Gimeno-Segovia}, \citenamefont
  {Stace},\ and\ \citenamefont {Browne}}]{auger_fault-tolerance_2017}%
  \BibitemOpen
  \bibfield  {author} {\bibinfo {author} {\bibfnamefont {James~M.}\
  \bibnamefont {Auger}}, \bibinfo {author} {\bibfnamefont {Hussain}\
  \bibnamefont {Anwar}}, \bibinfo {author} {\bibfnamefont {Mercedes}\
  \bibnamefont {Gimeno-Segovia}}, \bibinfo {author} {\bibfnamefont {Thomas~M.}\
  \bibnamefont {Stace}}, \ and\ \bibinfo {author} {\bibfnamefont {Dan~E.}\
  \bibnamefont {Browne}},\ }\bibfield  {title} {\enquote {\bibinfo {title}
  {Fault-tolerance thresholds for the surface code with fabrication errors},}\
  }\href {\doibase 10.1103/PhysRevA.96.042316} {\bibfield  {journal} {\bibinfo
  {journal} {Physical Review A}\ }\textbf {\bibinfo {volume} {96}},\ \bibinfo
  {pages} {042316} (\bibinfo {year} {2017})}\BibitemShut {NoStop}%
\bibitem [{\citenamefont {Delfosse}\ and\ \citenamefont
  {Nickerson}(2021)}]{delfosse_almost-linear_2021}%
  \BibitemOpen
  \bibfield  {author} {\bibinfo {author} {\bibfnamefont {Nicolas}\ \bibnamefont
  {Delfosse}}\ and\ \bibinfo {author} {\bibfnamefont {Naomi~H.}\ \bibnamefont
  {Nickerson}},\ }\bibfield  {title} {\enquote {\bibinfo {title} {Almost-linear
  time decoding algorithm for topological codes},}\ }\href {\doibase
  10.22331/q-2021-12-02-595} {\bibfield  {journal} {\bibinfo  {journal}
  {Quantum}\ }\textbf {\bibinfo {volume} {5}},\ \bibinfo {pages} {595}
  (\bibinfo {year} {2021})}\BibitemShut {NoStop}%
\bibitem [{\citenamefont {Goto}\ and\ \citenamefont
  {Uchikawa}(2014)}]{goto_soft-decision_2014}%
  \BibitemOpen
  \bibfield  {author} {\bibinfo {author} {\bibfnamefont {Hayato}\ \bibnamefont
  {Goto}}\ and\ \bibinfo {author} {\bibfnamefont {Hironori}\ \bibnamefont
  {Uchikawa}},\ }\bibfield  {title} {\enquote {\bibinfo {title} {Soft-decision
  decoder for quantum erasure and probabilistic-gate error models},}\ }\href
  {\doibase 10.1103/PhysRevA.89.022322} {\bibfield  {journal} {\bibinfo
  {journal} {Physical Review A}\ }\textbf {\bibinfo {volume} {89}},\ \bibinfo
  {pages} {022322} (\bibinfo {year} {2014})}\BibitemShut {NoStop}%
\bibitem [{\citenamefont {McLauchlan}\ \emph {et~al.}(2024)\citenamefont
  {McLauchlan}, \citenamefont {Gehér},\ and\ \citenamefont
  {Moylett}}]{mclauchlan_accommodating_2024}%
  \BibitemOpen
  \bibfield  {author} {\bibinfo {author} {\bibfnamefont {Campbell}\
  \bibnamefont {McLauchlan}}, \bibinfo {author} {\bibfnamefont {György~P.}\
  \bibnamefont {Gehér}}, \ and\ \bibinfo {author} {\bibfnamefont
  {Alexandra~E.}\ \bibnamefont {Moylett}},\ }\href
  {http://arxiv.org/abs/2405.15854} {\enquote {\bibinfo {title} {Accommodating
  {Fabrication} {Defects} on {Floquet} {Codes} with {Minimal} {Hardware}
  {Requirements}},}\ } (\bibinfo {year} {2024}),\ \bibinfo {note}
  {arXiv:2405.15854}\BibitemShut {NoStop}%
\bibitem [{\citenamefont {Nagayama}\ \emph {et~al.}(2017)\citenamefont
  {Nagayama}, \citenamefont {Fowler}, \citenamefont {Horsman}, \citenamefont
  {Devitt},\ and\ \citenamefont {Meter}}]{nagayama_surface_2017}%
  \BibitemOpen
  \bibfield  {author} {\bibinfo {author} {\bibfnamefont {Shota}\ \bibnamefont
  {Nagayama}}, \bibinfo {author} {\bibfnamefont {Austin~G.}\ \bibnamefont
  {Fowler}}, \bibinfo {author} {\bibfnamefont {Dominic}\ \bibnamefont
  {Horsman}}, \bibinfo {author} {\bibfnamefont {Simon~J.}\ \bibnamefont
  {Devitt}}, \ and\ \bibinfo {author} {\bibfnamefont {Rodney~Van}\ \bibnamefont
  {Meter}},\ }\bibfield  {title} {\enquote {\bibinfo {title} {Surface code
  error correction on a defective lattice},}\ }\href {\doibase
  10.1088/1367-2630/aa5918} {\bibfield  {journal} {\bibinfo  {journal} {New
  Journal of Physics}\ }\textbf {\bibinfo {volume} {19}},\ \bibinfo {pages}
  {023050} (\bibinfo {year} {2017})}\BibitemShut {NoStop}%
\bibitem [{\citenamefont {Siegel}\ \emph {et~al.}(2023)\citenamefont {Siegel},
  \citenamefont {Strikis}, \citenamefont {Flatters},\ and\ \citenamefont
  {Benjamin}}]{siegel_adaptive_2023}%
  \BibitemOpen
  \bibfield  {author} {\bibinfo {author} {\bibfnamefont {Adam}\ \bibnamefont
  {Siegel}}, \bibinfo {author} {\bibfnamefont {Armands}\ \bibnamefont
  {Strikis}}, \bibinfo {author} {\bibfnamefont {Thomas}\ \bibnamefont
  {Flatters}}, \ and\ \bibinfo {author} {\bibfnamefont {Simon}\ \bibnamefont
  {Benjamin}},\ }\bibfield  {title} {\enquote {\bibinfo {title} {Adaptive
  surface code for quantum error correction in the presence of temporary or
  permanent defects},}\ }\href {\doibase 10.22331/q-2023-07-25-1065} {\bibfield
   {journal} {\bibinfo  {journal} {Quantum}\ }\textbf {\bibinfo {volume} {7}},\
  \bibinfo {pages} {1065} (\bibinfo {year} {2023})}\BibitemShut {NoStop}%
\bibitem [{\citenamefont {Steinberg}\ \emph {et~al.}(2024)\citenamefont
  {Steinberg}, \citenamefont {Fan}, \citenamefont {Harris}, \citenamefont
  {Elkouss}, \citenamefont {Feld},\ and\ \citenamefont
  {Jahn}}]{steinberg_far_2024}%
  \BibitemOpen
  \bibfield  {author} {\bibinfo {author} {\bibfnamefont {Matthew}\ \bibnamefont
  {Steinberg}}, \bibinfo {author} {\bibfnamefont {Junyu}\ \bibnamefont {Fan}},
  \bibinfo {author} {\bibfnamefont {Robert~J.}\ \bibnamefont {Harris}},
  \bibinfo {author} {\bibfnamefont {David}\ \bibnamefont {Elkouss}}, \bibinfo
  {author} {\bibfnamefont {Sebastian}\ \bibnamefont {Feld}}, \ and\ \bibinfo
  {author} {\bibfnamefont {Alexander}\ \bibnamefont {Jahn}},\ }\href {\doibase
  10.48550/arXiv.2407.11926} {\enquote {\bibinfo {title} {Far from {Perfect}:
  {Quantum} {Error} {Correction} with ({Hyperinvariant}) {Evenbly} {Codes}},}\
  } (\bibinfo {year} {2024}),\ \bibinfo {note} {arXiv:2407.11926}\BibitemShut
  {NoStop}%
\bibitem [{\citenamefont {Vodola}\ \emph {et~al.}(2022)\citenamefont {Vodola},
  \citenamefont {Rispler}, \citenamefont {Kim},\ and\ \citenamefont
  {Müller}}]{vodola_fundamental_2022}%
  \BibitemOpen
  \bibfield  {author} {\bibinfo {author} {\bibfnamefont {Davide}\ \bibnamefont
  {Vodola}}, \bibinfo {author} {\bibfnamefont {Manuel}\ \bibnamefont
  {Rispler}}, \bibinfo {author} {\bibfnamefont {Seyong}\ \bibnamefont {Kim}}, \
  and\ \bibinfo {author} {\bibfnamefont {Markus}\ \bibnamefont {Müller}},\
  }\bibfield  {title} {\enquote {\bibinfo {title} {Fundamental thresholds of
  realistic quantum error correction circuits from classical spin models},}\
  }\href {\doibase 10.22331/q-2022-01-05-618} {\bibfield  {journal} {\bibinfo
  {journal} {Quantum}\ }\textbf {\bibinfo {volume} {6}},\ \bibinfo {pages}
  {618} (\bibinfo {year} {2022})}\BibitemShut {NoStop}%
\bibitem [{\citenamefont {Stricker}\ \emph {et~al.}(2020)\citenamefont
  {Stricker}, \citenamefont {Vodola}, \citenamefont {Erhard}, \citenamefont
  {Postler}, \citenamefont {Meth}, \citenamefont {Ringbauer}, \citenamefont
  {Schindler}, \citenamefont {Monz}, \citenamefont {Müller},\ and\
  \citenamefont {Blatt}}]{stricker_experimental_2020}%
  \BibitemOpen
  \bibfield  {author} {\bibinfo {author} {\bibfnamefont {Roman}\ \bibnamefont
  {Stricker}}, \bibinfo {author} {\bibfnamefont {Davide}\ \bibnamefont
  {Vodola}}, \bibinfo {author} {\bibfnamefont {Alexander}\ \bibnamefont
  {Erhard}}, \bibinfo {author} {\bibfnamefont {Lukas}\ \bibnamefont {Postler}},
  \bibinfo {author} {\bibfnamefont {Michael}\ \bibnamefont {Meth}}, \bibinfo
  {author} {\bibfnamefont {Martin}\ \bibnamefont {Ringbauer}}, \bibinfo
  {author} {\bibfnamefont {Philipp}\ \bibnamefont {Schindler}}, \bibinfo
  {author} {\bibfnamefont {Thomas}\ \bibnamefont {Monz}}, \bibinfo {author}
  {\bibfnamefont {Markus}\ \bibnamefont {Müller}}, \ and\ \bibinfo {author}
  {\bibfnamefont {Rainer}\ \bibnamefont {Blatt}},\ }\bibfield  {title}
  {\enquote {\bibinfo {title} {Experimental deterministic correction of qubit
  loss},}\ }\href {\doibase 10.1038/s41586-020-2667-0} {\bibfield  {journal}
  {\bibinfo  {journal} {Nature}\ }\textbf {\bibinfo {volume} {585}},\ \bibinfo
  {pages} {207--210} (\bibinfo {year} {2020})}\BibitemShut {NoStop}%
\bibitem [{\citenamefont {Delfosse}\ and\ \citenamefont
  {Zémor}(2020)}]{delfosse_linear-time_2020}%
  \BibitemOpen
  \bibfield  {author} {\bibinfo {author} {\bibfnamefont {Nicolas}\ \bibnamefont
  {Delfosse}}\ and\ \bibinfo {author} {\bibfnamefont {Gilles}\ \bibnamefont
  {Zémor}},\ }\bibfield  {title} {\enquote {\bibinfo {title} {Linear-time
  maximum likelihood decoding of surface codes over the quantum erasure
  channel},}\ }\href {\doibase 10.1103/PhysRevResearch.2.033042} {\bibfield
  {journal} {\bibinfo  {journal} {Physical Review Research}\ }\textbf {\bibinfo
  {volume} {2}},\ \bibinfo {pages} {033042} (\bibinfo {year}
  {2020})}\BibitemShut {NoStop}%
\bibitem [{\citenamefont {Bravyi}\ \emph {et~al.}(2014)\citenamefont {Bravyi},
  \citenamefont {Suchara},\ and\ \citenamefont
  {Vargo}}]{bravyi_efficient_2014}%
  \BibitemOpen
  \bibfield  {author} {\bibinfo {author} {\bibfnamefont {Sergey}\ \bibnamefont
  {Bravyi}}, \bibinfo {author} {\bibfnamefont {Martin}\ \bibnamefont
  {Suchara}}, \ and\ \bibinfo {author} {\bibfnamefont {Alexander}\ \bibnamefont
  {Vargo}},\ }\bibfield  {title} {\enquote {\bibinfo {title} {Efficient
  algorithms for maximum likelihood decoding in the surface code},}\ }\href
  {\doibase 10.1103/PhysRevA.90.032326} {\bibfield  {journal} {\bibinfo
  {journal} {Physical Review A}\ }\textbf {\bibinfo {volume} {90}},\ \bibinfo
  {pages} {032326} (\bibinfo {year} {2014})}\BibitemShut {NoStop}%
\bibitem [{\citenamefont {Chubb}(2021)}]{chubb_general_2021}%
  \BibitemOpen
  \bibfield  {author} {\bibinfo {author} {\bibfnamefont {Christopher~T.}\
  \bibnamefont {Chubb}},\ }\href {\doibase 10.48550/arXiv.2101.04125} {\enquote
  {\bibinfo {title} {General tensor network decoding of {2D} {Pauli} codes},}\
  } (\bibinfo {year} {2021}),\ \bibinfo {note} {arXiv:2101.04125}\BibitemShut
  {NoStop}%
\bibitem [{\citenamefont {Laflamme}\ \emph {et~al.}(1996)\citenamefont
  {Laflamme}, \citenamefont {Miquel}, \citenamefont {Paz},\ and\ \citenamefont
  {Zurek}}]{laflamme_perfect_1996}%
  \BibitemOpen
  \bibfield  {author} {\bibinfo {author} {\bibfnamefont {Raymond}\ \bibnamefont
  {Laflamme}}, \bibinfo {author} {\bibfnamefont {Cesar}\ \bibnamefont
  {Miquel}}, \bibinfo {author} {\bibfnamefont {Juan~Pablo}\ \bibnamefont
  {Paz}}, \ and\ \bibinfo {author} {\bibfnamefont {Wojciech~Hubert}\
  \bibnamefont {Zurek}},\ }\href {\doibase 10.48550/arXiv.quant-ph/9602019}
  {\enquote {\bibinfo {title} {Perfect {Quantum} {Error} {Correction}
  {Code}},}\ } (\bibinfo {year} {1996}),\ \bibinfo {note}
  {arXiv:quant-ph/9602019}\BibitemShut {NoStop}%
\bibitem [{\citenamefont {Sahay}\ \emph {et~al.}(2023)\citenamefont {Sahay},
  \citenamefont {Jin}, \citenamefont {Claes}, \citenamefont {Thompson},\ and\
  \citenamefont {Puri}}]{sahay_high-threshold_2023}%
  \BibitemOpen
  \bibfield  {author} {\bibinfo {author} {\bibfnamefont {Kaavya}\ \bibnamefont
  {Sahay}}, \bibinfo {author} {\bibfnamefont {Junlan}\ \bibnamefont {Jin}},
  \bibinfo {author} {\bibfnamefont {Jahan}\ \bibnamefont {Claes}}, \bibinfo
  {author} {\bibfnamefont {Jeff~D.}\ \bibnamefont {Thompson}}, \ and\ \bibinfo
  {author} {\bibfnamefont {Shruti}\ \bibnamefont {Puri}},\ }\bibfield  {title}
  {\enquote {\bibinfo {title} {High-{Threshold} {Codes} for {Neutral}-{Atom}
  {Qubits} with {Biased} {Erasure} {Errors}},}\ }\href {\doibase
  10.1103/PhysRevX.13.041013} {\bibfield  {journal} {\bibinfo  {journal}
  {Physical Review X}\ }\textbf {\bibinfo {volume} {13}},\ \bibinfo {pages}
  {041013} (\bibinfo {year} {2023})}\BibitemShut {NoStop}%
\bibitem [{\citenamefont {Quinn}\ \emph {et~al.}(2024)\citenamefont {Quinn},
  \citenamefont {Gregory}, \citenamefont {Moore}, \citenamefont {Brudney},
  \citenamefont {Metzner}, \citenamefont {Ritchie}, \citenamefont {O'Reilly},
  \citenamefont {Wineland},\ and\ \citenamefont
  {Allcock}}]{quinn_high-fidelity_2024}%
  \BibitemOpen
  \bibfield  {author} {\bibinfo {author} {\bibfnamefont {A.}~\bibnamefont
  {Quinn}}, \bibinfo {author} {\bibfnamefont {G.~J.}\ \bibnamefont {Gregory}},
  \bibinfo {author} {\bibfnamefont {I.~D.}\ \bibnamefont {Moore}}, \bibinfo
  {author} {\bibfnamefont {S.}~\bibnamefont {Brudney}}, \bibinfo {author}
  {\bibfnamefont {J.}~\bibnamefont {Metzner}}, \bibinfo {author} {\bibfnamefont
  {E.~R.}\ \bibnamefont {Ritchie}}, \bibinfo {author} {\bibfnamefont
  {J.}~\bibnamefont {O'Reilly}}, \bibinfo {author} {\bibfnamefont {D.~J.}\
  \bibnamefont {Wineland}}, \ and\ \bibinfo {author} {\bibfnamefont {D.~T.~C.}\
  \bibnamefont {Allcock}},\ }\href {\doibase 10.48550/arXiv.2411.12727}
  {\enquote {\bibinfo {title} {High-fidelity entanglement of metastable
  trapped-ion qubits with integrated erasure conversion},}\ } (\bibinfo {year}
  {2024}),\ \bibinfo {note} {arXiv:2411.12727}\BibitemShut {NoStop}%
\bibitem [{\citenamefont {Pastawski}\ and\ \citenamefont
  {Yoshida}(2015)}]{pastawski_fault-tolerant_2015}%
  \BibitemOpen
  \bibfield  {author} {\bibinfo {author} {\bibfnamefont {Fernando}\
  \bibnamefont {Pastawski}}\ and\ \bibinfo {author} {\bibfnamefont {Beni}\
  \bibnamefont {Yoshida}},\ }\bibfield  {title} {\enquote {\bibinfo {title}
  {Fault-tolerant logical gates in quantum error-correcting codes},}\ }\href
  {\doibase 10.1103/PhysRevA.91.012305} {\bibfield  {journal} {\bibinfo
  {journal} {Physical Review A}\ }\textbf {\bibinfo {volume} {91}},\ \bibinfo
  {pages} {012305} (\bibinfo {year} {2015})}\BibitemShut {NoStop}%
\bibitem [{\citenamefont {Pastawski}\ \emph {et~al.}(2015)\citenamefont
  {Pastawski}, \citenamefont {Yoshida}, \citenamefont {Harlow},\ and\
  \citenamefont {Preskill}}]{pastawski_holographic_2015}%
  \BibitemOpen
  \bibfield  {author} {\bibinfo {author} {\bibfnamefont {Fernando}\
  \bibnamefont {Pastawski}}, \bibinfo {author} {\bibfnamefont {Beni}\
  \bibnamefont {Yoshida}}, \bibinfo {author} {\bibfnamefont {Daniel}\
  \bibnamefont {Harlow}}, \ and\ \bibinfo {author} {\bibfnamefont {John}\
  \bibnamefont {Preskill}},\ }\bibfield  {title} {\enquote {\bibinfo {title}
  {Holographic quantum error-correcting codes: toy models for the bulk/boundary
  correspondence},}\ }\href {\doibase 10.1007/JHEP06(2015)149} {\bibfield
  {journal} {\bibinfo  {journal} {Journal of High Energy Physics}\ }\textbf
  {\bibinfo {volume} {2015}},\ \bibinfo {pages} {149} (\bibinfo {year}
  {2015})}\BibitemShut {NoStop}%
\bibitem [{\citenamefont {Dumer}\ \emph {et~al.}(2015)\citenamefont {Dumer},
  \citenamefont {Kovalev},\ and\ \citenamefont
  {Pryadko}}]{dumer_thresholds_2015}%
  \BibitemOpen
  \bibfield  {author} {\bibinfo {author} {\bibfnamefont {Ilya}\ \bibnamefont
  {Dumer}}, \bibinfo {author} {\bibfnamefont {Alexey~A.}\ \bibnamefont
  {Kovalev}}, \ and\ \bibinfo {author} {\bibfnamefont {Leonid~P.}\ \bibnamefont
  {Pryadko}},\ }\bibfield  {title} {\enquote {\bibinfo {title} {Thresholds for
  {Correcting} {Errors}, {Erasures}, and {Faulty} {Syndrome} {Measurements} in
  {Degenerate} {Quantum} {Codes}},}\ }\href {\doibase
  10.1103/PhysRevLett.115.050502} {\bibfield  {journal} {\bibinfo  {journal}
  {Physical Review Letters}\ }\textbf {\bibinfo {volume} {115}},\ \bibinfo
  {pages} {050502} (\bibinfo {year} {2015})}\BibitemShut {NoStop}%
\bibitem [{\citenamefont {Dennis}\ \emph {et~al.}(2002)\citenamefont {Dennis},
  \citenamefont {Kitaev}, \citenamefont {Landahl},\ and\ \citenamefont
  {Preskill}}]{dennis_topological_2002}%
  \BibitemOpen
  \bibfield  {author} {\bibinfo {author} {\bibfnamefont {Eric}\ \bibnamefont
  {Dennis}}, \bibinfo {author} {\bibfnamefont {Alexei}\ \bibnamefont {Kitaev}},
  \bibinfo {author} {\bibfnamefont {Andrew}\ \bibnamefont {Landahl}}, \ and\
  \bibinfo {author} {\bibfnamefont {John}\ \bibnamefont {Preskill}},\
  }\bibfield  {title} {\enquote {\bibinfo {title} {Topological quantum
  memory},}\ }\href {\doibase 10.1063/1.1499754} {\bibfield  {journal}
  {\bibinfo  {journal} {Journal of Mathematical Physics}\ }\textbf {\bibinfo
  {volume} {43}},\ \bibinfo {pages} {4452--4505} (\bibinfo {year}
  {2002})}\BibitemShut {NoStop}%
\bibitem [{\citenamefont {Delfosse}\ and\ \citenamefont
  {Zémor}(2013)}]{delfosse_upper_2013}%
  \BibitemOpen
  \bibfield  {author} {\bibinfo {author} {\bibfnamefont {Nicolas}\ \bibnamefont
  {Delfosse}}\ and\ \bibinfo {author} {\bibfnamefont {Gilles}\ \bibnamefont
  {Zémor}},\ }\bibfield  {title} {\enquote {\bibinfo {title} {Upper bounds on
  the rate of low density stabilizer codes for the quantum erasure channel},}\
  }\href {https://dl.acm.org/doi/10.5555/2535680.2535684} {\bibfield  {journal}
  {\bibinfo  {journal} {Quantum Info. Comput.}\ }\textbf {\bibinfo {volume}
  {13}},\ \bibinfo {pages} {793--826} (\bibinfo {year} {2013})}\BibitemShut
  {NoStop}%
\bibitem [{\citenamefont {Kovalev}\ and\ \citenamefont
  {Pryadko}(2013)}]{kovalev_fault_2013}%
  \BibitemOpen
  \bibfield  {author} {\bibinfo {author} {\bibfnamefont {Alexey~A.}\
  \bibnamefont {Kovalev}}\ and\ \bibinfo {author} {\bibfnamefont {Leonid~P.}\
  \bibnamefont {Pryadko}},\ }\bibfield  {title} {\enquote {\bibinfo {title}
  {Fault tolerance of quantum low-density parity check codes with sublinear
  distance scaling},}\ }\href {\doibase 10.1103/PhysRevA.87.020304} {\bibfield
  {journal} {\bibinfo  {journal} {Physical Review A}\ }\textbf {\bibinfo
  {volume} {87}},\ \bibinfo {pages} {020304} (\bibinfo {year}
  {2013})}\BibitemShut {NoStop}%
\bibitem [{\citenamefont {Woolls}\ and\ \citenamefont
  {Pryadko}(2022)}]{woolls_homology-changing_2022}%
  \BibitemOpen
  \bibfield  {author} {\bibinfo {author} {\bibfnamefont {Michael}\ \bibnamefont
  {Woolls}}\ and\ \bibinfo {author} {\bibfnamefont {Leonid~P.}\ \bibnamefont
  {Pryadko}},\ }\bibfield  {title} {\enquote {\bibinfo {title}
  {Homology-changing percolation transitions on finite graphs},}\ }\href
  {\doibase 10.1063/5.0036418} {\bibfield  {journal} {\bibinfo  {journal}
  {Journal of Mathematical Physics}\ }\textbf {\bibinfo {volume} {63}},\
  \bibinfo {pages} {023301} (\bibinfo {year} {2022})}\BibitemShut {NoStop}%
\bibitem [{\citenamefont {Ohzeki}(2012)}]{ohzeki_error_2012}%
  \BibitemOpen
  \bibfield  {author} {\bibinfo {author} {\bibfnamefont {Masayuki}\
  \bibnamefont {Ohzeki}},\ }\bibfield  {title} {\enquote {\bibinfo {title}
  {Error threshold estimates for surface code with loss of qubits},}\ }\href
  {\doibase 10.1103/PhysRevA.85.060301} {\bibfield  {journal} {\bibinfo
  {journal} {Physical Review A}\ }\textbf {\bibinfo {volume} {85}},\ \bibinfo
  {pages} {060301} (\bibinfo {year} {2012})}\BibitemShut {NoStop}%
\bibitem [{\citenamefont {Calderbank}\ and\ \citenamefont
  {Shor}(1996)}]{calderbank_good_1996}%
  \BibitemOpen
  \bibfield  {author} {\bibinfo {author} {\bibfnamefont {A.~R.}\ \bibnamefont
  {Calderbank}}\ and\ \bibinfo {author} {\bibfnamefont {Peter~W.}\ \bibnamefont
  {Shor}},\ }\bibfield  {title} {\enquote {\bibinfo {title} {Good quantum
  error-correcting codes exist},}\ }\href {\doibase 10.1103/PhysRevA.54.1098}
  {\bibfield  {journal} {\bibinfo  {journal} {Physical Review A}\ }\textbf
  {\bibinfo {volume} {54}},\ \bibinfo {pages} {1098--1105} (\bibinfo {year}
  {1996})}\BibitemShut {NoStop}%
\bibitem [{\citenamefont {Steane}(1997)}]{steane_multiple-particle_1997}%
  \BibitemOpen
  \bibfield  {author} {\bibinfo {author} {\bibfnamefont {Andrew}\ \bibnamefont
  {Steane}},\ }\bibfield  {title} {\enquote {\bibinfo {title}
  {Multiple-particle interference and quantum error correction},}\ }\href
  {\doibase 10.1098/rspa.1996.0136} {\bibfield  {journal} {\bibinfo  {journal}
  {Proceedings of the Royal Society of London. Series A: Mathematical, Physical
  and Engineering Sciences}\ }\textbf {\bibinfo {volume} {452}},\ \bibinfo
  {pages} {2551--2577} (\bibinfo {year} {1997})}\BibitemShut {NoStop}%
\bibitem [{\citenamefont {Stace}\ and\ \citenamefont
  {Barrett}(2010)}]{stace_error_2010}%
  \BibitemOpen
  \bibfield  {author} {\bibinfo {author} {\bibfnamefont {Thomas~M.}\
  \bibnamefont {Stace}}\ and\ \bibinfo {author} {\bibfnamefont {Sean~D.}\
  \bibnamefont {Barrett}},\ }\bibfield  {title} {\enquote {\bibinfo {title}
  {Error correction and degeneracy in surface codes suffering loss},}\ }\href
  {\doibase 10.1103/PhysRevA.81.022317} {\bibfield  {journal} {\bibinfo
  {journal} {Physical Review A}\ }\textbf {\bibinfo {volume} {81}},\ \bibinfo
  {pages} {022317} (\bibinfo {year} {2010})}\BibitemShut {NoStop}%
\bibitem [{\citenamefont {Chubb}\ and\ \citenamefont
  {Flammia}(2021)}]{chubb_statistical_2021}%
  \BibitemOpen
  \bibfield  {author} {\bibinfo {author} {\bibfnamefont {Christopher~T.}\
  \bibnamefont {Chubb}}\ and\ \bibinfo {author} {\bibfnamefont {Steven~T.}\
  \bibnamefont {Flammia}},\ }\bibfield  {title} {\enquote {\bibinfo {title}
  {Statistical mechanical models for quantum codes with correlated noise},}\
  }\href {\doibase 10.4171/aihpd/105} {\bibfield  {journal} {\bibinfo
  {journal} {Annales de l’Institut Henri Poincaré D}\ }\textbf {\bibinfo
  {volume} {8}},\ \bibinfo {pages} {269--321} (\bibinfo {year}
  {2021})}\BibitemShut {NoStop}%
\bibitem [{\citenamefont {Old}\ \emph {et~al.}(2024)\citenamefont {Old},
  \citenamefont {Rispler},\ and\ \citenamefont
  {Müller}}]{old_lift-connected_2024}%
  \BibitemOpen
  \bibfield  {author} {\bibinfo {author} {\bibfnamefont {Josias}\ \bibnamefont
  {Old}}, \bibinfo {author} {\bibfnamefont {Manuel}\ \bibnamefont {Rispler}}, \
  and\ \bibinfo {author} {\bibfnamefont {Markus}\ \bibnamefont {Müller}},\
  }\bibfield  {title} {\enquote {\bibinfo {title} {Lift-connected surface
  codes},}\ }\href {\doibase 10.1088/2058-9565/ad5eb6} {\bibfield  {journal}
  {\bibinfo  {journal} {Quantum Science and Technology}\ }\textbf {\bibinfo
  {volume} {9}},\ \bibinfo {pages} {045012} (\bibinfo {year} {2024})},\
  \bibinfo {note} {publisher: IOP Publishing}\BibitemShut {NoStop}%
\bibitem [{\citenamefont {Wang}\ \emph {et~al.}(2023)\citenamefont {Wang},
  \citenamefont {Wu},\ and\ \citenamefont {Wang}}]{wang_intrinsic_2023}%
  \BibitemOpen
  \bibfield  {author} {\bibinfo {author} {\bibfnamefont {Zijian}\ \bibnamefont
  {Wang}}, \bibinfo {author} {\bibfnamefont {Zhengzhi}\ \bibnamefont {Wu}}, \
  and\ \bibinfo {author} {\bibfnamefont {Zhong}\ \bibnamefont {Wang}},\ }\href
  {\doibase 10.48550/arXiv.2307.13758} {\enquote {\bibinfo {title} {Intrinsic
  {Mixed}-state {Topological} {Order} {Without} {Quantum} {Memory}},}\ }
  (\bibinfo {year} {2023}),\ \bibinfo {note} {arXiv:2307.13758}\BibitemShut
  {NoStop}%
\bibitem [{\citenamefont {Hauser}\ \emph {et~al.}(2024)\citenamefont {Hauser},
  \citenamefont {Bao}, \citenamefont {Sang}, \citenamefont {Lavasani},
  \citenamefont {Agrawal},\ and\ \citenamefont
  {Fisher}}]{hauser_information_2024}%
  \BibitemOpen
  \bibfield  {author} {\bibinfo {author} {\bibfnamefont {Jacob}\ \bibnamefont
  {Hauser}}, \bibinfo {author} {\bibfnamefont {Yimu}\ \bibnamefont {Bao}},
  \bibinfo {author} {\bibfnamefont {Shengqi}\ \bibnamefont {Sang}}, \bibinfo
  {author} {\bibfnamefont {Ali}\ \bibnamefont {Lavasani}}, \bibinfo {author}
  {\bibfnamefont {Utkarsh}\ \bibnamefont {Agrawal}}, \ and\ \bibinfo {author}
  {\bibfnamefont {Matthew P.~A.}\ \bibnamefont {Fisher}},\ }\href {\doibase
  10.48550/arXiv.2407.07882} {\enquote {\bibinfo {title} {Information dynamics
  in decohered quantum memory with repeated syndrome measurements: a dual
  approach},}\ } (\bibinfo {year} {2024}),\ \bibinfo {note} {arXiv:2407.07882
  [quant-ph]}\BibitemShut {NoStop}%
\bibitem [{\citenamefont {Su}\ \emph {et~al.}(2024)\citenamefont {Su},
  \citenamefont {Yang},\ and\ \citenamefont {Jian}}]{su_tapestry_2024}%
  \BibitemOpen
  \bibfield  {author} {\bibinfo {author} {\bibfnamefont {Kaixiang}\
  \bibnamefont {Su}}, \bibinfo {author} {\bibfnamefont {Zhou}\ \bibnamefont
  {Yang}}, \ and\ \bibinfo {author} {\bibfnamefont {Chao-Ming}\ \bibnamefont
  {Jian}},\ }\href {\doibase 10.48550/arXiv.2401.17359} {\enquote {\bibinfo
  {title} {Tapestry of dualities in decohered quantum error correction
  codes},}\ } (\bibinfo {year} {2024}),\ \bibinfo {note}
  {arXiv:2401.17359}\BibitemShut {NoStop}%
\bibitem [{\citenamefont {Lyons}(2024)}]{lyons_understanding_2024}%
  \BibitemOpen
  \bibfield  {author} {\bibinfo {author} {\bibfnamefont {Anasuya}\ \bibnamefont
  {Lyons}},\ }\href {\doibase 10.48550/arXiv.2403.03955} {\enquote {\bibinfo
  {title} {Understanding {Stabilizer} {Codes} {Under} {Local} {Decoherence}
  {Through} a {General} {Statistical} {Mechanics} {Mapping}},}\ } (\bibinfo
  {year} {2024}),\ \bibinfo {note} {arXiv:2403.03955}\BibitemShut {NoStop}%
\bibitem [{\citenamefont {Zhao}\ and\ \citenamefont
  {Liu}(2023)}]{zhao_extracting_2023}%
  \BibitemOpen
  \bibfield  {author} {\bibinfo {author} {\bibfnamefont {Yuanchen}\
  \bibnamefont {Zhao}}\ and\ \bibinfo {author} {\bibfnamefont {Dong~E.}\
  \bibnamefont {Liu}},\ }\href {\doibase 10.48550/arXiv.2312.16991} {\enquote
  {\bibinfo {title} {Extracting {Error} {Thresholds} through the {Framework} of
  {Approximate} {Quantum} {Error} {Correction} {Condition}},}\ } (\bibinfo
  {year} {2023}),\ \bibinfo {note} {arXiv:2312.16991}\BibitemShut {NoStop}%
\bibitem [{\citenamefont {Lee}\ \emph {et~al.}(2023)\citenamefont {Lee},
  \citenamefont {Jian},\ and\ \citenamefont {Xu}}]{lee_quantum_2023}%
  \BibitemOpen
  \bibfield  {author} {\bibinfo {author} {\bibfnamefont {Jong~Yeon}\
  \bibnamefont {Lee}}, \bibinfo {author} {\bibfnamefont {Chao-Ming}\
  \bibnamefont {Jian}}, \ and\ \bibinfo {author} {\bibfnamefont {Cenke}\
  \bibnamefont {Xu}},\ }\bibfield  {title} {\enquote {\bibinfo {title} {Quantum
  {Criticality} {Under} {Decoherence} or {Weak} {Measurement}},}\ }\href
  {\doibase 10.1103/PRXQuantum.4.030317} {\bibfield  {journal} {\bibinfo
  {journal} {PRX Quantum}\ }\textbf {\bibinfo {volume} {4}},\ \bibinfo {pages}
  {030317} (\bibinfo {year} {2023})}\BibitemShut {NoStop}%
\bibitem [{\citenamefont {Bao}\ \emph {et~al.}(2023)\citenamefont {Bao},
  \citenamefont {Fan}, \citenamefont {Vishwanath},\ and\ \citenamefont
  {Altman}}]{bao_mixed-state_2023}%
  \BibitemOpen
  \bibfield  {author} {\bibinfo {author} {\bibfnamefont {Yimu}\ \bibnamefont
  {Bao}}, \bibinfo {author} {\bibfnamefont {Ruihua}\ \bibnamefont {Fan}},
  \bibinfo {author} {\bibfnamefont {Ashvin}\ \bibnamefont {Vishwanath}}, \ and\
  \bibinfo {author} {\bibfnamefont {Ehud}\ \bibnamefont {Altman}},\ }\href
  {\doibase 10.48550/arXiv.2301.05687} {\enquote {\bibinfo {title} {Mixed-state
  topological order and the errorfield double formulation of
  decoherence-induced transitions},}\ } (\bibinfo {year} {2023}),\ \bibinfo
  {note} {arXiv:2301.05687}\BibitemShut {NoStop}%
\bibitem [{\citenamefont {Sang}\ and\ \citenamefont
  {Hsieh}(2024)}]{sang_stability_2024}%
  \BibitemOpen
  \bibfield  {author} {\bibinfo {author} {\bibfnamefont {Shengqi}\ \bibnamefont
  {Sang}}\ and\ \bibinfo {author} {\bibfnamefont {Timothy~H.}\ \bibnamefont
  {Hsieh}},\ }\href {\doibase 10.48550/arXiv.2404.07251} {\enquote {\bibinfo
  {title} {Stability of mixed-state quantum phases via finite {Markov}
  length},}\ } (\bibinfo {year} {2024}),\ \bibinfo {note}
  {arXiv:2404.07251}\BibitemShut {NoStop}%
\bibitem [{\citenamefont {Eckstein}\ \emph {et~al.}(2024)\citenamefont
  {Eckstein}, \citenamefont {Han}, \citenamefont {Trebst},\ and\ \citenamefont
  {Zhu}}]{eckstein_robust_2024}%
  \BibitemOpen
  \bibfield  {author} {\bibinfo {author} {\bibfnamefont {Finn}\ \bibnamefont
  {Eckstein}}, \bibinfo {author} {\bibfnamefont {Bo}~\bibnamefont {Han}},
  \bibinfo {author} {\bibfnamefont {Simon}\ \bibnamefont {Trebst}}, \ and\
  \bibinfo {author} {\bibfnamefont {Guo-Yi}\ \bibnamefont {Zhu}},\ }\bibfield
  {title} {\enquote {\bibinfo {title} {Robust {Teleportation} of a {Surface}
  {Code} and {Cascade} of {Topological} {Quantum} {Phase} {Transitions}},}\
  }\href {\doibase 10.1103/PRXQuantum.5.040313} {\bibfield  {journal} {\bibinfo
   {journal} {PRX Quantum}\ }\textbf {\bibinfo {volume} {5}},\ \bibinfo {pages}
  {040313} (\bibinfo {year} {2024})}\BibitemShut {NoStop}%
\bibitem [{\citenamefont {Bombin}\ \emph {et~al.}(2012)\citenamefont {Bombin},
  \citenamefont {Andrist}, \citenamefont {Ohzeki}, \citenamefont {Katzgraber},\
  and\ \citenamefont {Martin-Delgado}}]{bombin_strong_2012}%
  \BibitemOpen
  \bibfield  {author} {\bibinfo {author} {\bibfnamefont {H.}~\bibnamefont
  {Bombin}}, \bibinfo {author} {\bibfnamefont {Ruben~S.}\ \bibnamefont
  {Andrist}}, \bibinfo {author} {\bibfnamefont {Masayuki}\ \bibnamefont
  {Ohzeki}}, \bibinfo {author} {\bibfnamefont {Helmut~G.}\ \bibnamefont
  {Katzgraber}}, \ and\ \bibinfo {author} {\bibfnamefont {M.~A.}\ \bibnamefont
  {Martin-Delgado}},\ }\bibfield  {title} {\enquote {\bibinfo {title} {Strong
  {Resilience} of {Topological} {Codes} to {Depolarization}},}\ }\href
  {\doibase 10.1103/PhysRevX.2.021004} {\bibfield  {journal} {\bibinfo
  {journal} {Physical Review X}\ }\textbf {\bibinfo {volume} {2}},\ \bibinfo
  {pages} {021004} (\bibinfo {year} {2012})}\BibitemShut {NoStop}%
\bibitem [{\citenamefont {Stauffer}\ and\ \citenamefont
  {Aharony}(1992)}]{stauffer_introduction_1992}%
  \BibitemOpen
  \bibfield  {author} {\bibinfo {author} {\bibfnamefont {Dietrich}\
  \bibnamefont {Stauffer}}\ and\ \bibinfo {author} {\bibfnamefont {Amnon}\
  \bibnamefont {Aharony}},\ }\href@noop {} {\emph {\bibinfo {title}
  {Introduction to {Percolation} {Theory}}}}\ (\bibinfo  {publisher} {Taylor \&
  Francis},\ \bibinfo {year} {1992})\BibitemShut {NoStop}%
\bibitem [{\citenamefont {Gottesman}(1997)}]{gottesman_stabilizer_1997}%
  \BibitemOpen
  \bibfield  {author} {\bibinfo {author} {\bibfnamefont {Daniel}\ \bibnamefont
  {Gottesman}},\ }\href {\doibase 10.48550/arXiv.quant-ph/9705052} {\enquote
  {\bibinfo {title} {Stabilizer {Codes} and {Quantum} {Error} {Correction}},}\
  } (\bibinfo {year} {1997}),\ \bibinfo {note}
  {arXiv:quant-ph/9705052}\BibitemShut {NoStop}%
\bibitem [{\citenamefont {Kitaev}(2006)}]{kitaev_anyons_2006}%
  \BibitemOpen
  \bibfield  {author} {\bibinfo {author} {\bibfnamefont {Alexei}\ \bibnamefont
  {Kitaev}},\ }\bibfield  {title} {\enquote {\bibinfo {title} {Anyons in an
  exactly solved model and beyond},}\ }\href {\doibase
  10.1016/j.aop.2005.10.005} {\bibfield  {journal} {\bibinfo  {journal} {Annals
  of Physics}\ }\bibinfo {series} {January {Special} {Issue}},\ \textbf
  {\bibinfo {volume} {321}},\ \bibinfo {pages} {2--111} (\bibinfo {year}
  {2006})}\BibitemShut {NoStop}%
\bibitem [{\citenamefont {Kitaev}(1997)}]{kitaev_quantum_1997}%
  \BibitemOpen
  \bibfield  {author} {\bibinfo {author} {\bibfnamefont {A.~Yu}\ \bibnamefont
  {Kitaev}},\ }\bibfield  {title} {\enquote {\bibinfo {title} {Quantum
  computations: algorithms and error correction},}\ }\href {\doibase
  10.1070/RM1997v052n06ABEH002155} {\bibfield  {journal} {\bibinfo  {journal}
  {Russian Mathematical Surveys}\ }\textbf {\bibinfo {volume} {52}},\ \bibinfo
  {pages} {1191} (\bibinfo {year} {1997})}\BibitemShut {NoStop}%
\bibitem [{\citenamefont {Bombin}\ and\ \citenamefont
  {Martin-Delgado}(2007{\natexlab{a}})}]{bombin_topological_2007}%
  \BibitemOpen
  \bibfield  {author} {\bibinfo {author} {\bibfnamefont {H.}~\bibnamefont
  {Bombin}}\ and\ \bibinfo {author} {\bibfnamefont {M.~A.}\ \bibnamefont
  {Martin-Delgado}},\ }\bibfield  {title} {\enquote {\bibinfo {title}
  {Topological {Computation} without {Braiding}},}\ }\href {\doibase
  10.1103/PhysRevLett.98.160502} {\bibfield  {journal} {\bibinfo  {journal}
  {Physical Review Letters}\ }\textbf {\bibinfo {volume} {98}},\ \bibinfo
  {pages} {160502} (\bibinfo {year} {2007}{\natexlab{a}})}\BibitemShut
  {NoStop}%
\bibitem [{\citenamefont {Bennett}\ \emph {et~al.}(1997)\citenamefont
  {Bennett}, \citenamefont {DiVincenzo},\ and\ \citenamefont
  {Smolin}}]{bennett_capacities_1997}%
  \BibitemOpen
  \bibfield  {author} {\bibinfo {author} {\bibfnamefont {Charles~H.}\
  \bibnamefont {Bennett}}, \bibinfo {author} {\bibfnamefont {David~P.}\
  \bibnamefont {DiVincenzo}}, \ and\ \bibinfo {author} {\bibfnamefont
  {John~A.}\ \bibnamefont {Smolin}},\ }\bibfield  {title} {\enquote {\bibinfo
  {title} {Capacities of {Quantum} {Erasure} {Channels}},}\ }\href {\doibase
  10.1103/PhysRevLett.78.3217} {\bibfield  {journal} {\bibinfo  {journal}
  {Physical Review Letters}\ }\textbf {\bibinfo {volume} {78}},\ \bibinfo
  {pages} {3217--3220} (\bibinfo {year} {1997})}\BibitemShut {NoStop}%
\bibitem [{\citenamefont {Lloyd}(1997)}]{lloyd_capacity_1997}%
  \BibitemOpen
  \bibfield  {author} {\bibinfo {author} {\bibfnamefont {Seth}\ \bibnamefont
  {Lloyd}},\ }\bibfield  {title} {\enquote {\bibinfo {title} {Capacity of the
  noisy quantum channel},}\ }\href {\doibase 10.1103/PhysRevA.55.1613}
  {\bibfield  {journal} {\bibinfo  {journal} {Physical Review A}\ }\textbf
  {\bibinfo {volume} {55}},\ \bibinfo {pages} {1613--1622} (\bibinfo {year}
  {1997})}\BibitemShut {NoStop}%
\bibitem [{\citenamefont {Nielsen}\ and\ \citenamefont
  {Chuang}(2010)}]{nielsen_quantum_2010}%
  \BibitemOpen
  \bibfield  {author} {\bibinfo {author} {\bibfnamefont {Michael~A.}\
  \bibnamefont {Nielsen}}\ and\ \bibinfo {author} {\bibfnamefont {Isaac~L.}\
  \bibnamefont {Chuang}},\ }\href@noop {} {\emph {\bibinfo {title} {Quantum
  {Computation} and {Quantum} {Information}: 10th {Anniversary} {Edition}}}}\
  (\bibinfo  {publisher} {Cambridge University Press},\ \bibinfo {year}
  {2010})\BibitemShut {NoStop}%
\bibitem [{\citenamefont {Huang}\ \emph
  {et~al.}(2024{\natexlab{b}})\citenamefont {Huang}, \citenamefont
  {Colmenarez}, \citenamefont {Müller},\ and\ \citenamefont
  {Diehl}}]{huang_coherent_2024}%
  \BibitemOpen
  \bibfield  {author} {\bibinfo {author} {\bibfnamefont {Ze-Min}\ \bibnamefont
  {Huang}}, \bibinfo {author} {\bibfnamefont {Luis}\ \bibnamefont
  {Colmenarez}}, \bibinfo {author} {\bibfnamefont {Markus}\ \bibnamefont
  {Müller}}, \ and\ \bibinfo {author} {\bibfnamefont {Sebastian}\ \bibnamefont
  {Diehl}},\ }\href {\doibase 10.48550/arXiv.2412.12279} {\enquote {\bibinfo
  {title} {Coherent information as a mixed-state topological order parameter of
  fermions},}\ } (\bibinfo {year} {2024}{\natexlab{b}}),\ \bibinfo {note}
  {arXiv:2412.12279}\BibitemShut {NoStop}%
\bibitem [{\citenamefont {Breuckmann}\ and\ \citenamefont
  {Eberhardt}(2021)}]{breuckmann_quantum_2021}%
  \BibitemOpen
  \bibfield  {author} {\bibinfo {author} {\bibfnamefont {Nikolas~P.}\
  \bibnamefont {Breuckmann}}\ and\ \bibinfo {author} {\bibfnamefont
  {Jens~Niklas}\ \bibnamefont {Eberhardt}},\ }\bibfield  {title} {\enquote
  {\bibinfo {title} {Quantum {Low}-{Density} {Parity}-{Check} {Codes}},}\
  }\href {\doibase 10.1103/PRXQuantum.2.040101} {\bibfield  {journal} {\bibinfo
   {journal} {PRX Quantum}\ }\textbf {\bibinfo {volume} {2}},\ \bibinfo {pages}
  {040101} (\bibinfo {year} {2021})}\BibitemShut {NoStop}%
\bibitem [{\citenamefont {Placke}\ and\ \citenamefont
  {Breuckmann}(2023)}]{placke_random-bond_2023}%
  \BibitemOpen
  \bibfield  {author} {\bibinfo {author} {\bibfnamefont {Benedikt}\
  \bibnamefont {Placke}}\ and\ \bibinfo {author} {\bibfnamefont {Nikolas~P.}\
  \bibnamefont {Breuckmann}},\ }\bibfield  {title} {\enquote {\bibinfo {title}
  {Random-bond {Ising} model and its dual in hyperbolic spaces},}\ }\href
  {\doibase 10.1103/PhysRevE.107.024125} {\bibfield  {journal} {\bibinfo
  {journal} {Physical Review E}\ }\textbf {\bibinfo {volume} {107}},\ \bibinfo
  {pages} {024125} (\bibinfo {year} {2023})}\BibitemShut {NoStop}%
\bibitem [{\citenamefont {Jiang}\ \emph {et~al.}(2019)\citenamefont {Jiang},
  \citenamefont {Dumer}, \citenamefont {Kovalev},\ and\ \citenamefont
  {Pryadko}}]{jiang_duality_2019}%
  \BibitemOpen
  \bibfield  {author} {\bibinfo {author} {\bibfnamefont {Yi}~\bibnamefont
  {Jiang}}, \bibinfo {author} {\bibfnamefont {Ilya}\ \bibnamefont {Dumer}},
  \bibinfo {author} {\bibfnamefont {Alexey~A.}\ \bibnamefont {Kovalev}}, \ and\
  \bibinfo {author} {\bibfnamefont {Leonid~P.}\ \bibnamefont {Pryadko}},\
  }\bibfield  {title} {\enquote {\bibinfo {title} {Duality and free energy
  analyticity bounds for few-body {Ising} models with extensive homology
  rank},}\ }\href {\doibase 10.1063/1.5039735} {\bibfield  {journal} {\bibinfo
  {journal} {Journal of Mathematical Physics}\ }\textbf {\bibinfo {volume}
  {60}},\ \bibinfo {pages} {083302} (\bibinfo {year} {2019})}\BibitemShut
  {NoStop}%
\bibitem [{\citenamefont {Wang}\ \emph {et~al.}(2003)\citenamefont {Wang},
  \citenamefont {Harrington},\ and\ \citenamefont
  {Preskill}}]{wang_confinement-higgs_2003}%
  \BibitemOpen
  \bibfield  {author} {\bibinfo {author} {\bibfnamefont {Chenyang}\
  \bibnamefont {Wang}}, \bibinfo {author} {\bibfnamefont {Jim}\ \bibnamefont
  {Harrington}}, \ and\ \bibinfo {author} {\bibfnamefont {John}\ \bibnamefont
  {Preskill}},\ }\bibfield  {title} {\enquote {\bibinfo {title}
  {Confinement-{Higgs} transition in a disordered gauge theory and the accuracy
  threshold for quantum memory},}\ }\href {\doibase
  10.1016/S0003-4916(02)00019-2} {\bibfield  {journal} {\bibinfo  {journal}
  {Annals of Physics}\ }\textbf {\bibinfo {volume} {303}},\ \bibinfo {pages}
  {31--58} (\bibinfo {year} {2003})}\BibitemShut {NoStop}%
\bibitem [{\citenamefont {Tomita}\ and\ \citenamefont
  {Svore}(2014)}]{tomita_low-distance_2014}%
  \BibitemOpen
  \bibfield  {author} {\bibinfo {author} {\bibfnamefont {Yu}~\bibnamefont
  {Tomita}}\ and\ \bibinfo {author} {\bibfnamefont {Krysta~M.}\ \bibnamefont
  {Svore}},\ }\bibfield  {title} {\enquote {\bibinfo {title} {Low-distance
  surface codes under realistic quantum noise},}\ }\href {\doibase
  10.1103/PhysRevA.90.062320} {\bibfield  {journal} {\bibinfo  {journal}
  {Physical Review A}\ }\textbf {\bibinfo {volume} {90}},\ \bibinfo {pages}
  {062320} (\bibinfo {year} {2014})}\BibitemShut {NoStop}%
\bibitem [{\citenamefont {Bombin}\ and\ \citenamefont
  {Martin-Delgado}(2006)}]{bombin_topological_2006}%
  \BibitemOpen
  \bibfield  {author} {\bibinfo {author} {\bibfnamefont {H.}~\bibnamefont
  {Bombin}}\ and\ \bibinfo {author} {\bibfnamefont {M.~A.}\ \bibnamefont
  {Martin-Delgado}},\ }\bibfield  {title} {\enquote {\bibinfo {title}
  {Topological {Quantum} {Distillation}},}\ }\href {\doibase
  10.1103/PhysRevLett.97.180501} {\bibfield  {journal} {\bibinfo  {journal}
  {Physical Review Letters}\ }\textbf {\bibinfo {volume} {97}},\ \bibinfo
  {pages} {180501} (\bibinfo {year} {2006})}\BibitemShut {NoStop}%
\bibitem [{\citenamefont {Kuno}\ \emph {et~al.}(2024)\citenamefont {Kuno},
  \citenamefont {Orito},\ and\ \citenamefont {Ichinose}}]{kuno_intrinsic_2024}%
  \BibitemOpen
  \bibfield  {author} {\bibinfo {author} {\bibfnamefont {Yoshihito}\
  \bibnamefont {Kuno}}, \bibinfo {author} {\bibfnamefont {Takahiro}\
  \bibnamefont {Orito}}, \ and\ \bibinfo {author} {\bibfnamefont {Ikuo}\
  \bibnamefont {Ichinose}},\ }\href {\doibase 10.48550/arXiv.2410.14258}
  {\enquote {\bibinfo {title} {Intrinsic mixed state topological order in a
  stabilizer system under stochastic decoherence},}\ } (\bibinfo {year}
  {2024}),\ \bibinfo {note} {arXiv:2410.14258}\BibitemShut {NoStop}%
\bibitem [{cro()}]{crossings}%
  \BibitemOpen
  \href@noop {} {}\bibinfo {note} {Crossings are computed by finding the root
  of the interpolated function $y=I_{d+2}-I_{d}$. The sub-index refers to the
  code distance. The root finder routine used was scipy.optimize.brent, which
  delivers close-to machine precision roots. Error bars are estimated by the
  resolution in error rate indicated in the main text and prior to
  interpolation. It should be noted that this is likely overestimating the
  interpolation error in finding the crossing}\BibitemShut {NoStop}%
\bibitem [{\citenamefont {Katzgraber}\ \emph {et~al.}(2009)\citenamefont
  {Katzgraber}, \citenamefont {Bombin},\ and\ \citenamefont
  {Martin-Delgado}}]{katzgraber_error_2009}%
  \BibitemOpen
  \bibfield  {author} {\bibinfo {author} {\bibfnamefont {Helmut~G.}\
  \bibnamefont {Katzgraber}}, \bibinfo {author} {\bibfnamefont
  {H.}~\bibnamefont {Bombin}}, \ and\ \bibinfo {author} {\bibfnamefont {M.~A.}\
  \bibnamefont {Martin-Delgado}},\ }\bibfield  {title} {\enquote {\bibinfo
  {title} {Error {Threshold} for {Color} {Codes} and {Random} {Three}-{Body}
  {Ising} {Models}},}\ }\href {\doibase 10.1103/PhysRevLett.103.090501}
  {\bibfield  {journal} {\bibinfo  {journal} {Physical Review Letters}\
  }\textbf {\bibinfo {volume} {103}},\ \bibinfo {pages} {090501} (\bibinfo
  {year} {2009})}\BibitemShut {NoStop}%
\bibitem [{\citenamefont {Panteleev}\ and\ \citenamefont
  {Kalachev}(2021)}]{panteleev_degenerate_2021}%
  \BibitemOpen
  \bibfield  {author} {\bibinfo {author} {\bibfnamefont {Pavel}\ \bibnamefont
  {Panteleev}}\ and\ \bibinfo {author} {\bibfnamefont {Gleb}\ \bibnamefont
  {Kalachev}},\ }\bibfield  {title} {\enquote {\bibinfo {title} {Degenerate
  {Quantum} {LDPC} {Codes} {With} {Good} {Finite} {Length} {Performance}},}\
  }\href {\doibase 10.22331/q-2021-11-22-585} {\bibfield  {journal} {\bibinfo
  {journal} {Quantum}\ }\textbf {\bibinfo {volume} {5}},\ \bibinfo {pages}
  {585} (\bibinfo {year} {2021})}\BibitemShut {NoStop}%
\bibitem [{\citenamefont {Panteleev}\ and\ \citenamefont
  {Kalachev}(2022)}]{panteleev_quantum_2022}%
  \BibitemOpen
  \bibfield  {author} {\bibinfo {author} {\bibfnamefont {Pavel}\ \bibnamefont
  {Panteleev}}\ and\ \bibinfo {author} {\bibfnamefont {Gleb}\ \bibnamefont
  {Kalachev}},\ }\bibfield  {title} {\enquote {\bibinfo {title} {Quantum {LDPC}
  {Codes} {With} {Almost} {Linear} {Minimum} {Distance}},}\ }\href {\doibase
  10.1109/TIT.2021.3119384} {\bibfield  {journal} {\bibinfo  {journal} {IEEE
  Transactions on Information Theory}\ }\textbf {\bibinfo {volume} {68}},\
  \bibinfo {pages} {213--229} (\bibinfo {year} {2022})}\BibitemShut {NoStop}%
\bibitem [{\citenamefont {Placke}\ \emph {et~al.}(2024)\citenamefont {Placke},
  \citenamefont {Rakovszky}, \citenamefont {Breuckmann},\ and\ \citenamefont
  {Khemani}}]{placke_topological_2024}%
  \BibitemOpen
  \bibfield  {author} {\bibinfo {author} {\bibfnamefont {Benedikt}\
  \bibnamefont {Placke}}, \bibinfo {author} {\bibfnamefont {Tibor}\
  \bibnamefont {Rakovszky}}, \bibinfo {author} {\bibfnamefont {Nikolas~P.}\
  \bibnamefont {Breuckmann}}, \ and\ \bibinfo {author} {\bibfnamefont {Vedika}\
  \bibnamefont {Khemani}},\ }\href {\doibase 10.48550/arXiv.2412.13248}
  {\enquote {\bibinfo {title} {Topological {Quantum} {Spin} {Glass} {Order} and
  its realization in {qLDPC} codes},}\ } (\bibinfo {year} {2024}),\ \bibinfo
  {note} {arXiv:2412.13248}\BibitemShut {NoStop}%
\bibitem [{\citenamefont {Roeck}\ \emph {et~al.}(2024)\citenamefont {Roeck},
  \citenamefont {Khemani}, \citenamefont {Li}, \citenamefont {O'Dea},\ and\
  \citenamefont {Rakovszky}}]{roeck_ldpc_2024}%
  \BibitemOpen
  \bibfield  {author} {\bibinfo {author} {\bibfnamefont {Wojciech~De}\
  \bibnamefont {Roeck}}, \bibinfo {author} {\bibfnamefont {Vedika}\
  \bibnamefont {Khemani}}, \bibinfo {author} {\bibfnamefont {Yaodong}\
  \bibnamefont {Li}}, \bibinfo {author} {\bibfnamefont {Nicholas}\ \bibnamefont
  {O'Dea}}, \ and\ \bibinfo {author} {\bibfnamefont {Tibor}\ \bibnamefont
  {Rakovszky}},\ }\href {\doibase 10.48550/arXiv.2411.02384} {\enquote
  {\bibinfo {title} {{LDPC} stabilizer codes as gapped quantum phases:
  stability under graph-local perturbations},}\ } (\bibinfo {year} {2024}),\
  \bibinfo {note} {arXiv:2411.02384}\BibitemShut {NoStop}%
\bibitem [{\citenamefont {Roffe}(2022)}]{Roffe_LDPC_Python_tools_2022}%
  \BibitemOpen
  \bibfield  {author} {\bibinfo {author} {\bibfnamefont {Joschka}\ \bibnamefont
  {Roffe}},\ }\href {https://pypi.org/project/ldpc/} {\enquote {\bibinfo
  {title} {{LDPC: Python tools for low density parity check codes}},}\ }
  (\bibinfo {year} {2022})\BibitemShut {NoStop}%
\bibitem [{\citenamefont {Harris}\ \emph {et~al.}(2020)\citenamefont {Harris},
  \citenamefont {Millman}, \citenamefont {van~der Walt}, \citenamefont
  {Gommers}, \citenamefont {Virtanen}, \citenamefont {Cournapeau},
  \citenamefont {Wieser}, \citenamefont {Taylor}, \citenamefont {Berg},
  \citenamefont {Smith}, \citenamefont {Kern}, \citenamefont {Picus},
  \citenamefont {Hoyer}, \citenamefont {van Kerkwijk}, \citenamefont {Brett},
  \citenamefont {Haldane}, \citenamefont {del R{\'{i}}o}, \citenamefont
  {Wiebe}, \citenamefont {Peterson}, \citenamefont {G{\'{e}}rard-Marchant},
  \citenamefont {Sheppard}, \citenamefont {Reddy}, \citenamefont {Weckesser},
  \citenamefont {Abbasi}, \citenamefont {Gohlke},\ and\ \citenamefont
  {Oliphant}}]{harris2020array}%
  \BibitemOpen
  \bibfield  {author} {\bibinfo {author} {\bibfnamefont {Charles~R.}\
  \bibnamefont {Harris}}, \bibinfo {author} {\bibfnamefont {K.~Jarrod}\
  \bibnamefont {Millman}}, \bibinfo {author} {\bibfnamefont {St{\'{e}}fan~J.}\
  \bibnamefont {van~der Walt}}, \bibinfo {author} {\bibfnamefont {Ralf}\
  \bibnamefont {Gommers}}, \bibinfo {author} {\bibfnamefont {Pauli}\
  \bibnamefont {Virtanen}}, \bibinfo {author} {\bibfnamefont {David}\
  \bibnamefont {Cournapeau}}, \bibinfo {author} {\bibfnamefont {Eric}\
  \bibnamefont {Wieser}}, \bibinfo {author} {\bibfnamefont {Julian}\
  \bibnamefont {Taylor}}, \bibinfo {author} {\bibfnamefont {Sebastian}\
  \bibnamefont {Berg}}, \bibinfo {author} {\bibfnamefont {Nathaniel~J.}\
  \bibnamefont {Smith}}, \bibinfo {author} {\bibfnamefont {Robert}\
  \bibnamefont {Kern}}, \bibinfo {author} {\bibfnamefont {Matti}\ \bibnamefont
  {Picus}}, \bibinfo {author} {\bibfnamefont {Stephan}\ \bibnamefont {Hoyer}},
  \bibinfo {author} {\bibfnamefont {Marten~H.}\ \bibnamefont {van Kerkwijk}},
  \bibinfo {author} {\bibfnamefont {Matthew}\ \bibnamefont {Brett}}, \bibinfo
  {author} {\bibfnamefont {Allan}\ \bibnamefont {Haldane}}, \bibinfo {author}
  {\bibfnamefont {Jaime~Fern{\'{a}}ndez}\ \bibnamefont {del R{\'{i}}o}},
  \bibinfo {author} {\bibfnamefont {Mark}\ \bibnamefont {Wiebe}}, \bibinfo
  {author} {\bibfnamefont {Pearu}\ \bibnamefont {Peterson}}, \bibinfo {author}
  {\bibfnamefont {Pierre}\ \bibnamefont {G{\'{e}}rard-Marchant}}, \bibinfo
  {author} {\bibfnamefont {Kevin}\ \bibnamefont {Sheppard}}, \bibinfo {author}
  {\bibfnamefont {Tyler}\ \bibnamefont {Reddy}}, \bibinfo {author}
  {\bibfnamefont {Warren}\ \bibnamefont {Weckesser}}, \bibinfo {author}
  {\bibfnamefont {Hameer}\ \bibnamefont {Abbasi}}, \bibinfo {author}
  {\bibfnamefont {Christoph}\ \bibnamefont {Gohlke}}, \ and\ \bibinfo {author}
  {\bibfnamefont {Travis~E.}\ \bibnamefont {Oliphant}},\ }\bibfield  {title}
  {\enquote {\bibinfo {title} {Array programming with {NumPy}},}\ }\href
  {\doibase 10.1038/s41586-020-2649-2} {\bibfield  {journal} {\bibinfo
  {journal} {Nature}\ }\textbf {\bibinfo {volume} {585}},\ \bibinfo {pages}
  {357--362} (\bibinfo {year} {2020})}\BibitemShut {NoStop}%
\bibitem [{\citenamefont {Virtanen}\ \emph {et~al.}(2020)\citenamefont
  {Virtanen}, \citenamefont {Gommers}, \citenamefont {Oliphant}, \citenamefont
  {Haberland}, \citenamefont {Reddy}, \citenamefont {Cournapeau}, \citenamefont
  {Burovski}, \citenamefont {Peterson}, \citenamefont {Weckesser},
  \citenamefont {Bright}, \citenamefont {{van der Walt}}, \citenamefont
  {Brett}, \citenamefont {Wilson}, \citenamefont {Millman}, \citenamefont
  {Mayorov}, \citenamefont {Nelson}, \citenamefont {Jones}, \citenamefont
  {Kern}, \citenamefont {Larson}, \citenamefont {Carey}, \citenamefont {Polat},
  \citenamefont {Feng}, \citenamefont {Moore}, \citenamefont {{VanderPlas}},
  \citenamefont {Laxalde}, \citenamefont {Perktold}, \citenamefont {Cimrman},
  \citenamefont {Henriksen}, \citenamefont {Quintero}, \citenamefont {Harris},
  \citenamefont {Archibald}, \citenamefont {Ribeiro}, \citenamefont
  {Pedregosa}, \citenamefont {{van Mulbregt}},\ and\ \citenamefont {{SciPy 1.0
  Contributors}}}]{2020SciPy-NMeth}%
  \BibitemOpen
  \bibfield  {author} {\bibinfo {author} {\bibfnamefont {Pauli}\ \bibnamefont
  {Virtanen}}, \bibinfo {author} {\bibfnamefont {Ralf}\ \bibnamefont
  {Gommers}}, \bibinfo {author} {\bibfnamefont {Travis~E.}\ \bibnamefont
  {Oliphant}}, \bibinfo {author} {\bibfnamefont {Matt}\ \bibnamefont
  {Haberland}}, \bibinfo {author} {\bibfnamefont {Tyler}\ \bibnamefont
  {Reddy}}, \bibinfo {author} {\bibfnamefont {David}\ \bibnamefont
  {Cournapeau}}, \bibinfo {author} {\bibfnamefont {Evgeni}\ \bibnamefont
  {Burovski}}, \bibinfo {author} {\bibfnamefont {Pearu}\ \bibnamefont
  {Peterson}}, \bibinfo {author} {\bibfnamefont {Warren}\ \bibnamefont
  {Weckesser}}, \bibinfo {author} {\bibfnamefont {Jonathan}\ \bibnamefont
  {Bright}}, \bibinfo {author} {\bibfnamefont {St{\'e}fan~J.}\ \bibnamefont
  {{van der Walt}}}, \bibinfo {author} {\bibfnamefont {Matthew}\ \bibnamefont
  {Brett}}, \bibinfo {author} {\bibfnamefont {Joshua}\ \bibnamefont {Wilson}},
  \bibinfo {author} {\bibfnamefont {K.~Jarrod}\ \bibnamefont {Millman}},
  \bibinfo {author} {\bibfnamefont {Nikolay}\ \bibnamefont {Mayorov}}, \bibinfo
  {author} {\bibfnamefont {Andrew R.~J.}\ \bibnamefont {Nelson}}, \bibinfo
  {author} {\bibfnamefont {Eric}\ \bibnamefont {Jones}}, \bibinfo {author}
  {\bibfnamefont {Robert}\ \bibnamefont {Kern}}, \bibinfo {author}
  {\bibfnamefont {Eric}\ \bibnamefont {Larson}}, \bibinfo {author}
  {\bibfnamefont {C~J}\ \bibnamefont {Carey}}, \bibinfo {author} {\bibfnamefont
  {{\.I}lhan}\ \bibnamefont {Polat}}, \bibinfo {author} {\bibfnamefont
  {Yu}~\bibnamefont {Feng}}, \bibinfo {author} {\bibfnamefont {Eric~W.}\
  \bibnamefont {Moore}}, \bibinfo {author} {\bibfnamefont {Jake}\ \bibnamefont
  {{VanderPlas}}}, \bibinfo {author} {\bibfnamefont {Denis}\ \bibnamefont
  {Laxalde}}, \bibinfo {author} {\bibfnamefont {Josef}\ \bibnamefont
  {Perktold}}, \bibinfo {author} {\bibfnamefont {Robert}\ \bibnamefont
  {Cimrman}}, \bibinfo {author} {\bibfnamefont {Ian}\ \bibnamefont
  {Henriksen}}, \bibinfo {author} {\bibfnamefont {E.~A.}\ \bibnamefont
  {Quintero}}, \bibinfo {author} {\bibfnamefont {Charles~R.}\ \bibnamefont
  {Harris}}, \bibinfo {author} {\bibfnamefont {Anne~M.}\ \bibnamefont
  {Archibald}}, \bibinfo {author} {\bibfnamefont {Ant{\^o}nio~H.}\ \bibnamefont
  {Ribeiro}}, \bibinfo {author} {\bibfnamefont {Fabian}\ \bibnamefont
  {Pedregosa}}, \bibinfo {author} {\bibfnamefont {Paul}\ \bibnamefont {{van
  Mulbregt}}}, \ and\ \bibinfo {author} {\bibnamefont {{SciPy 1.0
  Contributors}}},\ }\bibfield  {title} {\enquote {\bibinfo {title} {{{SciPy}
  1.0: Fundamental Algorithms for Scientific Computing in Python}},}\ }\href
  {\doibase 10.1038/s41592-019-0686-2} {\bibfield  {journal} {\bibinfo
  {journal} {Nature Methods}\ }\textbf {\bibinfo {volume} {17}},\ \bibinfo
  {pages} {261--272} (\bibinfo {year} {2020})}\BibitemShut {NoStop}%
\bibitem [{\citenamefont {Harris}\ \emph {et~al.}(2018)\citenamefont {Harris},
  \citenamefont {McMahon}, \citenamefont {Brennen},\ and\ \citenamefont
  {Stace}}]{harris_calderbank-shor-steane_2018}%
  \BibitemOpen
  \bibfield  {author} {\bibinfo {author} {\bibfnamefont {Robert~J.}\
  \bibnamefont {Harris}}, \bibinfo {author} {\bibfnamefont {Nathan~A.}\
  \bibnamefont {McMahon}}, \bibinfo {author} {\bibfnamefont {Gavin~K.}\
  \bibnamefont {Brennen}}, \ and\ \bibinfo {author} {\bibfnamefont {Thomas~M.}\
  \bibnamefont {Stace}},\ }\bibfield  {title} {\enquote {\bibinfo {title}
  {Calderbank-{Shor}-{Steane} holographic quantum error-correcting codes},}\
  }\href {\doibase 10.1103/PhysRevA.98.052301} {\bibfield  {journal} {\bibinfo
  {journal} {Physical Review A}\ }\textbf {\bibinfo {volume} {98}},\ \bibinfo
  {pages} {052301} (\bibinfo {year} {2018})}\BibitemShut {NoStop}%
\bibitem [{\citenamefont {Bombin}(2013)}]{bombin_introduction_2013}%
  \BibitemOpen
  \bibfield  {author} {\bibinfo {author} {\bibfnamefont {H.}~\bibnamefont
  {Bombin}},\ }\href {\doibase 10.48550/arXiv.1311.0277} {\enquote {\bibinfo
  {title} {An {Introduction} to {Topological} {Quantum} {Codes}},}\ } (\bibinfo
  {year} {2013}),\ \bibinfo {note} {arXiv:1311.0277}\BibitemShut {NoStop}%
\bibitem [{\citenamefont {Bombin}\ and\ \citenamefont
  {Martin-Delgado}(2007{\natexlab{b}})}]{bombin_optimal_2007}%
  \BibitemOpen
  \bibfield  {author} {\bibinfo {author} {\bibfnamefont {H.}~\bibnamefont
  {Bombin}}\ and\ \bibinfo {author} {\bibfnamefont {M.~A.}\ \bibnamefont
  {Martin-Delgado}},\ }\bibfield  {title} {\enquote {\bibinfo {title} {Optimal
  resources for topological two-dimensional stabilizer codes: {Comparative}
  study},}\ }\href {\doibase 10.1103/PhysRevA.76.012305} {\bibfield  {journal}
  {\bibinfo  {journal} {Physical Review A}\ }\textbf {\bibinfo {volume} {76}},\
  \bibinfo {pages} {012305} (\bibinfo {year} {2007}{\natexlab{b}})}\BibitemShut
  {NoStop}%
\bibitem [{\citenamefont {Landahl}\ \emph {et~al.}(2011)\citenamefont
  {Landahl}, \citenamefont {Anderson},\ and\ \citenamefont
  {Rice}}]{landahl_fault-tolerant_2011}%
  \BibitemOpen
  \bibfield  {author} {\bibinfo {author} {\bibfnamefont {Andrew~J.}\
  \bibnamefont {Landahl}}, \bibinfo {author} {\bibfnamefont {Jonas~T.}\
  \bibnamefont {Anderson}}, \ and\ \bibinfo {author} {\bibfnamefont
  {Patrick~R.}\ \bibnamefont {Rice}},\ }\href {\doibase
  10.48550/arXiv.1108.5738} {\enquote {\bibinfo {title} {Fault-tolerant quantum
  computing with color codes},}\ } (\bibinfo {year} {2011}),\ \bibinfo {note}
  {arXiv:1108.5738}\BibitemShut {NoStop}%
\end{thebibliography}%

\newpage
\clearpage
\onecolumngrid
\appendix

\section{Method for computing coherent information under erasure errors}\label{appendix:ci_erasure}

Given a CSS stabilizer code $[[n,k,d]]$, there are $k$ pairs of logical operators $\{O^z_i,O^x_i\}$ which are the generators of the logical group. Each operator can be interpreted as a binary vector of size $n$, where 0 (1) on position $j$ means empty (occupied) by either $X$ or $Z$ Pauli. Since the code is CSS we treat $X$ and $Z$ operators independently, although the same formalism can be applied to non-CSS codes.  
We also have parity check matrices $H^x$ and $H^z$ whose rows are the stabilizer generators, hence they have dimension $n\times(n-k)$. We define a loss vector $r_l$ of dimension $n$ which is all zeros except on the positions of the erased qubit. The configuation of qubit erasures is denoted by the index $l=1,2,...,2^n$. 
In Refs. \cite{amaro_analytical_2020,harris_calderbank-shor-steane_2018}, the authors show an algorithm for determining whether the erasure $r_l$ has support on at least one representative of the logical operators $\{O^z_i,O^x_i\}$ based only on $H^z$, $H^x$ and $r_l$. In other words, this indicates whether the logical operator was measured by the environment or not.
If for the $i-$th logical qubit one of the two logical operators is measured then it becomes a logical bit. If both logical operators are measured then we call it a lost qubit. Computing the coherent information for a fixed erasure configuration $l$ means computing how many logical qubits, logical bits and lost logical qubits there are.
Another way to interpret the appearance of logical bits and lost logical qubits is as a reduction in the number of generators of the logical group. If one or more logical operators cannot be recovered, then the logical group will have fewer generators than the original $2k$ logical generators. This affects the size of the code space, thereby decreasing the amount of information that can be protected by the QEC code.

The whole procedure for computing the coherent information for a fixed error configuration $l$ looks as follows: 
\begin{enumerate}
    \item Run the algorithms in Refs.~\cite{amaro_analytical_2020,harris_calderbank-shor-steane_2018} and determine the set of generators of the logical group $\tilde{O}^x$ and $\tilde{O}^z$, namely the minimum set of operators that generates all possible logical operators . Sometimes different generators $O^x$ might not be well-defined anymore but their product might still be. For instance $O^x_1 O^x_2$ maybe well-defined but $O^x_1$ and $O^x_2$ are not. Therefore, in the worst case, a total of $2^k$ combinations of $X$ and $Z$ logical operators must be checked.  
    \item The number of remaining logical qubits $k'_l$ is the number of anti-commuting operators $[\tilde{O}^x_i,\tilde{O}^z_i]\neq 0$ with $i=1,..,k'_l$. The number of logical bits $b_l$ is equal to the number of unpaired logical operators $\tilde{O}^x$ and $\tilde{O}^z$, i.e that commute with all other logical operators.  Hence the number of lost qubits is $c = k-k'_l-b_l$.
    \item The coherent information for an erasure configuration $l$ is then calculated as $I_l=k-b_l-2c_l$.
\end{enumerate}

As pointed out in Ref. \cite{harris_calderbank-shor-steane_2018} computing $I_l$ for each of the $2^n$ erasure configurations may not be necessary. First one has to realize that
\begin{eqnarray}\label{eq:ci_sampling}
I & = & \sum_{m=0}^{n} e^m (1-e)^{n-m} \sum_{\mathcal{C}(l)=m} I_l \nonumber \\
& = & \sum_{m=0}^{n} e^m (1-e)^{n-m} \mathcal{I}_m , 
\end{eqnarray}
where $\mathcal{C}(l)$ denotes the number of lost qubits in the erasure configuration $l$. Hence the term $\mathcal{I}_m$ is the summation of all $I_l$ with the same number of erased qubits $m$. This fact considerably simplifies the calculation. First, for $m<d$, where $d$ is the code distance, we know that $\mathcal{I}_m = {n\choose m} k \log 2 $ because one needs to erase at least $d$ qubits for measuring a logical operator. Second, for $m \geq d$, logical operators are measured in some configurations, however their contribution to $\mathcal{I}_m$ can be negligible. That's why we can approximate it by a Monte Carlo average of the following kind
\begin{eqnarray}
\mathcal{I}_m \approx \sum_{i=1}^{N_s} \dfrac{I^m_i}{N_s}.
\end{eqnarray}
Where $I^m_i$ is the CI for a random configuration $i$ of $m$ erased qubits and $N_s$ is the number of samples. The erasure configurations $i$ are randomly sampled, therefore $\mathcal{I}_m$ does not depend on the erasure probability $e$. Let us note that Eq.~\eqref{eq:ci_sampling} also applies in the presence of computational errors, then $I_l$ picks the contributions of the computational errors discussed in Sec.~\ref{sec:summary} Eq.~\eqref{eq:ci_comp_and_erasure}. Random sampling of error configurations with fixed number of erased qubits suffices for computing $I$ in the present work, but there is still room for improvement in approximating $\mathcal{I}_m$. For instance,  some type of importance sampling that takes into account the distribution of $I_l$ could help to speed up convergence.

\section{QEC codes}\label{app:qec_codes}

\begin{figure}[h]
    \centering
    \includegraphics[width=0.46\textwidth]{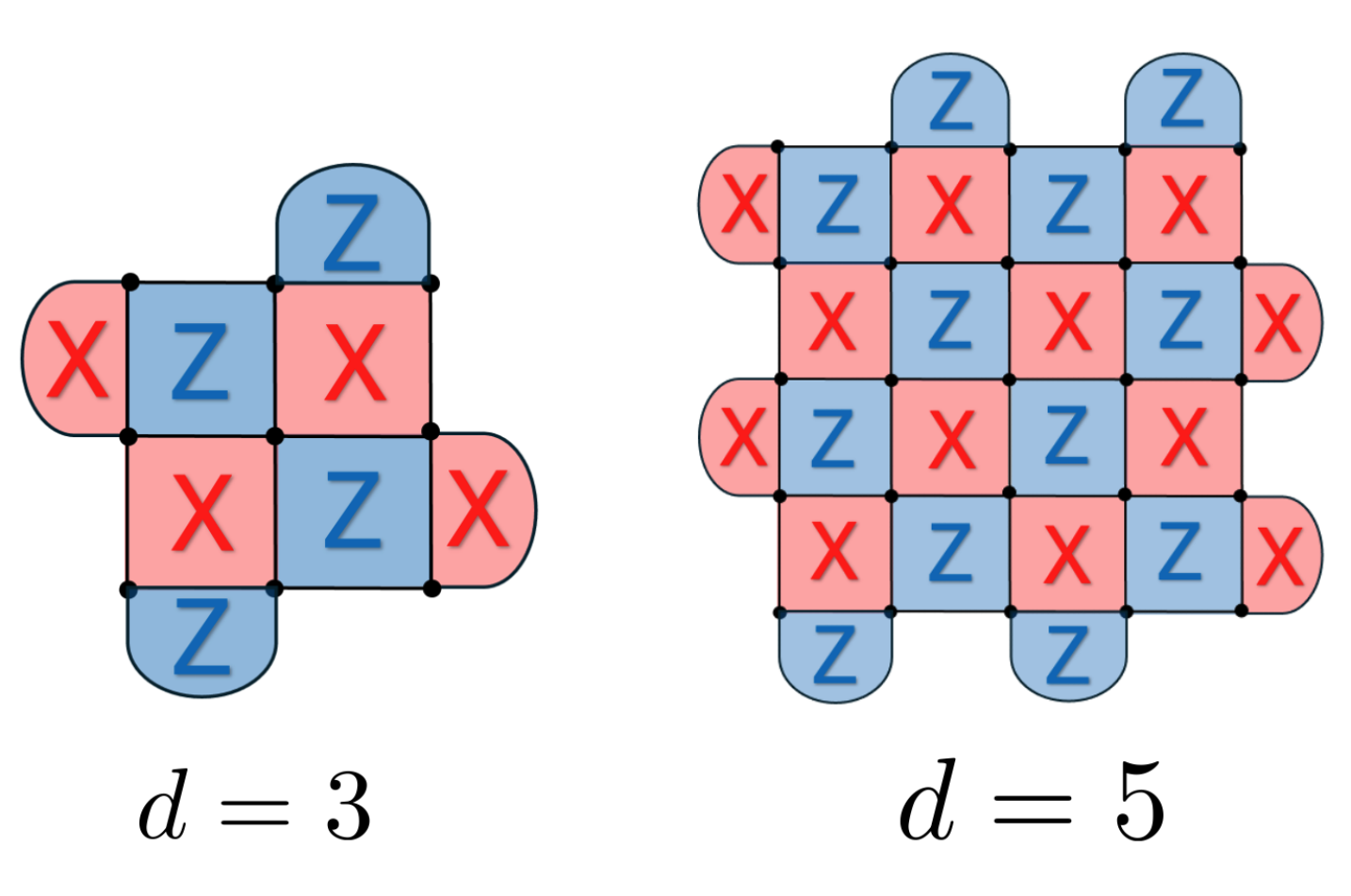}
    \caption{Rotated surface codes with $d=3$ and $d=5$. The code distance goes as $d=\sqrt{n}$, and  $[[n=d^2,1,d]]$. Blue (Red) plaquettes correspond $Z$ ($X$) stabilizers. Rounded shapes denote the weight-2 stabilizers, the square are the weight-4 stabilizers in the bulk. Physical qubits (black dots) are located on the vertices of the square lattice.}
    \label{fig:rotated_surface_codes}
\end{figure}

\begin{figure}[h]
    \centering
    \includegraphics[width=0.485\textwidth]{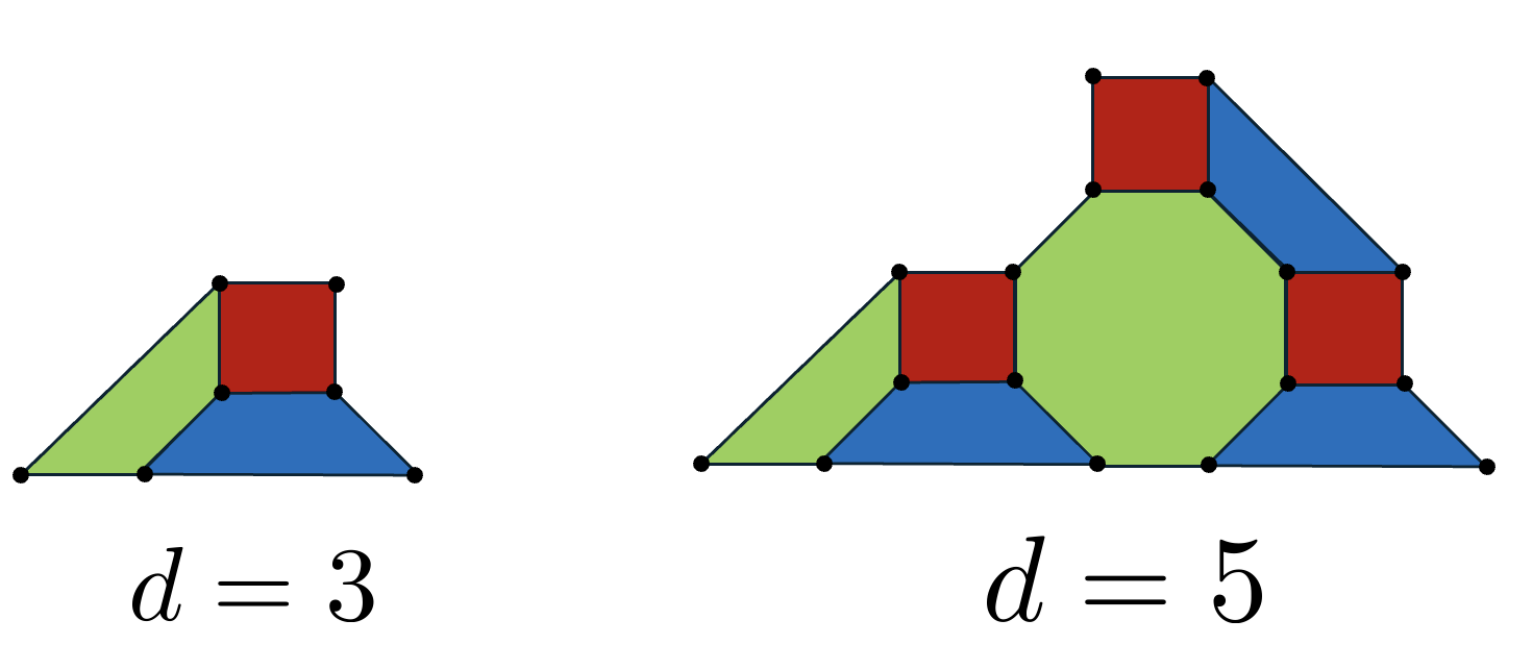}
    \caption{$d=3,5$ instances of the 4.8.8 color code. The number of physical qubits is given by $n=(d^2-1)/2+d$ yielding $[[(d^2-1)/2+d,1,d]]$. Both $X$ and $Z$ stabilizers are defined on the plaquettes of the lattice. Physical qubits (black dots) are placed on the vertices such that each qubit is shared by three plaquettes (one or two at the boundaries). }
    \label{fig:octogonal_color_codes}
\end{figure}

In this section, we describe in more detail the two topological codes used in the numerical calculations of Sec.~\ref{sec:results_topo_codes}. Toric and color codes belong to the family of topological codes \cite{bombin_introduction_2013}. They store quantum information in topologically ordered many-body states.
The toric code \cite{kitaev_quantum_1997, kitaev_fault-tolerant_2003, dennis_topological_2002} protects $k = 2$ logical qubits, while 2D color codes on a torus hold $k = 4$ logical qubits \cite{bombin_topological_2006, bombin_topological_2007}. Both codes require $n = 2d^2$ physical qubits to realize the corresponding topological code of distance $d$.

In order to reduce the overhead in physical qubits for the numerical calculations, we have chosen to work with the planar version of these codes. As QEC codes, they share the same features as their counterparts with periodic boundaries, but with a reduced number of logical qubits and a milder physical qubit overhead.
The rotated surface code \cite{bombin_optimal_2007, tomita_low-distance_2014,lidar_quantum_2013} protects only one logical qubit and requires $n = d^2$ physical qubits. The code distance $d$ is equal to the linear length of the square lattice (see Fig.~\ref{fig:rotated_surface_codes}). The $X$ and $Z$ stabilizers are located in alternating plaquettes of the lattice. In addition, there are weight-2 stabilizers on the boundaries. The logical operator $Z_L$ ($X_L$) is defined along the boundary where the weight-2 $X$ ($Z$) stabilizers are placed.

Regarding the 2D color code, we focus on the $4.8.8$ color code \cite{landahl_fault-tolerant_2011}. This code protects one logical qubit and has a qubit overhead of $n = (d^2 - 1)/2 + d$, where $d$ is the code distance and the linear size of the lattice (see the smallest representatives of the code in Fig.~\ref{fig:octogonal_color_codes}). Indeed, the 7-qubit Steane code \cite{steane_multiple-particle_1997} is the smallest representative for this and other types of triangular color codes. Each plaquette of the $4.8.8$ tiling is filled with one color, such that each color has neighboring plaquettes of the other two colors. $X$ and $Z$ stabilizers are defined on each plaquette, so they have the same support. The logical operators $X_L$ and $Z_L$ are defined on the boundaries of the lattice and they also share the same support.

\section{Exact statistical mechanics model for [[15,3,3]] lift-connected surface code}\label{sec:lcs_appendix}

In this section, we describe the exact statistical mechanics model for the lift-connected surface (LCS) code with parameters $L = 3$ and $\ell = 1$. To be concrete, we write down the exact form of the spin interactions; thus, the model applies to any choice of $p_x$, $p_z$, $p_y$, and $e$ (see Eq.~\eqref{eq:hamiltonian_xyz_first}). Moreover, this serves as the starting point for constructing analogous models for arbitrary values of $L$ and $\ell$.
First, let us note that there are two $\sigma$ and two $\tau$ spins per surface code sheet (see Fig.~\ref{fig:lcs_d3}). Starting from Eq.~\eqref{eq:hamiltonian_xyz_first}, we now write down the terms $P_\ell^X$ for each qubit:

\begin{eqnarray}\label{eq:single_terms}
    P_0^X = \sigma_{L,0}; \quad P_1^X = \sigma_{L,1}; \quad P_2^X = \sigma_{L,2}; \quad P_3^X = \sigma_{R,0}; \quad P_4^X = \sigma_{R,1}; \quad P_5^X = \sigma_{R,2};
\end{eqnarray}

\begin{eqnarray}\label{eq:double_terms}
    P_6^X = \sigma_{L,0} \sigma_{L,2}, \quad P_7^X = \sigma_{L,0} \sigma_{L,1}; \quad P_8^X = \sigma_{L,1}\sigma_{L,2}; \quad P_9^X = \sigma_{R,0} \sigma_{R,2}; \quad P_{10}^X = \sigma_{R,0} \sigma_{R,1}; \quad P_{11}^X = \sigma_{R,1} \sigma_{R,2};
\end{eqnarray}

\begin{eqnarray}\label{eq:triple_terms}
    P_{12}^X = \sigma_{L,0} \sigma_{R,0} \sigma_{R,1}; \quad P_{13}^X = \sigma_{L,1} \sigma_{R,1} \sigma_{R,2}; \quad P_{14}^X = \sigma_{L,2} \sigma_{R,2}\sigma_{R,0}; 
\end{eqnarray}

We denote $\sigma_{L(R),q}$ as the spin on the left (right) belonging to the $q$-th surface code sheet. We identify three classes of interaction terms: 

\begin{itemize}
    \item[i)] \textbf{Single-spin terms} [Eq.~\eqref{eq:single_terms}] arising from the upper qubits on each surface code sheet.
    \item[ii)] \textbf{Two-spin terms} [Eq.~\eqref{eq:double_terms}] between spins on consecutive surface code sheets at the same in-sheet position, originating from the lower qubits on each surface code sheet.
    \item[iii)] \textbf{Three-spin terms} [Eq.~\eqref{eq:triple_terms}] arising from the bulk qubits, i.e., those in the middle of the surface code sheet, where the two spins on the $q$-th sheet are coupled to the right spin on the $(q+1)$-th sheet.
\end{itemize}

At this point, it is worth noting that the spin interactions exhibit directionality along the vertical axis; that is, the interactions depend on vertical position, unlike the more symmetric interactions found in the Ising model derived from toric or surface codes.

Now we write down the $P_\ell^Z$ in terms of the $\tau$ spins in Fig.~\ref{fig:lcs_d3}: 

\begin{eqnarray}\label{eq:single_terms_z}
    P_0^Z = \tau_{U,0}; \quad P_1^Z = \tau_{U,1}; \quad P_2^Z = \tau_{U,2}; \quad P_6^Z = \tau_{L,0}; \quad P_7^Z = \tau_{L,1}; \quad P_8^Z = \tau_{L,2};
\end{eqnarray}

\begin{eqnarray}\label{eq:double_terms_z}
    P_3^Z = \tau_{U,0} \tau_{U,2}, \quad P_4^Z = \tau_{U,0} \tau_{U,1}; \quad P_5^Z = \tau_{U,1}\tau_{U,2}; \quad P_9^Z = \tau_{L,3} \tau_{L,5}; \quad P_{10}^Z = \tau_{L,3} \tau_{L,4}; \quad P_{11}^Z = \tau_{L,4} \tau_{L,5};
\end{eqnarray}

\begin{eqnarray}\label{eq:triple_terms_z}
    P_{12}^Z = \tau_{U,0} \tau_{L,0} \tau_{L,1}; \quad P_{13}^Z = \tau_{U,1} \tau_{L,1} \tau_{L,2}; \quad P_{14}^Z = \tau_{U,2} \tau_{L,2} \tau_{L,0}.
\end{eqnarray}

We define $\tau_{U(L),q}$ as the spin at the upper (lower) part of the $q$-th surface code sheet. The spin model obtained for the $Z$ stabilizers is simply a $90^\circ$ rotation of the spin model for the $X$ stabilizers. As a result, there is also directionality along the horizontal direction. To summarize, the three types of terms identified, namely the boundary terms [Eqs.~\eqref{eq:single_terms}, \eqref{eq:double_terms}] and the bulk terms [Eq.~\eqref{eq:triple_terms}]; serve as the building blocks for larger codes. 
The spin interactions exhibit a form of ``chirality,'' as they do not posses mirror symmetry along either spatial direction, even in the absence of disorder.

\section{The stabilizer configuration picture}\label{appendix:stab_conf}

\begin{figure}[h]
    \centering
    \includegraphics[width=0.5\textwidth]{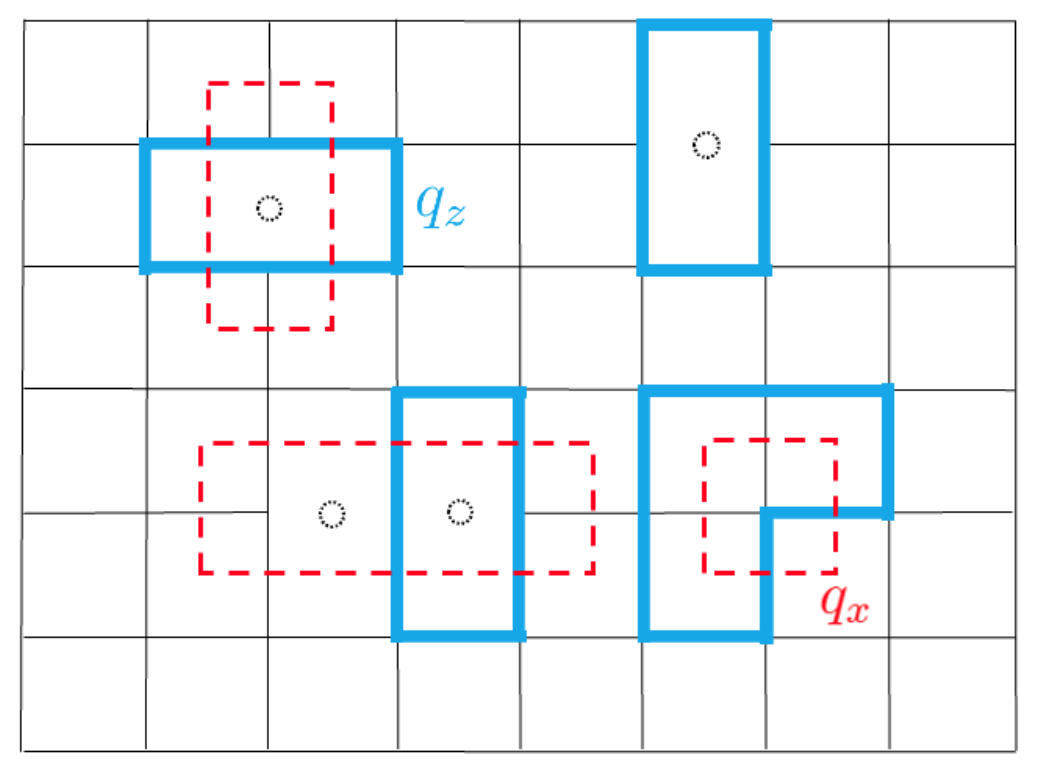}
    \caption{Example of $q_x$ (red) and $q_z$ (blue) stabilizer operators in the toric code. In the presence of erasure errors (open circles), super-plaquettes and super-star surrounding the erased qubits become stabilizer generators. We note that $q_x$ and $q_z$ are loop operators on a square lattice but always avoiding the position of the erased qubits.}
    \label{fig:loop_operators}
\end{figure}

\begin{figure}[h]
    \centering
    \includegraphics[width=0.6\textwidth]{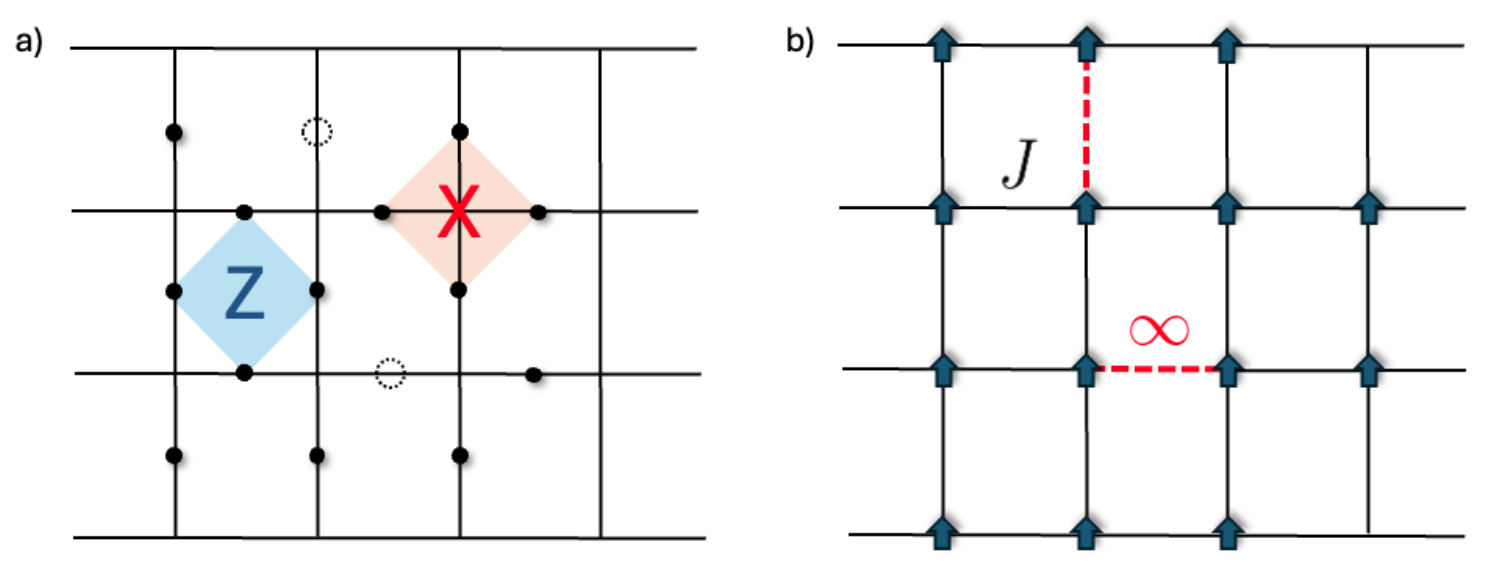}
    \caption{Erasure configuration in the toric code (a) and the corresponding $r-1$ flavour Ising model for $r=2$ (b). Only the $X$ part of the statistical mechanics model is shown.  Each site with no erasures has a coupling $J=-\log \sqrt{1-2p}$, while the sites where erasures have taken place have infinite coupling, effectively locking two neighbouring spins rigidly together.}
    \label{fig:ising-1flavor}
\end{figure}

As studied in detail in Refs.~\cite{fan_diagnostics_2024, lyons_understanding_2024, hauser_information_2024}, there is a family of statistical mechanics models that are dual to the models derived from the procedure shown in Sec.~\ref{sec:results}. The equivalence stems from the fact that both are representations of the moments $\Tr(\rho^r)$. Here, we show how this family of mappings is modified by introducing erasure errors. Let us start with the mixed state after erasures, as shown in Sec.~\ref{sec:results}:

\begin{eqnarray}\label{eq:rho_rq_lost}
\rho^l_{RQ} =  \dfrac{\mathbb{1}}{2^{h+2m}} \prod_{i=1}^{2k-h}\left(\frac{1+O_{R_i}O_{L_i}}{2}\right)\prod_{i=1}^{n-2m-k} \left(\frac{1+g'_i}{2}\right) .
\end{eqnarray}
Here, $h$ is the number of lost logical operators (see Appendix \ref{appendix:ci_erasure}) and $m$ is the number of erased qubits. The first product in Eq.~\eqref{eq:rho_rq_lost} runs over all the remaining logical operators. $g'_i$ are the well-defined stabilizer generators, which now contain plaquettes and super-plaquettes, see Fig.~\ref{fig:super_plaquette}.  After tracing out the reference system we obtain 
\begin{eqnarray}
\rho^l_{Q} & = &  \dfrac{\mathbb{1}}{2^{k+2m}} \prod_{i=1}^{n-2m-k} \left(\frac{1+g'_i}{2}\right) \nonumber \\
& = &  \dfrac{1}{2^n} \sum_{\{q_z,q_x\}} q_x q_z, 
\end{eqnarray}
where $\{q_x\}$ and $\{q_z\}$ are all possible combinations of the \emph{well-defined} $X$ and $Z$ stabilizer generators. In the toric code, the plaquettes and super-plaquettes discussed in Ref.~\cite{stace_thresholds_2009} are now stabilizer generators, so $q_x$ and $q_z$ are still loop operators on a square lattice (see Fig.~\ref{fig:loop_operators}). For simplicity, let us consider uncorrelated bit and phase flip errors, both with probability $p$. The same intuition for the erasure errors applies equally to depolarizing asymmetric channels. We then summarize the action of the error channel as

\begin{eqnarray}\label{eq:flip_on_loop}
\mathcal{N}_{X,i}\left[q_z\right] & = & (1-2p) q_z  \quad i \in q_z \nonumber \\
                 & =  &  q_z \quad \quad \quad \quad \quad  i \notin q_z
\end{eqnarray}
\begin{eqnarray}\label{eq:flip_on_loop}
\mathcal{N}_{Z,i}\left[q_x\right] & = & (1-2p) q_x  \quad i \in q_x \nonumber \\
                 & =  &  q_x \quad \quad \quad \quad \quad  i \notin q_x. 
\end{eqnarray}
After the noise map is applied we obtain the state
\begin{eqnarray}
\rho_{Q} = \mathcal{N}(\rho^l_{Q}) = \dfrac{1}{2^n} \sum_{\{q_z,q_x\}} e^{-\mu |q_z|-\mu |q_x|} q_x q_z ,
\end{eqnarray}
where $\mu = -\log(1-2p)$ and $|q_{z,x}|$ is the length of the respective stabilizer operator. In the toric code these are closed loops whose generators are plaquettes and super-plaquettes. Now we compute $\Tr(\rho^r_{Q})$ and $\Tr(\rho^r_{RQ})$, for simplicity we do it for the 2D toric code but highlighting the steps that must be modified for studying other QEC codes and error models. 

\subsection{Calculation of $\Tr(\rho^r_{Q})$}

After taking the trace we get
\begin{eqnarray}
\Tr\left(\rho^r_{Q}\right) = \dfrac{1}{2^{nr}} \sum_{\{q_z^{(s)},q_x^{(s)}\}} \Tr\left( \prod_{s=1}^{r} q_x^{(s)} q_z^{(s)} \right) e^{-\mu \sum_s |q^{(s)}_z| - \mu \sum_s |q^{(s)}_x|  }.
\end{eqnarray}
The non-vanishing contributions to trace force the constraint $\prod_{s=1}^r q^{(s)}_{x,z} = 1$, which is convenient to write as
\begin{eqnarray}
q^{(r)}_{x,z} = \prod_{s=1}^{r-1} q^{(s)}_{x,z}.
\end{eqnarray}
Then we are left with
\begin{eqnarray}
\Tr\left(\rho^r_{Q}\right) =   \dfrac{1}{2^{n(r-1)}} \sum_{\{q_z^{(s)}, q_x^{(s)}\}} e^{-\mu \sum_{s=1}^{r-1} |q^{(s)}_z| -\mu |\prod_{s=1}^{r-1} q^{(s)}_z| } e^{-\mu \sum_{s=1}^{r-1} |q^{(s)}_x| -\mu |\prod_{s=1}^{r-1} q^{(s)}_x| }.
\end{eqnarray}
The task is now to sum over all loops $q_z^{(s)}$. We can write $|q_z^{(s)}| = \sum_{\ell} |q_{z,\ell}^{(s)}| $ with
\begin{eqnarray}
|q_{z,\ell}^{(s)}| = \dfrac{1-\sigma^{(s)}_i \sigma^{(s)}_j}{2},
\end{eqnarray}
where $i,j$ denotes the plaquette locations that share the edge/qubit at $\ell$. In general one assigns spin variables to the positions of the Tanner graph and groups them according to the physical qubit they share, just like the $P^{X,Z}_{\ell}$ terms used in Sec.~\ref{sec:summary}. To faithfully count the super-plaquettes one should flip the spins denoting the former plaquettes together, this will be interpreted later as an infinite coupling between spins. Moreover, unlike for the erasure-free case, the loop operators for $X$ and $Z$ stabilizers might not be equivalent to one another because super-plaquettes and superstars are not necessarily equal up to a lattice shift and global rotation. 
For the second term $|\prod_{s=1}^{r-1} q^{(s)}|$ we compare the edges on all replicas such that $|\prod_{s=1}^{r-1} q^{(s)}| = \sum_{<i,j>}\left(1-\prod_{s=1}^{r-1} \sigma^{(s)}_i\sigma^{(s)}_j\right)/2$. We can then rewrite everything as
\begin{eqnarray}
\Tr\left(\rho^r_{Q}\right) =  \dfrac{e^{-J(n-m)r}}{2^{n(r-1)}} \sum_{\{\sigma^{(s)}\}} e^{-JH^{(r)}_X } \sum_{\{\sigma^{(s)}\}} e^{-JH^{(r)}_Z } =  \dfrac{e^{-J(n-m)r}}{2^{n(r-1)}} \mathcal{Z}_X(r) \mathcal{Z}_Z(r) 
\end{eqnarray}
with $J = \mu/2 = - \log(\sqrt{1-2p})$, $\mathcal{Z}_\alpha(r) =  \sum_{\{\sigma^{(s)}\}} e^{-JH^{(r)}_\alpha }$, $\alpha=X,Z$ and 

\begin{eqnarray}
H^{(r)} = \sum_{<i,j>} \left(\sum_{s=1}^{r-1} \sigma^{(s)}_i \sigma^{(s)}_j+ \prod_{s=1}^{r-1}\sigma^{(s)}_i \sigma^{(s)}_j\right).
\end{eqnarray}

The above expression represents an \emph{$r-1$-flavour Ising model}, as described in Ref.~\cite{fan_diagnostics_2024}. The coupling $J$ ranges from $J = 0$ (infinite temperature) for $p = 0$ to $J = \infty$ (zero temperature) for $p = 1/2$. The key difference from the erasure-free case is that the Hamiltonian excludes the qubit positions where qubits have been erased (see Fig.~\ref{fig:ising-1flavor}).
As a result, we can think of the super-plaquettes as being identified as single classical spins in a lattice with defects or as two spins locked together in the original lattice \cite{stace_thresholds_2009}.

\subsection{Calculation of $S(\rho_{RQ})$}

We write now the state $\rho^l_{RQ}$ in the stabilizer configuration picture:
\begin{eqnarray}
\rho^l_{RQ}  = \dfrac{1}{2^{n+k}} \Gamma^{x}_0 \Gamma^z_0 ,
\end{eqnarray}
with 
\begin{eqnarray}
\Gamma^{\alpha}_0 =  \sum_{\{q_\alpha\}}  q_\alpha  \prod_{i=1}^{k_\alpha} (1+O^\alpha_{R_i}O^\alpha_{L_i}) ,
\end{eqnarray}
as all possible $X$ or $Z$ well-defined stabilizer operators and the well-defined logical and reference operators $O^\alpha_{R_i}O^\alpha_{L_i}$. Let us note that $k_x$ and $k_z$ are the number of well-defined logical $X$ and $Z$ operators, respectively, which need not to be equal in the presence of erasures.  The noisy mixed state is then written as 
\begin{eqnarray}
\rho_{RQ} = \mathcal{N}(\rho^l_{RQ}) = \dfrac{1}{2^{n+k}} \Gamma^z \Gamma^x.
\end{eqnarray}
Here, we defined
\begin{eqnarray}
\Gamma^{\alpha} =  \sum_{\{q_\alpha\}} \sum_{\vec{d}_\alpha} e^{-\mu|q_\alpha O^{\vec{d}_\alpha}_\alpha|} q_\alpha O^{\vec{d}_\alpha}_\alpha,  
\end{eqnarray}
with $O^{\vec{d}_\alpha}_\alpha = O^{d_1}_1 O_2^{d_2} ... O_{k_\alpha}^{d_\alpha}$ as products of logical and reference operators $O_\alpha = O^{\alpha}_{R_i} O^{\alpha}_{L_i}$,  $\vec{d}_\alpha$ is binary vector of length $k_\alpha$ that denotes which logical/reference operators are included. For instance, for $k_x=2$ we have $d_x \in \{(0,0), (0,1), (1,0), (1,1)\}$. Now we study the $r$-th power of the state
\begin{eqnarray}
\Tr(\rho^r_{RQ}) = \dfrac{1}{2^{(n+k)r}} \Tr(\prod_{s=1}^{r}\Gamma^z(s) \Gamma^x(s)).
\end{eqnarray}
Similarly to the previous section, the non-vanishing terms of the trace are those for which the product of operators in different replicas equals the identity, then we get 
\begin{eqnarray}
\Tr(\rho^r_{RQ}) = \dfrac{1}{2^{(n+k)(r-1)}}  \sum_{\{q^{(s)}_\alpha,\vec{d}(s)\}} e^{-\mu\sum_{s=1}^{r-1} |q_\alpha^{(s)} O_\alpha^{\vec{d}_\alpha(s)}|-\mu|\prod_{s=1}^{r-1} q_\alpha^{(s)} O_\alpha^{\vec{d}_\alpha(s)}|}.
\end{eqnarray}
Similarly to the case without logical operators we can take $|q_\alpha^{(s)} O_\alpha^{\vec{d}_\alpha(s)}| = \sum_\ell |q_{\alpha,\ell}^{(s)} O_\alpha^{\vec{d}_\alpha(s)}| $ and think in therms of spins located at the center of the plaquettes (or Tanner graph in general),
\begin{eqnarray}
|q_{\alpha,\ell}^{(s)} O_{\alpha,\ell}^{\vec{d}_\alpha(s)}| = \dfrac{1-(-1)^{\lambda_\ell(s)}\sigma^{(s)}_i \sigma^{(s)}_j}{2},
\end{eqnarray}
where $\lambda_\ell$ denotes whether one of the well-defined logical operators occupies that link. We omit the $x,z$ indices of the spin variables $\sigma$. Thus the Hamiltonian describing this systems reads: 
\begin{eqnarray}
H^{(r)}_{\vec{d}_\alpha} = \sum_{<i,j>} \left(\sum_{s=1}^{r-1} (-1)^{\lambda_\ell(s)} \sigma^{(s)}_i \sigma^{(s)}_j+ \prod_{s=1}^{r-1} (-1)^{\lambda_\ell(s)}\sigma^{(s)}_i \sigma^{(s)}_j\right).
\end{eqnarray}
And the whole expression is now
\begin{eqnarray}
\Tr\left(\rho^r_{RQ}\right) =  \dfrac{e^{-J(n-m)r}}{2^{(n+k)(r-1)}} \sum_{\{\sigma^{(s)}\}} \sum_{\{\vec{d}_X(s)\}} e^{-JH^{(r)}_{\vec{d}_X}} \sum_{\{\sigma^{(s)}\}} \sum_{\{\vec{d}_Z(s)\}} e^{-JH^{(r)}_{\vec{d}_Z}} = \dfrac{e^{-J(n-m)r}}{2^{(n+k)(r-1)}} \sum_{\{\vec{d}_X(s)\}} \mathcal{Z}_{\vec{d}_X}(r) \sum_{\{\vec{d}_Z(s)\}} \mathcal{Z}_{\vec{d}_Z}(r).\nonumber \\
\end{eqnarray}

We have thereby obtained a family of models with defects along the lines of the \emph{well-defined} logical operators. This is in stark contrast to the error configuration picture used in the main text, where the defects run over the support of the logical operators that anti-commute with all well-defined logical operators. This fact is independent of the code and error model considered, so it can be directly translated to depolarizing noise and other QEC codes. 

\subsection{Renyi coherent information}

Let us recall the Renyi entropies $S^{(r)}(\rho) = \log [\Tr(\rho^r)]/(1-r)$. The trick shown in Eq.~\eqref{eq:entropy_identity} is not useful anymore for computing the von Neumann entropy because the $r$ dependency in the partition functions $\mathcal{Z}_{\vec{d}_\alpha}(r)$ is highly non-trivial. Therefore a straight-forward computation of $I$ is not possible in the present picture. However we can compute the \emph{Renyi coherent information} for $r>1$
\begin{eqnarray}
I^{(r)}_l= S^{(r)}(\rho_Q) - S^{(r)}(\rho_{RQ})  =  \dfrac{1}{r-1} \log \left[  \dfrac{\Tr \left(\rho^{r}_{RQ}\right)}{\Tr \left( \rho^{r}_{Q}\right)}\right].
\end{eqnarray}
We can then express the Renyi coherent information in terms of the statistical mechanics mappings:
\begin{eqnarray}
I^{(r)}_l & = &  \dfrac{1}{r-1} \log \left[ \dfrac{\sum_{\vec{d}} \mathcal{Z}_{\vec{d}}(r)}{2^{k(r-1)}\mathcal{Z}(r)}\right] \\
        & = & - k\log 2 + \log \left[ \dfrac{\sum_{\vec{d}} \mathcal{Z}_{\vec{d}}(r)}{\mathcal{Z}(r)}\right],
\end{eqnarray}
where $\vec{d}=(\vec{d}_x,\vec{d}_z)$ encompasses both $X$ and $Z$. The average Renyi coherent information $I^{(r)} = \sum_l P(l) I^{(r)}_l$ is then
\begin{eqnarray}
I^{(r)} & = & - k\log 2 +  \Big\langle  \log \left[ \dfrac{\sum_{\vec{d}} \mathcal{Z}_{\vec{d}}(r)}{\mathcal{Z}(r)}\right] \Big\rangle,
\end{eqnarray}
where $\langle...\rangle$ denotes erasure average. As expected, the disorder average over error chains is absent, however we can not get rid of the erasure average. 
Furthermore the erasure errors manifest themselves as infinite couplings that lock neighboring spins rigidly together. We can see this as fully depolarizing channels, i.e.~$p=1/2$, producing $J\rightarrow \infty$ on the location of the erasures. 
Therefore one can just extend the existing mappings in the stabilizer configuration picture by adding fully depolarizing channels with some probability $e$. 
Besides, the contributions of erasure and computational errors are entangled in the average over different partition functions. In the absence of computational errors, $p=0$, the Hamiltonians have zero coupling yielding an infinite temperature state. Hence we get $2^{2k-h}$ inside the log and recover $I^{(r)}_l=(k-h_l)\log 2$ for any $r$.

\end{document}